\documentclass[a4paper,11pt]{article}

\usepackage{jheppubnew}
\usepackage{amsmath}
\usepackage{amsfonts}
\usepackage{amssymb}
\usepackage{amsthm}
\usepackage{mathrsfs}
\usepackage{graphicx}
\usepackage[all]{xy}
\usepackage{bm}
\usepackage{xcolor}
\usepackage{comment}
\usepackage[numbers,sort&compress]{natbib}

\newcommand{\be}{\begin{equation}}
\newcommand{\ee}{\end{equation}}

\newcommand{\I}{\mathrm{i}}
\newcommand{\D}{\mathrm{d}}

\newcommand{\R}{\mathbb{R}}
\newcommand{\Z}{\mathbb{Z}}

\renewcommand{\O}{\mathcal{O}}

\newcommand{\<}{\langle}
\renewcommand{\>}{\rangle}

\newcommand{\nn}{\nonumber}
\newcommand{\lla}{\langle \! \langle}
\newcommand{\rra}{\rangle \! \rangle}
\renewcommand{\a}{\alpha}

\DeclareMathOperator{\re}{Re}

\DeclareMathOperator{\tr}{tr}
\DeclareMathOperator{\Tr}{Tr}

\newcommand{\bs}[1]{\boldsymbol{#1}}

\DeclareMathOperator{\K}{K}
\DeclareMathOperator{\Lo}{L}
\DeclareMathOperator{\Ro}{R}

\numberwithin{equation}{section}

\title{\Large Implications of conformal invariance in momentum space}
\author[a,d]{Adam Bzowski,}
\author[c]{Paul McFadden}
\author[a,b,d,e]{and Kostas Skenderis.}
\affiliation[a]{Mathematical Sciences, University of Southampton, UK.}
\affiliation[b]{STAG research centre, University of Southampton, UK.}
\affiliation[c]{Perimeter Institute for Theoretical Physics, Waterloo, Canada.}
\affiliation[d]{Korteweg-de~Vries Institute for Mathematics, Amsterdam, Netherlands.}
\affiliation[e]{Institute of Physics, Amsterdam, Netherlands.}
\emailAdd{a.w.bzowski@uva.nl} 
\emailAdd{pmcfadden@perimeterinstitute.ca} 
\emailAdd{k.skenderis@soton.ac.uk}

\begin{document}

\abstract{
	We present a comprehensive analysis of the implications of conformal invariance for 3-point functions of the stress-energy tensor, conserved currents and scalar operators in general dimension and {\it in momentum space}. Our starting point is a novel and very effective decomposition of tensor
	correlators which reduces their computation to that of a number of scalar form factors. For example, the most general 3-point function of a conserved and traceless stress-energy tensor is determined by only five form factors. Dilatations and special conformal Ward identities then impose additional conditions on these form factors. The special conformal Ward identities become a set of first and second order differential equations, whose general solution is given in terms of integrals involving a product of three Bessel functions (`triple-$K$ integrals'). All in all, the correlators are completely determined up to a number of
	constants, in agreement with well-known position space results. In odd dimensions 3-point functions are finite without renormalisation while in even dimensions non-trivial renormalisation in required. In this paper we restrict ourselves to odd dimensions. A comprehensive analysis of renormalisation will be discussed elsewhere.
	
	This paper contains two parts that can be read independently of each other. In the first part, we explain the method that leads to the solution for the correlators in terms of triple-$K$ integrals
	while the second part contains
	a self-contained presentation of all results. Readers interested only in results may directly consult the second part of the paper.
} 

\setcounter{tocdepth}{2}
\maketitle

\section{Introduction and summary of results}

It is well known that conformal invariance imposes strong constraints on
correlation functions. In particular, 2- and 3-point functions
of the stress-energy tensor, conserved currents and scalar primary operators are
completely determined up to a few constants. The result for the 3-point function
of scalar primary operators already appeared in \cite{Polyakov:1970xd}, while
the 3-point function of currents for $d=4$ was determined a few years later in
\cite{Schreier:1971um}. A complete analysis of all such 3-point functions, and
in general dimension, was carried out in \cite{Osborn:1993cr,Erdmenger:1996yc};
for a sample of more recent work on this topic see also \cite{Giombi:2011,
Costa:2011a,Costa:2011b,Maldacena:2011, Maldacena:2012sf, Stanev:2012nq,
Zhiboedov:2012bm}. All of these papers obtain their results through the use
position space techniques.

The purpose of this paper is to present the analogous set of results in momentum
space. In principle, the results in momentum space can be obtained
from those in position space by Fourier transform. Typically, however, the
position space expressions (in the form often quoted) are only valid at
separated points, and do not possess a Fourier transform prior to
renormalisation.\footnote{ For example, the 2-point function
$\langle \mathcal{O}(\bs{x}) \mathcal{O}(0)\rangle \sim 1/|\bs{x}|^{2 \Delta}$
of
a scalar operator $\mathcal{O}$ of dimension $\Delta$ does not  possess a
Fourier transform due to short distance singularities
when $\Delta = d/2 + k,\ k=0,1,2, \ldots\, $. \label{2pt}
Only after renormalisation (using, for example, differential renormalisation
\cite{Freedman:1991tk}) can one Fourier transform this correlator to momentum
space.}
Even after renormalising, it is technically rather
difficult to carry out explicitly the Fourier transforms, see for example
\cite{Coriano:2012wp}.
Here we will present a complete analysis from first principles of the
constraints due to conformal
symmetry directly in momentum space. We believe such an analysis gives
considerably more insight into the results and is interesting in its own right.

A momentum space analysis is natural from the perspective of Feynman diagram
computations, which are usually performed in momentum space.
Furthermore, a number of recent works have exemplified the need for CFT results
in momentum space. Our original motivation for analysing this question
was the requirement for these results in our work on holographic cosmology
\cite{McFadden:2009fg,McFadden:2010na,McFadden:2010vh,Bzowski:2011ab,
Bzowski:2012ih},
and similar applications of the conformal/de Sitter symmetry in cosmology have
been discussed in \cite{Antoniadis:2011ib, Maldacena:2011nz, Creminelli:2012ed,
Kehagias:2012pd,Kehagias:2012td,Coriano:2012hd,Mata:2012bx}. Other recent works
that contain explicit computations of CFT correlation functions in momentum
space include \cite{Giannotti:2008cv, Armillis:2009sm, Armillis:2009pq,
Coriano:2012wp,Chowdhury:2012km}. Our results may also be useful in the context
of work on
an a-theorem in diverse dimensions, see \cite{Cappelli:2001pz} for a relevant
discussion in $d=4$.

There are two main issues that complicate the analysis of the implications of
conformal invariance in momentum space.
While conformal transformations act naturally in position space, they lead to
differential operators in momentum space.
Dilatations, $\delta x^\mu = \lambda x^\mu$, being linear in $x^\mu$ lead to a
Ward identity
that is a first-order differential equation, and as such, it is easy to solve in
complete generality. Special conformal transformations however are non-linear,
$\delta x^\mu = b^\mu x^2 -2 x^\mu b \cdot x$, so after Fourier transform we
obtain a Ward identity that is a second-order differential equation, which makes
the analysis more complicated.

The second main issue is the complicated tensorial decomposition required for
correlators involving vectors and tensors. We will return to this below, but let
us begin by illustrating the first issue, focusing on the case of the 3-point
function of scalar operators $\mathcal{O}_j$ of dimension $\Delta_j$, $\langle
\mathcal{O}_1(\bs{p}_1) \mathcal{O}_2(\bs{p}_2) \mathcal{O}_3(\bs{p}_3)
\rangle$. We will discuss this computation in detail in section \ref{theory2_P},
so here we simply summarise the main points. As usual, translational invariance
implies that we can pull out a momentum-conserving delta function,
\begin{equation} \label{delta}
\langle \mathcal{O}_1(\bs{p}_1) \mathcal{O}_2(\bs{p}_2) \mathcal{O}_3(\bs{p}_3)
\rangle = (2 \pi)^d \delta( \bs{p}_1 + \bs{p}_2 + \bs{p}_3 ) \lla
\mathcal{O}_1(\bs{p}_1) \mathcal{O}_2(\bs{p}_2) \mathcal{O}_3(\bs{p}_3) \rra,
\end{equation}
thereby defining the reduced matrix element which we denote with double
brackets.
Lorentz invariance then implies that $\lla \mathcal{O}_1(\bs{p}_1)
\mathcal{O}_2(\bs{p}_2) \mathcal{O}_3(\bs{p}_3) \rra$ is only a function of the
magnitude of the momenta $p_j = | \bs{p}_j |$, while dilatation invariance
implies that it is a homogeneous function of total degree $(\Delta_t - 2 d)$,
where
$\Delta_t =\sum \Delta_j$.

Finally, we impose invariance under special conformal transformations. The
corresponding Ward identities are second-order differential equations which can
be manipulated into the form
\begin{equation}
0=\K_{12} \lla \mathcal{O}_1(\bs{p}_1) \mathcal{O}_2(\bs{p}_2)
\mathcal{O}_3(\bs{p}_3) \rra = \K_{23} \lla \mathcal{O}_1(\bs{p}_1)
\mathcal{O}_2(\bs{p}_2) \mathcal{O}_3(\bs{p}_3) \rra,
\end{equation}
where
\begin{align} \label{scalarCWI}
\K_{ij} & = \K_i - \K_j, \qquad
\K_j = \frac{\partial^2}{\partial p_j^2} + \frac{d + 1 - 2 \Delta_j}{p_j}
\frac{\partial}{\partial p_j}, \qquad i,j = 1, 2, 3.
\end{align}
This system of differential equations is precisely that defining Appell's $F_4$
generalised hypergeometric function of two variables. There are four linearly
independent solutions of these equations but three of
them have unphysical singularities at certain values of the momenta leaving one
physically acceptable solution.
This solution has the following integral representation, which we will refer to
as a \emph{triple-$K\,$integral}:\footnote{This expression can also obtained
by direct Fourier transform of the well-known position space result, see
\textit{e.g.}, \cite{Barnes:2010}.}
\begin{align}
& \lla \mathcal{O}_1(\bs{p}_1) \mathcal{O}_2(\bs{p}_2) \mathcal{O}_3(\bs{p}_3)
\rra \nn\\
& \qquad = C_{123}  p_1^{\Delta_1 - \frac{d}{2}} p_2^{\Delta_2 - \frac{d}{2}}
p_3^{\Delta_3 - \frac{d}{2}} \int_0^\infty \D x \: x^{\frac{d}{2} - 1}
K_{\Delta_1 - \frac{d}{2}} (p_1 x) K_{\Delta_2 - \frac{d}{2}} (p_2 x)
K_{\Delta_3 - \frac{d}{2}} (p_3 x), \label{scalar_3pt}
\end{align}
where $K_\nu(p)$ is a modified Bessel function of the second kind (or Bessel $K$
function, for short) and $C_{123}$ is an overall undetermined constant. We thus
arrive at the conclusion that scalar 3-point functions are uniquely determined
up to one constant.

This result is still formal, however, since the integral in (\ref{scalar_3pt})
may not converge. Depending on the conformal dimensions involved there are three
cases: (i) the integral converges; (ii) the integral diverges but it can be
defined via analytic continuation in the spacetime dimension and the conformal
dimensions $\Delta_i$; (iii) the integral diverges and explicit
subtractions are necessary. In the last case, after renomalisation, the
correlators exhibit anomalous scaling transformations: the theory suffers from
conformal anomalies. This is analogous to the discussion of 2-point functions
(see footnote \ref{2pt}): renormalising the 2-point functions results in
conformal anomalies, see, \textit{e.g.}, the discussion in \cite{Petkou:1999fv}.

We now turn to discuss tensorial correlation functions, such as those involving
stress-energy tensors
and conserved currents. Lorentz invariance implies that the tensor structure
will be carried by tensors
constructed from the momenta $p^\mu$ and the metric $\delta_{\mu \nu}$
(throughout this paper we work with Euclidean signature). The standard procedure
consists of writing down all possible such independent tensor structures and
expressing the correlators as a sum of these structures, each multiplied by
scalar form factor. In the case of correlators involving conserved currents
and/or stress-energy tensors one then imposes the restrictions enforced by
conservation (and tracelessness of the stress-energy tensor in the case of
CFTs). Recent works discussing such a tensor decomposition include
\cite{Armillis:2009pq,Cappelli:2001pz,Armillis:2009sm,Giannotti:2008cv,
Coriano:2012wp}.
This methodology is in principle straightforward, but an inefficient
parametrisation can produce unwieldy expressions. Here we present a new
parametrisation that appears to yield a minimal number of form factors.

Before proceeding with this, let us briefly discuss the transverse and trace
Ward identities (also known as the diffeomorphism and Weyl Ward identities,
respectively). The fact that classically a current or stress-energy tensor is
conserved implies that
$n$-point functions involving insertions of $\partial_\alpha J^\alpha$ or
$\partial_\alpha T^{\alpha \beta}$
are semi-local (\textit{i.e.}, at least two points are coincident) and can be
expressed in terms of lower-point functions without such insertions. Similarly,
the trace Ward identity implies that correlation functions with insertions of
the trace of the stress-energy tensor are also semi-local and are related to
lower-point functions. The first step in our analysis is to implement these Ward
identities. We do this by providing reconstruction formulae that yield the full
3-point functions involving stress-energy tensors/currents/scalar operators
starting from expressions that are exactly conserved/traceless. These 3-point
functions automatically satisfy the transverse and trace Ward identities.

To determine the general form of correlators consistent with the transverse and
trace Ward identities, it thus suffices to start from an expression that is
exactly conserved/traceless in all relevant indices.  Such an expression may be
obtained by means of projection operators. Recall that in momentum
space the operator
\begin{equation} \label{e:pi1}
\pi^{\mu}_{\alpha}(\bs{p}) = \delta^{\mu}_{\alpha} - \frac{p^{\mu}
p_{\alpha}}{p^2}
\end{equation}
is a projector onto tensors transverse to $\bs{p}$, \textit{i.e.}, $p_\mu
\pi^{\mu}_{\alpha}(\bs{p}) = 0$. Similarly, in $d$ dimensions, the operator
\begin{equation} \label{e:Pi1}
\Pi^{\mu \nu}_{\alpha \beta}(\bs{p}) = \frac{1}{2} \left(
\pi^{\mu}_{\alpha}(\bs{p}) \pi^{\nu}_{\beta}(\bs{p}) + \pi^{\mu}_{\beta}(\bs{p})
\pi^{\nu}_{\alpha}(\bs{p}) \right) - \frac{1}{d - 1} \pi^{\mu \nu}(\bs{p})
\pi_{\alpha \beta}(\bs{p})
\end{equation}
is a projector onto transverse to $\bs{p}$, traceless, symmetric tensors of rank
two.

To illustrate our discussion we will use as an example the 3-point function of
the stress-energy tensor,
$\langle T^{\mu_1 \nu_1}(\bs{p}_1) T^{\mu_2 \nu_2}(\bs{p}_2) T^{\mu_3
\nu_3}(\bs{p}_3) \rangle$. This is the
most complicated case, but also perhaps the most interesting one. In the main
text we will explain the method
using the simpler example of $\langle T^{\mu_1 \nu_1}(\bs{p}_1) T^{\mu_2
\nu_2}(\bs{p}_2) \mathcal{O}(\bs{p}_3) \rangle$,
and in part II we present the corresponding results for all 3-point functions.

To obtain the most general 3-point function of the stress-energy tensor
satisfying the transverse and trace Ward identities  it suffices to  start from
the following transverse-traceless expression,
\be \label{ttt}
\lla t^{\mu_1 \nu_1}(\bs{p}_1) t^{\mu_2 \nu_2}(\bs{p}_2) t^{\mu_3
\nu_3}(\bs{p}_3) \rra
= \Pi^{\mu_1 \nu_1}_{\alpha_1 \beta_1}(\bs{p}_1) \Pi^{\mu_2 \nu_2}_{\alpha_2
\beta_2}(\bs{p}_2) \Pi^{\mu_3 \nu_3}_{\alpha_3 \beta_3}(\bs{p}_3) X^{\alpha_1
\beta_1 \alpha_2 \beta_2 \alpha_3 \beta_3},
\ee
where $X^{\alpha_1 \beta_1 \alpha_2 \beta_2 \alpha_3 \beta_3}$ is a rank six
tensor built from the momenta and the metric. The complete 3-point function may
then be obtained using the reconstruction formula in (\ref{re_TTT}).

 At this point it seems that we have not gained much since we traded a rank six
tensor, the left-hand side of (\ref{ttt}), with another rank six tensor,
$X^{\alpha_1 \beta_1 \alpha_2 \beta_2 \alpha_3 \beta_3}$. This is deceptive,
however, as the explicit projection operators annihilate many of the possible
terms that can appear in $X^{\alpha_1 \beta_1 \alpha_2 \beta_2 \alpha_3
\beta_3}$. To obtain the most economic parametrisation we would still like to
impose
 one more requirement. The 3-point function is invariant under permutations of
the labels:
\begin{equation} \label{perm}
\lla t^{\mu_1 \nu_1}(\bs{p}_1) t^{\mu_2 \nu_2}(\bs{p}_2) t^{\mu_3
\nu_3}(\bs{p}_3) \rra = \lla t^{\mu_{\sigma(1)}
\nu_{\sigma(1)}}(\bs{p}_{\sigma(1)}) t^{\mu_{\sigma(2)}
\nu_{\sigma(2)}}(\bs{p}_{\sigma(2)}) t^{\mu_{\sigma(3)}
\nu_{\sigma(3)}}(\bs{p}_{\sigma(3)}) \rra,
\end{equation}
where $\sigma$ denotes any element of the permutation group $S_3$ of the set
$\{1,2,3\}$. We would like to find a parametrisation making this invariance as
manifest as possible.

Recall that the reduced matrix elements are multiplied by a momentum-conserving
delta function, as in (\ref{delta}). Often one uses momentum
conservation to solve for one momentum in terms of the others, say $\bs{p}_3 = -
(\bs{p}_1 + \bs{p}_2)$, and then the right-hand side of (\ref{ttt}) contains
only
$\bs{p}_1$ and $\bs{p}_2$. In doing so, however, one obscures the relation (\ref{perm}).
Here, we will instead choose the independent momenta differently for
different Lorentz indices:
\begin{equation} \label{e:momenta_choice_intro}
\bs{p}_1, \bs{p}_2 \text{ for } \mu_1, \nu_1; \ \bs{p}_2, \bs{p}_3 \text{ for }
\mu_2, \nu_2 \text{  and  } \bs{p}_3, \bs{p}_1 \text{ for }\mu_3, \nu_3.
\end{equation}
With this choice, it is straightforward to show that $X^{\alpha_1 \beta_1
\alpha_2 \beta_2 \alpha_3 \beta_3}$ is determined by five form factors:
\begin{align}
X^{\alpha_1 \beta_1 \alpha_2 \beta_2 \alpha_3 \beta_3} &=A_1 p_2^{\alpha_1}
p_2^{\beta_1} p_3^{\alpha_2} p_3^{\beta_2} p_1^{\alpha_3} p_1^{\beta_3}
\nonumber \\
&+ A_2 \delta^{\beta_1 \beta_2} p_2^{\alpha_1} p_3^{\alpha_2} p_1^{\alpha_3}
p_1^{\beta_3} + A_2(p_1 \leftrightarrow p_3) \delta^{\beta_2 \beta_3}
p_2^{\alpha_1} p_2^{\beta_1} p_3^{\alpha_2} p_1^{\alpha_3} \nonumber \\
& \qquad \qquad + A_2(p_2 \leftrightarrow p_3) \delta^{\beta_1 \beta_3}
p_2^{\alpha_1} p_3^{\alpha_2} p_3^{\beta_2} p_1^{\alpha_3} \nonumber \\
& + A_3 \delta^{\alpha_1 \alpha_2} \delta^{\beta_1 \beta_2} p_1^{\alpha_3}
p_1^{\beta_3} + A_3(p_1 \leftrightarrow p_3) \delta^{\alpha_2 \alpha_3}
\delta^{\beta_2 \beta_3} p_2^{\alpha_1} p_2^{\beta_1} \nonumber \\
& \qquad \qquad + A_3(p_2 \leftrightarrow p_3) \delta^{\alpha_1 \alpha_3}
\delta^{\beta_1 \beta_3} p_3^{\alpha_2} p_3^{\beta_2} \nonumber \\
& + A_4 \delta^{\alpha_1 \alpha_3} \delta^{\alpha_2 \beta_3} p_2^{\beta_1}
p_3^{\beta_2} + A_4(p_1 \leftrightarrow p_3) \delta^{\alpha_1 \alpha_3}
\delta^{\alpha_2 \beta_1} p_3^{\beta_2} p_1^{\beta_3} \nonumber \\
 & \qquad \qquad
 + A_4(p_2 \leftrightarrow p_3) \delta^{\alpha_1 \alpha_2} \delta^{\alpha_3
\beta_2} p_2^{\beta_1} p_1^{\beta_3} \nonumber \\
& + A_5 \delta^{\alpha_1 \beta_2} \delta^{\alpha_2 \beta_3} \delta^{\alpha_3
\beta_1}\, , \label{e:decompTTT_Intro}
\end{align}
where the $A_i$ are functions of three variables, $A_i(p_1, p_2, p_3)$,  and
$A_2(p_1 \leftrightarrow p_3)$ denotes the
same function but with $p_1$ interchanged with $p_3$, \textit{i.e.}, $A_2(p_1
\leftrightarrow p_3) = A_2(p_3, p_2, p_1)$, {\it etc}.
In this paper we assume that there are no parity-violating terms in the tensorial decomposition
of the various correlators. It
would be interesting to incorporate such terms in our analysis. We leave this for
future work.

The form factors
$A_1$ and $A_5$ are $S_3$-invariant,
\begin{equation}
A_j(p_1, p_2, p_3) = A_j(p_{\sigma(1)}, p_{\sigma(2)}, p_{\sigma(3)}), \quad j
\in \{1,5\},
\end{equation}
while the remaining ones are symmetric under $p_1 \leftrightarrow p_2$,
\textit{i.e.}, they satisfy
\begin{equation}
A_j(p_2, p_1, p_3) = A_j(p_1, p_2, p_3), \qquad j \in \{2,3,4\}.
\end{equation}
Under the action of $S_3$ the first and last terms are invariant on their own,
while the three terms with $A_2$ are mapped to each other and similar for $A_3$
and $A_4$.

To illustrate the mechanics behind the decomposition \eqref{e:decompTTT_Intro}
let us explain why there no terms with either $p_1^{\alpha_1}$ or
$p_3^{\alpha_1}$. First, note that the index $\alpha_1$ is linked with the
indices
$\mu_1$ and $\nu_1$ via (\ref{ttt}) and from (\ref{e:momenta_choice_intro})
follows that we should only use $p_1^{\alpha_1}$ and $p_2^{\alpha_1}$ when
working out the possible terms (a possible $p_3^{\alpha_1}$ is converted into
$-(p_1^{\alpha_1}+p_2^{\alpha_1})$ using momentum conservation). However, terms
containing $p_1^{\alpha_1}$ vanish due to project operator in (\ref{ttt}).
Repeating this argument for the other indices leads to
(\ref{e:decompTTT_Intro}).

Thus we find that the most general 3-point function of the stress-energy tensor
in any dimension satisfying the transverse and trace Ward identities is
specified by five form factors. If we relax the condition of Weyl invariance
(\textit{i.e.}, if we consider a stress-energy tensor with non-vanishing trace)
then the number of form factors becomes ten,
see appendix \ref{ch:decTTT}. These results should be compared with those of
\cite{Cappelli:2001pz} in $d=4$.
There, the expansion of the 3-point function for the conformal case (traceless
$T_{\mu \nu}$) is done in terms
of 13 tensors, of which eight are transverse traceless, two are transverse but
not traceless, and three are neither transverse nor traceless.
Our reconstruction formula (\ref{re_TTT}) eliminates the need to construct a
basis for the non-transverse-traceless part of the correlator,
while the difference between the number of transverse traceless tensors is due
to the fact that in \cite{Cappelli:2001pz}
the expression is manifestly invariant under $\bs{p}_1 \leftrightarrow \bs{p}_2$
and not under the full $S_3$ permutation group.
In the non-conformal case, \cite{Cappelli:2001pz} uses 22 tensor structures
while we need only ten.

One may be concerned that the form factors so defined are difficult to extract
from explicit computations of correlators, which directly give the
entire correlator rather than the transverse traceless piece with the projection
operators extracted. It turns out this is not the case. The form factors can be
simply extracted by looking at the coefficients of certain tensor structures in
the full correlator.
For example, in the case of the 3-point function of the stress-energy tensor,
one can extract the form factors
from the following coefficients of the complete 3-point functions $\lla T^{\mu_1
\nu_1}(\bs{p}_1) T^{\mu_2 \nu_2}(\bs{p}_2) T^{\mu_3 \nu_3}(\bs{p}_3) \rra$:
\begin{align}
A_1 & = \text{coefficient of } p_2^{\mu_1} p_2^{\nu_1} p_3^{\mu_2} p_3^{\nu_2}
p_1^{\mu_3} p_1^{\nu_3}, \\
A_2 & = 4 \cdot \text{coefficient of } \delta^{\nu_1 \nu_2} p_2^{\mu_1}
p_3^{\mu_2} p_1^{\mu_3} p_1^{\nu_3}, \\
A_3 & = 2 \cdot \text{coefficient of } \delta^{\mu_1 \mu_2} \delta^{\nu_1 \nu_2}
p_1^{\mu_3} p_1^{\nu_3}, \\
A_4 & = 8 \cdot \text{coefficient of } \delta^{\mu_1 \mu_3} \delta^{\mu_2 \nu_3}
p_2^{\nu_1} p_3^{\nu_2}, \\
A_5 & = 8 \cdot \text{coefficient of } \delta^{\mu_1 \nu_2} \delta^{\mu_2 \nu_3}
\delta^{\mu_3 \nu_1}\,.
\end{align}

We are now ready to impose the dilatation and conformal Ward identities. Let us
first define the tensorial dimension $N_j$ of $A_j$ to be equal to the number of
momenta in the tensor structure that this form factor multiplies. We use the
convention that $A_j$ are ordered according to the tensorial dimension, with the
form factor of highest tensorial dimension being the first one, $A_1$, {\it etc}.
In the example above, $N_1=6, N_2=4, N_3=N_4=2, N_5=0$. The dilatation Ward
identity then implies that the form factors are homogeneous functions of the
momenta of degree
\begin{equation}
\deg (A_j) = \Delta_t - 2d - N_j,
\end{equation}
where $\Delta_t$ is the sum of the conformal dimensions of the three operators.

All that remains to be discussed are the special conformal Ward identities
(CWIs). These split into a set of second-order differential equations, which we
call primary CWIs, and a set of first-order partial differential equations,
which we call secondary CWIs.  The primary CWIs are very similar to the
conformal Ward identity found in the case of scalar operators,
(\ref{scalarCWI}). In particular, $A_1$, the term of highest tensorial dimension
always satisfies (\ref{scalarCWI}), while the terms with lower tensorial
dimension satisfy similar equations with linear inhomogeneous terms on the
right-hand side. In the case
of the 3-point function of $T_{\mu \nu}$, these read:
\begin{equation}
\begin{array}{ll}
\K_{12} A_1 = 0, &\qquad \K_{13} A_1 = 0, \\
\K_{12} A_2 = 0, &\qquad \K_{13} A_2 = 8 A_1, \\
\K_{12} A_3 = 0, &\qquad \K_{13} A_3 = 2 A_2, \\
\K_{12} A_4 = 4 \left[ A_2(p_1 \leftrightarrow p_3) - A_2(p_2 \leftrightarrow
p_3) \right], &\qquad \K_{13} A_4 = - 4 A_2(p_2 \leftrightarrow p_3), \\
\K_{12} A_5 = 2 \left[ A_4(p_2 \leftrightarrow p_3) - A_4(p_1 \leftrightarrow
p_3) \right], &\qquad \K_{13} A_5 = 2 \left[ A_4 - A_4(p_1 \leftrightarrow p_3)
\right].
\end{array}
\end{equation}
It turns out that each pair of primary CWIs has a unique solution up to one
arbitrary constant, and that these solutions can be expressed in terms of
triple-$K$ integrals, similar to the solution in (\ref{scalar_3pt}). We will
call these integration constants the primary constants.
It follows that the 3-point functions are determined up to a number of constants
that are
at most equal to the number of form factors.

Finally, we need to impose the secondary CWIs. These are first order partial
differential equations that depend in particular on the specific 2-point
functions that appear in the transverse Ward identities. (As discussed above,
the transverse Ward identities relate 3-point functions involving the divergence
of the stress-energy tensor/conserved current to 2-point functions.) When
inserting the solutions of the primary CWI to the secondary
CWIs these become algebraic relations among the primary constants and the
normalisation of the operators (through the coefficient of the 2-point
functions). Solving these relations we obtain the final number of constants that
determine the 3-point function. In the case of the 3-point function of the
stress-energy tensor we find that the final answer depends on three constants,
which may be taken to be two of the primary constants and the coefficient of the
2-point function. This agrees exactly with the position space analysis in
\cite{Osborn:1993cr}.

As in the case of scalar operators, the 3-point functions involving the stress energy tensor,
symmetry currents and scalar operators sometimes diverge and must be regularised and renormalised.
This is the case in all even dimensions. We will discuss in detail regularisation and renormalisation
in a companion paper \cite{II}. In this paper we restrict ourselves to CFTs in odd dimensions where
the 3-point functions are finite. Even in these cases the individual triple-$K$ integrals that enter
in the 3-point function may diverge, with only the total combination being finite. In such cases
one needs to regulate the integrals and only remove the regulator after the divergences cancel.
Using appropriate regulator, the Bessel $K$ functions
that appear in the triple-$K$ integrals reduce to elementary functions and one
can compute the integrals by standard methods. On the other hand, when $d$ is even the evaluation of the integrals is less trivial. As will be discussed in \cite{II}, these integrals can be computed using a reduction scheme, generalising Davydychev's recursion relations \cite{Davydychev:1992xr}.

We will now discuss in detail the structure of the paper. We split this paper in
two parts. In the first part we explain the method we use, while in the second
part we give a complete list of all results.  The second part is completely
self-contained and can be used without reference
to the first part of the paper. This second part starts with a collection of all
basic definitions and a summary of conventions, followed by a list of all
results for all 3-point functions involving the stress-energy tensor, conserved
currents and scalar operators. For each such correlator we list:
(i) the relevant trace and transverse Ward identities,
(ii) the reconstruction
formula that yield the complete correlator from its transverse(-traceless) part,
(iii) the tensorial decomposition of the transverse(-traceless) part,
(iv) how to extract the form factors from the complete correlator,
(v) the primary
CWIs and their solution in terms of triple-$K$ integrals,
(vi) the secondary Ward
identities and the relations they impose on the integration constants of the
primary Ward identities (the primary constants),
(vii) evaluation of the triple-$K$
integrals in dimensions $d=3,5$.
 Special effort was made so that each section
describing any given correlator can be read in isolation
so the reader interested only in the results for a specific correlator may
directly consult that section.

In the first part we strive to keep a balance between describing the method and
computations in enough detail so that all subtleties entering the derivation of
the results in Part II are covered, while still keeping the level of
technicalities to a minimum. We start
in section \ref{sec:2pt} with a brief review of 2-point functions in momentum space.
In section \ref{theory2_P}, we then discuss
in detail the case of the 3-point function of scalar operators.  This case
serves as a good warm-up exercise, as one has to face many of the subtleties of
the conformal Ward identities in momentum space without having to deal with
complications  due to the tensorial structure of other correlators.
Here we see that the solution of the special conformal Ward identities may be
expressed in terms of the Appell $F_4$ function, and discuss its representation
in terms of triple-$K$ integrals.
This is
also a case where the answer can easily be obtained by a Fourier transform so we
can directly check our results.

In section \ref{sec:decomp}, we discuss the tensorial decomposition of
correlation functions. We illustrate our approach using
the case of $\langle T^{\mu_1 \nu_1} T^{\mu_2 \nu_2} \mathcal{O} \rangle$. We
also compare in this section our decomposition with previous state-of-the-art
results in \cite{Cappelli:2001pz}. In section \ref{sec:CWIs}, we derive the form
of the conformal Ward identities in momentum space. In particular, we show that
the special conformal Ward identities
can be split into two classes, the primary and secondary conformal Ward identities.

Section \ref{section:solnofCWIs} is devoted to the solution of the CWIs. We show
first that the primary CWIs can be solved using triple-$K$ integrals and we then
substitute these into the secondary CWIs. This results in a number of relations
between the primary constants and the normalisation of the 2-point functions, as
described earlier. In this section we pay attention to the issues of
regularisation and we discuss how to deal with special cases (when the dimension
of the scalar operators takes special values, {\it etc}.).

In section \ref{sec:TJJworkedexample} we work out completely the case of
$\langle T^{\mu_1 \nu_1} J^{\mu_2} J^{\mu_3}\rangle$,
where $J^\mu$ is a conserved current. This example is more complex than the case of
$\langle T^{\mu_1 \nu_1} T^{\mu_2 \nu_2} \mathcal{O} \rangle$ used to
illustrate the method in earlier sections, yet simpler than correlators with
more stress-energy tensors. We also discuss the
evaluation of 
integrals in $d=3$ 
and present a concrete model, free fermions, where
these correlators can be explicitly computed by standard Feynman diagrams.

Part I finishes with a brief discussion of possible extensions in section
\ref{sec:extensions}. One such extension is to develop a helicity formalism. In
this formalism one uses helicity projected operators, so we trade the
transverse(-traceless) tensor for the two helicity components, so this approach
may simplify the tensor decomposition of the 3-point functions. Section
\ref{ch:helicity} contains a short introduction to this method. Another
extension is to higher-point functions and we briefly discuss this in section
\ref{ch:higher}.

Finally, this paper contains a number of appendices. Appendix \ref{ch:decTTT}
contains a discussion of the tensorial decomposition of the 3-point function of
a conserved but not traceless stress-energy tensor. In appendix
\ref{ch:degeneracy}, we explain that the general tensorial decomposition is
degenerate when $d=3$. In this case, the 3-point function of the stress-energy
tensor is determined by only two form factors. In appendix \ref{ch:toKKK}, we
discuss how to Fourier transform the scalar 3-point functions. In appendix,
\ref{ch:prop} we list useful properties of the triple-$K$ integrals that are
used in the derivations in the main text. Appendix \ref{ch:F4} contains a
collection of facts about the Appell $F_4$ function: its definition, the
differential equations it satisfies and useful integral representations.
In appendix \ref{ch:vanish}, we
prove the triviality of the $\langle T^{\mu_1 \nu_1} J^{\mu_2} \mathcal{O}
\rangle$ correlator, and
in appendix \ref{ch:identities} we present a set of useful identities involving projection
operators appearing in the main text.

We finish this introduction with a few comments about how to read this paper.
Readers only interested in an overview and specific results may read this
introduction and then directly move to the relevant section of Part II. Section 3 provides
a longer and more technical introduction to CFT in momentum space, but without the complication of tensorial decompositions.

\bigskip

{\bf Note added:} As this paper was finalised \cite{Coriano:2013jba} appeared
which has significant overlap with the discussion in section 3 of this paper.
We also understand that our discussion of the tensor decomposition may have overlap
with upcoming work by A.~Dymarsky \cite{Dymarsky}.

\newpage
\part*{Theory}
\setcounter{part}{1}

\section{2-point functions in momentum space} \label{sec:2pt}

In this section, we present a review of 2-point functions in momentum
space. This material is well known but is included here as it allows
us to discuss in the simplest possible setup many of the issues
that we will meet later on in our discussion of 3-point functions.
Furthermore, as 2-point functions appear in the Ward identities of
3-point functions, the results we present here will also be needed in
later sections.

\subsection{Scalar 2-point function} \label{sec:2ptsubsec}

The standard position space expressions for CFT 2- and 3-point
functions found in
\cite{Polyakov:1970xd,Schreier:1971um,Osborn:1993cr,Erdmenger:1996yc}
and other works are valid at non-coincident points only. When two or
more points coincide, these expressions become singular. Correlation
functions should however be well-defined distributions, and in
particular they should have well-defined Fourier transforms. We can
therefore analyse the coincident limit and regularise the position
space correlation functions by requiring that the Fourier transform
exists.
In this paper we deal primarily with cases where the
correlators are finite. This is the case in odd dimensions. Even in
such cases, however, infinities may appear at intermediate steps so it
is important to regularise. In even dimensions, one must carry out the
full renormalisation procedure and we will provide a comprehensive
discussion in \cite{II}.

Let us consider first the 2-point function of scalar operators,
\begin{equation} \label{e:twopoint}
\< \O(\bs{x}) \O(0) \> = \frac{C_{\O}}{x^{2 \Delta}}.
\end{equation}
The Fourier transform may be obtained explicitly using
\begin{equation} \label{e:2ptFourier}
\int \D^d \bs{x} \: e^{-\I \bs{p} \cdot \bs{x}} \frac{1}{x^{2 \Delta}}
= \frac{\pi^{d/2} 2^{d - 2 \Delta} \Gamma \left( \frac{d - 2
\Delta}{2} \right)}{\Gamma ( \Delta )} p^{2 \Delta - d},
\end{equation}
where the integral converges for $0 < 2 \Delta < d$.
Extracting the Dirac delta function associated with momentum
conservation, we will use double brackets to denote the reduced
correlation function
\begin{equation}
\< \mathcal{O}(\bs{p}_1) \mathcal{O}(\bs{p}_2) \> = (2 \pi)^d
\delta(\bs{p}_1 + \bs{p}_2) \lla \mathcal{O}(\bs{p}_1)
\mathcal{O}(-\bs{p}_1) \rra,
\end{equation}
where
\begin{equation} \label{e:twopointp}
\lla \O(\bs{p}) \O(-\bs{p}) \rra = \frac{C_{\O} \pi^{d/2} 2^{d - 2
\Delta} \Gamma \left( \frac{d - 2 \Delta}{2} \right)}{\Gamma ( \Delta
)} p^{2 \Delta - d}.
\end{equation}
In unitary theories $\Delta > 0$, but the upper bound $2\Delta<d$ for
the convergence of \eqref{e:2ptFourier} may be violated.  Since the
right-hand side of \eqref{e:2ptFourier} is nevertheless a well-defined
analytic function of $d$ and $\Delta$, we can extend the result by
analytic continuation to any $2 \Delta \neq d + 2 n$, where $n$ is a
non-negative integer.

The cases $2 \Delta = d + 2 n$ are special: \eqref{e:twopointp} is
singular and a non-trivial regularisation and renormalisation is
required. We will proceed by dimensional regularisation. As we will
see later on in section \ref{ch:tripleKand2pt}, it is convenient to
simultaneously shift the dimension of the operator:
\begin{equation} \label{e:dimreg00}
d \mapsto d - \epsilon, \qquad\qquad \Delta \mapsto \Delta - \epsilon.
\end{equation}
Keeping fixed the form of the coupling between the source and the operator,
\be
\int \D^{d-\epsilon}\bs{x}\, \phi_0 \O,
\ee
the dimension of the source $\phi_0$ then remains $d/2-n$.
In the case of 3-point functions, this prescription amounts to
regularising the momentum space integrals by shifting the dimension in
the integration measure while keeping the integrand fixed.

Summarising, in the general case the regulated 2-point function of
scalar operators reads
\begin{equation}
\lla \mathcal{O}(\bs{p}) \mathcal{O}(-\bs{p}) \rra = c_{\mathcal{O}}
\Gamma \left( \frac{d}{2} - \Delta + \frac{1}{2} \epsilon \right) p^{2
\Delta - d - \epsilon}, \label{e:OO}
\end{equation}
where $c_{\mathcal{O}}$ (which is proportional to $C_{\O}$) is the
momentum space 2-point function normalisation constant.
Using such a parametrisation we cover both the finite case of $2
\Delta - d \neq 2 n$ and the singular one $2 \Delta - d = 2 n$.
In the singular case there is a $1/\epsilon$ pole which must be removed by a counterterm.
This subtraction in turn introduces a
scale into the problem leading to a conformal anomaly, as will be
discussed in detail in \cite{II}.

\subsection{Tensorial 2-point functions}

Let us now consider the form of 2-point functions of spin-1 conserved
currents $J^\mu$ and the stress-energy tensor $T^{\mu \nu}$ for
some general QFT. One could start from the position space expressions
and Fourier transform them directly. It is more convenient, however, and also a
warm up exercise for the tensorial 3-point functions
that we will discuss later, to derive them from scratch.

To do so, observe that the operator
\begin{equation} \label{e:pi0}
\pi^{\mu}_{\alpha}(\bs{p}) = \delta^{\mu}_{\alpha} - \frac{p^{\mu}
p_{\alpha}}{p^2}
\end{equation}
is a projector onto tensors transverse to $\bs{p}$, \textit{i.e.},
$p_\mu \pi^{\mu}_{\alpha}(\bs{p}) = 0$. Similarly, in $d$ dimensions,
the operator
\begin{equation} \label{e:Pi0}
\Pi^{\mu \nu}_{\alpha \beta}(\bs{p}) = \frac{1}{2} \left[
\pi^{\mu}_{\alpha}(\bs{p}) \pi^{\nu}_{\beta}(\bs{p}) +
\pi^{\mu}_{\beta}(\bs{p}) \pi^{\nu}_{\alpha}(\bs{p}) \right] -
\frac{1}{d - 1} \pi^{\mu \nu}(\bs{p}) \pi_{\alpha \beta}(\bs{p})
\end{equation}
is a projector onto transverse to $\bs{p}$, traceless, symmetric
tensors of rank two. In particular,
\begin{equation}
p_\mu \pi^{\mu}_{\alpha}(\bs{p}) = 0, \qquad \delta_{\mu \nu} \Pi^{\mu
\nu}_{\alpha \beta}(\bs{p}) = 0, \qquad p_{\mu} \Pi^{\mu \nu}_{\alpha
\beta}(\bs{p}) = 0.
\end{equation}
Any  traceless, symmetric tensor of rank two that is transverse to
$\bs{p}$ may therefore be written as $t^{\mu \nu} = \Pi^{\mu
\nu}_{\alpha \beta}(\bs{p}) X^{\alpha \beta}$, where $X^{\alpha
\beta}$ is an arbitrary tensor. More properties of the projectors are
listed in appendix \ref{ch:identities}.

The transverse Ward identities (to be discussed in section
\ref{ch:transverseWI})
imply that the divergence of any 2-point function of conserved
currents is proportional to 1-point functions.  Assuming the latter
vanish, the 2-point function is then transverse, leading to the
general decompositions
\begin{align}
\lla T^{\mu\nu}(\bs{p}) T^{\rho\sigma}(-\bs{p}) \rra & =
\Pi^{\mu\nu\rho\sigma}(\bs{p}) A(p) + \pi^{\mu\nu}(\bs{p})
\pi^{\rho\sigma}(\bs{p}) B(p), \label{e:TTform} \\
\lla J^\mu(\bs{p}) J^\nu(-\bs{p}) \rra & = \pi^{\mu\nu}(\bs{p}) C(p).
\label{e:JJform}
\end{align}
Due to Lorentz invariance, $A(p)$, $B(p)$, and $C(p)$ are arbitrary
functions of $p$, the magnitude of the momentum. Provided  $\< J^\mu
\> = \< T^{\mu \nu} \> = 0$, these general expressions are then valid
in any quantum field theory.

Specialising now to the conformal  case, for which 1-point functions
necessarily vanish, the trace Ward identity (to be discussed in
section \ref{ch:trace}) implies that $\< T T^{\rho\sigma} \> = 0$,
setting $B = 0$ in \eqref{e:TTform}.
We also know that the conformal dimensions of $T^{\mu\nu}$ and $J^\mu$
are $d$ and $d-1$ respectively.

As discussed in section \ref{ch:dilatation} the dilation Ward
identity implies that the form factors should be homogeneous functions
of degree $2 \Delta -d$. This implies that
\begin{equation} \label{e:cJcT1}
A(p)= \tilde{c}_T p^d, \qquad C(p) = \tilde{c}_J p^{d-2}
\end{equation}
where $\tilde{c}_T$ and $\tilde{c}_J$ are arbitrary constants.
This result is valid in the absence of scale anomalies,
\textit{i.e.}, when the dilatation Ward identity holds without any
anomalous terms. In even spacetime dimensions scale anomalies must appear
leading to logarithmic terms in correlation functions. The reason is that in
even dimensions (\ref{e:cJcT1}) is analytic in momenta and thus represents contact terms only.
We cover all cases by dimensionally regularising (\ref{e:cJcT1}) and making the constants
$\tilde{c}_T$ and $\tilde{c}_J$ singular in even dimensions.\footnote{The same argument also holds  for the 2-point function of scalar operators: the dilatation Ward identity implies that the 2-point function should be proportional to $p^{2 \Delta -d}$, but when $2 \Delta -d$ is a non-negative integer this represents a contact term and conformal anomalies must be present.} A convenient choice is
\begin{align}
\lla T^{\mu \nu}(\bs{p}) T^{\rho \sigma}(-\bs{p}) \rra & = c_{T}
\Pi^{\mu \nu \rho \sigma}(\bs{p}) \Gamma \left( - \frac{d}{2} +
\frac{\epsilon}{2} \right) p^{d - \epsilon}, \label{e:TT1} \\
\lla J^{\mu}(\bs{p}) J^{\nu}(-\bs{p}) \rra & = c_{J} \pi^{\mu
\nu}(\bs{p}) \Gamma \left( 1 - \frac{d}{2} + \frac{\epsilon}{2}
\right) p^{d-2-\epsilon}, \label{e:JJ1}
\end{align}
where $c_T$ and $c_J$ are now regular in all cases.
In odd spacetime dimensions $d$ the limit $\epsilon \rightarrow 0$ is finite
and we recover \eqref{e:cJcT1}. In even dimensions
the correlation functions become singular in the limit $\epsilon\rightarrow 0$, and the expressions
\eqref{e:TT1} and \eqref{e:JJ1} represent the 2-point functions
regularised in the dimensional regularisation scheme
\eqref{e:dimreg00}. After removing the singularity by a counterterm the renormalised expression now contains a logarithm.

Expressions \eqref{e:TT1} and \eqref{e:JJ1} can be obtained directly
by the Fourier transform of the position-space expressions. Alternatively,
one may extract and solve the momentum-space Ward identities
associated with special conformal transformations. A systematic
approach will be presented in detail in section \ref{ch:to_momentum}.
Using equations \eqref{e:ward_cwi0} and \eqref{e:ward_cwi1}, one can
write down the differential equations satisfied by the 2-point
functions,
\begin{align}
0 & = \left( - 2 p^\alpha \delta^{\kappa \beta} + p^\kappa
\delta^{\alpha \beta} \right) \frac{\partial^2}{\partial p^\alpha
\partial p^\beta} \lla T^{\mu \nu}(\bs{p}) T^{\rho \sigma}(-\bs{p})
\rra \nn\\
& \qquad + \: 2 \left[ ( \delta^{\kappa \mu} \delta^{\alpha \gamma} -
\delta^{\kappa \alpha} \delta^{\mu \gamma} ) \delta^{\nu \beta} +
\delta^{\mu \alpha} ( \delta^{\kappa \mu} \delta^{\beta \gamma} -
\delta^{\kappa \beta} \delta^{\mu \gamma} ) \right]
\frac{\partial}{\partial p^\gamma} \lla T_{\alpha \beta}(\bs{p})
T^{\rho \sigma}(-\bs{p}) \rra, \label{e:TTeq}\\
0 & = \left[ \left( - 2 p^\alpha \delta^{\kappa \beta} + p^\kappa
\delta^{\alpha \beta} \right) \frac{\partial^2}{\partial p^\alpha
\partial p^\beta} - 2 \frac{\partial}{\partial p_\kappa} \right] \lla
J^{\mu}(\bs{p}) J^{\nu}(-\bs{p}) \rra \nn\\
& \qquad + \: 2 ( \delta^{\kappa \mu} \delta^{\alpha \gamma} -
\delta^{\kappa \alpha} \delta^{\mu \gamma} ) \frac{\partial}{\partial
p^\gamma} \lla J_{\alpha}(\bs{p}) J^{\nu}(-\bs{p}) \rra. \label{e:JJeq}
\end{align}
These equations, representing the special conformal Ward identities,
constitute a set of tensorial equations with a free Lorentz index $\kappa$.
Each equation consists of two distinct parts:
a universal second-order differential
operator whose form is independent of the Lorentz structure of the
correlation function, and a first-order
differential operator whose precise form depends on the Lorentz structure,
as will be discussed in more detail in section \ref{ch:to_momentum}.

Having imposed the transverse and dilatation Ward identities to arrive at
expressions \eqref{e:TTform}, \eqref{e:JJform} and \eqref{e:cJcT1}, by
substituting these results into the differential equations above, one
finds that the equation for $\lla J^\mu J^{\nu} \rra$ is identically
satisfied while the equation for $\lla T^{\mu \nu} T^{\rho \sigma}
\rra$ requires $B = 0$ in \eqref{e:TTform}. This confirms that the
analysis based on the trace Ward identity coincides with the
constraints imposed by the special conformal Ward identities.

Alternatively, one may substitute \eqref{e:TTform} and \eqref{e:JJform} into \eqref{e:TTeq} and \eqref{e:JJeq} respectively, without imposing the dilatation Ward identity. In this case, the coefficient of each independent tensor in \eqref{e:TTform} and \eqref{e:JJform} becomes a differential equation for the form factors $A(p)$, $B(p)$ and $C(p)$. By looking at the coefficient of $\delta^{\kappa \mu} \delta^{\rho \sigma} p^\nu$ on the right-hand side of \eqref{e:TTeq} we find $B(p) = 0$. The coefficient of $\delta^{\kappa \rho} \delta^{\mu \nu} p^\sigma$ then implies the differential equation
\begin{equation}
\left( d - p \frac{\partial}{\partial p} \right) A(p) = 0,
\end{equation}
which is precisely the same constraint as that derived from the dilatation Ward identity, leading us back to \eqref{e:cJcT1}. Similarly, for the 2-point function of the conserved current, the coefficient of $\delta^{\kappa \mu} p^\nu$ in \eqref{e:JJeq} requires
\begin{equation}
\left( d - 2 - p \frac{\partial}{\partial p} \right) C(p) = 0,
\end{equation}
which is precisely the dilatation Ward identity. Furthermore, one can check that the differential equations extracted from the other tensors in \eqref{e:TTform} and \eqref{e:JJform} do not impose any additional constraints.
As we can see, in the case of 2-point functions, the dilatation and special conformal Ward identities are equivalent.

\section{3-point function of scalar operators} \label{theory2_P}

\subsection{From position to momentum space}

We start with
a revision of some well-known facts regarding 3-point functions of conformal primary scalar operators in any CFT. Considering three scalar operators $\mathcal{O}_j$ of dimensions $\Delta_j$, $j = 1, 2, 3$, the 3-point function is unique up to an overall constant $c_{123}$  and in position space takes the form \cite{DiFrancesco:1997nk}
\begin{equation} \label{e:3ptx}
\langle \mathcal{O}_1(\bs{x}_1) \mathcal{O}_2(\bs{x}_2) \mathcal{O}_3(\bs{x}_3) \rangle = \frac{c_{123}}{|\bs{x}_1 - \bs{x}_2|^{\Delta_1 + \Delta_2 - \Delta_3} |\bs{x}_2 - \bs{x}_3|^{\Delta_2 + \Delta_3 - \Delta_1} |\bs{x}_3 - \bs{x}_1|^{\Delta_3 + \Delta_1 - \Delta_2}},
\end{equation}
where we work in Euclidean signature.
This expression, in principle, can be Fourier transformed to obtain the corresponding result in momentum space.
Extracting the Dirac delta function encoding overall momentum conservation, we
define the reduced matrix element (denoted with double brackets) as in \eqref{delta},
\begin{equation}
\langle \mathcal{O}_1(\bs{p}_1) \mathcal{O}_2(\bs{p}_2) \mathcal{O}_3(\bs{p}_3) \rangle = (2 \pi)^d \delta( \bs{p}_1 + \bs{p}_2 + \bs{p}_3 ) \lla \mathcal{O}_1(\bs{p}_1) \mathcal{O}_2(\bs{p}_2) \mathcal{O}_3(\bs{p}_3) \rra.
\end{equation}
Assuming $d \geq 3$, since $\bs{p}_1 + \bs{p}_2 + \bs{p}_3 = 0$ there are two independent momenta. Defining
\begin{equation} \label{e:deltaDelta}
\Delta_t = \Delta_1 + \Delta_2 + \Delta_3, \quad \qquad \delta_j = \frac{d - \Delta_t}{2} + \Delta_j, 
\end{equation}
a useful representation of the Fourier transform of \eqref{e:3ptx} is
\begin{align}
& \lla \mathcal{O}_1(\bs{p}_1) \mathcal{O}_2(\bs{p}_2) \mathcal{O}_3(\bs{p}_3) \rra = \nn\\
& \qquad = c_{123} \pi^{\frac{3d}{2}} 2^{3d - \Delta_t} \prod_{j=1}^3 \frac{\Gamma (\delta_j)}{\Gamma \left( \frac{d}{2} - \delta_j \right)} \cdot \int \frac{\D^d \bs{k}}{(2 \pi)^d} \frac{1}{|\bs{k}|^{2 \delta_3} | \bs{p}_1 - \bs{k} |^{2 \delta_2} | \bs{p}_2 + \bs{k} |^{2 \delta_1}} \nn\\
& \qquad = \frac{c_{123} \pi^d 2^{4 + \frac{3 d}{2} - \Delta_t}}{\Gamma \left( \frac{\Delta_t - d}{2} \right) \Gamma \left( \frac{\Delta_1 + \Delta_2 - \Delta_3}{2} \right) \Gamma \left( \frac{\Delta_2 + \Delta_3 - \Delta_1}{2} \right) \Gamma \left( \frac{\Delta_3 + \Delta_1 - \Delta_2}{2} \right)}  \nn\\
& \qquad \qquad \times p_1^{\Delta_1 - \frac{d}{2}} p_2^{\Delta_2 - \frac{d}{2}} p_3^{\Delta_3 - \frac{d}{2}} \int_0^\infty \D x \: x^{\frac{d}{2} - 1} K_{\Delta_1 - \frac{d}{2}}(p_1 x) K_{\Delta_2 - \frac{d}{2}}(p_2 x) K_{\Delta_3 - \frac{d}{2}}(p_3 x), \label{e:Otriple-$K$}
\end{align}
where $K_\nu(z)$ is a Bessel $K$ function, \textit{i.e.}, a modified Bessel function of the second kind.  A derivation of this result is given in appendix \ref{ch:toKKK}.
As mentioned in the introduction, we will refer to integrals of the form above featuring three Bessel $K$ functions and a power of $x$ as \emph{triple-$K$} integrals. This form of the 3-point function is familiar in the context of the AdS/CFT correspondence, where every bulk-to-boundary propagator for the field dual to the conformal operator $\mathcal{O}_j$ contains one Bessel $K$ function \cite{Freedman:1998tz}.

The expression \eqref{e:Otriple-$K$} may be severely divergent and requires regularisation. This stems from the fact that the original position space expression \eqref{e:3ptx} is valid at non-coincident points only and must itself be regularised. As we will discuss in more detail in section \ref{section:solnofCWIs} for generic values\footnote{
That is for values with
$
\frac{d}{2} \neq \pm (\Delta_1 - \frac{d}{2}) \pm (\Delta_2 - \frac{d}{2})\pm (\Delta_3 - \frac{d}{2})-2n
$, with $n=0,1,2, \ldots$.}
of $d$ and $\Delta_j$ the divergent integrals can be computed by analytic continuation in $d$ and $\Delta_j$.
In the special cases non-trivial renormalisation may required and we will discuss this in detail in \cite{II}.

As a simple example, consider the 3-point function of the operator $\O = \phi^2$ in the theory of a free real massless scalar in $d=3$ spacetime dimensions. By the standard Feynman diagrams one finds
\begin{equation} \label{e:3ptO1}
\lla \O(\bs{p}_1) \O(\bs{p}_2) \O(\bs{p}_3) \rra = 8 \int \frac{\D^3 \bs{k}}{(2 \pi)^3} \frac{1}{k^2 | \bs{p}_1 - \bs{k} |^2 | \bs{p}_2 + \bs{k} |^2} = \frac{1}{p_1 p_2 p_3}.
\end{equation}
The integral can be evaluated by means of the subtitution $\tilde{\bs{k}} = \bs{k}/k^2$. We can compare this result with a direct evaluation of the triple-$K$ integral \eqref{e:Otriple-$K$}. In this case $\Delta_j = 1$, $j = 1,2,3$ and the Bessel functions can be expressed in terms of elementary functions (see \eqref{e:Khalf}). However, since the integral in \eqref{e:Otriple-$K$} has a logarithmic divergence, we must regularise it. One way to regularise this result is to substitute
\begin{equation} \label{e:Oscheme}
d \mapsto d + 2 \epsilon, \qquad \qquad \Delta_j \mapsto \Delta_j + \epsilon.
\end{equation}
This regularisation scheme is extremely useful in context of triple-$K$ integrals since it preserves the indices of the Bessel functions in \eqref{e:Otriple-$K$}. In the present case, we find
\begin{align}
\lla \mathcal{O}(\bs{p}_1) \mathcal{O}(\bs{p}_2) \mathcal{O}(\bs{p}_3) \rra &=
\frac{16 c_{123} \pi^{3}}{\Gamma \left( \frac{\epsilon}{2} \right) p_1 p_2 p_3} \cdot \int_0^\infty \D x \: x^{-1 + \epsilon} e^{-x(p_1 + p_2 + p_3)} \nn\\
& = \frac{(2 \pi)^3 c_{123}}{p_1 p_2 p_3} + O(\epsilon).
\end{align}
As we can see, one recovers \eqref{e:3ptO1} with $c_{123} = (2 \pi)^{-3}$.

In summary, the Fourier transform of the position space expression \eqref{e:3ptx} for the 3-point function of scalar operators in any CFT may be expressed, at least formally, and up to an overall multiplicative constant, in terms of the triple-$K$ integral \eqref{e:Otriple-$K$}. In the next section we will show that this representation in terms of a triple-$K$ integral is very natural in the context of the conformal Ward identities. In fact, we will be able to re-derive the expression \eqref{e:Otriple-$K$} by solving the conformal Ward identities directly in momentum space, without any reference to position space.

\subsection{Conformal Ward identities} \label{ch:scalar_CWIs}

The conformal Ward identities (CWIs) in position space may be found in any standard reference text, \textit{e.g.}, \cite{DiFrancesco:1997nk}. In momentum space, the Ward identities for scalar operators have been partially analysed in \cite{Maldacena:2011nz,Creminelli:2012ed}, and we will use these results here before generalising them in the following sections. First, observe that due to Lorentz invariance any 3-point function may be expressed in terms of the magnitudes of the momenta,
\begin{equation}
p_j = | \bs{p}_j | = \sqrt{ \bs{p}_j^2 }, \qquad j = 1, 2, 3.
\end{equation}
The expression \eqref{e:Otriple-$K$} is in accord with this fact. Regarding the 3-point function as a function of the momentum magnitudes, the dilatation Ward identity then reads
\begin{equation} \label{e:Oward_dil_scal}
0= \left[ 2d + \sum_{j=1}^{3} \left( p_j \frac{\partial}{\partial p_j} - \Delta_j \right) \right] \lla \mathcal{O}_1(\bs{p}_1) \mathcal{O}_2(\bs{p}_2) \mathcal{O}_3(\bs{p}_3) \rra.
\end{equation}
Similarly, the Ward identity associated with special conformal transformations is
\begin{equation}
0 = \sum_{j=1}^3 p_j^\kappa \left[ \frac{\partial^2}{\partial p_j^2} + \frac{d + 1 - 2 \Delta_j}{p_j} \frac{\partial}{\partial p_j} \right] \lla \mathcal{O}_1(\bs{p}_1) \mathcal{O}_2(\bs{p}_2) \mathcal{O}_3(\bs{p}_3) \rra,
\end{equation}
where $\kappa$ is a free Lorentz index. Choosing $\bs{p}_1$ and $\bs{p}_2$ as independent momenta, we may split this vector equation into two independent scalar equations
\begin{align} \label{e:Oward_sp_scal}
& 0 = \left[ \left( \frac{\partial^2}{\partial p_1^2} + \frac{d + 1 - 2 \Delta_1}{p_1} \frac{\partial}{\partial p_1} \right) - \left( \frac{\partial^2}{\partial p_3^2} + \frac{d + 1 - 2 \Delta_3}{p_3} \frac{\partial}{\partial p_3} \right) \right] \lla \mathcal{O}_1(\bs{p}_1) \mathcal{O}_2(\bs{p}_2) \mathcal{O}_3(\bs{p}_3) \rra, \\
& 0 = \left[ \left( \frac{\partial^2}{\partial p_2^2} + \frac{d + 1 - 2 \Delta_2}{p_2} \frac{\partial}{\partial p_2} \right) - \left( \frac{\partial^2}{\partial p_3^2} + \frac{d + 1 - 2 \Delta_3}{p_3} \frac{\partial}{\partial p_3} \right) \right] \lla \mathcal{O}_1(\bs{p}_1) \mathcal{O}_2(\bs{p}_2) \mathcal{O}_3(\bs{p}_3) \rra. \label{e:Oward_sp_scal2}
\end{align}
As an immediate check, we may verify that the expression \eqref{e:Otriple-$K$}
satisfies (\ref{e:Oward_sp_scal}, \ref{e:Oward_sp_scal2}) using the well-known Bessel function relations \cite{Abramowitz}
\begin{align}
\frac{\partial}{\partial a} \left[ a^\nu K_\nu(a x) \right] & = - x a^\nu K_{\nu - 1}(a x), \\
K_{\nu-1}(x) + \frac{2 \nu}{x} K_{\nu}(x) & = K_{\nu + 1}(x).
\end{align}

As we will see shortly, equations of the form (\ref{e:Oward_sp_scal}, \ref{e:Oward_sp_scal2}) also arise in the case of 3-point correlators of general tensor operators.

\subsection{Uniqueness of the solution} \label{ch:Ounique}

To frame our analysis purely in momentum space, we need to show that there is a unique physically acceptable solution, up to an overall multiplicative constant, of the system
(\ref{e:Oward_dil_scal}, \ref{e:Oward_sp_scal}, \ref{e:Oward_sp_scal2}) of dilatation and special CWIs.
To accomplish this, it suffices to transform these equations into generalised hypergeometric form by writing
\begin{equation} \label{e:OsubstoF4}
\lla \mathcal{O}_1(\bs{p}_1) \mathcal{O}_2(\bs{p}_2) \mathcal{O}_3(\bs{p}_3) \rra = p_3^{\Delta_t - 2d} \left( \frac{p_1^2}{p_3^2} \right)^\mu \left( \frac{p_2^2}{p_3^2} \right)^\lambda F \left( \frac{p_1^2}{p_3^2}, \frac{p_2^2}{p_3^2} \right),
\end{equation}
where the overall power of momenta on the right-hand side is fixed by the dilatation Ward identity  (\ref{e:Oward_dil_scal}), and we have chosen to multiply the arbitrary function $F$ by the prefactor\\
$(p_1^2/p_3^2)^\mu (p_2^2/p_3^2)^\lambda$, where $\mu$ and $\lambda$ are arbitrary constants.
Substituting this parametrisation into (\ref{e:Oward_sp_scal}, \ref{e:Oward_sp_scal2}) then yields a pair of differential equations satisfied by $F$. Taking $\mu$ and $\lambda$ to be any of the four combinations obtainable from the values
\begin{equation} \label{e:OF4param}
\mu = 0, \ \Delta_1 - \frac{d}{2}, \qquad \qquad \lambda = 0, \ \Delta_2 - \frac{d}{2},
\end{equation}
these equations for $F$ read
\begin{align}
& 0 = \left[ \xi (1 - \xi) \frac{\partial^2}{\partial \xi^2} - \eta^2 \frac{\partial^2}{\partial \eta^2} - 2 \xi \eta \frac{\partial^2}{\partial \xi \partial \eta} \right. \nn \\
& \qquad \left. + \: \left( \gamma - ( \alpha + \beta + 1) \xi \right) \frac{\partial}{\partial \xi} - (\alpha + \beta + 1) \eta \frac{\partial}{\partial \eta} - \alpha \beta \right] F(\xi, \eta), \label{e:OsysF4a} \\
& 0 = \left[ \eta (1 - \eta) \frac{\partial^2}{\partial \eta^2} - \xi^2 \frac{\partial^2}{\partial \xi^2} - 2 \xi \eta \frac{\partial^2}{\partial \xi \partial \eta} \right. \nn \\
& \qquad \left. + \: \left( \gamma' - ( \alpha + \beta + 1) \eta \right) \frac{\partial}{\partial \eta} - (\alpha + \beta + 1) \xi \frac{\partial}{\partial \xi} - \alpha \beta \right] F(\xi, \eta), \label{e:OsysF4b}
\end{align}
where
\begin{equation}
\xi = \frac{p_1^2}{p_3^2}, \qquad \qquad \eta = \frac{p_2^2}{p_3^2},
\end{equation}
and the values of the parameters $\alpha, \beta, \gamma, \gamma'$ depend on the choice of $\mu$ and $\lambda$. Specifically, parametrising the four choices for $\mu$ and $\lambda$ by two variables $\epsilon_1, \epsilon_2 \in \{ -1, +1 \}$ according to
\begin{equation}
\mu= \frac{1}{2}\big(\Delta_1-\frac{d}{2}\big)(\epsilon_1+1),\qquad \lambda =\frac{1}{2}\big(\Delta_2-\frac{d}{2}\big)(\epsilon_2+1),
\end{equation}
we have
\begin{align}
& \alpha = \frac{1}{2} \left[ \epsilon_1 \left( \Delta_1 - \frac{d}{2} \right) + \epsilon_2 \left( \Delta_2 - \frac{d}{2} \right) + \Delta_3 \right], \qquad \beta = \alpha - \left( \Delta_3 - \frac{d}{2} \right), \nn \\
& \gamma = 1 + \epsilon_1 \left( \Delta_1 - \frac{d}{2} \right), \qquad \gamma' = 1 + \epsilon_2 \left( \Delta_2 - \frac{d}{2} \right). \label{e:Ochoices}
\end{align}

The system of equations (\ref{e:OsysF4a}, \ref{e:OsysF4b}) defines the generalised hypergeometric function of two variables Appell $F_4$. This function has been extensively studied by mathematicians (see, \textit{e.g.}, \cite{Appell,Erdelyi}), and its important properties are summarised in appendix \ref{ch:F4}. In particular, the system of equations (\ref{e:OsysF4a}, \ref{e:OsysF4b}) has at most four linearly independent solutions, each of which may be expressed in terms of the $F_4$ function \cite{Erdelyi, Exton:1995}. The four possible choices for $\mu$ and $\lambda$ reproduce these four solutions exactly.

In a physical context only one linear combination of these four solutions is acceptable: all the others contain divergences for collinear momentum configurations, for example when $p_1 + p_2 = p_3$. To see this, consider the integral representation \cite{Prudnikov}
\begin{align}
F_4 \left( \alpha, \beta; \gamma, \gamma'; \frac{p_1^2}{p_3^2}, \frac{p_2^2}{p_3^2} \right) & = \frac{\Gamma(\gamma) \Gamma(\gamma')}{2^{\alpha+\beta-\gamma-\gamma'} \Gamma(\alpha) \Gamma(\beta)} \cdot \frac{p_3^{\alpha+\beta}}{p_1^{\gamma-1} p_2^{\gamma'-1}} \times \nn \\
& \times \int_0^\infty \D x x^{\alpha + \beta - \gamma - \gamma'+ 1} I_{\gamma - 1}(p_1 x) I_{\gamma' - 1}(p_2 x) K_{\beta - \alpha}(p_3 x), \label{e:F4toIIK}
\end{align}
where $I_\nu(x)$ is the Bessel $I$ function. This expression is formal in the sense that the integral converges only for $\alpha$, $\beta$, $\gamma$, $\gamma'$ in certain ranges, see appendix \ref{ch:F4} for details. For the remaining parameter values the integral is defined by the analytic continuation \eqref{e:Oscheme}. Using \eqref{e:Ochoices}, one can then write the four solutions for the 3-point functions in the form
\begin{equation} \label{e:IIK}
p_1^{\Delta_1 - \frac{d}{2}} p_2^{\Delta_2 - \frac{d}{2}} p_3^{\Delta_3 - \frac{d}{2}} \int_0^\infty \D x \: x^{\frac{d}{2} - 1} I_{\pm(\Delta_1 - \frac{d}{2})} (p_1 x) I_{\pm(\Delta_2 - \frac{d}{2})} (p_2 x) K_{\Delta_3 - \frac{d}{2}} (p_3 x).
\end{equation}
For large $x$ we have the asymptotic expansions
\begin{equation} \label{e:A}
I_\nu(x) = \frac{1}{\sqrt{2 \pi}} \frac{e^x}{\sqrt{x}} + \ldots, \qquad K_\nu(x) = \sqrt{\frac{\pi}{2}} \frac{e^{-x}}{\sqrt{x}} + \ldots,
\end{equation}
from which we see that the integral \eqref{e:IIK} converges at infinite $x$ only for non-triangle (\textit{i.e.}, unphysical) momentum configurations where $p_1 + p_2 < p_3$.
Moreover, for the physical collinear momentum configuration $p_1 + p_2 = p_3$, the integral diverges for dimensions $d\ge 3$. However, the 3-point function itself is a linear combination of these four solutions and may be continued to the physical regime by choosing the linear combination for which the collinear divergences cancel.  This may be accomplished by subtracting two integrals with the same asymptotics, \textit{i.e.}, we sum the four terms of the form \eqref{e:IIK} with signs chosen so as to obtain Bessel $K$ functions
\begin{equation}
K_\nu(x) = \frac{\pi}{2 \sin ( \nu \pi)} \left[ I_\nu(x) - I_{-\nu}(x) \right].
\end{equation}
Therefore we arrive at the unique solution
\begin{align}
& \lla \mathcal{O}_1(\bs{p}_1) \mathcal{O}_2(\bs{p}_2) \mathcal{O}_3(\bs{p}_3) \rra \label{e:Osol}\\
& \qquad = C_{123} \cdot p_1^{\Delta_1 - \frac{d}{2}} p_2^{\Delta_2 - \frac{d}{2}} p_3^{\Delta_3 - \frac{d}{2}} \int_0^\infty \D x \: x^{\frac{d}{2} - 1} K_{\Delta_1 - \frac{d}{2}} (p_1 x) K_{\Delta_2 - \frac{d}{2}} (p_2 x) K_{\Delta_3 - \frac{d}{2}} (p_3 x), \nn
\end{align}
where $C_{123}$ is an overall undetermined constant.  From the asymptotic expansion \eqref{e:A}, it is clear that this triple-$K$
integral converges at infinite $x$ for physical momentum configurations $p_1+p_2+p_3>0$.  Depending on the values of the parameters $\Delta_j$ and $d$, however, the triple-$K$ integral may still diverge at $x=0$. This divergence may be regularised using \eqref{e:Oscheme} or the dimensional regularisation \eqref{e:dimreg00}. We will analyse such cases in sections \ref{ch:regul} and \ref{ch:tripleKand2pt}.

In summary then,
we have shown that the conformal Ward identities may be solved directly in momentum space leading to a unique result \eqref{e:Osol}. As we will see shortly, a similar procedure also holds for tensorial correlation functions: solving the momentum space Ward identities will lead to a unique solution for 3-point correlators without any reference to the position space analysis.

\section{Decomposition of tensors} \label{sec:decomp}

In this section, we present a natural decomposition of tensorial correlation functions.
Correlation functions of conserved currents are transverse and/or traceless -- up to local terms -- and we would like to find a decomposition which reflects these properties. At this point, we will not yet impose conformal invariance.

The problem of decomposition has already been tackled in a number of papers, see for example \cite{Rosenberg:1962pp,Cappelli:2001pz,Giannotti:2008cv,Armillis:2009sm,Armillis:2009pq,Coriano:2012wp}. The usual approach consists of writing down the most general tensor structure before imposing the constraints following from symmetries and Ward identities.
Here we refine this approach to take account of the permutation symmetries of operator insertions inside correlators, obtaining a convenient and natural decomposition applicable for any correlation function. In particular, our decomposition contains the minimal number of tensor structures, leading to the simplest form for the conformal Ward identities.

We remind the reader we will always be working in $d$-dimensional Euclidean field theory with a flat metric $\delta_{\mu \nu}$ for which indices are raised and lowered trivially.

\subsection{Representations of tensor structures} \label{ch:Tensor_structure}

The operator
\begin{equation} \label{e:pi}
\pi^{\mu}_{\alpha}(\bs{p}) = \delta^{\mu}_{\alpha} - \frac{p^{\mu} p_{\alpha}}{p^2}
\end{equation}
is a projector onto tensors transverse to $\bs{p}$, \textit{i.e.}, $p_\mu \pi^{\mu}_{\alpha}(\bs{p}) = 0$. Similarly, in $d$ dimensions, the operator
\begin{equation} \label{e:Pi}
\Pi^{\mu \nu}_{\alpha \beta}(\bs{p}) = \frac{1}{2} \left( \pi^{\mu}_{\alpha}(\bs{p}) \pi^{\nu}_{\beta}(\bs{p}) + \pi^{\mu}_{\beta}(\bs{p}) \pi^{\nu}_{\alpha}(\bs{p}) \right) - \frac{1}{d - 1} \pi^{\mu \nu}(\bs{p}) \pi_{\alpha \beta}(\bs{p})
\end{equation}
is a projector onto transverse to $\bs{p}$, traceless, symmetric tensors of rank two. In particular
\begin{equation} \label{e:properties_of_Pi}
\delta_{\mu \nu} \Pi^{\mu \nu}_{\alpha \beta}(\bs{p}) = 0, \qquad p_{\mu} \Pi^{\mu \nu}_{\alpha \beta}(\bs{p}) = 0.
\end{equation}
Therefore, any transverse to $\bs{p}$, traceless, symmetric tensor $t^{\mu \nu}$ of rank two may be written as $t^{\mu \nu} = \Pi^{\mu \nu}_{\alpha \beta}(\bs{p}) X^{\alpha \beta}$, where $X^{\alpha \beta}$ is an arbitrary tensor.

As we are interested in correlation functions, we must consider tensor functions that depend on a number of momenta. Let $\mathcal{T}_j^{\mu_{j1} \mu_{j2} \ldots \mu_{j r_j}}$, $j = 1, 2, \ldots, n$ be a given set of QFT operators. Due to the momentum conservation, the $n$-point function contains a delta function which may be written explicitly by introducing the reduced matrix element which we denote with double brackets $\lla \ldots \rra$,
\begin{align}
& \langle \mathcal{T}_1^{\mu_{11} \mu_{12} \ldots \mu_{1 r_1}}(\bs{p}_1) \mathcal{T}_2^{\mu_{21} \mu_{22} \ldots \mu_{2 r_2}}(\bs{p}_2) \ldots \mathcal{T}_n^{\mu_{n1} \mu_{n2} \ldots \mu_{n r_n}}(\bs{p}_n) \rangle \nn\\
& = (2 \pi)^d \delta \left( \sum_{k=1}^n \bs{p}_k \right) \lla \mathcal{T}_1^{\mu_{11} \mu_{12} \ldots \mu_{1 r_1}}(\bs{p}_1) \mathcal{T}_2^{\mu_{21} \mu_{22} \ldots \mu_{2 r_2}}(\bs{p}_2) \ldots \mathcal{T}_n^{\mu_{n1} \mu_{n2} \ldots \mu_{n r_n}}(\bs{p}_n) \rra.
\end{align}
The $n$-point function thus depends on at most $n - 1$ vectors, say $\bs{p}_1, \ldots, \bs{p}_{n-1}$. If
$n-1<d$, then all $n-1$ momenta are independent. If $n - 1 \geq d$, however, then only $d$ generic momenta are independent. In this case we can write \cite{Denner:2005nn}
\begin{equation} \label{e:metric_as_momenta}
\delta^{\mu \nu} = \sum_{j,k = 1}^d p_j^\mu p_k^\nu ( Z^{-1} )_{kj},
\end{equation}
where $Z$ is the Gram matrix, $Z = [ \bs{p}_k \cdot \bs{p}_l ]_{k,l=1}^d$, hence the metric $\delta^{\mu \nu}$ is no longer an independent tensor.

From now on we assume $d \geq 3$. Since we are primarily interested in 3-point functions, the degeneracy does not occur.
Nevertheless, the case $d=3$ is still special since the existence of the cross-product allows the metric tensor to be re-expressed purely in terms of the momenta. This degeneracy serves to reduce the number of independent form factors for certain correlators, as we discuss in appendix \ref{ch:degeneracy}.  In the following discussion we will ignore this degeneracy however and concentrate on the general case.  We will therefore choose two out of the three  $\bs{p}_1, \bs{p}_2, \bs{p}_3$ as independent momenta, and treat the metric $\delta^{\mu \nu}$ as an independent tensor.

As an example consider a 3-point function of two transverse, traceless, symmetric rank two operators $t^{\mu \nu}$ and a scalar operator $\mathcal{O}$. Using the projectors \eqref{e:Pi} one can write the most general form
\begin{equation} \label{e:genformTTO}
\lla t^{\mu_1 \nu_1}(\bs{p}_1) t^{\mu_2 \nu_2}(\bs{p}_2) \mathcal{O}(\bs{p}_3) \rra
= \Pi^{\mu_1 \nu_1}_{\alpha_1 \beta_1}(\bs{p}_1) \Pi^{\mu_2 \nu_2}_{\alpha_2 \beta_2}(\bs{p}_2) X^{\alpha_1 \beta_1 \alpha_2 \beta_2},
\end{equation}
where $X^{\alpha_1 \beta_1 \alpha_2 \beta_2}$ is a general tensor of rank four built from the metric and momenta. Usually one chooses two independent momenta once and for all. On the other hand, there is no obstacle to choosing different independent momenta for different Lorentz indices. In this paper we always choose
\begin{equation} \label{e:momenta_choice}
\bs{p}_1, \bs{p}_2 \text{ for } \mu_1, \nu_1; \ \bs{p}_2, \bs{p}_3 \text{ for } \mu_2, \nu_2 \text{  and  } \bs{p}_3, \bs{p}_1 \text{ for }\mu_3, \nu_3.
\end{equation}
Such a choice enhances the symmetry properties of the decomposition, as we will discuss at length in the next section.

Let us now enumerate all possible tensors that can appear in $X^{\alpha_1 \beta_1 \alpha_2 \beta_2}$. Observe that whenever a simple tensor contains at least one of the following tensors
\begin{equation}
\delta^{\alpha_1 \beta_1}, \delta^{\alpha_2 \beta_2}, p_1^{\alpha_1}, p_1^{\beta_1}, p_2^{\alpha_2}, p_2^{\beta_2},
\end{equation}
then the contraction with the projectors in \eqref{e:genformTTO} vanishes. Therefore, in accordance with the choice \eqref{e:momenta_choice}, the only tensors giving a non-zero result after contraction with the projectors are
\begin{equation}
\delta^{\alpha_1 \alpha_2}, \delta^{\alpha_1 \beta_2}, \delta^{\beta_1 \alpha_2}, \delta^{\beta_1 \beta_2}, p_2^{\alpha_1}, p_2^{\beta_1}, p_3^{\alpha_2}, p_3^{\beta_2}.
\end{equation}
Since the projector \eqref{e:Pi} is symmetric in $\mu \leftrightarrow \nu$ and $\alpha \leftrightarrow \beta$, the most general form of our 3-point function is then
\begin{align}
\lla t^{\mu_1 \nu_1}(\bs{p}_1) t^{\mu_2 \nu_2}(\bs{p}_2) \mathcal{O}(\bs{p}_3) \rra
& = \Pi^{\mu_1 \nu_1}_{\alpha_1 \beta_1}(\bs{p}_1) \Pi^{\mu_2 \nu_2}_{\alpha_2 \beta_2}(\bs{p}_2) \left[
A_1 p_2^{\alpha_1} p_2^{\beta_1} p_3^{\alpha_2} p_3^{\beta_2} \right. \nn\\
& \left. \qquad \qquad + \: A_2 p_2^{\alpha_1}  p_3^{\alpha_2} \delta^{\beta_1 \beta_2} + A_3 \delta^{\alpha_1 \alpha_2} \delta^{\beta_1 \beta_2} \right], \label{e:TTOform}
\end{align}
where the coefficients $A_1$, $A_2$ and $A_3$ are scalar functions of momenta. We will refer to the coefficients $A_j$, and their analogous counterparts in more general correlation functions, as \emph{form factors}. By Lorentz invariance, these form factors are functions of the momentum magnitudes
\begin{equation}
p_j = | \bs{p}_j| = \sqrt{ \bs{p}_j^2}, \qquad j = 1,2,3,
\end{equation}
\textit{i.e.}, $A_j = A_j(p_1, p_2, p_3)$.  In particular, any scalar product of two momenta can be written as a combination of momentum magnitudes, for example
\begin{equation} \label{e:scal_prod_to_squares}
\bs{p}_1 \cdot \bs{p}_2 = \frac{1}{2} ( p_3^2 - p_1^2 - p_2^2 ).
\end{equation}
For brevity, we will henceforth suppress the dependence of form factors on the momentum magnitudes, writing $A_j(p_1, p_2, p_3)$ as simply $A_j$.

Note that the correlation function on the left-hand side of \eqref{e:TTOform} is symmetric under a transposition $(\bs{p}_1, \mu_1, \nu_1) \leftrightarrow (\bs{p}_2, \mu_2, \nu_2)$. One can apply this symmetry to the right-hand side to find that all form factors $A_1$, $A_2$ and $A_3$ are symmetric under $p_1 \leftrightarrow p_2$. To prove this, observe that one has, for example, $\pi^{\mu_1}_{\alpha_1}(\bs{p}_1) p_3^{\alpha_1} = - \pi^{\mu_1}_{\alpha_1}(\bs{p}_1) p_2^{\alpha_1}$. Therefore $\bs{p}_2$ and $-\bs{p}_3$ can be exchanged under both $\pi^{\mu_1}_{\alpha_1}(\bs{p}_1)$ and $\Pi^{\mu_1 \nu_1}_{\alpha_1 \beta_1}(\bs{p}_1)$, and similarly for other momenta.

For any form factor $A_j$ we define an associated non-negative integer $N_j$, the \emph{tensorial dimension} of $A_j$, similar to that defined in \cite{Cappelli:2001pz}. Specifically, the tensorial dimension $N_j$ is the number of momenta that appear in the tensorial expression multiplying $A_j$ (excluding those in the transverse-traceless projectors) in the decomposition of the correlation function. As we will see later, this quantity will appear explicitly in the conformal Ward identities. In the example \eqref{e:TTOform}, we have the following tensorial dimensions: $N_1 = 4$, $N_2 = 2$ and $N_3 = 0$.

Decompositions for other correlation functions may be found in the second part of the paper. Observe that in each case the form factor $A_1$ stands in front of the unique tensor containing momenta only. The tensorial dimension $N_1$ is therefore always equal to the number of Lorentz indices in the 3-point function, and tensorial dimensions of all remaining form factors are smaller than $N_1$.

\subsection{\texorpdfstring{Decomposition of $\< t^{\mu_1 \nu_1} t^{\mu_2 \nu_2} t^{\mu_3 \nu_3} \>$}{Decomposition of <ttt>}}

In the previous section we introduced a natural decomposition of tensor structures. Rather than fixing two independent momenta (as is done for example in \cite{Rosenberg:1962pp,Cappelli:2001pz,Giannotti:2008cv,Armillis:2009sm,Armillis:2009pq,Coriano:2012wp}) we chose a different set of independent momenta for different Lorentz indices according to \eqref{e:momenta_choice}. Such a choice respects all symmetries of the correlation function, as we now discuss.

In \cite{Cappelli:2001pz}, it was shown that the transverse-traceless correlation function $\lla t^{\mu_1 \nu_1} t^{\mu_2 \nu_2} t^{\mu_3 \nu_3} \rra$ can be decomposed into eight tensor structures plus their $\bs{p}_1 \leftrightarrow \bs{p}_2$ symmetric versions. In our method, however, we arrive at only five tensor structures (for the general case $d \geq 3$, see appendix \ref{ch:degeneracy} for the case $d=3$)
according to the following decomposition
\begin{align}
& \lla t^{\mu_1 \nu_1}(\bs{p}_1) t^{\mu_2 \nu_2}(\bs{p}_2) t^{\mu_3 \nu_3}(\bs{p}_3) \rra \nonumber \\
& \qquad = \Pi^{\mu_1 \nu_1}_{\alpha_1 \beta_1}(\bs{p}_1) \Pi^{\mu_2 \nu_2}_{\alpha_2 \beta_2}(\bs{p}_2) \Pi^{\mu_3 \nu_3}_{\alpha_3 \beta_3}(\bs{p}_3) \left[
A_1 p_2^{\alpha_1} p_2^{\beta_1} p_3^{\alpha_2} p_3^{\beta_2} p_1^{\alpha_3} p_1^{\beta_3} \right. \nonumber \\
& \qquad \qquad + \: A_2 \delta^{\beta_1 \beta_2} p_2^{\alpha_1} p_3^{\alpha_2} p_1^{\alpha_3} p_1^{\beta_3} + A_2(p_1 \leftrightarrow p_3) \delta^{\beta_2 \beta_3} p_2^{\alpha_1} p_2^{\beta_1} p_3^{\alpha_2} p_1^{\alpha_3} \nonumber \\
& \qquad \qquad \qquad \qquad + \: A_2(p_2 \leftrightarrow p_3) \delta^{\beta_1 \beta_3} p_2^{\alpha_1} p_3^{\alpha_2} p_3^{\beta_2} p_1^{\alpha_3} \nonumber \\
& \qquad \qquad + \: A_3 \delta^{\alpha_1 \alpha_2} \delta^{\beta_1 \beta_2} p_1^{\alpha_3} p_1^{\beta_3} + A_3(p_1 \leftrightarrow p_3) \delta^{\alpha_2 \alpha_3} \delta^{\beta_2 \beta_3} p_2^{\alpha_1} p_2^{\beta_1} \nonumber \\
& \qquad \qquad \qquad \qquad + \: A_3(p_2 \leftrightarrow p_3) \delta^{\alpha_1 \alpha_3} \delta^{\beta_1 \beta_3} p_3^{\alpha_2} p_3^{\beta_2} \nonumber \\
& \qquad \qquad + \: A_4 \delta^{\alpha_1 \alpha_3} \delta^{\alpha_2 \beta_3} p_2^{\beta_1} p_3^{\beta_2} + A_4(p_1 \leftrightarrow p_3) \delta^{\alpha_1 \alpha_3} \delta^{\alpha_2 \beta_1} p_3^{\beta_2} p_1^{\beta_3} \nonumber \\
 & \qquad \qquad \qquad \qquad + \: A_4(p_2 \leftrightarrow p_3) \delta^{\alpha_1 \alpha_2} \delta^{\alpha_3 \beta_2} p_2^{\beta_1} p_1^{\beta_3} \nonumber \\
& \left. \qquad \qquad + \: A_5 \delta^{\alpha_1 \beta_2} \delta^{\alpha_2 \beta_3} \delta^{\alpha_3 \beta_1} \right]. \label{e:decompTTT}
\end{align}
By $p_1 \leftrightarrow p_3$ we denote the exchange of the arguments $p_1$ and $p_3$, $A_2(p_1 \leftrightarrow p_3) = A_2(p_3, p_2, p_1)$. If no arguments are specified, then the standard ordering is assumed, \textit{i.e.}, $A_2 = A_2(p_1, p_2, p_3)$.

First observe that this decomposition is manifestly invariant under the permutation group $S_3$ of the set $\{1,2,3\}$, \textit{i.e.}, for any $\sigma \in S_3$,
\begin{equation}
\lla t^{\mu_1 \nu_1}(\bs{p}_1) t^{\mu_2 \nu_2}(\bs{p}_2) t^{\mu_3 \nu_3}(\bs{p}_3) \rra = \lla t^{\mu_{\sigma(1)} \nu_{\sigma(1)}}(\bs{p}_{\sigma(1)}) t^{\mu_{\sigma(2)} \nu_{\sigma(2)}}(\bs{p}_{\sigma(2)}) t^{\mu_{\sigma(3)} \nu_{\sigma(3)}}(\bs{p}_{\sigma(3)}) \rra.
\end{equation}
In particular, the form factors $A_1$ and $A_5$ are $S_3$-invariant,
\begin{equation}
A_j(p_1, p_2, p_3) = A_j(p_{\sigma(1)}, p_{\sigma(2)}, p_{\sigma(3)}), \quad j \in \{1,5\},
\end{equation}
since the tensors they multiply are $S_3$-invariant. The action of the symmetry group on the remaining terms is then clearly visible. As an example, let us concentrate on the third line of \eqref{e:decompTTT} with the $A_2$ form factor. The $(\bs{p}_1, \mu_1, \nu_1) \leftrightarrow (\bs{p}_2, \mu_2, \nu_2)$ permutation leaves the tensor in the first term invariant, therefore the $A_2$ factor exhibits the $p_1 \leftrightarrow p_2$ symmetry. On the other hand, the $(\bs{p}_1, \mu_1, \nu_1) \leftrightarrow (\bs{p}_3, \mu_3, \nu_3)$ permutation swaps tensor structures of the first and the second term in the third line. This requires that the form factor of the second term is related to the form factor of the first term by the $p_1 \leftrightarrow p_3$ permutation, as indicated. Working out the remaining lines of \eqref{e:decompTTT} one finds that both remaining factors $A_3$ and $A_4$ are $p_1 \leftrightarrow p_2$ symmetric.

Let us comment then on the apparent disagreement between the number of tensor structures between \eqref{e:decompTTT} and the results of \cite{Cappelli:2001pz}. As already mentioned, the mismatch follows from the choice of two independent momenta in \cite{Cappelli:2001pz} to be $\bs{p}_1$ and $\bs{p}_2$, in our notation. Such a choice breaks the $S_3$ symmetry down to the $(\bs{p}_1, \mu_1, \nu_1) \leftrightarrow (\bs{p}_2, \mu_2, \nu_2)$ symmetry. One can easily recover eight tensor structures from \eqref{e:decompTTT} by substituting $\bs{p}_3 = - \bs{p}_1 - \bs{p}_2$ and writing the decomposition in terms of $\bs{p}_1$ and $\bs{p}_2$ only. One finds
\begin{align}
& \lla t^{\mu_1 \nu_1}(\bs{p}_1) t^{\mu_2 \nu_2}(\bs{p}_2) t^{\mu_3 \nu_3}(\bs{p}_3) \rra \nn\\
& \qquad = \Pi^{\mu_1 \nu_1}_{\alpha_1 \beta_1}(\bs{p}_1) \Pi^{\mu_2 \nu_2}_{\alpha_2 \beta_2}(\bs{p}_2) \Pi^{\mu_3 \nu_3}_{\alpha_3 \beta_3}(\bs{p}_3) \left[
\frac{1}{2} A_1 p_2^{\alpha_1} p_2^{\beta_1} p_1^{\alpha_2} p_1^{\beta_2} p_1^{\alpha_3} p_1^{\beta_3} \right. \nonumber \\
& \qquad \qquad - \: \frac{1}{2} A_2 \delta^{\beta_1 \beta_2} p_2^{\alpha_1} p_1^{\alpha_2} p_1^{\alpha_3} p_1^{\beta_3} - A_2(p_1 \leftrightarrow p_3) \delta^{\beta_2 \beta_3} p_2^{\alpha_1} p_2^{\beta_1} p_1^{\alpha_2} p_1^{\alpha_3} \nonumber \\
& \qquad \qquad + \: \frac{1}{2} A_3 \delta^{\alpha_1 \alpha_2} \delta^{\beta_1 \beta_2} p_1^{\alpha_3} p_1^{\beta_3} + A_3(p_1 \leftrightarrow p_3) \delta^{\alpha_2 \alpha_3} \delta^{\beta_2 \beta_3} p_2^{\alpha_1} p_2^{\beta_1} \nonumber \\
& \qquad \qquad - \: \frac{1}{2} A_4 \delta^{\alpha_1 \alpha_3} \delta^{\alpha_2 \beta_3} p_2^{\beta_1} p_1^{\beta_2} - A_4(p_1 \leftrightarrow p_3) \delta^{\alpha_1 \alpha_3} \delta^{\alpha_2 \beta_1} p_1^{\beta_2} p_1^{\beta_3} \nonumber \\
& \qquad \qquad + \: \frac{1}{2} A_5 \delta^{\alpha_1 \beta_2} \delta^{\alpha_2 \beta_3} \delta^{\alpha_3 \beta_1} \nonumber \\
& \left. \qquad \qquad + \: \text{everything with } ( \bs{p}_1, \alpha_1, \beta_1 ) \leftrightarrow ( \bs{p}_2, \alpha_2, \beta_2 ) \right].
\end{align}
As we can see, the number of tensor structures increases to exactly eight, as the symmetry group is diminished.

Summarising, our decomposition method based on \eqref{e:momenta_choice} gives the minimal number of tensor structures obeying the symmetries of the correlation function. It is an improvement over the standard method with two independent momenta fixed, since such a choice breaks symmetries and leads therefore to the larger number of tensor structures.

Finally, we should comment on the fact that we decompose the transverse-traceless part of the correlation function only. This is because the difference between the full 3-point function and its transverse-traceless part is semi-local, and hence may be entirely reconstructed from the Ward identities. We will discuss this method for recovering the full correlation function from its transverse-traceless part in the next section.

Let us note in passing that the decomposition method described here may also be used for correlation functions in non-conformal theories. For example, in cases where the stress-energy tensor is transverse but no longer traceless one should use the $\pi^\mu_\alpha$ projectors \eqref{e:pi} in place of $\Pi^{\mu \nu}_{\alpha \beta}$ in \eqref{e:decompTTT}. In this way one obtains ten tensor structures, five of which have nonzero trace. This decomposition is given in appendix \ref{ch:decTTT}.

\subsection{Finding the form factors} \label{ch:finding}

We would like to apply the results of the previous section to spin-$1$ and spin-$2$ conserved currents $J^\mu$ and a stress-energy tensor $T^{\mu \nu}$. These quantum operators, however, are only transverse and traceless on-shell, and in the quantum case, we need to analyse Ward identities. To proceed, we define transverse, transverse-traceless and local parts of $J^\mu$ and $T^{\mu \nu}$ by
\begin{align}
j^{\mu} \equiv \pi^{\mu}_{\alpha} J^{\alpha}, & \qquad\qquad j_{\text{loc}}^{\mu} \equiv J^{\mu} - j^{\mu}, \label{e:decompJ} \\
t^{\mu \nu} \equiv \Pi^{\mu \nu}_{\alpha \beta} T^{\alpha \beta}, & \qquad\qquad t_{\text{loc}}^{\mu \nu} \equiv T^{\mu \nu} - t^{\mu \nu}, \label{e:decompT}
\end{align}
as well as longitudinal and trace parts
\begin{equation} \label{e:locterms}
r = p_\mu J^\mu, \qquad R^\nu = p_\mu T^{\mu \nu}, \qquad R = p_\nu R^\nu, \qquad T = T^{\mu}_{\mu}.
\end{equation}
From these definitions, we then have
\begin{align}
j_{\text{loc}}^{\mu} & = \frac{p^\mu}{p^2} r, \label{e:jloc} \\
t_{\text{loc}}^{\mu \nu} & = \frac{p^\mu}{p^2} R^\nu + \frac{p^\nu}{p^2} R^\mu - \frac{p^\mu p^\nu}{p^4} R + \frac{1}{d-1} \pi^{\mu \nu} \left( T - \frac{R}{p^2} \right) \nn\\
& = \mathscr{T}^{\mu \nu}_{\alpha} R^\alpha + \frac{\pi^{\mu\nu}}{d-1} T, \label{e:tloc}
\end{align}
where the operator
\begin{equation} \label{e:curlyT}
\mathscr{T}^{\mu\nu}_{\alpha} (\bs{p}) = \frac{1}{p^2} \left[ 2 p^{(\mu} \delta^{\nu)}_\alpha - \frac{p_\alpha}{d-1} \left( \delta^{\mu\nu} + (d-2) \frac{p^\mu p^\nu}{p^2} \right) \right].
\end{equation}
In the following, we will also use $\mathscr{T}^{\mu\nu\alpha} = \delta^{\alpha \beta}\mathscr{T}^{\mu\nu}_{\beta}$.

We now observe that in a CFT, all terms in \eqref{e:jloc} and \eqref{e:tloc} are computable by means of the transverse and trace Ward identities. We can therefore divide a 3-point function into two parts: the \emph{transverse-traceless part} represented as in section \ref{ch:Tensor_structure}, and the \emph{semi-local part} (indicated by the subscript \emph{loc}) expressible through the transverse Ward identities. For simplicity we will use the term `transverse-traceless part' in all cases, even if the correlation function does not contain the stress-energy tensor.

As an example, consider
\begin{equation} \label{e:TOOdef}
\lla t^{\mu_1 \nu_1}(\bs{p}_1) t^{\mu_2 \nu_2}(\bs{p}_2) \mathcal{O}(\bs{p}_3) \rra = \Pi^{\mu_1 \nu_1}_{\alpha_1 \beta_1}(\bs{p}_1) \Pi^{\mu_2 \nu_2}_{\alpha_2 \beta_2}(\bs{p}_2) \lla T^{\alpha_1 \beta_1}(\bs{p}_1) T^{\alpha_2 \beta_2}(\bs{p}_2) \mathcal{O}(\bs{p}_3) \rra.
\end{equation}
One can recover the full 3-point function by writing
\begin{align}
& \lla T^{\mu_1 \nu_1} T^{\mu_2 \nu_2} \mathcal{O} \rra = \lla t^{\mu_1 \nu_1} t^{\mu_2 \nu_2} \mathcal{O} \rra + \lla t^{\mu_1 \nu_1} t_{\text{loc}}^{\mu_2 \nu_2} \mathcal{O} \rra + \lla t_{\text{loc}}^{\mu_1 \nu_1} t^{\mu_2 \nu_2} \mathcal{O} \rra + \lla t_{\text{loc}}^{\mu_1 \nu_1} t_{\text{loc}}^{\mu_2 \nu_2} \mathcal{O} \rra \nn\\
& \qquad = \lla t^{\mu_1 \nu_1} t^{\mu_2 \nu_2} \mathcal{O} \rra - \lla T^{\mu_1 \nu_1} t_{\text{loc}}^{\mu_2 \nu_2} \mathcal{O} \rra - \lla t_{\text{loc}}^{\mu_1 \nu_1} T^{\mu_2 \nu_2} \mathcal{O} \rra + \lla t_{\text{loc}}^{\mu_1 \nu_1} t_{\text{loc}}^{\mu_2 \nu_2} \mathcal{O} \rra \label{e:to_cwi}
\end{align}
All terms on the right-hand side apart from the first may be computed by means of Ward identities. The exact form of the Ward identities depends on the exact definition of the operators involved, but more importantly, all these terms depend on 2-point functions only.

Due to the complicated nature of contractions of the projectors \eqref{e:pi} and \eqref{e:Pi} one might fear that it is very difficult to calculate the form factors by means of Feynman rules, given some particular QFT.  Reassuringly, this is not the case, as we can see in the following example. First, we decompose the full 3-point function $\lla T^{\mu_1 \nu_1} T^{\mu_2 \nu_2} \mathcal{O} \rra$ into simple tensors using the choice of momenta \eqref{e:momenta_choice} and denote
\begin{equation} \label{e:TTOdecompab}
\lla T^{\alpha_1 \beta_1} T^{\alpha_2 \beta_2} \mathcal{O} \rra = \tilde{A}_1 p_2^{\alpha_1} p_2^{\beta_1} p_3^{\alpha_2} p_3^{\beta_2} + \tilde{A}_2 p_2^{\alpha_1} p_3^{\alpha_2} \delta^{\beta_1 \beta_2} + \tilde{A}_3 \delta^{\alpha_1 \alpha_2} \delta^{\beta_1 \beta_2} + \ldots,
\end{equation}
where the omitted terms do not contain the tensors we have listed explicitly. Next, we apply the projectors \eqref{e:Pi}. Obverse, for example, that the projector $\Pi^{\mu_1 \nu_1}_{\alpha_1 \beta_1}(\bs{p}_1)$ is constructed from the metric and the momentum $\bs{p}_1$ only, and therefore when applied to the 3-point function it cannot change the coefficient of any tensor containing $p_2^{\alpha_1} p_2^{\beta_1}$, \textit{i.e.},
\begin{equation}
\Pi^{\mu_1 \nu_1}_{\alpha_1 \beta_1}(\bs{p}_1) p_2^{\alpha_1} p_2^{\beta_1} = p_2^{\mu_1} p_2^{\nu_1} + \ldots,
\end{equation}
where the omitted terms do not contain $p_2^{\mu_1} p_2^{\nu_1}$. Using the same argument for $\Pi^{\mu_2 \nu_2}_{\alpha_2 \beta_2}(\bs{p}_2)$, we see that the coefficients of $p_2^{\alpha_1} p_2^{\beta_1} p_3^{\alpha_2} p_3^{\beta_2}$ in \eqref{e:TTOdecompab} and $p_2^{\mu_1} p_2^{\nu_1} p_3^{\mu_2} p_3^{\nu_2}$ in $\lla t^{\mu_1 \nu_1}(\bs{p}_1) t^{\mu_2 \nu_2}(\bs{p}_2) \mathcal{O}(\bs{p}_3) \rra$ in \eqref{e:TOOdef} are equal, \textit{i.e.}, $A_1 = \tilde{A}_1$. Similarly, we find that
\begin{align}
& \Pi^{\mu_1 \nu_1}_{\alpha_1 \beta_1}(\bs{p}_1) \Pi^{\mu_2 \nu_2}_{\alpha_2 \beta_2}(\bs{p}_2) \lla T^{\alpha_1 \beta_1}(\bs{p}_1) T^{\alpha_2 \beta_2}(\bs{p}_2) \mathcal{O}(\bs{p}_3) \rra = \nn\\
& \qquad = \tilde{A}_1 p_2^{\mu_1} p_2^{\nu_1} p_3^{\mu_2} p_3^{\nu_2} + \frac{1}{4} \tilde{A}_2 p_2^{\mu_1} p_3^{\mu_2} \delta^{\nu_1 \nu_2} + \frac{1}{2} \tilde{A}_3 \delta^{\mu_1 \mu_2} \delta^{\nu_1 \nu_2} + \ldots,
\end{align}
where the omitted terms do not contain the tensors we have listed explicitly. We therefore have
\begin{align}
A_1 & = \text{coefficient of } p_2^{\mu_1} p_2^{\nu_1} p_3^{\mu_2} p_3^{\nu_2} \text{ in } \lla T^{\mu_1 \nu_1}(\bs{p}_1) T^{\mu_2 \nu_2}(\bs{p}_2) \mathcal{O}(\bs{p}_3) \rra, \label{e:toA1} \\
A_2 & = 4 \cdot \text{coefficient of } p_2^{\mu_1} p_3^{\mu_2} \delta^{\nu_1 \nu_2} \text{ in } \lla T^{\mu_1 \nu_1}(\bs{p}_1) T^{\mu_2 \nu_2}(\bs{p}_2) \mathcal{O}(\bs{p}_3) \rra, \label{e:toA2}\\
A_3 & = 2 \cdot \text{coefficient of } \delta^{\mu_1 \mu_2} \delta^{\nu_1 \nu_2} \text{ in } \lla T^{\mu_1 \nu_1}(\bs{p}_1) T^{\mu_2 \nu_2}(\bs{p}_2) \mathcal{O}(\bs{p}_3) \rra. \label{e:toA3}
\end{align}
We list the analogous formulae for all other 3-point functions in the second part of the paper.

\subsection{Example} \label{ch:example_calc}

Let us consider a conformally coupled free scalar free massless field $\phi$ in $d$ Euclidean dimensions.
In the presence of a non-trivial source $g^{\mu \nu}$ for the stress-energy tensor, the action reads
\begin{equation}
S = \int \D^d x \: \sqrt{g} \left[ \frac{1}{2} g^{\mu \nu} \partial_\mu \phi \partial_\nu \phi + \frac{d-2}{8(d-1)} R \phi^2 \right],
\end{equation}
where $R$ is the Ricci scalar of $g_{\mu \nu}$. The stress-energy tensor in the presence of the sources is then
\begin{align}
T_{\mu \nu} & = \frac{2}{\sqrt{g}} \frac{\delta S}{\delta g^{\mu \nu}}
= \partial_\mu \phi \partial_\nu \phi - \frac{1}{2} g_{\mu \nu} \partial_\alpha \phi \partial^\alpha \phi \nn\\
& \qquad + \frac{d-2}{4(d-1)} \left( g_{\mu \nu} \nabla^2 - \nabla_\mu \nabla_\nu + R_{\mu \nu} - \frac{1}{2} g_{\mu \nu} R \right) \phi^2. \label{e:exampleT}
\end{align}
In this CFT, $\mathcal{O}(x) = \phi^2(x)$ is a conformal primary operator of dimension $\Delta_3 = d - 2$.

For later use we quote the result for the form factors
of $\lla T^{\mu_1 \nu_1} T^{\mu_2 \nu_2} \mathcal{O} \rra$ in this theory.
Writing down the expression for $\lla T^{\mu_1 \nu_1} T^{\mu_2 \nu_2} \mathcal{O} \rra$ using the regular Feynman rules, from (\ref{e:toA1}, \ref{e:toA2}, \ref{e:toA3}) we may then read off expressions for the form factors.  Explicitly evaluating these integrals for the case $d=3$, we find
\begin{align} \label{e:exA1}
A_1 & = \frac{3 (p_1^2 + p_2^2) + p_3^2 + 12 p_1 p_2 + 4 p_3 (p_1 + p_2) }{48 \: p_3 (p_1 + p_2 + p_3)^4}, \\
A_2 & = \frac{2 (p_1^2 + p_2^2) + p_3^2 + 6 p_1 p_2 + 3 p_3 (p_1 + p_2) }{12 \: (p_1 + p_2 + p_3)^3}, \label{e:exA2}\\
A_3 & = - \frac{2 (p_1^2 p_2 + p_1 p_2^2 + p_1 p_3^2 + p_3 p_1^2 + p_2 p_3^2 + p_3 p_2^2) + 2 p_1 p_2 p_3 + p_1^3 + p_2^3 + p_3^3}{24 \: (p_1 + p_2 + p_3)^2}, \label{e:exA3}
\end{align}
in agreeement with the direct evaluation of this correlator given in \cite{Bzowski:2011ab}.

\section{Conformal Ward identities in momentum space} \label{sec:CWIs}

In section \ref{ch:scalar_CWIs} we wrote down the Ward identities associated with dilatations and special conformal transformation for the case of correlators involving three scalars.  In this section, we discuss the corresponding Ward identities for 3-point correlators involving insertions of the stress-energy tensor and conserved currents.
First, in section \ref{ch:to_momentum}, we obtain the dilatation and special conformal Ward identities in momentum space by Fourier transforming the well-known position space expressions; in sections \ref{ch:dilatation} and \ref{ch:special} we then reduce these identities to a set of simple scalar equations using the tensor decomposition introduced in section \ref{ch:Tensor_structure}. Finally, in sections \ref{ch:transverseWI} and \ref{ch:trace}, we write down the transverse and trace Ward identities.

\subsection{From position space to momentum space} \label{ch:to_momentum}

Let $\mathcal{T}_1, \mathcal{T}_2, \ldots, \mathcal{T}_n$ represent quantum operators  of dimensions $\Delta_1, \Delta_2, \ldots, \Delta_n$ and of arbitrary Lorentz structure in some CFT. The dilatation Ward identity in position space is especially simple and reads \cite{DiFrancesco:1997nk}
\begin{equation} \label{e:ward_dil_x}
0 = \left[ \sum_{j=1}^n \Delta_j + \sum_{j=1}^n x_j^\alpha \frac{\partial}{\partial x_j^\alpha} \right] \langle \mathcal{T}_1(\bs{x}_1) \ldots \mathcal{T}_n(\bs{x}_n) \rangle.
\end{equation}
The Ward identity associated with special conformal transformations for the $n$-point function of scalar operators $\mathcal{O}_1, \mathcal{O}_2, \ldots, \mathcal{O}_n$ is
\begin{equation} \label{e:ward_cwi_x0}
0 = \left[ \sum_{j=1}^n \left( 2 \Delta_j x_j^\kappa + 2 x_j^\kappa x_j^\alpha \frac{\partial}{\partial x_j^\alpha} - x_j^2 \frac{\partial}{\partial x_{j \kappa}} \right) \right] \langle \mathcal{O}_1(\bs{x}_1) \ldots \mathcal{O}_n(\bs{x}_n) \rangle,
\end{equation}
where $\kappa$ is a free Lorentz index. For general tensors $\mathcal{T}_j$ one needs to add an additional term to the equation. This term depends on the Lorentz structure, and to write it down, we assume that the tensor $\mathcal{T}_j$ has $r_j$ Lorentz indices, \textit{i.e.}, $\mathcal{T}_j = \mathcal{T}_j^{\mu_{j1} \ldots \mu_{j r_j}}$, for $j = 1, 2, \ldots, n$. In this case, the contribution
\begin{align} \label{e:ward_cwi_x1}
& 2 \sum_{j=1}^n \sum_{k=1}^{r_j} \left[ (x_j)_{\alpha_{jk}} \delta^{\kappa \mu_{jk}} - x_j^{\mu_{jk}} \delta_{\alpha_{jk}}^\kappa \right] \times \nn\\
& \qquad \times \langle \mathcal{T}_1^{\mu_{11} \ldots \mu_{1 n_1}}(\bs{x}_1) \ldots \mathcal{T}_j^{\mu_{j1} \ldots \alpha_{jk} \ldots \mu_{j r_j}}(\bs{x}_j) \ldots \mathcal{T}_n^{\mu_{n1} \ldots \mu_{n r_n}}(\bs{x}_n) \rangle
\end{align}
must then be added to the right-hand side of \eqref{e:ward_cwi_x0}.

Expressions (\ref{e:ward_dil_x}, \ref{e:ward_cwi_x0}, \ref{e:ward_cwi_x1}) may be Fourier transformed in a similar manner to that discussed in \cite{Mehen:1999}. Due to the translation invariance the position space correlators depend only on the differences $\bs{x}_j - \bs{x}_n$.  Therefore, we can set $\bs{x}_n = 0$ and take
\begin{equation}\label{p3eqn}
\bs{p}_n = - \sum_{j=1}^{n-1} \bs{p}_j.
\end{equation}
The Ward identities \eqref{e:ward_dil_x} and \eqref{e:ward_cwi_x0} then transform to
\begin{align}
0 & = \left[ \sum_{j=1}^n \Delta_j - (n - 1) d - \sum_{j=1}^{n-1} p_j^\alpha \frac{\partial}{\partial p_j^\alpha} \right] \lla \mathcal{T}_1(\bs{p}_1) \ldots \mathcal{T}_n(\bs{p}_n) \rra, \label{e:ward_dil} \\
0 & = \left[ \sum_{j=1}^{n-1} \left( 2 (\Delta_j - d) \frac{\partial}{\partial p_j^\kappa} - 2 p_j^\alpha \frac{\partial}{\partial p_j^\alpha} \frac{\partial}{\partial p_j^\kappa} + (p_j)_\kappa \frac{\partial}{\partial p_j^\alpha} \frac{\partial}{\partial p_{j \alpha}} \right) \right] \lla \mathcal{O}_1(\bs{p}_1) \ldots \mathcal{O}_n(\bs{p}_n) \rra, \label{e:ward_cwi0}
\end{align}
while the additional contribution \eqref{e:ward_cwi_x1} transforms to
\begin{align} \label{e:ward_cwi1}
& 2 \sum_{j=1}^{n-1} \sum_{k=1}^{n_j} \left( \delta^{\mu_{jk} \kappa} \frac{\partial}{\partial p_j^{\alpha_{jk}}} - \delta^\kappa_{\alpha_{jk}} \frac{\partial}{\partial p_{j \mu_{jk}}} \right) \times \nn\\
& \qquad \times \lla \mathcal{T}_1^{\mu_{11} \ldots \mu_{1 r_1}}(\bs{p}_1) \ldots \mathcal{T}_j^{\mu_{j1} \ldots \alpha_{jk} \ldots \mu_{j r_j}}(\bs{p}_j) \ldots \mathcal{T}_n^{\mu_{n1} \ldots \mu_{n r_n}}(\bs{p}_n), \rra
\end{align}
and once again must be added to the right-hand side of \eqref{e:ward_cwi0}.
It will be useful to denote the differential operator obtained by summing the right-hand side of \eqref{e:ward_cwi0} and \eqref{e:ward_cwi1} as $\mathcal{K}^\kappa$, so that the CWIs may be compactly expressed as
\begin{equation} \label{e:cwiK}
\mathcal{K}^\kappa \lla \mathcal{T}_1(\bs{p}_1) \ldots \mathcal{T}_n(\bs{p}_n) \rra = 0.
\end{equation}
In view of \eqref{e:ward_cwi1}, note that $\mathcal{K}^\kappa$ acts non-trivially on Lorentz indices and so in fact is really of the form
\begin{equation}
\mathcal{K}^\kappa = \mathcal{K}^{\mu_{11} \ldots \mu_{n r_n}, \kappa}_{\alpha_{11} \ldots \alpha_{n r_n}},
\end{equation}
however for simplicity we will omit the tensor indices on $\mathcal{K}^\kappa$.

In the following analysis we will focus specifically on 3-point functions. The idea will be to take the tensor decomposition the 3-point function described in section \ref{ch:Tensor_structure}, then apply the operators \eqref{e:ward_cwi0} and \eqref{e:ward_cwi1} yielding differential equations for the form factors. Since by Lorentz invariance the form factors are purely functions of the momentum magnitudes, the action of momentum derivatives on form factors may be obtained using the chain rule,
\begin{align}
\frac{\partial}{\partial p_{1 \mu}} & = \frac{\partial p_1}{\partial p_{1 \mu}} \frac{\partial}{\partial p_1} + \frac{\partial p_2}{\partial p_{1 \mu}} \frac{\partial}{\partial p_2} + \frac{\partial p_3}{\partial p_{1 \mu}} \frac{\partial}{\partial p_3} \nn\\
& = \frac{p_1^\mu}{p_1} \frac{\partial}{\partial p_1} + \frac{p_1^\mu + p_2^\mu}{p_3} \frac{\partial}{\partial p_3}, \label{e:pdiff}
\end{align}
noting that $\bs{p}_3$ is fixed via (\ref{p3eqn}).  We may express derivatives with respect to $\bs{p}_2$ similarly, and the final results may then be re-expressed purely in terms of the momentum magnitudes.

\subsection{Dilatation Ward identity} \label{ch:dilatation}

Using \eqref{e:pdiff}, it is simple to rewrite the dilatation Ward identity \eqref{e:ward_dil} for any 3-point function $\lla \mathcal{T}_1 \mathcal{T}_2 \mathcal{T}_3 \rra$ in terms of its form factors as
\begin{equation} \label{e:dilA}
0 = \left[ 2d + N_n + \sum_{j=1}^{3} \left( p_j \frac{\partial}{\partial p_j} - \Delta_j \right) \right] A_n(p_1, p_2, p_3),
\end{equation}
where $N_n$ is the tensorial dimension of $A_n$, \textit{i.e.}, the number of momenta in the tensor multiplying the form factor $A_n$ and the transverse-traceless projectors. As previously, $\Delta_j$, $j=1,2,3$ denote the conformal dimensions of the operators $\mathcal{T}_j$ in the 3-point function: for a conserved current we thus have $\Delta = d - 1$ while for a stress-energy tensor $\Delta = d$.

The dilatation Ward identity determines the total degree of the 3-point function and hence of its form factors. In general, if a function $A$ satisfies
\begin{equation} \label{e:dilD}
0 = \left[ - D + \sum_{j=1}^{3} p_j \frac{\partial}{\partial p_j} \right] A(p_1, p_2, p_3)
\end{equation}
for some constant $D$ then we will refer to $D$ as the \emph{degree} of $A$, denoted $\deg(A) = D$. (A homogeneous polynomial in momenta of degree $D$ has dilatation degree $D$.) Therefore \eqref{e:dilA} implies that the form factor $A_n$ has degree
\begin{equation} \label{e:degree}
\deg (A_n) = \Delta_t - 2d - N_n,
\end{equation}
where $\Delta_t = \Delta_1 + \Delta_2 + \Delta_3$.

\subsection{Special conformal Ward identities} \label{ch:special}

In this section, we now extract scalar equations for the form factors by inserting our tensor decomposition for the transverse-traceless part of the 3-point functions into the special conformal Ward identities. While the details are somewhat involved, the procedure is nonetheless conceptually straightforward.
We will outline the method using as an example the 3-point function $\lla T^{\mu_1 \nu_1} T^{\mu_2 \nu_2} \mathcal{O} \rra$, which captures all the important features.

Consider then the action of the CWI operator $\mathcal{K}^\kappa$ defined in \eqref{e:cwiK} on the decomposition \eqref{e:to_cwi},
\begin{align}
0 & = \mathcal{K}^{\kappa} \lla T^{\mu_1 \nu_1} T^{\mu_2 \nu_2} \mathcal{O} \rra \nn \\
& = \mathcal{K}^{\kappa} \lla t^{\mu_1 \nu_1} t^{\mu_2 \nu_2} \mathcal{O} \rra + \mathcal{K}^{\kappa} \lla t^{\mu_1 \nu_1} t_{\text{loc}}^{\mu_2 \nu_2} \mathcal{O} \rra
+ \mathcal{K}^{\kappa} \lla t_{\text{loc}}^{\mu_1 \nu_1} t^{\mu_2 \nu_2} \mathcal{O} \rra + \mathcal{K}^{\kappa} \lla t_{\text{loc}}^{\mu_1 \nu_1} t_{\text{loc}}^{\mu_2 \nu_2} \mathcal{O} \rra, \label{e:to_cwi2}
\end{align}
recalling that our notation for $\mathcal{K}^\kappa$ suppresses Lorentz indices so that in reality, \textit{e.g.},
\begin{equation}
\mathcal{K}^{\kappa} \lla t^{\mu_1 \nu_1} t^{\mu_2 \nu_2} \mathcal{O} \rra = \mathcal{K}^{\mu_1 \nu_1 \mu_2 \nu_2, \kappa}_{\alpha_1 \beta_1 \alpha_2 \beta_2} \lla t^{\alpha_1 \beta_1} t^{\alpha_2 \beta_2} \mathcal{O} \rra.
\end{equation}
Through a direct but lengthy calculation we find that the first term on the right-hand side of \eqref{e:to_cwi2}, $\mathcal{K}^{\kappa} \lla t^{\mu_1 \nu_1} t^{\mu_2 \nu_2} \mathcal{O} \rra$, is transverse-traceless in $\mu_1, \nu_1$ and $\mu_2, \nu_2$ with respect to the corresponding momenta,
\begin{align}
& \delta_{\mu_1 \nu_1} \left[ \mathcal{K}^\kappa \lla t^{\mu_1 \nu_1}(\bs{p}_1) t^{\mu_2 \nu_2}(\bs{p}_2) \mathcal{O}(\bs{p}_3) \rra \right] = \delta_{\mu_2 \nu_2} \left[ \mathcal{K}^\kappa \lla t^{\mu_1 \nu_1}(\bs{p}_1) t^{\mu_2 \nu_2}(\bs{p}_2) \mathcal{O}(\bs{p}_3) \rra \right] = 0, \\
& p_{1 \mu_1} \left[ \mathcal{K}^\kappa \lla t^{\mu_1 \nu_1}(\bs{p}_1) t^{\mu_2 \nu_2}(\bs{p}_2) \mathcal{O}(\bs{p}_3) \rra \right] = p_{2 \mu_2} \left[ \mathcal{K}^\kappa \lla t^{\mu_1 \nu_1}(\bs{p}_1) t^{\mu_2 \nu_2}(\bs{p}_2) \mathcal{O}(\bs{p}_3) \rra \right] = 0,\label{label1}
\end{align}
where we used the definitions \eqref{e:ward_cwi0} and \eqref{e:ward_cwi1} for $\mathcal{K}^\kappa$ and the identities given in appendix \ref{ch:identities}. For correlators involving conserved currents, we find that the analogue of (\ref{label1}) similarly applies.

We are now free to apply transverse-traceless projectors \eqref{e:pi} and \eqref{e:Pi} to \eqref{e:to_cwi2}, in order to isolate equations for the form factors appearing in the decomposition of $\lla t^{\mu_1 \nu_1} t^{\mu_2 \nu_2} \mathcal{O} \rra$.
Evaluating the action of $\mathcal{K}^\kappa$ on the semi-local terms in \eqref{e:to_cwi2} via the formulae in appendix \ref{ch:identities}, we find
\begin{align} \label{e:loc}
\pi^\mu_\alpha \mathcal{K}^\kappa j_{\text{loc}}^\alpha & = \frac{2 (d - 2)}{p^2} \pi^{\kappa \mu} r, \\
\Pi^{\mu \nu}_{\alpha \beta} \mathcal{K}^\kappa t_{\text{loc}}^{\alpha \beta} & = \frac{4 d}{p^2} \Pi^{\mu \nu \kappa}_{\ \ \ \ \alpha} R^\alpha, \\
\pi^{\mu}_{\alpha} j^{\alpha}_{\text{loc}} = \Pi^{\mu \nu}_{\alpha \beta} t^{\alpha \beta}_{\text{loc}} & = 0. \label{e:pij0}
\end{align}
The last equation implies that any correlation function with more than one insertion of $t_{\text{loc}}^{\mu \nu}$ or $j_{\text{loc}}^{\mu}$ vanishes when the CWI operator $\mathcal{K}^\kappa$ and the projectors \eqref{e:pi} and \eqref{e:Pi} are applied. This is because the CWI operator $\mathcal{K}^\kappa$ can be written as a sum of two terms
\begin{equation}
\mathcal{K}^\kappa = \mathcal{K}_1^\kappa \left( \frac{\partial}{\partial p_1^\mu} \right) + \mathcal{K}_2^\kappa \left( \frac{\partial}{\partial p_2^\mu} \right),
\end{equation}
each depending only on derivatives with respect to the appropriate momenta, hence
\begin{equation}
\Pi^{\mu_1 \nu_1}_{\alpha_1 \beta_1}(\bs{p}_1) \Pi^{\mu_2 \nu_2}_{\alpha_2 \beta_2}(\bs{p}_2) \mathcal{K}^\kappa \lla t_{\text{loc}}^{\alpha_1 \beta_1}(\bs{p}_1) t_{\text{loc}}^{\alpha_2 \beta_2}(\bs{p}_2) \mathcal{O}(\bs{p}_3) \rra = 0.
\end{equation}

Substituting all results into \eqref{e:to_cwi2}, we find
\begin{align}
0 & = \Pi^{\mu_1 \nu_1}_{\alpha_1 \beta_1}(\bs{p}_1) \Pi^{\mu_2 \nu_2}_{\alpha_2 \beta_2}(\bs{p}_2) \mathcal{K}^\kappa \lla t^{\alpha_1 \beta_1}(\bs{p}_1) t^{\alpha_2 \beta_2}(\bs{p}_2) \mathcal{O}(\bs{p}_3) \rra \nn\\
& \qquad + \: \frac{4 d}{p_1^2} \Pi^{\mu_1 \nu_1 \kappa}_{\ \ \ \ \ \ \alpha_1}(\bs{p}_1) \left[ p_{1 \beta_1} \lla T^{\alpha_1 \beta_1}(\bs{p}_1) t^{\mu_2 \nu_2}(\bs{p}_2) \mathcal{O}(\bs{p}_3) \rra \right] \nn\\
& \qquad + \: \frac{4 d}{p_2^2} \Pi^{\mu_2 \nu_2 \kappa}_{\ \ \ \ \ \ \alpha_2}(\bs{p}_2) \left[ p_{2 \beta_2} \lla t^{\mu_1 \nu_1}(\bs{p}_1) T^{\alpha_2 \beta_2}(\bs{p}_2) \mathcal{O}(\bs{p}_3) \rra \right]. \label{e:semilocal_contribution}
\end{align}
Two last terms are semi-local and may be re-expressed in terms of 2-point functions via the transverse Ward identities. The remaining task is then to rewrite the first term of \eqref{e:semilocal_contribution} in terms of form factors and extract the CWIs. Via the method of section \ref{ch:Tensor_structure}, we can write the most general form of the result as
\begin{align}
0 & = \Pi_{\alpha_1 \beta_1}^{\mu_1 \nu_1}(\bs{p}_1) \Pi_{\alpha_2 \beta_2}^{\mu_2 \nu_2}(\bs{p}_2) \mathcal{K}^\kappa \lla T^{\alpha_1 \beta_1}(\bs{p}_1) T^{\alpha_2  \beta_2}(\bs{p}_2) \mathcal{O}(\bs{p}_3) \rra = \Pi_{\alpha_1 \beta_1}^{\mu_1 \nu_1}(\bs{p}_1) \Pi_{\alpha_2 \beta_2}^{\mu_2 \nu_2}(\bs{p}_2) \times \nn\\
& \qquad \times \left[ p^\kappa_1 \left( C_{11} p_2^{\alpha_1} p_2^{\beta_1} p_3^{\alpha_2} p_3^{\beta_2} + C_{12} p_2^{\alpha_1} p_3^{\alpha_2} \delta^{\beta_1 \beta_2} + C_{13} \delta^{\alpha_1 \alpha_2} \delta^{\beta_1 \beta_2} \right) \right. \nn\\
& \qquad \qquad + \: p^\kappa_2 \left( C_{21} p_2^{\alpha_1} p_2^{\beta_1} p_3^{\alpha_2} p_3^{\beta_2} + C_{22} p_2^{\alpha_1} p_3^{\alpha_2} \delta^{\beta_1 \beta_2} + C_{23} \delta^{\alpha_1 \alpha_2} \delta^{\beta_1 \beta_2} \right) \nn\\
& \qquad \qquad + \: \left( C_{31} \delta^{\kappa \alpha_1} p_2^{\beta_1} p_3^{\alpha_2} p_3^{\beta_2} + C_{32} \delta^{\kappa \alpha_1} \delta^{\alpha_2 \beta_1} p_3^{\beta_2} \right. \nn\\
& \left. \left. \qquad \qquad \qquad + \: C_{41} \delta^{\kappa \alpha_2} p_2^{\alpha_1} p_2^{\beta_1} p_3^{\beta_2} + C_{42} \delta^{\kappa \alpha_2} \delta^{\alpha_1 \beta_2} p_2^{\beta_1} \right) \right]. \label{e:TTOcwis}
\end{align}
In this expression, the coefficients $C_{jk}$ are differential equations involving the form factors $A_1$, $A_2$ and $A_3$ of \eqref{e:TTOform}. Each CWI can then be presented in terms of the momentum magnitudes $p_j = |\bs{p}_j|$.

As we can see, there are ten coefficients $C_{jk}$ in \eqref{e:TTOcwis}, so there are at most ten equations to consider. Usually not all of the CWIs, however, are independent. In this example, the $\bs{p}_1 \leftrightarrow \bs{p}_2$ symmetry implies that the equations following from $C_{1j}$ and $C_{2j}$, as well as $C_{3j}$ and $C_{4j}$, are pairwise equivalent.

For any 3-point function, the resulting equations can be divided into two groups: the \emph{primary} and the \emph{secondary} CWIs. The primary CWIs are second-order differential equations and appear as the coefficients of transverse or transverse-traceless tensors containing $p_1^\kappa$ or $p_2^\kappa$, where $\kappa$ is the `special' index in the CWI operator $\mathcal{K}^\kappa$. In the expression \eqref{e:TTOcwis} above, the primary CWIs are equivalent to the vanishing of the coefficients $C_{1j}$ and $C_{2j}$. The remaining equations, following from all other transverse or transverse-traceless terms, are then the secondary CWIs and are first-order differential equations. In the expression \eqref{e:TTOcwis}, the secondary CWIs are equivalent to the vanishing of the coefficients $C_{3j}$ and $C_{4j}$.

In the next two subsections we will examine the general form of the primary and secondary CWIs and discuss some of their properties.
In section \ref{section:solnofCWIs}, we will return to analyse their solution for the form factors.   In outline our strategy will be, first, to solve each of the primary CWIs up to an overall multiplicative constant, then second, to insert these solutions into the secondary CWIs typically allowing the number of undetermined constants to be further reduced. In the case of the correlator $\lla T^{\mu_1 \nu_1} T^{\mu_2 \nu_2} \mathcal{O} \rra$, for example, we will find that the final result is then uniquely determined up to one numerical constant, in agreement with the position space analysis of \cite{Osborn:1993cr}.

\subsubsection{Primary conformal Ward identities} \label{ch:primaryCWIs}

It turns out that in all cases the primary CWIs look very similar to the CWIs \eqref{e:Oward_sp_scal} for scalar operators. In order to write the primary CWIs in a simple way, we define the following fundamental differential operators
\begin{align}
\K_j & = \frac{\partial^2}{\partial p_j^2} + \frac{d + 1 - 2 \Delta_j}{p_j} \frac{\partial}{\partial p_j}, \quad j = 1, 2, 3, \label{e:K} \\
\K_{ij} & = \K_i - \K_j, \quad j = 1, 2, 3, \label{e:KK}
\end{align}
where $\Delta_j$ is the conformal dimension of the $j$-th operator in the 3-point function under consideration. (Observe that this same operator appeared earlier in (\ref{e:Oward_sp_scal}, \ref{e:Oward_sp_scal2}).)

In the case of our example $\lla T^{\mu_1 \nu_1} T^{\mu_2 \nu_2} \mathcal{O} \rra$, the primary CWIs for the form factors defined in \eqref{e:TTOform} read
\begin{equation}
\begin{array}{lll}
\K_{13} A_1 = 0, & \qquad \K_{13} A_2 = 8 A_1, & \qquad \K_{13} A_3 = 2 A_2, \\
\K_{23} A_1 = 0, & \qquad \K_{23} A_2 = 8 A_1, & \qquad \K_{23} A_3 = 2 A_2,
\end{array}
\label{e:preCWIsTTO}
\end{equation}
Note that, from the definition \eqref{e:KK}, we have
\begin{equation}
\K_{ii} = 0, \qquad \K_{ji} = - \K_{ij}, \qquad \K_{ij} + \K_{jk} = \K_{ik},
\end{equation}
for any $i,j,k \in \{1,2,3\}$.  One can therefore subtract corresponding pairs of equations and obtain the following system of independent partial differential equations
\begin{equation}
\begin{array}{lll}
\K_{12} A_1 = 0, & \qquad \K_{12} A_2 = 0, & \qquad \K_{12} A_3 = 0, \\
\K_{13} A_1 = 0, & \qquad \K_{13} A_2 = 8 A_1, & \qquad \K_{13} A_3 = 2 A_2. \\
\end{array}
\label{e:CWIsTTO}
\end{equation}
As we will prove, each equation has a unique solution up to one numerical constant. This means that at this point the 3-point function $\lla T^{\mu_1 \nu_1} T^{\mu_2 \nu_2} \mathcal{O} \rra$ is determined by three numerical constants. After the application of the secondary CWIs this number will decrease further.

The primary CWIs for all 3-point functions are listed explicitly in the second part of the paper. They share the following properties:
\begin{enumerate}

\item All primary CWIs are second-order linear differential equations.

\item The equations for the coefficient $A_1$ are always homogeneous and given by (\ref{e:Oward_sp_scal}, \ref{e:Oward_sp_scal2}) exactly, \textit{i.e.},
\begin{equation} \label{e:A1eq}
\K_{12} A_1 = 0, \qquad\qquad \K_{13} A_1 = 0.
\end{equation}

\item The equations for the remaining form factors are similar to (\ref{e:Oward_sp_scal}, \ref{e:Oward_sp_scal2}), but they may contain a linear inhomogeneous part. For a form factor $A_n$ multiplying a tensor of tensorial dimension $N_n$, the only form factors $A_j$ which can appear in the inhomogeneous part are those with $N_j = N_n + 2$. It is therefore always possible to solve the primary CWIs recursively, starting with $A_1$.

In the case of our example, the recursive structure of the equations \eqref{e:CWIsTTO} is clearly visible.

\item There is no semi-local contribution to any primary CWI. In our example, last two terms in \eqref{e:semilocal_contribution} do not contribute to the primary CWIs. This conclusion is valid in general and can be checked explicitly in all cases.

\item The solution to each pair of primary CWIs is unique up to one numerical constant, as we will prove in section \ref{section:solnofCWIs}.
\end{enumerate}

\subsubsection{Secondary conformal Ward identities} \label{ch:secondaryCWIs}

The secondary CWIs are first-order partial differential equations and in principle involve the semi-local information contained in $j^\mu_{\text{loc}}$ and $t^{\mu \nu}_{\text{loc}}$.
In order to write them compactly, we define the two differential operators
\begin{align}
\Lo_{N} & = p_1 (p_1^2 + p_2^2 - p_3^2) \frac{\partial}{\partial p_1} + 2 p_1^2 p_2 \frac{\partial}{\partial p_2} \nn\\
& + \: \left[ (2d - \Delta_1 - 2 \Delta_2 + N) p_1^2 + (2 \Delta_1 - d) (p_3^2 - p_2^2) \right], \label{e:L}\\
\Ro & = p_1 \frac{\partial}{\partial p_1} - (2 \Delta_1 - d), \label{e:R}
\end{align}
as well as their symmetric versions
\begin{align}
\Lo'_{N} & = \Lo_{N} \text{ with } p_1 \leftrightarrow p_2 \text{ and } \Delta_1 \leftrightarrow \Delta_2, \label{e:Lp} \\
\Ro' & = \Ro \text{ with } p_1 \mapsto p_2 \text{ and } \Delta_1 \mapsto \Delta_2. \label{e:Rp}
\end{align}
These operators depend on the conformal dimensions of the operators involved in the 3-point function under consideration, and additionally on a single parameter $N$ determined by the Ward identity in question.

In our example \eqref{e:TTOform} one finds two independent secondary CWIs following from the coefficients $C_{31}$ and $C_{32}$ in \eqref{e:TTOcwis}, namely
\begin{align}
& \Lo_{2} A_1 + \Ro A_2 = 2 d \cdot \text{coeff. of } p_2^{\mu_1} p_3^{\mu_2} p_3^{\nu_2} \text{ in } p_{1 \nu_1} \lla T^{\mu_1 \nu_1}(\bs{p}_1) T^{\mu_2 \nu_2}(\bs{p}_2) \mathcal{O}(\bs{p}_3) \rra, \label{e:secondaryCWIs1}\\
& \Lo_{2} A_2 + 4 \Ro A_3 = 8 d \cdot \text{coeff. of } \delta^{\mu_1 \mu_2} p_3^{\mu_2} \text{ in } p_{1 \nu_1} \lla T^{\mu_1 \nu_1}(\bs{p}_1) T^{\mu_2 \nu_2}(\bs{p}_2) \mathcal{O}(\bs{p}_3) \rra, \label{e:secondaryCWIs2}
\end{align}
with $\Delta_1 = \Delta_2 = d$. Note that in order to correctly extract the coefficient of a tensor, the rule \eqref{e:momenta_choice} regarding the momenta associated with a given Lorentz index must be observed. The semi-local terms on the right-hand sides may be computed by means of transverse Ward identities, to which we now turn our attention.

\subsection{Transverse Ward identities} \label{ch:transverseWI}

In this section we review briefly the transverse (or diffeomorphism) Ward identities in momentum space.
These Ward identities arise from the conservation law for currents. In particular we will need the precise form of all semi-local terms that appear in these Ward identities since these terms are required for the explicit evaluation of the right-hand sides of the secondary CWIs such as (\ref{e:secondaryCWIs1}, \ref{e:secondaryCWIs2}).

We assume the CFT contains the following data:
\begin{itemize}
\item A symmetry group $G$. The conserved current $J^{\mu a}$, $a = 1, \ldots, \dim G$, is then the Noether current associated with the symmetry and is sourced by a potential $A_\mu^a$.
\item Scalar primary operators $\mathcal{O}^I$ all of the same dimension $\Delta$, sourced by $\phi_0^I$.
\item A stress-energy tensor $T_{\mu \nu}$ sourced by the metric $g^{\mu \nu}$.
\end{itemize}
Under a symmetry transformation with parameter $\alpha^a$ the sources transform as
\begin{align}\label{e:gtrules}
\delta g^{\mu \nu} & = 0, \\
\delta A^a_\mu & = - D^{ac}_\mu \alpha^c = - \partial_\mu \alpha^a - f^{abc} A_\mu^b \alpha^c, \\
\delta \phi^I_0 & = - \I \alpha^a (T^a_R)^{IJ} \phi^J_0,
\end{align}
where $T_R^a$ are matrices of a representation $R$ and $f^{abc}$ are structure constants of the group $G$. The gauge field transforms in the adjoint representation while the $\phi^I$ may transform in any representation $R$. The covariant derivative is
\begin{equation}\label{e:covderivdef}
D^{IJ}_\mu = \delta^{IJ} \partial_\mu - \I A_\mu^a (T^a_R)^{IJ}.
\end{equation}

Similarly, under a diffeomorphism $\xi^\mu$ the sources transform as
\begin{align}
\delta g^{\mu \nu} & = - ( \nabla^\mu \xi^\nu + \nabla^\nu \xi^\mu ), \\
\delta A^a_\mu & = \xi^\nu \nabla_\nu A_\mu^a + \nabla_\mu \xi^\nu \cdot A_\nu^a, \\
\delta \phi^I_0 & = \xi^\nu \partial_\nu \phi^I_0,
\end{align}
where $\nabla$ is a Levi-Civita connection.

From the generating functional
\begin{equation} \label{e:Z}
Z[\phi_0^I, A_\mu^a, g^{\mu \nu}] = \int \mathcal{D} \Phi \: \exp \left( - S[A_\mu^a, g^{\mu \nu}] - \int \D^d \bs{x} \: \sqrt{g} \phi_0^I \mathcal{O}^I \right),
\end{equation}
we have the one-point functions in the presence of sources
\begin{align}
\< T_{\mu \nu}(\bs{x}) \> & = - \frac{2}{\sqrt{g(\bs{x})}} \frac{\delta}{\delta g^{\mu \nu}(\bs{x})} Z, \\
\< J^{\mu a}(\bs{x}) \> & = - \frac{1}{\sqrt{g(\bs{x})}} \frac{\delta}{\delta A^a_\mu(\bs{x})} Z, \\
\< \mathcal{O}^I(\bs{x}) \> & = - \frac{1}{\sqrt{g(\bs{x})}} \frac{\delta}{\delta \phi^I_0(\bs{x})} Z.
\end{align}
By taking more functional derivatives we can obtain higher-point correlation functions, \textit{e.g.},
\begin{align}
& \< T_{\mu_1 \nu_1}(\bs{x}) T_{\mu_2 \nu_2}(\bs{y}) T_{\mu_3 \nu_3}(\bs{z}) \> = \frac{-2}{\sqrt{g(\bs{z})}} \frac{\delta}{\delta g^{\mu_3 \nu_3}(\bs{z})} \frac{-2}{\sqrt{g(\bs{y})}} \frac{\delta}{\delta g^{\mu_2 \nu_2}(\bs{y})} \frac{-2}{\sqrt{g(\bs{x})}} \frac{\delta}{\delta g^{\mu_1 \nu_1}(\bs{x})} Z[g^{\mu \nu}] \nn\\
& \qquad + \: 2 \< \frac{\delta T_{\mu_1 \nu_1}(\bs{x})}{\delta g^{\mu_2 \nu_2}(\bs{y})} T_{\mu_3 \nu_3}(\bs{z}) \> + 2 \< \frac{\delta T_{\mu_2 \nu_2}(\bs{y})}{\delta g^{\mu_3 \nu_3}(\bs{z})} T_{\mu_1 \nu_1}(\bs{x}) \> + 2 \< \frac{\delta T_{\mu_3 \nu_3}(\bs{z})}{\delta g^{\mu_1 \nu_1}(\bs{x})} T_{\mu_2 \nu_2}(\bs{y}) \>. \label{e:defTTT}
\end{align}
Note that here we define the 3-point function of the stress-energy tensor to be the correlator of three separate stress-energy tensor insertions (and similarly for other correlators involving conserved currents), rather than the correlator obtained by functionally differentiating the generating functional with respect to the metric three times.
While the latter definition is used in \cite{Osborn:1993cr,Erdmenger:1996yc,Coriano:2012wp,Cappelli:2001pz}, our definition here is simpler for direct QFT computations. To convert between the two definitions simply requires the addition or subtraction of the semi-local terms in the formula above.

Requiring the partition function to be invariant under variation of the sources then leads to the transverse Ward identities
\begin{align}
0 & = D_\mu^{ac} \< J^{\mu a} \> - \I (T_R^a)^{IJ} \phi^J_0 \< \mathcal{O}^I \> \nn\\
& = \nabla_\mu \< J^{\mu a} \> + f^{abc} A_\mu^b \< J^{\mu c} \> - \I (T_R^a)^{IJ} \phi^J_0 \< \mathcal{O}^I \>, \label{e:toWardJ} \\
0 & = \nabla^\mu \< T_{\mu \nu} \> + \nabla_\nu A^a_\mu \cdot \< J^{\mu a} \> - \nabla_\mu A_\nu^a \cdot \< J^{\mu a} \> + \partial_\nu \phi^I_0 \cdot \< \mathcal{O}^I \> - A^a_\nu \nabla_\mu \< J^{\mu a} \> \nn\\
& = \nabla^\mu \< T_{\mu \nu} \>  - F_{\mu \nu}^a \< J^{\mu a} \> + D_\nu^{IJ} \phi^J_0 \cdot \< \mathcal{O}^I \>. \label{e:toWardT}
\end{align}
These equations may then be differentiated with respect to the sources to obtain the corresponding Ward identities for higher point functions.

Explicit expressions for all the higher-point transverse Ward identities we need are listed in the second part of the paper.  In obtaining these expressions we have used the assumptions:
\begin{enumerate}
\item $\mathcal{O}^I$ is independent of the sources, \textit{i.e.},
\begin{equation}
\frac{\delta \mathcal{O}^I}{\delta \phi_0^J} = 0, \qquad \frac{\delta \mathcal{O}^I}{\delta A_\mu^a} = 0, \qquad \frac{\delta \mathcal{O}^I}{\delta g^{\mu \nu}} = 0.
\end{equation}
\item The source $\phi_0^I$ appears only as in \eqref{e:Z}, so that
\begin{equation}
\frac{\delta T_{\mu \nu}(\bs{x})}{\delta \phi_0^I(\bs{y})} = -g_{\mu \nu}(\bs{x}) \mathcal{O}(\bs{x}) \delta(\bs{x} - \bs{y}), \qquad \frac{\delta J^{\mu a}}{\delta \phi_0^I} = 0.
\end{equation}
\item The gauge field $A_\mu^a$ couples either through covariant derivatives or acts as an external source for the current in the form of $A^a_{\mu} J^{\mu a}$. This means there are no kinetic terms for $A_\mu^a$, \textit{i.e.}, no derivatives acting on $A_\mu^a$ in the action, hence
\begin{equation}
\frac{\delta T_{\mu \nu}(\bs{x})}{\delta A_\rho^a(\bs{y})} = F_{\mu \nu}^{\rho a}(\bs{x}) \delta(\bs{x} - \bs{y}), \qquad \frac{\delta J^{\mu a}(\bs{x})}{\delta A_\nu^b(\bs{y})} = G^{\mu \nu a b}(\bs{x}) \delta(\bs{x} - \bs{y})
\end{equation}
where $F$ and $G$ are functions of the CFT fields.
\end{enumerate}
Of course, it may happen that renormalisation requires us to add counterterms violating one or more of the assumptions above, in which case the relevant Ward identities would need to be modified accordingly.

As a specific illustration of the general discussion above, let us consider the transverse Ward identity for $\lla T^{\mu_1 \nu_1} T^{\mu_2 \nu_2} \mathcal{O} \rra$ for a matter content consisting of conformal scalars, as defined in section \ref{ch:example_calc}.  We will take the operator $\mathcal{O} = \phi^2$.  The relevant Ward identity is
\begin{equation} \label{e:pTTO}
p_1^{\nu_1} \lla T_{\mu_1 \nu_1}(\bs{p}_1) T_{\mu_2 \nu_2}(\bs{p}_2) \mathcal{O}^I(\bs{p}_3) \rra = 2 p_1^{\nu_1} \lla \frac{\delta T_{\mu_1 \nu_1}}{\delta g^{\mu_2 \nu_2}}(\bs{p}_1, \bs{p}_2) \mathcal{O}^I(\bs{p}_3) \rra,
\end{equation}
where $\delta T_{\mu_1 \nu_1} / \delta g^{\mu_2 \nu_2}$ denotes taking the functional derivative of the stress-energy tensor with respect to the metric, after which we restore the metric to its background value $g_{\mu\nu}=\delta_{\mu\nu}$. Evaluating this functional derivative explicitly using \eqref{e:exampleT}, we find \cite{Bzowski:2011ab}
\begin{align}
\frac{\delta T_{\mu \nu}(\bs{x})}{\delta g^{\rho \sigma}(\bs{y})} & = - \frac{1}{2} \left[ \delta_{\mu \nu} \delta_\rho^\alpha \delta_\sigma^\beta + 2 \delta_{\mu (\rho} \delta_{\sigma) \nu} - \delta_{\mu \nu} \delta_{\rho \sigma} \delta^{\alpha \beta} \right] \delta(\bs{x} - \bs{y}) T_{\alpha \beta}(\bs{x}) \nn\\
& \qquad + \: \frac{1}{16} \left[ C_{\mu \nu \rho \sigma}^{(1) \alpha \beta} \delta(\bs{x} - \bs{y}) \partial_\alpha \partial_\beta + C_{\mu \nu \rho \sigma}^{(2) \alpha \beta} \partial_\alpha \delta(\bs{x} - \bs{y}) \partial_\beta \right. \nn\\
& \qquad \qquad \left. + \: C_{\mu \nu \rho \sigma}^{(3) \alpha \beta} \partial_\alpha \partial_\beta \delta(\bs{x} - \bs{y}) \right] \mathcal{O}(\bs{x}),
\end{align}
where partial derivatives are taken with respect to $\bs{x}$ and the prefactors are
\begin{align}
C_{\mu \nu \rho \sigma}^{(1) \alpha \beta} & = \delta_{\mu \nu} \delta_\rho^{(\alpha} \delta^{\beta)}_{\sigma} + 2 \delta_{\mu(\rho} \delta_{\sigma)\nu} \delta^{\alpha \beta} - \delta_{\mu \nu} \delta_{\rho \sigma} \delta^{\alpha \beta}, \\
C_{\mu \nu \rho \sigma}^{(2) \alpha \beta} & = 2 \delta_{\mu \nu} \delta_\rho^{(\alpha} \delta^{\beta)}_{\sigma} + \delta_{\mu(\rho} \delta_{\sigma)\nu} \delta^{\alpha \beta} - \delta_{\mu \nu} \delta_{\rho \sigma} \delta^{\alpha \beta} - 2 \delta^\alpha_{(\mu} \delta_{\nu)(\rho} \delta_{\sigma)}^\beta, \\
C_{\mu \nu \rho \sigma}^{(3) \alpha \beta} & = \delta_{\mu \nu} \delta_\rho^{(\alpha} \delta^{\beta)}_{\sigma} + \delta_{\mu(\rho} \delta_{\sigma)\nu} \delta^{\alpha \beta} - \delta_{\mu \nu} \delta_{\rho \sigma} \delta^{\alpha \beta} - 2 \delta^\alpha_{(\mu} \delta_{\nu)(\rho} \delta_{\sigma)}^\beta + \delta_{\rho \sigma} \delta_{\mu}^{(\alpha} \delta^{\beta)}_\nu.
\end{align}
After Fourier transforming and using the result for the 2-point function
\begin{equation} \label{e:exOO}
\lla \mathcal{O}(\bs{p}) \mathcal{O}(-\bs{p}) \rra = \frac{1}{4 p}
\end{equation}
we obtain
\begin{equation} \label{e:exWard}
p_{1 \nu_1} \lla T^{\mu_1 \nu_1}(\bs{p}_1) T^{\mu_2 \nu_2}(\bs{p}_2) \mathcal{O}(\bs{p}_3) \rra = - \frac{1}{32 \: p_3} p_2^{\mu_1} p_3^{\mu_2} p_3^{\nu_2} + \frac{p_3}{32} \delta^{\mu_1 \mu_2} p_3^{\nu_2} + \ldots,
\end{equation}
where we have retained only the terms appearing in the right-hand sides of the secondary CWIs
(\ref{e:secondaryCWIs1}) and (\ref{e:secondaryCWIs2}).  The omitted terms do not contain the tensors listed explicitly and will play no further role in our analysis.  As usual, we use the convention \eqref{e:momenta_choice} for the Lorentz indices.

\subsection{Trace Ward identities} \label{ch:trace}

Invariance of the generating functional \eqref{e:Z} under the Weyl transformations
\begin{equation}
\delta g^{\mu \nu} = 2 g^{\mu \nu} \delta \sigma, \qquad \delta \phi_0^I = (d - \Delta) \phi_0^I \delta \sigma, \qquad \delta A^a_\mu = 0.
\end{equation}
leads to the trace (or Weyl) Ward identity in the presence of the sources
\begin{equation} \label{e:toWardTr}
\langle T(\bs{x}) \rangle = (\Delta - d) \phi_0^I(\bs{x}) \langle \mathcal{O}^I(\bs{x}) \rangle,
\end{equation}
where $T = T^\mu_\mu$.  Functionally differentiating with respect to the sources then yields trace Ward identities for 3-point functions, \textit{e.g.},
\begin{align}
\lla T(\bs{p}_1) \mathcal{O}^I(\bs{p}_2) \mathcal{O}^J(\bs{p}_3) \rra & = - \Delta \left[ \lla \mathcal{O}^I(\bs{p}_2) \mathcal{O}^J(-\bs{p}_2) \rra + \lla \mathcal{O}^I(\bs{p}_3) \mathcal{O}^J(-\bs{p}_3) \rra \right], \\
\lla T(\bs{p}_1) T_{\mu \nu}(\bs{p}_2) \mathcal{O}^I(\bs{p}_3) \rra & = 2 \lla \frac{\delta T}{\delta g^{\mu \nu}}(\bs{p}_1, \bs{p}_2) \mathcal{O}^I(\bs{p}_3) \rra.
\end{align}
A complete list of all trace Ward identities is given in the second part of the paper.

As is well known, due to renormalisation the trace Ward identity may acquire an anomalous contribution.  The exact contribution depends strongly on the specifics of the theory, but its form is universal. In this paper we assume no anomalies in the transverse Ward identities \eqref{e:toWardJ} and \eqref{e:toWardT} can appear. The anomalous contributions are therefore still transverse.

\section{Solutions to conformal Ward identities}\label{section:solnofCWIs}

It is a rather pleasant fact that all the primary CWIs can be solved in terms of the triple-$K$ integrals similar to \eqref{e:Osol}. We will start by analysing some properties of the triple-$K$ integrals before proceeding to show how this class of integrals solves the primary CWIs.
In particular, we will find that the solution to each primary CWI is unique up to one numerical constant. Finally, we will analyse the structure and implications of the secondary CWIs.

\subsection{\texorpdfstring{Triple-$K$ integrals}{Triple-K integrals}}

All primary CWIs can be solved in terms of the general triple-$K$ integral
\begin{equation} \label{e:J}
I_{\alpha \{ \beta_1 \beta_2 \beta_3 \}}(p_1, p_2, p_3) = \int_0^\infty \D x \: x^\alpha \prod_{j=1}^3 p_j^{\beta_j} K_{\beta_j}(p_j x),
\end{equation}
where $K_\nu$ is a Bessel $K$ function. This integral depends on four parameters, namely the power $\alpha$ of the integration variable $x$, and the three Bessel function indices $\beta_j$. In the following we will generically refer to these as $\alpha$ and $\beta$ parameters respectively. Its arguments, $p_1, p_2, p_3$, are magnitudes of momenta $p_j = | \bs{p}_j |$, $j=1,2,3$.

It will be useful to define a reduced version $J_{N \{k_1 k_2 k_3\}}$ of the triple-$K$ integral by substituting
\begin{equation} \label{e:DeltaToAlpha}
\alpha = \frac{d}{2} - 1 + N, \qquad \qquad \beta_j = \Delta_j - \frac{d}{2} + k_j, \quad j = 1,2,3.
\end{equation}
Here we assume that we concentrate on some particular 3-point function and the conformal dimensions $\Delta_j$, $j=1,2,3$ are therefore fixed. In other words we define
\begin{equation} \label{e:Jred}
J_{N \{ k_j \}} = I_{\frac{d}{2} - 1 + N \{ \Delta_j - \frac{d}{2} + k_j \}},
\end{equation}
where we use a shortened notation $\{k_j\} = \{k_1 k_2 k_3\}$, {\it etc}. Finally we define
\begin{equation}
\Delta_t = \Delta_1 + \Delta_2 + \Delta_3, \qquad \beta_t = \beta_1 + \beta_2 + \beta_3, \qquad k_t = k_1 + k_2 + k_3.
\end{equation}

The main point of this section is to present relations showing that all primary CWIs for a given 3-point function can be solved by the triple-$K$ integrals \eqref{e:J}. The representation \eqref{e:Jred} turns out to be extremely useful, as the parameters $N$ and $k_j$ are fixed by the primary CWIs and have no dependence on either $\Delta_j$ or $d$.
If desired, these triple-$K$ integrals may also be re-expressed in terms of other familiar integrals such as Feynman or Schwinger parametrised integrals, as discussed in appendix \ref{ch:toKKK}.

\subsubsection{Region of validity and regularisation} \label{ch:regul}

We assume all parameters and variables in the triple-$K$ integral \eqref{e:J} are real. From the asymptotic expansion  \eqref{e:asymIK} the integral converges at large $x$, however in general there may still be a divergence at $x=0$.  From the series expansion \eqref{e:serI} and the definition \eqref{e:defK}, we see the triple-$K$ integral only converges if \cite{Erdelyi}
\begin{equation} \label{e:conv_cond}
\alpha > \sum_{j=1}^3 | \beta_j | - 1, \qquad p_1, p_2, p_3 > 0.
\end{equation}
If the parameters $\alpha$ and $\beta_j$ do not satisfy this inequality, we can use analytic continuation. We introduce two finite real parameters $u$ and $v$ and regulate the integral \eqref{e:J} by
continuing
\begin{equation} \label{e:genscheme}
I_{\alpha \{\beta_1 \beta_2 \beta_3 \}} \mapsto I_{\alpha + u \epsilon \{\beta_1 + v \epsilon, \beta_2  + v \epsilon, \beta_3  + v \epsilon \}}.
\end{equation}
In terms of the parameters $d$ and $\Delta_j$ this general regularisation corresponds to
\begin{equation} \label{e:genschemed}
d \mapsto d + 2 u \epsilon, \qquad\qquad \Delta_j \mapsto \Delta_j + (u + v) \epsilon,
\end{equation}
as one can infer using \eqref{e:DeltaToAlpha}.
The convenient scheme \eqref{e:Oscheme} preserving the indices of Bessel functions then corresponds to $u = 1$ and $v = 0$,
while the dimensional regularisation \eqref{e:dimreg00}
corresponds to the choice $u=v=-1/2$.
(Of course, strictly speaking, only the ratio $u/v$ is actually significant since we are always free to rescale $\epsilon$ by a constant.)
We will return to discuss the choice of $u$ and $v$ further in section \ref{ch:tripleKand2pt}, but for now we will keep the discussion general and treat $u$ and $v$ as arbitrary parameters.

Generically, the limit $\epsilon \rightarrow 0$ exists except for cases where
\begin{equation}\label{condition}
\alpha + 1 \pm \beta_1 \pm \beta_2 \pm \beta_3 = - 2 n,
\end{equation}
for some non-negative integer $n$.  If the limit does exist, then its value is independent of the specific choice of $u$ and $v$ due to the uniqueness of the analytic continuation. If, on the other hand, \eqref{condition} is satisfied then we obtain pole terms in the regulator $\epsilon$.
This singular behaviour arises from the divergence of the triple-$K$ integral at its lower limit $x=0$. To find the form of these poles more precisely, it suffices to expand the integrand about $x=0$ as follows.

First, let us consider the case where all the $\beta_j$ take non-integer values.
Expanding the integrand of the triple-$K$ integral using \eqref{e:serI} and \eqref{e:defK}, we obtain a sum of powers $x^a$ for various $a \in \R$.
Each such term makes a contribution to the integral about $x=0$ of the form
\begin{equation} \label{e:analcont1}
\int_0 x^a \D x = \frac{\text{const}}{a + 1},
\end{equation}
where the upper limit of the integral determines the value of the constant but is otherwise unimportant.  The right-hand side is an analytic function of $a$ with a single pole at $a=-1$,
and defines the integral by analytic continuation for $a<-1$.
Thus, while the triple-$K$ integral naively diverges if the expansion of its integrand contains terms of the form $x^a$ with $a < -1$, its value in such cases is in fact uniquely defined through the analytic continuation \eqref{e:analcont1}.
When \eqref{condition} is satisfied, the expansion of the integrand of the unregulated triple-$K$ integral has terms with $a=-1$, leading to single poles in $\epsilon$ in the regulated integral.

In cases where some of the $\beta_j$ take integer values, the regulated triple-$K$ integral may also contain higher-order poles in $\epsilon$.
This is because when expanding the integrand of the triple-$K$ integral about $x=0$ using \eqref{e:expKn} we now obtain logarithms and their powers.  From terms containing a single logarithm we obtain a contribution
\begin{equation}
\int_0 x^a \log x \: \D x = - \frac{\text{const}_1}{(a+1)^2} + \frac{\text{const}_2}{a+1},
\end{equation}
while terms containing powers of logarithms contribute still higher-order poles.  (In all cases these poles are located at $a=-1$ however.)
Using the series expansions \eqref{e:serI} and \eqref{e:expKn}, we can then confirm that in order for terms with $a=-1$ to arise in the expansion of the integrand of the unregulated triple-$K$ integral, the condition \eqref{condition} must be satisfied.  Thus, only when \eqref{condition} is satisfied do we obtain poles in $\epsilon$ for the regulated triple-$K$ integral, while in all other cases the limit $\epsilon\rightarrow 0$ is well defined.

\subsubsection{Basic properties} \label{ch:Basic_properties}

Let us now examine briefly some of the basic properties of triple-$K$ integrals.
The most obvious of these is the permutation symmetry
\begin{equation} \label{e:Jsym}
I_{\alpha \{ \beta_{\sigma(1)} \beta_{\sigma(2)} \beta_{\sigma(3)} \}}(p_1, p_2, p_3) = I_{\alpha \{ \beta_1 \beta_2 \beta_3 \}}(p_{\sigma^{-1}(1)}, p_{\sigma^{-1}(2)}, p_{\sigma^{-1}(3)}),
\end{equation}
where $\sigma$ is any permutation of the set $\{1,2,3\}$.
We also have the relations
\begin{align}
\frac{\partial}{\partial p_n} I_{\alpha \{ \beta_j \}} & = - p_n I_{\alpha + 1 \{ \beta_j - \delta_{jn} \}}, \label{e:Jid1} \\
I_{\alpha \{ \beta_j + \delta_{jn} \}} & = p_n^2 I_{\alpha \{ \beta_j - \delta_{jn} \}} + 2 \beta_n I_{\alpha - 1 \{ \beta_j \}}, \label{e:Jid2} \\
I_{\alpha \{\beta_1 \beta_2, - \beta_3\}} & = p_3^{-2 \beta_3} I_{\alpha \{\beta_1 \beta_2 \beta_3\}}, \label{e:Jid4}
\end{align}
for any $n=1,2,3$, as follows from the elementary Bessel function relations
\begin{align}
\frac{\partial}{\partial a} \left[ a^\nu K_\nu(a x) \right] & = - x a^\nu K_{\nu - 1}(a x), \label{e:Kid1a} \\
K_{\nu-1}(x) + \frac{2 \nu}{x} K_{\nu}(x) & = K_{\nu + 1}(x), \label{e:Kid2a} \\
K_{-\nu}(x) & = K_{\nu}(x). \label{e:Kid4a}
\end{align}
Some additional properties of Bessel functions and triple-$K$ integrals are listed in appendix \ref{ch:prop}.

\subsubsection{\texorpdfstring{Dilatation degree of the triple-$K$ integral}{Dilatation degree of the triple-K integral}}

As the triple-$K$ integral solves the CWIs, it should also solve the dilatation Ward identity \eqref{e:dilD} and hence must have a specific dilatation dimension.
Using first \eqref{e:Kid1a} then \eqref{e:Jid1}, we can write
\begin{align} \label{e:bnd_term}
 \int_0^\infty \D x \frac{\partial}{\partial x} \left( x^{\alpha+1} \prod_{j=1}^3 p_j^{\beta_j} K_{\beta_j}(p_j x) \right)
&= ( \alpha + 1 - \beta_t ) I_{\alpha \{ \beta_k \}} - \sum_{j=1}^3 p_j^2 I_{\alpha + 1 \{ \beta_k - \delta_{jk} \}}
\nn\\
& = ( \alpha + 1 - \beta_t ) I_{\alpha \{ \beta_k \}} + \sum_{j=1}^3 p_j \frac{\partial}{\partial p_j} I_{\alpha \{ \beta_k \}}.
\end{align}
The expression on the left-hand side leads to a boundary term at $x = 0$. In the region of convergence \eqref{e:conv_cond}, all integrals in this expression are well-defined and the boundary term vanishes. We can now use the analytic continuation \eqref{e:genscheme} to argue that the analytically continued left-hand side vanishes generically, except in cases where \eqref{condition} is satisfied. Indeed, if we regard both sides of \eqref{e:bnd_term} as analytic functions of $\alpha$, with all other parameters and momenta fixed, then the validity of \eqref{e:bnd_term} in the region \eqref{e:conv_cond} implies its validity in the entire domain of analyticity. The vanishing of the left-hand side then implies that
\begin{equation} \label{e:degJ}
\deg I_{\alpha \{\beta_j\}} = \beta_t - \alpha - 1, \qquad\qquad \deg J_{N \{ k_j \}} = \Delta_t + k_t - 2d - N,
\end{equation}
provided $\alpha + 1 \pm \beta_1 \pm \beta_2 \pm \beta_3 \neq - 2 n$ for some non-negative integer $n$ and independent choice of signs.

If instead \eqref{condition} is satisfied then we expect scaling anomalies in $I_{\alpha \{\beta_j\}}$. Using the power series expansion \eqref{e:serI} of the Bessel $I$ functions, we see that the series expansion about $x=0$ of the boundary term $x^{\alpha+1} \prod_{j=1}^3 p_j^{\beta_j} K_{\beta_j}(p_j x)$ in \eqref{e:bnd_term} contains a constant piece exactly when \eqref{condition} holds. This indicates that the dilatation Ward identity for the $I_{\alpha \{\beta_j\}}$ is not satisfied in such cases. Note however that this is not a strict argument since the regulator cannot be removed from the integrals appearing in \eqref{e:bnd_term}: one should instead expand both sides in the regulator $\epsilon$ and match terms order by order.

\subsubsection{\texorpdfstring{Triple-$K$ integrals and 2-point functions}{Triple-K integrals and 2-point functions}} \label{ch:tripleKand2pt}

Before discussing the consequences of the primary and secondary Ward identities, we first need to analyse the possible singularities associated with the 2- and 3-point functions. This is because the secondary CWIs connect triple-$K$ integrals with semi-local terms expressible in terms of 2-point functions.  When suitably regulated, the singularities in the triple-$K$ integrals must then match the corresponding singularities in the 2-point functions.

An initial obstacle is that our convenient regularisation scheme \eqref{e:Oscheme} does not work for 2-point functions. The Fourier transform of the position space expression for a generic 2-point function was discussed in section \ref{sec:2pt} and there is a singularity when $2 \Delta = d + 2n$ for a non-negative integer $n$. This singularity however is not regularised by the scheme \eqref{e:Oscheme}.

For these reasons it is convenient to choose a different regularisation scheme which is applicable to all correlation functions.  Here, we choose the standard dimensional regularisation
\begin{equation} \label{e:dimreg}
d \mapsto d - \epsilon, \qquad\qquad \Delta_j \mapsto \Delta_j - \epsilon.
\end{equation}
Re-expressed in terms of the parameters $\alpha$ and $\beta$ in \eqref{e:J}, this is equivalent to
\begin{equation} \label{e:scheme}
\alpha \mapsto \alpha -\frac{\epsilon}{2}, \qquad\qquad \beta_j \mapsto \beta_j - \frac{\epsilon}{2}, \quad j = 1,2,3\, .
\end{equation}
To see this, consider for example the momentum space integral appearing in the 3-point function of three scalars, namely
\begin{equation} \label{e:mom_space_int}
\int \frac{\D^d \bs{k}}{(2 \pi)^d} \frac{k^{\mu_1} \ldots k^{\mu_r}}{\bs{k}^{2 \delta_{3}} |\bs{k} - \bs{p}_1|^{2 \delta_{2}} |\bs{k} + \bs{p}_2|^{2 \delta_{1}}}
\end{equation}
as given in (\ref{e:Otriple-$K$}).
The parameters $\delta_i$ are related to the conformal dimensions via (\ref{e:deltaDelta}), and
keeping $\delta_i$ fixed is then equivalent to \eqref{e:scheme}.

Note that the conformal dimensions are continued in \eqref{e:dimreg}.  This is necessary, firstly, to preserve the conservation of $J^\mu$ and $T^{\mu\nu}$,
and secondly,
to preserve the dimensions of the associated sources (so that, for example, the metric, representing the source for
$T^{\mu\nu}$, remains dimensionless).
The operators $J^\mu$ and $T^{\mu\nu}$ must moreover saturate the unitarity bound in a CFT, $\Delta = d - 2 + s$, where $s$ denotes the spin.

The dimensional regularisation scheme \eqref{e:scheme} has one important disadvantage, however, which is that it does not regularise all triple-$K$ integrals: starting from the general scheme \eqref{e:genscheme}, some triple-$K$ integrals become singular upon setting $u = v$.
As actual physical correlation functions are well-defined in dimensional regularisation, however, these singularities must cancel out when correlation functions are written as linear combinations of triple-$K$ integrals.
In light of this, we will first use the general regularisation scheme \eqref{e:genscheme} to solve the primary Ward identities for arbitrary $u$ and $v$.
In the combination of triple-$K$ integrals that solve the primary Ward identities the $(u-v)$-poles should cancel out so that one may then set $u=v$ \cite{II}. This imposes additional conditions on the Taylor expansion of the undetermined primary constants $\alpha_j$ around $u=v$ . One can extract these conditions, but it turns out that the entire information is already contained in the secondary Ward identities. Eventually one can set $u=v=-\frac{1}{2}$ to obtain the form factors in  dimensional regularisation \eqref{e:scheme}. All remaining singularities now appear as $\epsilon$-poles and these should be removed by adding local covariant counterterms. We will discuss in detail this procedure, the full set of counterterms, {\it etc.}, for all cases in \cite{II}.

An additional minor drawback of dimensional regularisation is the fact that it shifts the orders of the Bessel functions in triple-$K$ integrals, potentially making them harder to evaluate. If, for example, a given triple-$K$ integral with half-integer Bessel indices diverges, in dimensional regularisation one cannot simply use the analytic formula \eqref{e:Khalf} as one could in the scheme with $u=1$ and $v=0$. Such difficulties can however be avoided if the analytic continuation of the triple-$K$ integral exists and is finite. In this case any regularisation must lead to the same value of the triple-$K$ integral due to uniqueness of analytic continuation, and so we may simply use the most convenient scheme to evaluate it. This allows us, for example, to evaluate correlation functions such as $\< T^{\mu_1 \nu_1} J^{\mu_2} J^{\mu_3} \>$ or $\< T^{\mu_1 \nu_1} T^{\mu_2 \nu_2} T^{\mu_3 \nu_3} \>$ in odd spacetime dimensions exactly. In odd dimensions these correlation functions must be finite, since no covariant
counterterms of appropriate dimension exist,
allowing us to use the scheme \eqref{e:Oscheme} preserving the indices of the Bessel functions.

\subsection{Solutions to the primary conformal Ward identities} \label{ch:sol_to_pri_CWIs}

In the previous section, we defined the triple-$K$ integral and analysed its basic properties. We now want to use this information in order to write a solution to the CWIs. For this, we need the following fundamental identity. For any $m,n = 1,2,3$,
\begin{equation} \label{e:fund_eqK}
\K_{mn} J_{N \{ k_j \}} = - 2 k_m J_{N + 1 \{ k_j - \delta_{jm} \}} + 2 k_n J_{N + 1 \{ k_j - \delta_{jn} \}},
\end{equation}
for $k_1, k_2, k_3, N \in \R$.  The operator $\K_{mn}$ is the CWI operator defined in \eqref{e:KK}. This relation is a direct consequence of the identities \eqref{e:Jid1} and \eqref{e:Jid2}.

Let us first consider the pair of primary CWIs for the form factor $A_1$. As discussed in section \ref{ch:primaryCWIs}, such CWIs are always homogeneous and take the form \eqref{e:A1eq}. Observe that if we set all $k_j = 0$ in \eqref{e:fund_eqK} then $A_1 = \alpha_1 J_{N \{000\}}$ is a solution for arbitrary $N \in \R$ and an integration constant $\alpha_1 \in \R$. Furthermore, observe that if we impose only one homogeneous equation, say $\K_{12} A = 0$, then the most general solution in terms of the triple-$K$ integrals is $\alpha J_{N \{0 0 k_3\}}$ for any $\a, N, k_3 \in \R$. In general, the equation \eqref{e:fund_eqK} is sufficient to write down solutions to all primary CWIs.

The remaining piece of information is the value of $N$. In general, if $A_n = \alpha_n J_{N \{k_j \}}$ is a form factor of tensorial dimension $N_n$, then \eqref{e:degree} and \eqref{e:degJ} determine the value of $N = N(A_n)$ to be
\begin{equation} \label{e:Nval}
N(A_n) = N_n + k_t.
\end{equation}

Let us see how the procedure works for our example $\lla T^{\mu_1 \nu_1} T^{\mu_2 \nu_2} \mathcal{O} \rra$. The primary CWIs are given by \eqref{e:CWIsTTO} and, in particular,
\begin{align}
& \K_{12} A_1 = 0, \qquad \K_{13} A_1 = 0, \nn\\
& \K_{12} A_2 = 0, \qquad \K_{12} \K_{13} A_2 = 0, \qquad \K^2_{13} A_2 = 0, \nn\\
& \K_{12} A_3 = 0, \qquad \K_{12} \K_{13} A_3 = 0, \qquad \K_{12} \K^2_{13} A_3 = 0, \qquad \K^3_{13} A_3 = 0, \label{e:presolTTO}
\end{align}
Therefore, using \eqref{e:fund_eqK} and \eqref{e:Nval}, we can write the most general solution given in terms of the triple-$K$ integrals,
\begin{align}
A_1 & = \alpha_1 J_{4 \{000\}}, \label{e:preATTO0} \\
A_2 & = \alpha_{21} J_{3 \{001\}} + \alpha_2 J_{2 \{000\}}, \label{e:preATTO2} \\
A_3 & = \alpha_{31} J_{2 \{002\}} + \alpha_{32} J_{1 \{001\}} + \alpha_3 J_{0 \{000\}}, \label{e:preATTO}
\end{align}
where all the $\alpha$ are numerical constants. Finally, the inhomogeneous parts of \eqref{e:CWIsTTO} fix some of these constants. When the solution above is substituted into the primary CWIs, \eqref{e:fund_eqK} requires that
\begin{equation} \label{e:cTTO}
\alpha_{21} = 4 \alpha_1, \qquad \alpha_{31} = 2 \alpha_1, \qquad \alpha_{32} = \alpha_2.
\end{equation}
The three remaining undetermined constants $\alpha_1, \alpha_2, \alpha_3 \in \R$ multiply integrals of the form $J_{N \{000\}}$. Such integrals solve the homogeneous parts of the CWIs. Therefore the remaining constants, undetermined by the primary CWIs, will be called \emph{primary constants}.

Let us summarise our results. We have analysed the primary CWIs for the $\lla T^{\mu_1 \nu_1} T^{\mu_2 \nu_2} \mathcal{O} \rra$ correlation function and we found a solution
\begin{align}
A_1 & = \alpha_1 J_{4 \{000\}}, \label{e:ATTO0} \\
A_2 & = 4 \alpha_{1} J_{3 \{001\}} + \alpha_2 J_{2 \{000\}}, \label{e:ATTO2} \\
A_3 & = 2 \alpha_{1} J_{2 \{002\}} + \alpha_{2} J_{1 \{001\}} + \alpha_3 J_{0 \{000\}}, \label{e:ATTO}
\end{align}
with three undetermined constants $\alpha_1, \alpha_2, \alpha_3 \in \R$. We will show shortly that this solution to the primary CWIs is indeed unique, specifying $\lla T^{\mu_1 \nu_1} T^{\mu_2 \nu_2} \mathcal{O} \rra$ in momentum space up to three constants.
Following application of the secondary CWIs, we will find that the number of undetermined constants is reduced to just one. The method we have described is based purely in momentum space and is applicable to all 3-point functions.  Explicit solutions for all primary CWIs are listed in the second part of the paper.

The triple-$K$ integrals we discuss here also arise in AdS/CFT calculations of momentum space 3-point functions using a dual gravitational theory (recent papers include, \textit{e.g.}, \cite{Raju:2012zr,Raju:2012zs,Chowdhury:2012km}). As such, these calculations apply only to the specific CFTs dual to particular gravitational theories.   In contrast, our approach here is completely general, showing that all 3-point functions of conserved currents, stress-energy tensors and scalar operators in {\it any} CFT can be expressed in terms of triple-$K$ integrals.

\subsubsection{\texorpdfstring{More on  $\< T^{\mu_1 \nu_1} T^{\mu_2 \nu_2} \mathcal{O} \>$}{More on <TTO>}} \label{ch:Jint3}

In this section we wish to illustrate that the solution to the primary CWIs in terms of the triple-$K$ integrals can be evaluated explicitly with ease.
For concreteness, consider $\lla T^{\mu_1 \nu_1} T^{\mu_2 \nu_2} \mathcal{O} \rra$ with a scalar operator $\mathcal{O}$ of dimension $\Delta_3 = 1$ in $d = 3$ dimensional CFT. The solution to the primary CWIs is given by \eqref{e:ATTO0} - \eqref{e:ATTO} with constants fixed according to \eqref{e:cTTO}. In order to write the solution explicitly, we can use expressions \eqref{e:Khalf} and \eqref{e:Khalf1}, after which all integrals turn out to be elementary. The following integrals converge and can be easily computed
\begin{align}
J_{4 \{000\}} & = I_{\frac{9}{2} \{\frac{3}{2}\frac{3}{2},-\frac{1}{2}\}} = \left( \frac{\pi}{2} \right)^{3/2} \cdot \frac{3 (p_1^2 + p_2^2) + p_3^2 + 12 p_1 p_2 + 4 p_3 (p_1 + p_2) }{p_3 (p_1 + p_2 + p_3)^4}, \\
J_{3 \{001\}} & = I_{\frac{7}{2} \{\frac{3}{2}\frac{3}{2}\frac{1}{2}\}} = \left( \frac{\pi}{2} \right)^{3/2} \cdot \frac{2 (p_1^2 + p_2^2) + p_3^2 + 6 p_1 p_2 + 3 p_3 (p_1 + p_2) }{(p_1 + p_2 + p_3)^3},
\end{align}
assuming $\Delta_1 = \Delta_2 = 3$ and $\Delta_3 = 1$. The remaining integrals diverge and require a regularisation. As discussed in section \ref{ch:regul} and \ref{ch:tripleKand2pt}, we can consider the integrals $J_{N + \epsilon \{ k_j \}}$, since this particular 3-point function is finite and any regularisation must yield the same result. In this manner, we find
\begin{align}
J_{2 + \epsilon \{002\}} & = I_{\frac{5}{2} + \epsilon \{\frac{3}{2}\frac{3}{2}\frac{3}{2}\}} = - \left( \frac{\pi}{2} \right)^{3/2} \frac{1}{(p_1 + p_2 + p_3)^2} \left[ 2 p_1 p_2 p_3 + p_1^3 + p_2^3 + p_3^3 \right.\nn\\
& \qquad\qquad \left. +\:2 (p_1^2 p_2 + p_1 p_2^2 + p_1^2 p_3 + p_1 p_3^2 + p_2^2 p_3 + p_2 p_3^2) \right] + O(\epsilon), \\
J_{2 + \epsilon \{000\}} & = I_{\frac{5}{2} + \epsilon \{\frac{3}{2}\frac{3}{2},-\frac{1}{2}\}} = \left( \frac{\pi}{2} \right)^{3/2} \cdot \frac{1}{p_3 \epsilon} + O(\epsilon^0), \label{e:exj2000} \\
J_{1 + \epsilon \{001\}} & = I_{\frac{3}{2} + \epsilon \{\frac{3}{2}\frac{3}{2}\frac{1}{2}\}} = - \left( \frac{\pi}{2} \right)^{3/2} \cdot \frac{p_3}{\epsilon} + O(\epsilon^0), \label{e:exj1001} \\
J_{0 + \epsilon \{000\}} & = I_{\frac{1}{2} + \epsilon \{\frac{3}{2}\frac{3}{2},-\frac{1}{2}\}} = - \left( \frac{\pi}{2} \right)^{3/2} \frac{p_1^2 + p_2^2 - p_3^2}{2 p_3 \epsilon} + O(\epsilon^0). \label{e:exj0000}
\end{align}
As we will see the omitted terms make no contribution in our subsequent analysis. In order to further constrain the primary constants $\alpha_1, \alpha_2, \alpha_3$ we must consider the secondary CWIs. We will return to this task in section \ref{ch:exseccwi}.

At this point we can compare the result given by \eqref{e:ATTO0} - \eqref{e:ATTO} with the direct calculations of the 3-point function for the free scalar field carried out in section \ref{ch:example_calc}. We see that the form of the integrals $J_{4 \{000\}}$, $J_{3 \{001\}}$ and $J_{2 \{002\}}$ match the form factors $A_1$, $A_2$ and $A_3$ in the equations (\ref{e:exA1}) - (\ref{e:exA3}). Therefore,
working in the regularisation scheme \eqref{e:genscheme} with $u = 1$ and $v = 0$, the primary constants for this particular model are
\begin{equation} \label{e:excj}
\alpha_1 = \frac{1}{48} \left( \frac{2}{\pi} \right)^{\frac{3}{2}}, \qquad \alpha_2 = 0, \qquad \alpha_3 = 0.
\end{equation}
The scheme-dependence of the primary constants  here simply reflects the fact that while the divergent triple-$K$ integrals are scheme-dependent, the physical form factors themselves are finite and hence unique.
Note also that the relations \eqref{e:cTTO} provide a cross-check on our solution. Later, we will see that the secondary Ward identities impose two additional constraints on the primary constants that are not yet visible.

\subsubsection{Uniqueness of the solution}

In the previous sections we argued that all CWIs may be solved in terms of triple-$K$ integrals \eqref{e:J}. A case-by-case analysis confirms this and the list of complete solutions is given in the second part of the paper. Here we want to establish that these solutions are unique. To be more precise, we want to argue that each pair of the primary CWIs determines a form factor $A_n$ uniquely up to one numerical constant. This may be achieved by essentially the same reasoning as in section \ref{ch:Ounique}.

First, we assume that $A_n$ satisfies a pair of homogeneous primary CWIs
\begin{equation}
\K_{12} A_n = 0, \qquad \qquad \K_{13} A_n = 0,
\end{equation}
together with the dilatation Ward identity \eqref{e:dilA} with tensorial dimension $N_n$. We can then use the substitution
\begin{equation} \label{e:substoF4}
A_n(p_1, p_2, p_3) = p_3^{\Delta_t - 2d - N_n} \left( \frac{p_1^2}{p_3^2} \right)^\mu \left( \frac{p_2^2}{p_3^2} \right)^\lambda F \left( \frac{p_1^2}{p_3^2}, \frac{p_2^2}{p_3^2} \right)
\end{equation}
and proceed with the analysis analogous to that following equation \eqref{e:OsubstoF4}. The substitution leads to the system of equations (\ref{e:OsysF4a}, \ref{e:OsysF4b}) with four possible choices of parameters
\begin{align}
& \alpha = \frac{1}{2} \left[ N_n + \epsilon_1 \left( \Delta_1 - \frac{d}{2} \right) + \epsilon_2 \left( \Delta_2 - \frac{d}{2} \right) + \Delta_3 \right], \qquad \beta = \alpha - \left( \Delta_3 - \frac{d}{2} \right), \nn\\
& \gamma = 1 + \epsilon_1 \left( \Delta_1 - \frac{d}{2} \right), \qquad \gamma' = 1 + \epsilon_2 \left( \Delta_2 - \frac{d}{2} \right),
\end{align}
parametrised by $\epsilon_1, \epsilon_2 = \pm 1$. We can now use equation \eqref{e:F4toIIK} and the analysis that follows. This leads to the conclusion that the only physically acceptable solution to the homogenous part of the CWIs is given by the triple-$K$ integral $\alpha_n J_{N_n \{000\}}(p_1, p_2, p_3)$, where $\alpha_n$ is a single undetermined constant.

In general, the primary CWIs for a form factor $A_n$ contain inhomogeneous parts. The recursive nature of the primary CWIs discussed in section \ref{ch:primaryCWIs} then allows us to solve these equations one by one. Since the inhomogeneous part is linear in the other form factors, every two solutions to a given pair of equations differ by a solution to the homogeneous part of the equation. The full solution to the pair of primary CWIs and the dilatation Ward identity is therefore unique up to one numerical constant.

It is important to emphasise that while the solution to each pair of primary CWIs is unique up to one primary constant, the representation in terms of triple-$K$ integrals may not be. For example, for generic parameter values the equation \eqref{e:bnd_term} shows that
\begin{equation} \label{e:Jid3}
(\alpha + \beta_t) I_{\alpha - 1 \{\beta_1 \beta_2 \beta_3\}} = I_{\alpha \{\beta_1 + 1, \beta_2, \beta_3\}} + I_{\alpha \{\beta_1, \beta_2 + 1, \beta_3\}} + I_{\alpha \{\beta_1, \beta_2, \beta_3 + 1\}}.
\end{equation}
One can therefore re-write one triple-$K$ integral as a combination of others and hence the representation is not unique.

\subsection{Solutions to the secondary conformal Ward identities} \label{ch:sol_sec_CWI}

In this section we will finalise our theoretical considerations by solving the secondary CWIs. In general, the secondary CWIs lead to linear algebraic equations between the various primary constants appearing in solutions to the primary CWIs. The precise form of the secondary CWIs depends on the semi-local information provided by transverse Ward identities, which may be written in terms of the 2-point functions.

First, we will return to our example from section \ref{ch:Jint3} and show how the two secondary CWIs (\ref{e:secondaryCWIs1}, \ref{e:secondaryCWIs2}) constrain the values of the three primary constants appearing in the solution \eqref{e:ATTO0} - \eqref{e:ATTO} to the primary CWIs. As expected, we will find two algebraic linear equations between the three primary constants. From this we may conclude that the 3-point function $\lla T^{\mu_1 \nu_1} T^{\mu_2 \nu_2} \mathcal{O} \rra$ depends on a single undetermined primary constant.

Next, we will discuss how the secondary CWIs lead, in the general case, to a set of algebraic equations for the primary constants.  This set of equations may be extracted through an analysis of the zero-momentum limit of the secondary CWI.
In this limit the triple-$K$ integrals simplify, although the precise details of the analysis depend on whether or not the triple-$K$ integrals involved require regularisation.
When the regulator can be removed from all triple-$K$ integrals the procedure is relatively simple, however when the regulator cannot be removed special care must be taken when regulating the 2-point functions that appear in the right-hand side of the secondary CWIs.

\subsubsection{\texorpdfstring{$\< T^{\mu_1 \nu_1} T^{\mu_2 \nu_2} \mathcal{O} \>$ for free scalars}{<TTO> for free scalars}} \label{ch:exseccwi}

Let us begin by discussing our example correlation function $\lla T^{\mu_1 \nu_1} T^{\mu_2 \nu_2} \mathcal{O} \rra$.  We derived the secondary CWIs earlier in \eqref{e:secondaryCWIs1} and \eqref{e:secondaryCWIs2}, where the terms on the right-hand side of these equations are given by  the transverse Ward identity \eqref{e:exWard}. We now want to show that these data fix two out of three primary constants in the solution \eqref{e:ATTO0} - \eqref{e:ATTO} of the primary CWIs.
To fix the final remaining constant then requires additional physical input in the form of the specific field content.

Since the regulator $\epsilon$ in the triple-$K$ integrals \eqref{e:exj2000} - \eqref{e:exj0000} cannot be removed, we must assume that the primary constants $\alpha_2$ and $\alpha_3$ depend on the regulator $\epsilon$ as well. As remarked earlier in section \ref{ch:regul}, while each individual component may depend on the regulator, the full expression for the form factors $A_j$ cannot. Let us therefore define the power series expansions
\begin{equation} \label{e:an}
\alpha_j = \sum_{n = -\infty}^\infty \alpha_j^{(n)} \epsilon^n, \quad j = 2,3.
\end{equation}
Since the integral $J_{4 \{000\}}$ is finite, we can assume that the constant $\alpha_1$ does not depend on the regulator, \textit{i.e.}, $\a_1 = \a_1^{(0)}$.

We start by substituting the solutions (\ref{e:ATTO0}, \ref{e:ATTO2}) together with the series expansions \eqref{e:an} into the secondary CWI \eqref{e:secondaryCWIs1}, with right-hand side given by \eqref{e:exWard}.  Organising the equations according to powers of $\epsilon$,
upon sending $\epsilon \rightarrow 0$ all equations associated with negative powers of $\epsilon$ must vanish. In the present case, this yields $\alpha_2^{(n)} = 0$ for all $n \leq 0$. The equation coming from the $\epsilon^0$ terms then reads
\begin{equation} \label{e:secondaryCWIs1example}
- \frac{3}{p_3} \left(\frac{\pi}{2}\right)^{\frac{3}{2}} ( \alpha_2^{(1)} + 3 \alpha^{(0)}_1 ) = - \frac{3}{16 p_3}.
\end{equation}
The same procedure may now be applied to the remaining secondary CWI \eqref{e:secondaryCWIs2}, yielding $\alpha_3^{(n)} = 0$ for all $n \leq 0$ and
\begin{equation}
\left( \frac{\pi}{2} \right)^{\frac{3}{2}} \left[ 2 \alpha_3^{(1)} + 3 \alpha_1^{(0)} - \frac{1}{16} \left( \frac{2}{\pi} \right)^{\frac{3}{2}} \right] \frac{p_1^2 + 3 p_2^2 - 3 p_3^2}{p_3} + \frac{3}{4} p_3 = \frac{3}{4} p_3.
\end{equation}
Putting everything together, we have
\begin{align}
\alpha_2 & = \left[ - 3 \alpha_1 + \frac{1}{16} \left( \frac{2}{\pi} \right)^{\frac{3}{2}} \right] \epsilon + O(\epsilon^2), \label{e:exa2a1} \\
\alpha_3 & = \frac{1}{2} \left[ - 3 \alpha_1 + \frac{1}{16} \left( \frac{2}{\pi} \right)^{\frac{3}{2}} \right] \epsilon + O(\epsilon^2), \label{e:exa3a1}
\end{align}
where the constant $\alpha_1$ remains undetermined by Ward identities.
When we take the limit $\epsilon\rightarrow 0$, the leading terms of order $\epsilon$ in these expressions then multiply $1/\epsilon$ poles in the $J_{2+\epsilon \{000\}}$, $J_{1+\epsilon \{001\}}$ and $J_{0+\epsilon \{000\}}$ integrals yielding the correct finite result.
The omitted higher order terms in \eqref{e:exa2a1} and\eqref{e:exa3a1} make no contribution.

Finally, we can check the results of this section against the result \eqref{e:excj} for the specific theory discussed in section \ref{ch:Jint3}.  Inserting the value of $\alpha_1$ from \eqref{e:excj} into \eqref{e:exa2a1} and \eqref{e:exa3a1} we indeed recover the correct result $\alpha_2 = \alpha_3 = 0$ up to insignificant $O(\epsilon^2)$ terms.

\subsubsection{Simplifications in the generic case} \label{ch:seccwi1}

In the previous section we substituted the full solutions of the primary CWIs into the secondary CWIs in order to extract more information about the primary constants. At first sight this procedure might appear hard to carry out in general since the triple-$K$ integrals usually cannot be expressed in terms of elementary functions. It turns out, however, that examining the zero-momentum limit leads to simple algebraic equations for the primary constants.

In this section, for reasons of simplicity, we will assume that each triple-$K$ integral in a solution to the primary CWIs can be defined by an analytic continuation and the regulator can be completely removed. We will refer to this as the `generic case' in the present and following sections.  We will then analyse the remaining cases later.

In the zero-momentum limit
\begin{equation}
\bs{p}_3 \rightarrow 0, \qquad \bs{p}_1 = - \bs{p}_2 = \bs{p},
\end{equation}
we have
\begin{align} \label{e:Kexp}
p_3^{\beta_3} K_{\beta_3}(p_3 x) & = \left[ \frac{2^{\beta_3 - 1} \Gamma(\beta_3)}{x^{\beta_3}} + O(p_3^2) \right] \nn\\
& \qquad + p_3^{2 \beta_3} \left[ 2^{-\beta_3 - 1} \Gamma(- \beta_3) x^{\beta_3} + O(p_3^2) \right], \quad \beta_3 \notin \Z, \\
K_0(p_3 x) & = - \log p_3 - \log x + \log 2 - \gamma_E + O(p_3^2), \label{e:K0exp}
\end{align}
\begin{align}
p_3^n K_n(p_3 x) & = \left[ \frac{2^{n-1} \Gamma(n)}{x^n} + O(p_3^2) \right] + p_3^{2 n} \left[ \frac{(-1)^{n+1}}{2^n \Gamma(n+1)} x^n \log p_3 + \right. \nn\\
& \left. \qquad + \text{ultralocal} + O(p_3^2) \right], \quad n = 1,2,3,\ldots \label{e:Knexp}
\end{align}
From these expressions one can see that the zero momentum limit of $p_3^{\beta_3} K_{\beta_3} (p_3 x)$ exists if $\beta_3 > 0$. Since for any correlation function and any form factor $\beta_3 = \Delta_3 - \frac{d}{2} + k_3$ with non-negative $k_3$, this condition is satisfied if $\Delta_3 > \frac{d}{2}$. (For conserved currents and for the stress-energy tensor we thus have $\beta_3>0$ automatically.)  We will return to discuss the case where $\beta_3 \leq 0$ later in the text.

Assuming $\beta_3 > 0$ then, we can calculate the limit of the triple-$K$ integrals in the generic case
\begin{equation} \label{e:limJ}
\lim_{p_3 \rightarrow 0} I_{\alpha \{\beta_j\}}(p, p, p_3) = l_{\alpha \{\beta_j\}} p^{\beta_t - \alpha - 1},
\end{equation}
where, using the result \eqref{e:I2K}, we find
\begin{equation} \label{e:l}
l_{\alpha \{ \beta_k \}} = \frac{2^{\alpha - 3} \Gamma(\beta_3)}{\Gamma(\alpha - \beta_3 + 1)} \prod_{\epsilon_1, \epsilon_2 \in \{-1, 1\}} \Gamma \left( \frac{\alpha - \beta_3 + 1 + \epsilon_1 \beta_1 + \epsilon_2 \beta_2}{2} \right),
\end{equation}
which is valid away from poles of the gamma function.

Since the derivatives in the $\Lo$ and $\Ro$ operators defined in \eqref{e:L} and \eqref{e:R} acting on triple-$K$ integrals can also be expressed via \eqref{e:Jid1} in terms of triple-$K$ integrals, this procedure leads to algebraic constraints on the primary constants.

\subsubsection{Derivation of the equations in the generic case} \label{ch:seccwi2}

Let us illustrate the considerations above in the case of the correlator $\lla T^{\mu_1 \nu_1} T^{\mu_2 \nu_2} \mathcal{O} \rra$. The secondary CWIs are given by \eqref{e:secondaryCWIs1} and \eqref{e:secondaryCWIs2},
\begin{align}
& \Lo_{2} A_1 + \Ro A_2 = 2 d \cdot \text{coeff. of } p_2^{\mu_1} p_3^{\mu_2} p_3^{\nu_2} \text{ in } p_{1 \nu_1} \lla T^{\mu_1 \nu_1}(\bs{p}_1) T^{\mu_2 \nu_2}(\bs{p}_2) \mathcal{O}(\bs{p}_3) \rra, \label{e:secondaryCWIs1a} \\
& \Lo_{2} A_2 + 4 \Ro A_3 = 8 d \cdot \text{coeff. of } \delta^{\mu_1 \mu_2} p_3^{\mu_2} \text{ in } p_{1 \nu_1} \lla T^{\mu_1 \nu_1}(\bs{p}_1) T^{\mu_2 \nu_2}(\bs{p}_2) \mathcal{O}(\bs{p}_3) \rra, \label{e:secondaryCWIs2a}
\end{align}
with $\Delta_1 = \Delta_2 = d$ and $\Lo$ and $\Ro$ defined by \eqref{e:L} and \eqref{e:R}. The right-hand sides are semi-local and can be expressed in terms of 2-point functions by means of the transverse Ward identities. In section \ref{ch:transverseWI} we found the Ward identity \eqref{e:pTTO}, which reads
\begin{equation} \label{e:pTTO1}
p_1^{\nu_1} \lla T_{\mu_1 \nu_1}(\bs{p}_1) T_{\mu_2 \nu_2}(\bs{p}_2) \mathcal{O}(\bs{p}_3) \rra = 2 p_1^{\nu_1} \lla \frac{\delta T_{\mu_1 \nu_1}}{\delta g^{\mu_2 \nu_2}}(\bs{p}_1, \bs{p}_2) \mathcal{O}(\bs{p}_3) \rra.
\end{equation}
We omit the group index on $\mathcal{O}$ as we consider only one scalar operator. First, we want to argue that the right-hand side of \eqref{e:pTTO1} vanishes if $\beta_3 > 0$, unless some specific conditions on conformal dimensions are met. Therefore, in this section we will assume that the right-hand sides of \eqref{e:secondaryCWIs1a} and \eqref{e:secondaryCWIs2a} vanish, leaving a discussion of the various special cases to the following sections. Indeed, the only possibility for a non-vanishing right-hand side of \eqref{e:pTTO1} is if the functional derivative $\delta T_{\mu_1 \nu_1} / \delta g^{\mu_2 \nu_2}$ contains the operator $\mathcal{O}$ or its descendants. Since the dilatation degree of $\delta T_{\mu_1 \nu_1} / \delta g^{\mu_2 \nu_2}$ is equal to $d$, this requires $d = \Delta_3 + 2n$ where $n$ is a non-negative integer. Consider first the case $\Delta_3 = d$. In this case, we can write the most general form of $\delta T_{\mu_1 \nu_1} / \delta g^{\mu_2 \nu_2}$ which contains $\mathcal{O}$ as
\begin{equation}
\frac{\delta T_{\mu_1 \nu_1}(\bs{x})}{\delta g^{\mu_2 \nu_2}(\bs{y})} = \left[ c_1 \delta_{\mu_1 \nu_1} \delta_{\mu_2 \nu_2}  + c_2 \delta_{(\mu_1 (\mu_2} \delta_{\nu_2) \nu_1)} \right] \delta(\bs{x}-\bs{y}) \mathcal{O}(\bs{x}) + \ldots
\end{equation}
where $c_1$ and $c_2$ are numerical constants. If, on the other hand, $d = \Delta_3 + 2n$ with $n > 0$ then derivatives acting on both $\mathcal{O}$ and $\delta(\bs{x}-\bs{y})$ may also appear. In all cases, the Fourier transform reads
\begin{equation} \label{e:pTTOc}
\lla \frac{\delta T_{\mu_1 \nu_1}}{\delta g^{\mu_2 \nu_2}}(\bs{p}_1, \bs{p}_2) \mathcal{O}(\bs{p}_3) \rra = P_{\mu_1 \nu_1 \mu_2 \nu_2}(p_1^2, p_2^2, p_3^2) \lla \mathcal{O}(\bs{p}_3) \mathcal{O}(-\bs{p}_3) \rra,
\end{equation}
where $P$ is some polynomial built from momenta and the metric $\delta_{\mu\nu}$, with kinematic dependence on squares of momenta only. This form arises from the Fourier transform of the position space expression containing derivatives acting on delta functions and on the 2-point function. Since $\lla \mathcal{O}(\bs{p}_3) \mathcal{O}(-\bs{p}_3) \rra \sim p_3^{2 \beta_3}$, the expression vanishes in the $p_3 \rightarrow 0$ limit as long as $\beta_3 > 0$.

We now substitute the solutions of the primary CWIs (\ref{e:ATTO0}, \ref{e:ATTO2}) into the left-hand side of \eqref{e:secondaryCWIs1a} and take the zero-momentum limit. Assuming the regulator can be removed (see section \ref{ch:explicit} if not), the result is
\begin{equation}  \label{e:toexseccwi}
- \frac{l_{\frac{d}{2} + 1 \{ \frac{d}{2}, \frac{d}{2}, \Delta_3 - \frac{d}{2} \}}}{2} p^{\Delta_3 - 2} (2 + 2d - \Delta_3) \left[ \alpha_2 + \alpha_1 ( \Delta_3 + 2) (\Delta_3 - d + 2) \right] = 0,
\end{equation}
with $l_{\alpha\{\beta_k\}}$ as defined in \eqref{e:l}.
We then find
\begin{equation} \label{e:a2a1}
\alpha_2 = - (\Delta_3 + 2)(\Delta_3 + 2 - d) \alpha_1.
\end{equation}
Applying the same reasoning as above to \eqref{e:secondaryCWIs2a}, we  likewise find
\begin{equation} \label{e:a3a1}
\alpha_3 = \frac{1}{4} \Delta_3 (\Delta_3 + 2) (\Delta_3 - d) (\Delta_3 + 2 - d) \alpha_1.
\end{equation}

Summarising, in this and the previous section we presented a method for extracting algebraic dependencies between the primary constants following from the secondary CWIs. The analysis was performed in the generic case, where the regulator can be removed from all triple-$K$ integrals involved. Note that the results \eqref{e:a2a1} and \eqref{e:a3a1} agree with our example \eqref{e:exa2a1} and \eqref{e:exa3a1} in the leading term in $\epsilon$ only, \textit{i.e.}, they correctly predict $\alpha_2 = \alpha_3 = 0 + O(\epsilon)$. This is due to the fact that in our example the regulator cannot be removed from each triple-$K$ integral separately. Therefore, it does not satisfy the assumption of this section. Note, however, that the analysis of the generic case is sufficient if one is merely interested in finding the solution up to semi-local terms. This is because the possible non-generic cases arise due to the regularisation procedure, correcting the generic solution by at most semi-local terms.

\subsubsection{Secondary conformal Ward identities in all cases} \label{ch:allcases}

Let us now return to the discussion of the secondary CWIs in the case where the regulator cannot be removed in certain triple-$K$ integrals. In principle, the procedure is simple. One must keep the explicit dependence on $\epsilon$, both in the triple-$K$ integrals as well as in the primary constants, and carry out the analysis of sections \ref{ch:seccwi1} and \ref{ch:seccwi2} order by order in the regulator. Note that if the index of a Bessel function is integral, then the expansions \eqref{e:K0exp} and \eqref{e:Knexp} should be used instead of \eqref{e:Kexp}.

The only difference with section \ref{ch:seccwi2} is that looking at the zero-momentum limit may not be enough. We should look at both terms following from the first and second brackets in \eqref{e:Kexp}, \textit{i.e.}, the coefficients of $p_3^0$ and $p_3^{2 \beta_3}$ in the expansion in powers of $p_3$ with $p_1 = p_2 = p$. If the Bessel index is integral, then we should use \eqref{e:K0exp} and \eqref{e:Knexp} and look for the coefficients of $p_3^0$ and $p_3^{2 \beta_3} \log p_3$. This procedure will provide a set of algebraic equations relating the primary constants.

Let us now explain why this procedure is valid for $\beta_3 < 0$ and why the remaining terms in the expansions \eqref{e:Kexp} - \eqref{e:Knexp} are irrelevant.
First, the unitarity bound requires $-1 \leq \beta_3$. The unitarity bound can only be saturated by a non-composite scalar operator in a free field theory, \cite{Maldacena:2011} and \cite{Weinberg:2012cd}.  We can therefore assume $-1 < \beta_3 < 0$. It turns out that the considerations encountered in the case $\beta_3 > 0$ remain valid here. Since the zero-momentum limit does not exist in this case, we are going to look for the coefficient of $p_3^0$ in the expansion in $p_3$. The key observation is that on the left-hand sides of the secondary CWIs such as (\ref{e:secondaryCWIs1a}, \ref{e:secondaryCWIs2a}), the differential operators $\Lo$ and $\Ro$ defined by \eqref{e:L} and \eqref{e:R} do not contain derivatives with respect to $p_3$, and can only increase powers of $p_3$ by two. Therefore, the coefficient of $p_3^0$ in the series expansion in $p_3$ remains unaltered provided $-1 < \beta_3$. A similar analysis applies to the right-hand sides of the secondary CWIs.

Let us now examine why it is sufficient to look at the leading coefficients in \eqref{e:Kexp} - \eqref{e:Knexp} only. From \eqref{e:serI}, we know that in each successive term the power of the integration variable $x$ increases by two. After taking the zero-momentum limit, the integral \eqref{e:I2K} therefore leads to essentially the same expression as \eqref{e:l} with $\beta_3 \mapsto \beta_3 + 2n$, $n$ being a non-negative integer, plus some finite pre-factor following from the series expansion of the Bessel $K$ function. Since the singularities manifest themselves as poles of the gamma functions, we see that the result cannot be more singular than the original $l_{\alpha \{\beta_j\}}$.

\subsubsection{Back to the example} \label{ch:explicit}

Finally, let us see how the general considerations of the previous section apply in the case of the correlator $\lla T^{\mu_1 \nu_1} T^{\mu_2 \nu_2} \mathcal{O} \rra$. First, we carry out the same analysis as in sections \ref{ch:seccwi1} and \ref{ch:seccwi2} but keep the regulator explicitly. In place of \eqref{e:toexseccwi}, we then find
\begin{align}
0 & = l_{\frac{d}{2} + 1 + u \epsilon \{ \frac{d}{2} + v \epsilon, \frac{d}{2} + v \epsilon, \Delta_3 - \frac{d}{2} + v \epsilon \}} p^{\Delta_3 - 2 + \epsilon (3 v - u)} \left\{ - \frac{1}{2} (2 + 2d - \Delta_3 + (u + v)\epsilon ) \alpha_2^I  \right.\nn\\
& \qquad + \alpha^I_1\,\Big( \frac{2 + 2d - \Delta_3 + (u+v)\epsilon}{3 + d - \Delta_3 + (u-v) \epsilon}\Big) \left[ 2(2 + \Delta_3)(3 + d - \Delta_3)(d - 2 - \Delta_3) \right.\nn\\
& \qquad\qquad + \epsilon \left( 4 v + 3 v(4 + 2d - 3 \Delta_3)(d - \Delta_3) - u( 28+8d+2 d^2-8 \Delta_3 - 7 \Delta_3 d + 5 \Delta_3^2) \right) \nn \\
& \qquad\qquad + \epsilon^2 (u - v) \left( v (10 + 9 d - 10 \Delta_3) + u(4 \Delta_3 - 3 d - 10) \right) \nn\\
& \qquad\qquad \left. + O( (u-v)^2 \epsilon^3 ) \right] \Big\}, \label{e:eqexcwi}
\end{align}
with $l_{\alpha\{\beta_k\}}$ as defined in \eqref{e:l}.
If the $\epsilon \rightarrow 0$ limit exists, we recover \eqref{e:toexseccwi}. If $\Delta_3$ satisfies certain non-generic relations, however, the limit does not exist. In this paper we will not solve for all possible special cases, postponing this task to the follow-up paper \cite{II}.

In the case of $\lla T^{\mu_1 \nu_1} T^{\mu_2 \nu_2} \mathcal{O} \rra$, we  substitute $d = 3$ and $\Delta_3 = 1$. This leads to $\alpha_2 = O(\epsilon)$. Since the integrals building the form factor $A_2$ are at most linearly divergent, this equation does not specify the form of subleading orders. Instead, the subleading $O(\epsilon)$ term is determined by analysing the coefficient of $p_3$ following from the expansion of \eqref{e:Kexp}. To write this equation, we define two additional constants $c_1^K$ and $c_2^K$ by
\begin{align}
& \left. p_1^{\nu_1} \lla \frac{\delta T_{\mu_1 \nu_1}}{\delta g^{\mu_2 \nu_2}}(\bs{p}_1, \bs{p}_2) \mathcal{O}^I(\bs{p})_3 \rra \right|_{p_1 = p_2 = p} = c^K_1 p_2^{\mu_1} p_3^{\mu_2} p_3^{\nu_2} p^{d - \Delta_3 - 2} \lla \mathcal{O}^I(\bs{p}_3) \mathcal{O}^K(-\bs{p}_3) \rra \nn\\
& \qquad\qquad + c^K_2 \delta^{\mu_1 \mu_2} p_3^{\nu_2} p^{d - \Delta_3} \lla \mathcal{O}^I(\bs{p}_3) \mathcal{O}^K(-\bs{p}_3) \rra + \ldots \label{e:TOOc1c2}
\end{align}
By $p_1 = p_2 = p$, we mean here the following procedure: first, the correlation function on the left-hand side is expanded in terms of simple tensors according to the convention \eqref{e:momenta_choice}, then second, one applies $p_1 = p_2 = p$ to each coefficient separately. In this case \eqref{e:eqexcwi} simplifies to
\begin{align}
& l_{\frac{d}{2} + 3 + u \epsilon, \{ \frac{d}{2} + v \epsilon, \frac{d}{2} + v \epsilon, \frac{d}{2} - \Delta - v \epsilon \}} \left[ - \left(6 + 2 \Delta + (u+3v) \epsilon \right) \alpha_1^I \right.\nn\\
& \qquad\qquad \left. + \: \frac{2(3+\Delta + (u+v)\epsilon)}{(-2+d-\Delta-(u-v)\epsilon)(2+\Delta+(u+v)\epsilon)} \alpha_2^I \right] = \nn\\
& \qquad = 4 \delta_{d, \Delta-2-2n} (d + 2 u\epsilon) \Gamma \left( \frac{d}{2} - \Delta - v \epsilon \right) c_1^I c_{\O}.
\end{align}
In our case $d = 3$, $\Delta_3 = 1$ and we find
\begin{equation} \label{e:a2new}
\alpha_2 = - (u - v) \epsilon \left[ 3 \alpha_1 + \frac{8 \sqrt{2}}{\pi} c_1 c_{\O} \right] + O(\epsilon^2).
\end{equation}

A similar analysis may be carried out for the second CWI, \eqref{e:secondaryCWIs2a}. Putting all the ingredients together, we can now write the most general form of the correlator $\lla T^{\mu_1 \nu_1} T^{\mu_2 \nu_2} \mathcal{O}^I \rra$ for $d=3$ and $\Delta_2=\Delta_3=1$. Using the results for the triple-$K$ integrals from section \ref{ch:Jint3} we have
\begin{align}
A_1^I &= \frac{\a_1^I}{p_3 a_{123}^4} \left[ p_3^2 + 4 p_3 a_{12} + 3 (a_{12}^2 + 2 b_{12}) \right] \label{e:exsolA1}, \\
A_2^I &= \frac{\a_1^I}{p_3 a_{123}^3} \left[ p_3^3 + 3 p_2^2 a_{12} + p_3 (-a_{12}^2 + 8 b_{12}) - 3 a_{12}^3 \right] - \frac{4 \sqrt{\pi} c_1^I c_{\mathcal{O}}}{p_3}, \\
A_3^I &= \frac{\a_1^I (a_{12} - p_3)}{4 p_3 a_{123}^2} \left[ - p_3^3 - 3 p_3^2 a_{12} + p_3 (a_{12}^2 - 10 b_{12}) + 3 a_{12} ( a_{12}^2 - 2 b_{12}) \right] \nn\\
& \qquad + \frac{c_\mathcal{O}}{p_3} \sqrt{\pi} \left[ (c_1^I - 3 c_2^I) (p_1^2 + p_2^2) + 3 (c_1^I + c_2^I) p_3^2 \right], \label{e:exsolA3}
\end{align}
where we have defined the symmetric polynomials in momentum magnitudes
\begin{align}
& a_{123} = p_1 + p_2 + p_3, \qquad b_{123} = p_1 p_2 + p_1 p_3 + p_2 p_3, \qquad c_{123} = p_1 p_2 p_3, \nn\\
& a_{ij} = p_i + p_j, \qquad\qquad b_{ij} = p_i p_j,
\end{align}
where $i,j = 1,2,3$. The solution for this correlator is thus uniquely determined up to one numerical constant $\a_1^I$.  The remaining constants in the solution, namely $c_\mathcal{O}$, $c_1^I$ and $c_2^I$, are determined by the 2-point function normalisations: $c_\mathcal{O}$ is given in \eqref{e:OO} while $c_1^I$ and $c_2^I$ are given in \eqref{e:TOOc1c2}.

One can check this result against our example in section \ref{ch:exseccwi}. From  \eqref{e:exOO} and \eqref{e:exWard}, the solution for the parameters is
\begin{equation}
c_{\mathcal{O}} = \frac{1}{4 \sqrt{\pi}}, \qquad\qquad c_1^I = - \frac{1}{16}, \qquad\qquad c_2^I = 0.
\end{equation}
As we can see, \eqref{e:a2new} simplifies to \eqref{e:exa2a1} exactly in the regularisation scheme $u = 1$ and $v = 0$. As discussed in section \eqref{ch:regul}, we are free to choose such a scheme since the resulting form factor $A_2$ is finite, and therefore independent of the regularisation scheme. A similar analysis can be carried out for the second secondary CWI \eqref{e:secondaryCWIs2}.

\section{\texorpdfstring{Worked example: $\< T^{\mu_1 \nu_1} J^{\mu_2} J^{\mu_3} \>$}{Worked example: <TJJ>}} \label{sec:TJJworkedexample}

Now that our general method is complete, in this section we present a full worked example, the $\lla T^{\mu_1 \nu_1} J^{\mu_2} J^{\mu_3} \rra$ correlation function. Here we will take $J^{\mu}$ to be a conserved $U(1)$ current; more general results are listed in Part II. This correlator provides a useful test case as, while more complex than the $\langle T^{\mu_1 \nu_1} T^{\mu_2 \nu_2} \mathcal{O} \rangle$ correlator we used to illustrate the method in earlier sections, it is nonetheless simpler than correlators with more stress-energy tensors.

We will also discuss the complete evaluation of all integrals in 
$d=3$
and present a concrete model, free fermions, where these correlators can be explicitly computed by standard Feynman diagrams.  These results provide a nontrivial consistency check on our method.

\subsection{Primary conformal Ward identities}

We start with the analysis of primary CWIs for $\lla T^{\mu_1 \nu_1} J^{\mu_2} J^{\mu_3} \rra$ in general Euclidean dimension $d$. For the decomposition of the transverse-traceless part of $\lla T^{\mu_1 \nu_1} J^{\mu_2} J^{\mu_3} \rra$ we follow the analysis of section \ref{ch:Tensor_structure}. The decomposition consists of four form factors,
\begin{align}
& \lla t^{\mu_1 \nu_1}(\bs{p}_1) j^{\mu_2}(\bs{p}_2) j^{\mu_3}(\bs{p}_3) \rra \nn\\
& \qquad = \Pi^{\mu_1 \nu_1}_{\alpha_1 \beta_1}(\bs{p}_1) \pi^{\mu_2}_{\alpha_2}(\bs{p}_2) \pi^{\mu_3}_{\alpha_3}(\bs{p}_3) \left[ A_1 p_2^{\alpha_1} p_2^{\beta_1} p_3^{\alpha_2} p_1^{\alpha_3} + A_2 \delta^{\alpha_2 \alpha_3} p_2^{\alpha_1} p_2^{\beta_1} \right. \nn\\
& \qquad \qquad + \: A_3 \delta^{\alpha_1 \alpha_2} p_2^{\beta_1} p_1^{\alpha_3} + A_3(p_2 \leftrightarrow p_3) \delta^{\alpha_1 \alpha_3} p_2^{\beta_1} p_3^{\alpha_2} \nonumber \\
& \left. \qquad \qquad + \: A_4 \delta^{\alpha_1 \alpha_3} \delta^{\alpha_2 \beta_1} \right]. \label{e:tjjdecomp}
\end{align}
Here, $p_2 \leftrightarrow p_3$ denotes exchange of the arguments $p_2$ and $p_3$, \textit{i.e.}, $A_3(p_2 \leftrightarrow p_3) = A_3(p_1, p_3, p_2)$. If on the other hand no arguments are given then the standard ordering is assumed, \textit{i.e.}, $A_3 = A_3(p_1, p_2, p_3)$. Note that the form factors $A_1$, $A_2$ and $A_4$ are symmetric under $p_2 \leftrightarrow p_3$,
\begin{equation}
A_j(p_1, p_3, p_2) = A_j(p_1, p_2, p_3), \qquad j \in \{1,2,4\},
\end{equation}
while the form factor $A_3$ does not exhibit any symmetry properties.

Next, the primary CWIs can be extracted by means of the procedure described in section \ref{ch:special}. These CWIs are
\begin{equation}
\begin{array}{ll}
\K_{12} A_1 = 0, & \qquad\qquad \K_{13} A_1 = 0, \\
\K_{12} A_2 = -2 A_1, & \qquad\qquad \K_{13} A_2 = - 2 A_1, \\
\K_{12} A_3 = 0, & \qquad\qquad \K_{13} A_3 = 4 A_1, \\
\K_{12} A_4 = 2 A_3, & \qquad\qquad \K_{13} A_4 = 2 A_3(p_2 \leftrightarrow p_3).
\end{array}
\end{equation}
The solution follows from the analysis of section \ref{ch:sol_to_pri_CWIs},
\begin{align}
\label{e:fermAform0}
A_1 & = \alpha_1 J_{4 \{000\}}, \\
A_2 & = \alpha_1 J_{3 \{100\}} + \alpha_2 J_{2 \{000\}}, \\
\label{e:fermAform3}
A_3 & = 2 \alpha_1 J_{3 \{001\}} + \alpha_3 J_{2 \{000\}}, \\
A_4 & = 2 \alpha_1 J_{2 \{011\}} + \alpha_3 \left( J_{1 \{010\}} + J_{1 \{001\}} \right) + \alpha_4 J_{0 \{000\}}. \label{e:fermAform}
\end{align}

\subsection{Evaluation of secondary conformal Ward identities} \label{ch:tran_tjj}

The independent secondary CWIs for $\lla T^{\mu_1 \nu_1} J^{\mu_2} J^{\mu_3} \rra$ are listed in the second part of the paper and read
\begin{align}
& (*) \ \Lo_{4} A_1 + \Ro \left[ A_3 - A_3(p_2 \leftrightarrow p_3) \right] \nn\\
& \qquad = 2 d \cdot \text{coefficient of } p_2^{\mu_1} p_3^{\mu_2} p_1^{\mu_3} \text{ in } p_{1 \nu_1} \lla T^{\mu_1 \nu_1}(\bs{p}_1) J^{\mu_2}(\bs{p}_2) J^{\mu_3}(\bs{p}_3) \rra, \label{e:fermsec1} \\
& \Lo'_{3} A_1 + 2 \Ro' \left[ A_3 - A_2 \right] \nn\\
& \qquad = 2 d \cdot \text{coefficient of } p_2^{\mu_1} p_2^{\nu_1} p_1^{\mu_3} \text{ in } p_{2 \mu_2} \lla T^{\mu_1 \nu_1}(\bs{p}_1) J^{\mu_2}(\bs{p}_2) J^{\mu_3}(\bs{p}_3) \rra, \label{e:fermsec2} \\
& \Lo_{2} A_2 - p_1^2 \left[ A_3 - A_3(p_2 \leftrightarrow p_3) \right] \nn\\
& \qquad = 2 d \cdot \text{coefficient of } \delta^{\mu_2 \mu_3} p_2^{\mu_1} \text{ in } p_{1 \nu_1} \lla T^{\mu_1 \nu_1}(\bs{p}_1) J^{\mu_2}(\bs{p}_2) J^{\mu_3}(\bs{p}_3) \rra, \label{e:fermsec3} \\
& \Lo_{4} A_3 - 2 \Ro A_4 \nn\\
& \qquad = 4 d \cdot \text{coefficient of } \delta^{\mu_1 \mu_2} p_1^{\mu_3} \text{ in } p_{1 \nu_1} \lla T^{\mu_1 \nu_1}(\bs{p}_1) J^{\mu_2}(\bs{p}_2) J^{\mu_3}(\bs{p}_3) \rra, \label{e:fermsec4}
\end{align}
where $\Lo$ and $\Ro$ operators are defined in \eqref{e:L} and \eqref{e:R}. They can be obtained by the procedure outlined in section \ref{ch:secondaryCWIs}. Note that there are four primary constants and four secondary CWIs.  As some of the secondary CWIs are trivially satisfied, however, not all four of the primary constants are fixed, as we expect from the position space analysis \cite{Osborn:1993cr}.  Secondary CWIs that are trivially satisfied are denoted by asterisks in the second part of the paper (for example \eqref{e:fermsec1} above is of this type).

Before solving the secondary CWIs, we must simplify the semi-local terms appearing on their right-hand sides. Differentiating (\ref{e:toWardJ}, \ref{e:toWardT}, \ref{e:toWardTr}) we find the following transverse and trace Ward identities,
\begin{align}
& p_1^{\nu_1} \lla T_{\mu_1 \nu_1}(\bs{p}_1) J^{\mu_2}(\bs{p}_2) J^{\mu_3}(\bs{p}_3) \rra = \nn\\
& \qquad = p_1^{\nu_1} \lla \frac{\delta T_{\mu_1 \nu_1}}{\delta A_{\mu_3}}(\bs{p}_1, \bs{p}_3) J^{\mu_2}(\bs{p}_2) \rra + p_1^{\nu_1} \lla \frac{\delta T_{\mu_1 \nu_1}}{\delta A_{\mu_2}}(\bs{p}_1, \bs{p}_2) J^{\mu_3}(\bs{p}_3) \rra \nn\\
& \qquad\qquad - \: p_3^{\mu_1} \lla J^{\mu_2}(\bs{p}_2) J^{\mu_3}(-\bs{p}_2) \rra - p_2^{\mu_1} \lla J^{\mu_2}(\bs{p}_3) J^{\mu_3}(-\bs{p}_3) \rra \nn\\
& \qquad\qquad + \: \delta^{\mu_3}_{\mu_1} p_{3 \alpha} \lla J^{\mu_2}(\bs{p}_2) J^{\alpha}(-\bs{p}_2) \rra
 + \delta^{\mu_2}_{\mu_1} p_{2 \alpha} \lla J^{\alpha}(\bs{p}_3) J^{\mu_3}(-\bs{p}_3) \rra, \label{e:p1TJJ}\\
& p_{2 \mu_2} \lla T_{\mu_1 \nu_1}(\bs{p}_1) J^{\mu_2}(\bs{p}_2) J^{\mu_3}(\bs{p}_3) \rra = \nn\\
& \qquad = 2 p_{2 \mu_2} \lla \frac{\delta J^{\mu_2}}{\delta g^{\mu_1 \nu_1}}(\bs{p}_2, \bs{p}_1) J^{\mu_3}(\bs{p}_3) \rra + \delta_{\mu_1 \nu_1} p_{1 \alpha} \lla J^{\alpha}(\bs{p}_3) J^{\mu_3}(-\bs{p}_3) \rra, \label{e:p2TJJ} \\
& \lla T(\bs{p}_1) J^{\mu_2}(\bs{p}_2) J^{\mu_3}(\bs{p}_3) \rra = \lla \frac{\delta T}{\delta A_{\mu_2}}(\bs{p}_1, \bs{p}_2) J^{\mu_3}(\bs{p}_3) \rra + \lla \frac{\delta T}{\delta A_{\mu_3}}(\bs{p}_1, \bs{p}_3) J^{\mu_2}(\bs{p}_2) \rra. \label{e:trTJJ}
\end{align}
In the next section we will extract algebraic equations between the primary constants by taking the zero-momentum limit $p_3 \rightarrow 0$. The details of this procedure are described in  section \ref{ch:sol_sec_CWI}. We will find that in the zero-momentum limit the right-hand sides of the secondary CWIs \eqref{e:fermsec1} - \eqref{e:fermsec3} are given by
\begin{align}
\lim_{\substack{p_3 \rightarrow 0\\p_1 = p_2 = p}} \text{coefficient of } p_2^{\mu_1} p_2^{\nu_1} p_1^{\mu_3} \text{ in } p_{2 \mu_2} \lla T^{\mu_1 \nu_1}(\bs{p}_1) J^{\mu_2}(\bs{p}_2) J^{\mu_3}(\bs{p}_3) \rra & = 0, \label{e:to0a} \\
\lim_{\substack{p_3 \rightarrow 0\\p_1 = p_2 = p}} \text{coefficient of } p_2^{\mu_1} p_3^{\mu_2} p_1^{\mu_3} \text{ in } p_{1 \nu_1} \lla T^{\mu_1 \nu_1}(\bs{p}_1) J^{\mu_2}(\bs{p}_2) J^{\mu_3}(\bs{p}_3) \rra & = 0, \label{e:to0b} \\
\lim_{\substack{p_3 \rightarrow 0\\p_1 = p_2 = p}} \text{coefficient of } \delta^{\mu_2 \mu_3} p_2^{\mu_1} \text{ in } p_{1 \nu_1} \lla T^{\mu_1 \nu_1}(\bs{p}_1) J^{\mu_2}(\bs{p}_2) J^{\mu_3}(\bs{p}_3) \rra & = \nn\\
\qquad = \text{coefficient of } \delta^{\mu_2 \mu_3} \text{ in } \lla J^{\mu_2}(\bs{p}) J^{\mu_3}(-\bs{p}) \rra. \label{e:to0c}
\end{align}

Let us start with the first result \eqref{e:to0a}. Due to conformal invariance, the only operators in $\delta J^{\mu_2} / \delta g^{\mu_1 \nu_1}$ that can give a non-vanishing result under the expectation value with the current is another current $J^{\mu}$. In general,  the descendants of the current can also give a non-vanishing 2-point function with another current. In this case, however, the dilatation degree of $\delta J^{\mu_2} / \delta g^{\mu_1 \nu_1}$ is $d-1$, and so descendants cannot appear. The most general form of the functional derivative term is therefore
\begin{equation}
\frac{\delta J^{\mu_2}}{\delta g^{\mu_1 \nu_1}} = c_1 \delta_{\mu_1 \nu_1} J^{\mu_2} + c_2 \delta^{\mu_2}_{(\mu_1} J_{\nu_1)} + \ldots
\end{equation}
where $c_1$ and $c_2$ are numerical constants and the omitted terms may contain operators from different conformal families to that of $J^{\mu}$. The 2-point function then reads
\begin{equation}
\lla \frac{\delta J^{\mu_2}}{\delta g^{\mu_1 \nu_1}}(\bs{p}_2, \bs{p}_1) J^{\mu_3}(\bs{p}_3) \rra = \left[ c_1 \delta_{\mu_1 \nu_1} \delta_{\alpha}^{\mu_2} + c_2 \delta^{\mu_2}_{(\mu_1} \delta_{\nu_1) \alpha} \right] \lla J^{\alpha}(\bs{p}_3) J^{\mu_3}(-\bs{p}_3) \rra.
\end{equation}
In the limit $p_3 \rightarrow 0$, however, the 2-point function vanishes, since it behaves as $p_3^{d-2}$ and $d>2$. The same argument works for the second term in \eqref{e:p2TJJ} and so \eqref{e:to0a} also vanishes.

Let us now establish the remaining formulae \eqref{e:to0b} and \eqref{e:to0c}. Following the same argument for the limit $p_3 \rightarrow 0$, we can restrict consideration to the following terms in \eqref{e:p1TJJ}
\begin{equation}
p_1^{\nu_1} \lla \frac{\delta T_{\mu_1 \nu_1}}{\delta A_{\mu_3}}(\bs{p}_1, \bs{p}_3) J^{\mu_2}(\bs{p}_2) \rra - p_3^{\mu_1} \lla J^{\mu_2}(\bs{p}_2) J^{\mu_3}(-\bs{p}_2) \rra + \delta^{\mu_3}_{\mu_1} p_{3 \alpha} \lla J^{\mu_2}(\bs{p}_2) J^{\alpha}(-\bs{p}_2) \rra.
\end{equation}
Using the representation \eqref{e:JJ1} it is straightforward to expand the last two terms. As usual, we must use the convention \eqref{e:momenta_choice} for the momenta associated with Lorentz indices, leading to the right-hand sides of \eqref{e:to0b} and \eqref{e:to0c}. The remaining task is then to show that there are no contributions from the first term with the functional derivative.

Since the dimension of the stress-energy tensor is $d$, while that of the conserved current is $d - 1$ and that of the source $A_\mu$ is $1$, the only possible contributions to the first term in \eqref{e:p1TJJ} are
\begin{equation}
T_{\mu \nu} = c_3 \left[ A_{\mu} J_{\nu} + A_{\nu} J_{\mu} \right] + \ldots
\end{equation}
where $c_3$ is a numerical constant and the omitted terms do not contain the current or its descendants. This definition of $c_3$ applies if the $J^\mu$ operator is the unique spin-1 conserved current in theory. If not, we can instead define the constant $c_3$ through the 2-point function
\begin{equation} \label{e:TcAJ}
\lla \frac{\delta T_{\mu_1 \nu_1}}{\delta A_{\mu_2}}(\bs{p}_1, \bs{p}_2) J^{\mu_3}(\bs{p}_3) \rra = 2 c_3 \delta^{\mu_2}_{(\mu_1} \lla J_{\nu_1)}(\bs{p}_3) J^{\mu_3}(- \bs{p}_3) \rra.
\end{equation}
After taking the functional derivative one finds that tensors $p_2^{\mu_1} p_3^{\mu_2} p_1^{\mu_3}$ and $\delta^{\mu_2 \mu_3}$ are absent in \eqref{e:p1TJJ}.

Finally, with the definition of the $c_3$ constant as in \eqref{e:TcAJ}, the same method can be applied to work out the zero-momentum limit of the right-hand side of the final secondary CWI \eqref{e:fermsec4}, yielding the result
\begin{align}
\lim_{\substack{p_3 \rightarrow 0\\p_1 = p_2 = p}} \text{coefficient of } \delta^{\mu_1 \mu_2} p_1^{\mu_3} \text{ in } p_{1 \nu_1} \lla T^{\mu_1 \nu_1}(\bs{p}_1) J^{\mu_2}(\bs{p}_2) J^{\mu_3}(\bs{p}_3) \rra & = \nn\\
\qquad = c_3 \cdot \text{coefficient of } \delta^{\mu_1 \mu_2} \text{ in } \lla J^{\mu_1}(\bs{p}) J^{\mu_2}(-\bs{p}) \rra. \label{e:to0d}
\end{align}

\subsection{Solutions to secondary conformal Ward identities}

Our goal now is to analyse the additional constraints imposed by the secondary CWIs \eqref{e:fermsec2} - \eqref{e:fermsec4} on the solution \eqref{e:fermAform0}-\eqref{e:fermAform} of the primary CWIs. We proceed as in sections \ref{ch:seccwi1} and \ref{ch:seccwi2} by taking the zero-momentum limit $p_3 \rightarrow 0$ 
to  derive algebraic equations for the primary constants.

In odd spacetime dimensions, the integrals  appearing in the expressions \eqref{e:fermAform0}-\eqref{e:fermAform3} for the  form factors $A_1, A_2, A_3$ are automatically finite, since the condition \eqref{condition} cannot be satisfied for any choice of signs.
In even spacetime dimensions, however, these integrals have divergences corresponding to solutions of \eqref{condition} with three minus signs.
Upon closer inspection, however, these divergences are at most linear in $\epsilon$.  As discussed in section \ref{ch:tripleKand2pt}, for these form factors and the corresponding secondary CWIs \eqref{e:fermsec2} - \eqref{e:fermsec3} we can then use the simple dimensional regularisation scheme \eqref{e:dimreg} for which $u=v$.
The resulting constraints on primary constants are
\begin{align}
\alpha_3 & = \alpha_2, \label{e:sectjj1} \\
\alpha_2 & = - (d + 2 v \epsilon) \alpha_1 + \frac{2^{3 - \frac{d}{2} - v \epsilon} c_J}{\Gamma^2 \left( \frac{d}{2} + v \epsilon \right)}, \label{e:sectjj2}
\end{align}
where $c_{J}$ encodes the normalisation of the 2-point function as given in \eqref{e:JJ1}, and for the right-hand sides we used \eqref{e:to0a} - \eqref{e:to0c}.

The situation is more interesting for the remaining secondary CWI , which involves the form factor $A_4$.
The integrals $J_{0\{000\}}, J_{1\{010\}}, J_{1\{001\}}$ associated with this form factor (see \eqref{e:fermAform}) diverge quadratically in $\epsilon$, and hence are potentially singular in the dimensional regularisation  $u = v$.
Expanding \eqref{e:fermsec4} to first order in $(u-v)$, we find
\begin{align}
\alpha_4 & = - (d - 2 + 2 v \epsilon) \alpha_2 - \frac{1}{2} \epsilon (u - v) \left[ (d + 2 v \epsilon) \left( 2( d - 2 + 2 v \epsilon) \alpha_1 + \alpha_2 \right) \right.\nn\\
& \qquad\qquad\qquad \left. - \: \frac{2^{5 - \frac{d}{2} - v \epsilon} c_J c_3}{\Gamma \left( \frac{d}{2} - 1 + v \epsilon \right) \Gamma \left( \frac{d}{2} + v \epsilon \right)} \right] + O((u-v)^2 \epsilon^2). \label{e:sectjj3}
\end{align}
The physical form factor $A_4$, as given in \eqref{e:fermAform}, may then be calculated as
\begin{equation}
A_4 = 2 \alpha_1 J_{2 \{011\}} + \lim_{u \rightarrow v} \left[ \alpha_3 \left( J_{1 \{010\}} + J_{1 \{001\}} \right) + \alpha_4 J_{0 \{000\}} \right],
\end{equation}
where the singularities of $J_{2\{011\}}$ in dimensional regularisation cancel against those of the remaining terms after taking the limit $u\rightarrow v$.

Our solution of the primary and secondary CWIs above depends on one undetermined primary constant as well as two different 2-point function normalisations. This result is in fact consistent with the position space result of \cite{Osborn:1993cr} (which involves only a single 2-point function normalisation) by virtue of our different definition for the 3-point function, namely
\begin{align}
\< T_{\mu_1 \nu_1}(\bs{x}) J^{\mu_2}(\bs{y}) J^{\mu_3}(\bs{z}) \> & = \frac{-1}{\sqrt{g(\bs{z})}} \frac{\delta}{\delta A_{\mu_3}(\bs{z})} \frac{-1}{\sqrt{g(\bs{y})}} \frac{\delta}{\delta A_{\mu_2}(\bs{y})} \frac{-2}{\sqrt{g(\bs{x})}} \frac{\delta}{\delta g^{\mu_1 \nu_1}(\bs{x})} Z[g^{\mu \nu}, A_\rho] \nn\\
& \qquad + \: \< \frac{\delta T_{\mu_1 \nu_1}(\bs{x})}{\delta A_{\mu_2}(\bs{y})} J^{\mu_3}(\bs{z}) \> + \< \frac{\delta T_{\mu_1 \nu_1}(\bs{x})}{\delta A_{\mu_3}(\bs{z})} J^{\mu_2}(\bs{y}) \>. \label{e:defTJJ}
\end{align}
In \cite{Osborn:1993cr} (and similarly \cite{Giannotti:2008cv,Armillis:2009pq}) the semi-local terms on the right-hand side of this formula are absorbed into the definition of the $\< T_{\mu_1 \nu_1} J^{\mu_2} J^{\mu_3} \>$ correlator: it is these semi-local terms that are responsible, via \eqref{e:TcAJ}, for the dependence of our solution on the additional normalisation constant $c_3$.

\subsection{\texorpdfstring{General form of $\< T^{\mu_1 \nu_1} J^{\mu_2} J^{\mu_3} \>$ in $d = 3$}{General form of <TJJ> in d=3}}

Let us now focus on the special case of $d=3$. Since in odd dimensions the correlation function is automatically finite, we can use any regularisation scheme to achieve the goal. It is therefore most convenient to use the scheme \eqref{e:genscheme} with $u = 1$ and $v = 0$, since all triple-$K$ integrals can then be evaluated in terms of elementary integrals using \eqref{e:Khalf}. In this way, we find
\begin{align}
J_{4 \{000\}} & = I_{\frac{9}{2} \{ \frac{3}{2} \frac{1}{2} \frac{1}{2} \}} = 2 \left( \frac{\pi}{2} \right)^{\frac{3}{2}} \frac{4 p_1 + p_2 +p_3}{(p_1 + p_2 + p_3)^4}, \label{e:tjj3_integrals_first} \\
J_{3 \{100\}} & = I_{\frac{7}{2} \{ \frac{5}{2} \frac{1}{2} \frac{1}{2} \}} = \left( \frac{\pi}{2} \right)^{\frac{3}{2}} \frac{9 (p_1 p_2 + p_1 p_3) + 6 p_2 p_3 + 8p_1^2 + 3(p_2^2 + p_3^2)}{(p_1 + p_2 + p_3)^3}, \\
J_{2 \{000\}} & = I_{\frac{5}{2} \{ \frac{3}{2} \frac{1}{2} \frac{1}{2} \}} = \left( \frac{\pi}{2} \right)^{\frac{3}{2}} \frac{2 p_1 + p_2 + p_3}{(p_1 + p_2 + p_3)^2}, \\
J_{2 \{011\}} & = I_{\frac{5}{2} \{ \frac{3}{2} \frac{3}{2} \frac{3}{2} \}} = - \left( \frac{\pi}{2} \right)^{\frac{3}{2}} \frac{1}{(p_1 + p_2 + p_3)^2} \left[ 2 p_1 p_2 p_3 + p_1^3 + p_2^3 + p_3^3 \right.\nn\\
& \qquad\qquad\qquad\qquad \left. + \: 2 (p_1^2 p_2 + p_1 p_2^2 + p_1 p_3^2 + p_3 p_1^2 + p_2 p_3^2 + p_3 p_2^2) \right], \\
J_{1 + \epsilon \{010\}} & = I_{\frac{3}{2} + \epsilon \{ \frac{3}{2} \frac{3}{2} \frac{1}{2} \}} = \left( \frac{\pi}{2} \right)^{\frac{3}{2}} \left[ - \frac{p_3}{\epsilon} + p_3 \log(p_1 + p_2 + p_3) \right.\nn\\
& \qquad \left. + \: \frac{-p_1 p_2 + (\gamma_E - 2) ( p_1 p_3 + p_2 p_3 ) - p_1^2 - p_2^2 + (\gamma_E - 1) p_3^2}{p_1 + p_2 + p_3} + O(\epsilon) \right], \\
J_{0 + \epsilon \{000\}} & = I_{\frac{1}{2} + \epsilon \{ \frac{3}{2} \frac{1}{2} \frac{1}{2} \}} = \left( \frac{\pi}{2} \right)^{\frac{3}{2}} \left[ - \frac{p_2+p_3}{\epsilon} + (p_2 + p_3) \log(p_1 + p_2 + p_3) \right.\nn\\
& \qquad \left. + \: (\gamma_E - 1) (p_2 + p_3) - p_1 + O(\epsilon) \right], \label{e:tjj3_integrals_last}
\end{align}
with similar integrals following from the permutation formula \eqref{e:Jsym}.

Applying the secondary CWIs \eqref{e:sectjj1} - \eqref{e:sectjj3} we then obtain the final result
\begin{align}
A_1 & = \alpha_1 \frac{2(4 p_1 + p_2 + p_3)}{(p_1 + p_2 + p_3)^4}, \label{e:fintjj1} \\
A_2 & = \frac{2 \alpha_1 p_1^2}{(p_1 + p_2 + p_3)^3} + \frac{4 \sqrt{\pi} (2 p_1 + p_2 + p_3)}{(p_1 + p_2 + p_3)^2} c_{J},\\
A_3 & = \frac{\alpha_1}{(p_1 + p_2 + p_3)^3} \left[ - 2 p_1^2 - p_2^2+p_3^2 - 3 p_1 p_2 + 3 p_1 p_3 \right] + \frac{4 \sqrt{\pi} (2 p_1 + p_2 + p_3)}{(p_1 + p_2 + p_3)^2} c_{J}, \\
A_4 & = \alpha_1 \frac{(p_1 + p_2 - p_3)(p_1 - p_2 + p_3)(2 p_1 + p_2 + p_3)}{2 (p_1 + p_2 + p_3)^2} \nn\\
& \qquad \qquad - \:2 \sqrt{\pi} \left( \frac{2 p_1^2}{p_1 + p_2 + p_3} - p_2 - p_3 \right) c_{J} - 4 \sqrt{\pi} (p_2 + p_3) c_3 c_{J}. \label{e:fintjj4}
\end{align}
In these results we rescaled the coefficient $\alpha_1$ according to $\alpha_1 (\pi/2)^{3/2} \mapsto \alpha_1$, so as to remove the awkward factor of $(\pi/2)^{3/2}$.

The form factors build the transverse-traceless part of the correlation function according to \eqref{e:tjjdecomp}. The full correlation function can then be recovered by means of \eqref{e:jloc} and \eqref{e:tloc}. Using the transverse and trace Ward identities (\ref{e:p1TJJ}, \ref{e:p2TJJ}, \ref{e:trTJJ}), we find
\begin{align}
& \lla T^{\mu_1 \nu_1}(\bs{p}_1) J^{\mu_2}(\bs{p}_2) J^{\mu_3}(\bs{p}_3) \rra = \lla t^{\mu_1 \nu_1}(\bs{p}_1) j^{\mu_2}(\bs{p}_2) j^{\mu_3}(\bs{p}_3) \rra \nn\\
& \qquad + \left[ 2 \mathscr{T}^{\mu_1 \nu_1}_{\alpha}(\bs{p}_1) \pi^{\mu_3 [\alpha}(\bs{p}_3) p_3^{\beta]} + \frac{p_3^{\mu_3}}{p_3^2} \delta^{\mu_1 \nu_1} p_{1 \beta} \right] \lla J^{\mu_2}(\bs{p}_2) J_{\beta}(-\bs{p}_2) \rra \nn\\
& \qquad + \: \pi^{\mu_3}_{\alpha_3}(\bs{p}_3) \left[ \mathscr{T}^{\mu_1 \nu_1 \alpha_1}(\bs{p}_1) p_1^{\beta_1} + \frac{\pi^{\mu_1 \nu_1}(\bs{p}_1)}{d-1} \delta^{\alpha_1 \beta_1} \right] \lla \frac{\delta T_{\alpha_1 \beta_1}}{\delta A_{\alpha_3}}(\bs{p}_1, \bs{p}_3) J^{\mu_2}(\bs{p}_2) \rra \nn\\
& \qquad + \frac{2 p_3^{\mu_3} p_{3 \alpha_3}}{p_3^2} \delta^{\mu_1 \alpha_1} \delta^{\nu_1 \beta_1} \lla \frac{\delta J^{\alpha_3}}{\delta g^{\alpha_1 \beta_1}}(\bs{p}_3, \bs{p}_1) J^{\mu_2}(\bs{p}_2) \rra \nn\\
& \qquad + \: \text{everything with } (\bs{p}_2, \mu_2) \leftrightarrow (\bs{p}_3, \mu_3), \label{e:tjjrecover}
\end{align}
where $\mathscr{T}^{\mu \nu}_{\alpha}$ was given in \eqref{e:curlyT}. Here we assume no scale anomalies are present: if anomalies occur, the additional ultralocal contributions should be added to \eqref{e:tjjrecover}.

The result \eqref{e:tjjrecover} is the most general explicit expression for the $\lla T^{\mu_1 \nu_1} J^{\mu_2} J^{\mu_3} \rra$ correlation function in the momentum space. As we can see, it depends on one undetermined primary constant plus the normalisations of the 2-point functions.

\subsection{\texorpdfstring{Free fermions in $d=3$}{Free fermions in d=3}}

As a cross-check on our calculations we now consider free fermions in $d=3$ Euclidean dimensions given by the action
\begin{equation}
S = \int \D^3 \bs{x} \: e \left[ \bar{\psi} e^{\mu}_a \gamma^a \stackrel{\leftrightarrow}{D}_\mu \psi \right],
\end{equation}
where
\begin{equation}
D_\mu = \nabla_\mu - \I A_\mu, \qquad \nabla_\mu = \partial_\mu - \frac{\I}{2} \omega_\mu^{ab} \Sigma_{ab},
\end{equation}
and $\omega_\mu^{ab}$ is the spin connection
\begin{equation}
\omega_\mu^{ab} = e^a_\nu \partial_\mu e^{\nu b} + e^a_\nu e^{\sigma b} \Gamma^\nu_{\ \: \sigma \mu}, \qquad \Sigma^{ab} = \frac{\I}{4} [ \gamma^a, \gamma^b ].
\end{equation}
Here $\Gamma^\nu_{\ \sigma \mu}$ is the Christoffel symbol associated with the metric $g_{\mu \nu}$, while $e^a_\mu$ are vielbeins satisfying $e^a_{\mu} e_{\nu a} = g_{\mu \nu}$ and the gamma matrices $\gamma^a$ satisfy $\gamma^\mu = e^\mu_a \gamma^a$.  On flat space, we then have $\{\gamma^a, \gamma^b\} = - 2 \delta^{ab}$. In $d=3$, the spin-$\frac{1}{2}$ representation of the group $SO(3)$ is 2-dimensional and $\Tr(\gamma^a \gamma^b) = -2 \delta^{ab}$.

Notice that the gauge field $A_\mu$ is treated as a source for the conserved current and is not a degree of freedom. The stress-energy tensor and the conserved current in the presence of the sources are
\begin{align}
T_{\mu \nu} & = \frac{2}{\sqrt{g}} \frac{\delta S}{\delta g^{\mu \nu}} = \bar{\psi} \gamma_{(\mu} \stackrel{\leftrightarrow}{D}_{\nu)} \psi - g_{\mu \nu} \bar{\psi} \gamma^{\alpha} \stackrel{\leftrightarrow}{D}_{\alpha} \psi, \label{e:fermT} \\
J^\mu & = \frac{1}{\sqrt{g}} \frac{\delta S}{\delta A^{\mu}} = - \I \bar{\psi} \gamma^\mu \psi. \label{e:fermJ}
\end{align}
In this case the current is associated with the $U(1)$ symmetry, therefore we omit the group indices on $J^{\mu}$. By direct calculation we find
\begin{align}
\lla J^{\mu}(\bs{p}) J^{\nu}(-\bs{p}) \rra & = - \frac{1}{16} p \pi^{\mu \nu}(\bs{p}), \label{e:fermJJ} \\
\lla T^{\mu_1 \nu_1}(\bs{p}) T^{\mu_2 \nu_2}(-\bs{p}) \rra & = \frac{1}{128} p^3 \Pi^{\mu_1 \nu_1 \mu_2 \nu_2}(\bs{p}).
\end{align}
The transverse Ward identities can be obtained by differentiation of the equations (\ref{e:toWardJ}, \ref{e:toWardT}) and are listed in the second part of the paper. Some terms of the terms involve functional derivatives and may be evaluated directly from expressions (\ref{e:fermT}, \ref{e:fermJ}),
\begin{align}
\frac{\delta T_{\mu \nu}(\bs{x})}{\delta A_\rho(\bs{y})} & = \frac{1}{2} [ J_\mu \delta_\nu^\rho + J_\nu \delta_\mu^\rho - 2 J^\rho \delta_{\mu \nu} ] \delta(\bs{x} - \bs{y}), \label{e:fermTAJ} \\
\frac{\delta J^\mu(\bs{x})}{\delta g^{\alpha \beta}(\bs{y})} & = \frac{1}{4} [ J_\beta \delta^\mu_\alpha + J_\alpha \delta^\mu_\beta ] \delta(\bs{x} - \bs{y}),
\end{align}
where the sources are turned off after the derivative is taken. All together, for this particular CFT we find
\begin{equation} \label{e:ferm2pt}
c_J = \frac{1}{32 \sqrt{\pi}}, \qquad c_T = \frac{3}{512 \sqrt{\pi}}, \qquad c_3 = \frac{1}{2}.
\end{equation}
where the 2-point function normalisations $c_J$ and $c_T$, and the constant $c_3$, are as defined in \eqref{e:JJ1}, \eqref{e:TT1} and \eqref{e:TcAJ} respectively.

The 3-point function can be calculated by the usual Feynman rules. Using the results of section \ref{ch:finding}, one finds
\begin{align}
A_1 & = - \frac{4 p_1 + p_2 + p_3}{12 \: (p_1 + p_2 + p_3)^4}, \label{e:fermA1}\\
A_2 & = \frac{9 (p_1 p_2 + p_1 p_3) + 6 p_2 p_3 + 4 p_1^2 + 3 (p_2^2 + p_3^2)}{24 \: (p_1 + p_2 + p_3)^3}, \\
A_3 & = \frac{6 p_1 p_2 + 3 p_1 p_3 + 3 p_2 p_3 + 4 p_1^2 + 2 p_2^2 + p_3^2}{12 \: (p_1 + p_2 + p_3)^3}, \\
A_4 & = - \frac{4 p_1 p_2 p_3 + 7(p_1^2 p_2 + p_1^2 p_3) - 2 (p_1 p_2^2 + p_1 p_3^2) + p_2 p_3^2 + p_3 p_2^2 + 8 p_1^3 - (p_2^3 + p_3^3)}{48 \: (p_1 + p_2 + p_3)^2}. \label{e:fermA4}
\end{align}
The form factors $A_j$ are defined in the decomposition \eqref{e:tjjdecomp}.

We can compare this result directly with the solution \eqref{e:fintjj1} - \eqref{e:fintjj4}.
Since we know the 2-point function normalisations \eqref{e:ferm2pt} there is only one undetermined constant, $\alpha_1$.  The solution \eqref{e:fermA1} - \eqref{e:fermA4} then fits perfectly with $\alpha_1 = -\frac{1}{24}$. In fact, the secondary Ward identities provide quite a robust check on the standard QFT calculation of the 3-point function: for example, a mistake leading to the overall rescaling of all form factors in \eqref{e:fermA1} - \eqref{e:fermA4} by some factor would immediately lead to an inconsistency with the 2-point function normalisation constants \eqref{e:ferm2pt}.

\section{Extensions}
\label{sec:extensions}

In this final section of Part I, we briefly discuss two extensions of the present analysis: how to write the results for tensor correlators in terms of a helicity basis, and the issues that arise when we try to generalise to higher-point correlation functions.

\subsection{Helicity formalism} \label{ch:helicity}

In the helicity formalism, one writes down a basis for the space of transverse and transverse-traceless tensors in terms of polarisation tensors $\xi^{(s)}_\mu$ and $\epsilon^{(s)}_{\mu \nu}$ respectively, where the index $s$ ranges over helicities. The number of helicities depends on the tensor structure and is equal to the dimension of the corresponding representation of the little group in $D = d + 1$ dimensions. For the conserved current it is equal to $d - 1$ and for the symmetric, traceless tensors of rank $2$ it is equal to $(d+1)(d-2)/2$. Note that these numbers are equal to $\pi_{\mu\nu} \pi^{\mu \nu}$ and $\Pi_{\mu\nu\rho\sigma} \Pi^{\mu\nu\rho\sigma}$ respectively, where the projectors are defined in \eqref{e:pi} and \eqref{e:Pi}.

The polarisation tensors can be defined by the decomposition of the projectors,
\begin{align}
\pi_{\mu \nu}(\bs{p}) & = \sum_{s} \xi^{(s)}_\mu(\bs{p}) \bar{\xi}^{(s)}_\nu(\bs{p}), \label{e:xi}\\
\Pi_{\mu \nu \rho \sigma}(\bs{p}) & = \frac{1}{2} \sum_{s} \epsilon^{(s)}_{\mu \nu}(\bs{p}) \bar{\epsilon}^{(s)}_{\rho \sigma}(\bs{p}), \label{e:epsilon}
\end{align}
where the bar over a symbol denotes complex conjugation. Moreover, the helicity tensors satisfy
\begin{align}
& p^{\mu} \epsilon_{\mu \nu}^{(s)}(\bs{p}) = 0, && p^{\mu} \xi_{\mu}^{(s)}(\bs{p}) = 0, \nn\\
& \epsilon_{\mu \nu}^{(s)} = \epsilon_{\nu \mu}^{(s)}, && \delta^{\mu\nu} \epsilon_{\mu\nu}^{(s)} = 0, \nn\\
& \bar{\xi}_{\mu}^{(s)}(\bs{p}) = \xi_{\mu}^{(s)}(-\bs{p}), && \bar{\epsilon}_{\mu \nu}^{(s)}(\bs{p}) = \epsilon_{\mu \nu}^{(s)}(- \bs{p}).
\end{align}
Using the identities from appendix \ref{ch:identities} one finds
\begin{equation} \label{e:xixiee}
\xi_{\mu}^{(s)} \bar{\xi}^{(s') \mu} = \delta^{ss'}, \qquad\qquad \epsilon_{\mu\nu}^{(s)} \bar{\epsilon}^{(s') \mu\nu} = 2 \delta^{ss'}.
\end{equation}
The helicity-projected operators are then defined as
\begin{equation}
J^{(s)}(\bs{p}) = \bar{\xi}_\mu^{(s)}(\bs{p}) J^\mu(\bs{p}), \qquad  T^{(s)}(\bs{p}) = \frac{1}{2} \bar{\epsilon}_{\mu \nu}^{(s)}(\bs{p}) T^{\mu \nu}(\bs{p}).
\end{equation}

Correlation functions of the helicity-projected operators can easily be obtained from the transverse-traceless parts of the correlators. First observe that the semi-local parts of any correlation function vanish when contracted with polarisation tensors. Indeed, equations (\ref{e:xi}, \ref{e:epsilon}) together with \eqref{e:xixiee} imply that
\begin{equation} \label{e:helpi}
\pi_{\mu}^{\nu} \bar{\xi}_{\nu}^{(s)} = \bar{\xi}_\mu^{(s)}, \qquad\qquad \Pi_{\mu\nu}^{\rho \sigma} \bar{\epsilon}_{\rho\sigma}^{(s)} = \bar{\epsilon}_{\mu\nu}^{(s)}.
\end{equation}
Then, using equation \eqref{e:pij0} we can write
\begin{equation}
\bar{\xi}^{(s)}_\mu j^\mu_{\text{loc}} = \bar{\xi}^{(s)}_\nu \pi^\nu_\mu j^\mu_{\text{loc}} = 0
\end{equation}
and similarly $\bar{\epsilon}^{(s)}_{\mu\nu} t^{\mu\nu}_{\text{loc}} = 0$. To obtain correlation functions in the helicity formalism, one can therefore apply helicity projectors to the transverse-traceless parts of correlators only. Due to \eqref{e:helpi}, the projectors \eqref{e:pi} and \eqref{e:Pi} can then be removed as well. Finally, one needs to compute a small number of contractions of the helicity projectors with momenta and with the metric.

\subsubsection{\texorpdfstring{Examples in $d=3$}{Examples in d=3}}

As an example, consider the $\lla T^{\mu_1 \nu_1} T^{\mu_2 \nu_2} \mathcal{O} \rra$ correlation function in $d=3$ spacetime dimensions. Applying first the helicity projectors to its decomposition \eqref{e:TTOform}, we find
\begin{align}
& \lla T^{(s_1)}(\bs{p}_1) T^{(s_2)}(\bs{p}_2) \mathcal{O}(\bs{p}_3) \rra = \frac{1}{4} \bar{\epsilon}^{(s_1)}_{\mu_1 \nu_1}(\bs{p}_1) \bar{\epsilon}^{(s_2)}_{\mu_2 \nu_2}(\bs{p}_2) \lla t^{\mu_1 \nu_1}(\bs{p}_1) t^{\mu_2 \nu_2}(\bs{p}_2) \mathcal{O}(\bs{p}_3) \rra \nn\\
& = \frac{1}{4} \left[ A_1 \bar{\epsilon}^{(s_1)}_{\mu_1 \nu_1}(\bs{p}_1) p_2^{\mu_1} p_2^{\nu_1} \bar{\epsilon}^{(s_2)}_{\mu_2 \nu_2}(\bs{p}_2) p_3^{\mu_2} p_3^{\nu_2} + A_2 \bar{\epsilon}^{(s_1)}_{\mu_1 \alpha}(\bs{p}_1) \bar{\epsilon}^{(s_2)\alpha}_{\mu_2}(\bs{p}_2) p_2^{\mu_1} p_3^{\mu_2} \right.\nn\\
& \qquad\qquad \left. + \: A_3 \bar{\epsilon}^{(s_1)}_{\alpha \beta}(\bs{p}_1) \bar{\epsilon}^{(s_2)\alpha \beta}(\bs{p}_2) \right].
\end{align}
The contractions with helicity tensors depend on the precise definition of the latter and also the overall dimension. Let us consider, for example, the case $\Delta_2=\Delta_3=1$. In $d=3$ there are two helicities, which are usually denoted by $s = \pm$. The required contractions can be found in \cite{Bzowski:2011ab},
\begin{align}
\bar{\epsilon}^{(s_1)}_{\mu_1 \nu_1}(\bs{p}_1) p_2^{\mu_1} p_2^{\nu_1} & = \frac{J^2}{4 \sqrt{2} p_1^2}, \\
\bar{\epsilon}^{(s_1)}_{\mu_1 \alpha}(\bs{p}_1) \bar{\epsilon}^{(s_2)\alpha}_{\mu_2}(\bs{p}_2) p_2^{\mu_1} p_3^{\mu_2} & = \frac{J^2 S^{(s_1 s_2)}_3}{16 p_1^2 p_2^2}, \\
\bar{\epsilon}^{(s_1)}_{\alpha \beta}(\bs{p}_1) \bar{\epsilon}^{(s_2)\alpha \beta}(\bs{p}_2) & = \frac{(S_3^{(s_1 s_2)})^2}{8 p_1^2 p_2^2},
\end{align}
where $J^2$ and $S_3^{(s_1 s_2)}$ are defined as
\begin{align}
J^2 & = (p_1 + p_2 + p_3) (- p_1 + p_2 + p_3) (p_1 - p_2 + p_3) (p_1 + p_2 - p_3), \label{e:lambda} \\
S_3^{(s_1 s_2)} & = p_3^2 - (s_1 p_1 + s_2 p_2)^2.
\end{align}
Using \eqref{e:exsolA1} - \eqref{e:exsolA3} for the form factors, the most general solution is
\begin{align}
& \lla T^{(s_1)}(\bs{p}_1) T^{(s_2)}(\bs{p}_2) \mathcal{O}^I(\bs{p}_3) \rra = \alpha_1^I \frac{3 p_1 p_2}{4 p_3} \left( \frac{p_1 + p_2 - p_3}{p_1 + p_2 + p_3} \right)^2 \delta^{s_1 s_2} \nn\\
& \qquad\qquad + \: \frac{c_{\mathcal{O}} \sqrt{\pi} S_3^{(s_1 s_2)}}{32 p_1^2 p_2^2 p_3} \left[ -2 c_1^{I} J^2 + \left( (c_1^{I} - 3 c_2^{I}) (p_1^2 + p_2^2) + 3 (c_1^{I} + c_2^{I}) p_3^2 \right) S_3^{(s_1 s_2)} \right].
\end{align}
The constants $c_1^J$ and $c_2^J$ are defined in \eqref{e:TOOc1c2} and $c_{\mathcal{O}}$ is the normalisation constant of the 2-point function $\lla \mathcal{O}^J \mathcal{O}^I \rra$ defined in \eqref{e:OO}.

As a check on our results in \eqref{e:fermAform}, we compared our solution with that obtained in \cite{Chowdhury:2012km} for the $\lla T^{\mu_1 \nu_1} J^{\mu_2} J^{\mu_3} \rra$ correlator of free scalars and fermions finding perfect agreement.

The same method can be applied to the correlation function of three stress-energy tensors in $d=3$. This is an interesting example, since, according to the position space results of \cite{Osborn:1993cr}, there is one fewer independent conformal structure in $d=3$ than in dimensions $d>3$. Indeed, the application of the helicity formalism in $d=3$ to the correlation function of three stress-energy tensors given by \eqref{e:TTTA1} - \eqref{e:TTTA5} leads to the following result
\begin{align}
& \lla T^{(+)}(\bs{p}_1) T^{(+)}(\bs{p}_2) T^{(+)}(\bs{p}_3) \rra = \frac{30 \sqrt{2} \alpha_1 J^2 p_1 p_2 p_3}{a_{123}^4} \nn\\
& \qquad - \: c_T  \frac{\sqrt{\pi} J^2 a_{123}^2}{12 \sqrt{2} c_{123}^2} \left[ \left( 3 a_{123}^3 - 7 a_{123} b_{123} + 5 c_{123} \right) + 8 (p_1^3 + p_2^3 + p_3^3) c_g \right], \label{e:TTTppp} \\
& \lla T^{(+)}(\bs{p}_1) T^{(+)}(\bs{p}_2) T^{(-)}(\bs{p}_3) \rra = - c_T  \frac{\sqrt{\pi} J^2 (p_1 + p_2 - p_3)^2}{12 \sqrt{2} c^2_{123}} \nn\\
& \qquad \times \: \left[ \frac{1}{a_{123}^2} \left( 3 p_3^5 + 4 p_3^4 a_{12} + p_3^3 ( a_{12}^2 - b_{12} ) + p_3 a_{12} ( p_3 + 4 a_{12} ) ( a_{12}^2 - 3 b_{12} ) \right.\right.\nn\\
& \qquad\qquad \left. + \: a_{12}^3 ( 3 a_{12}^2 - 7 b_{12} ) \right) + 8 (p_1^3 + p_2^3 + p_3^3) c_g \Big], \label{e:TTTppm}
\end{align}
where $J^2$ is defined in \eqref{e:lambda} and all remaining variables are symmetric polynomials in magnitudes of momenta defined in \eqref{e:variables}. Notice that this solution depends on a single primary constant $\alpha_1$ and does not depend on $\alpha_2$, which features in the solution \eqref{e:TTTA1} - \eqref{e:TTTA5}. The same result can also be obtained directly in momentum space, as presented in appendix \ref{ch:degeneracy}. Note also that the $\lla T^{(+)} T^{(+)} T^{(-)} \rra$ part of the correlation function does not depend on $\alpha_1$, and hence is determined uniquely in terms of the 2-point function.

\subsection{Higher-point correlation functions} \label{ch:higher}

It would be interesting to apply the present formalism to higher-point correlation functions in momentum space. Unfortunately, this seems to be a much more difficult task. In general, the ideas of the tensor decomposition described in section \ref{ch:Tensor_structure} are valid, but much less constraining. For concreteness, consider the $\lla T^{\mu_1 \nu_1}(\bs{p}_1) T^{\mu_2 \nu_2}(\bs{p}_2) \mathcal{O}(\bs{p}_3) \mathcal{O}(\bs{p}_4) \rra$ correlation function.
As there are now three independent momenta, each transverse or transverse-traceless projector \eqref{e:pi} and \eqref{e:Pi} can be contracted with either of the two independent transverse momenta to yield a non-vanishing result. The decomposition of the transverse-traceless part of the correlation function under consideration is therefore
\begin{align}
& \lla t^{\mu_1 \nu_1}(\bs{p}_1) t^{\mu_2 \nu_2}(\bs{p}_2) \mathcal{O}(\bs{p}_3) \mathcal{O}(\bs{p}_4) \rra = \Pi^{\mu_1 \nu_1}_{\alpha_1 \beta_1}(\bs{p}_1) \Pi^{\mu_2 \nu_2}_{\alpha_2 \beta_2}(\bs{p}_2) \left[ \right. \nn\\
& \qquad \sum_{n_{1,2} \in \{2,3\}} \sum_{n_{3,4} \in \{3,4\}} A_{n_1 n_2 n_3 n_4} p_{n_1}^{\alpha_1} p_{n_2}^{\beta_1} p_{n_3}^{\alpha_2} p_{n_4}^{\beta_2} \nn\\
& \qquad \left. + \: \delta^{\alpha_1 \alpha_2} \sum_{n_1 \in \{2,3\}} \sum_{n_2 \in \{3,4\}} A_{n_1 n_2} p_{n_1}^{\beta_1} p_{n_2}^{\beta_2} + A_0 \: \delta^{\alpha_1 \alpha_2} \delta^{\beta_1 \beta_2} \right], \label{e:ttoo_decomp}
\end{align}
where $A_{n_1 n_2 n_3 n_4}$, $A_{n_1 n_2}$ and $A_0$ are form factors.  Initially, we thus have $2^4 = 16$ tensor structures following from the contractions of the transverse-traceless projectors with momenta, while in case of 3-point functions the corresponding tensor was unique.

The decomposition above is valid as long as $d \geq 4$. In case of $d=3$, the metric $\delta^{\mu \nu}$ is not an independent tensor according to \eqref{e:metric_as_momenta}. In this case, the decomposition \eqref{e:ttoo_decomp} can be truncated after the second line.

In the case of the correlator $\lla T^{\mu_1 \nu_1} T^{\mu_2 \nu_2} T^{\mu_3 \nu_3} T^{\mu_4 \nu_4} \rra$ our procedure reduces the number of independent tensor structures from the original $47\:868$ simple tensors built from the metric and three independent momenta down to $382$ transverse-traceless tensor structures. This number, however, should be further diminished when the full symmetry group $S_4$ is imposed. For the 3-point function, the full symmetry group was automatically encoded in the decomposition, since only one momentum could appear under a chosen Lorentz index. In case of the 4-point function, the permutation group mixes various momenta and the fully symmetric structure is not clearly visible. For comparison, in the case of the 3-point function $\lla T^{\mu_1 \nu_1} T^{\mu_2 \nu_2} T^{\mu_3 \nu_3} \rra$, $499$ simple tensors built up from the metric and two independent momenta were reduced down to five transverse-traceless tensor structures.

In addition, the form factors are no longer functions of momentum magnitudes only. If we consider an $n$-point function in a $d$-dimensional CFT with $d \geq n$, then the scalar products
\begin{equation} \label{e:scal_prod_high}
p_{ij} = \bs{p}_i \cdot \bs{p}_j, \quad i,j=1,2,\ldots,n, \quad i < j
\end{equation}
are independent variables. In this case, the form factors can be regarded as functions of $p_{ij}$.

It would be interesting to work out the form of the conformal Ward identities for higher-point correlation functions. It would be vital to understand the objects corresponding to conformal ratios in momentum space. So far, we can only count the number of degrees of freedom in an $n$-point function in a $d$-dimensional CFT and compare it to the number of independent conformal ratios which is $n(n-3)/2$, assuming $d \geq n$ \cite{DiFrancesco:1997nk}. Indeed, the number of independent scalar products \eqref{e:scal_prod_high} is reduced by the $n-1$ differential equations \eqref{e:ward_cwi_x0} and \eqref{e:ward_cwi_x1} following from the special conformal Ward identities, plus the additional constraint \eqref{e:ward_dil_x} from the dilatation Ward identity, leaving
\begin{equation}
\frac{n(n-1)}{2} - (n-1) - 1 = \frac{n(n-3)}{2}
\end{equation}
in accordance with the number of conformal ratios.


\newpage
\part*{List of results}
\addcontentsline{toc}{section}{List of results}
\setcounter{part}{2}

\appendix
\numberwithin{equation}{subsection}
\setcounter{section}{18}

\section*{Definitions}
\addcontentsline{toc}{subsection}{\numberline {1}Definitions}
\setcounter{subsection}{1}
\setcounter{equation}{0}

Here we collect together the necessary definitions and notation required to present our main results; further details may be found in the first part of the paper.

\subsection*{Basic conventions}

We assume $d \geq 3$ Euclidean dimensions. Vectors are denoted by bold letters, \textit{e.g.}, $\bs{p}_1$, but all results will be expressed in terms of the magnitudes of the momenta
\begin{equation}
p_j = | \bs{p}_j | = \sqrt{ \bs{p}_j^2 }, \qquad j = 1, 2, 3.
\end{equation}
In particular, the form factors $A_j = A_j(p_1, p_2, p_3)$ are functions of the momentum magnitudes. Arrows denote the exchange of arguments, \textit{e.g.}, $A_j(p_1 \leftrightarrow p_2) = A_j(p_2, p_1, p_3)$. If no arguments are given for a particular form factor then the standard ordering is assumed, $A_j(p_1, p_2, p_3)$.

To write the results in compact form, we frequently make use of the following symmetric polynomials in the momentum magnitudes
\begin{align}
& a_{123} = p_1 + p_2 + p_3, \qquad b_{123} = p_1 p_2 + p_1 p_3 + p_2 p_3, \qquad c_{123} = p_1 p_2 p_3, \nn\\
& a_{ij} = p_i + p_j, \qquad\qquad b_{ij} = p_i p_j, \label{e:variables}
\end{align}
where $i,j = 1,2,3$.

\subsection*{\texorpdfstring{Conformal Ward identities (CWIs) and triple-$K$ integrals}{Conformal Ward identities (CWIs) and triple-K integrals}}

The CWI operators $\K_j$ and $\K_{ij}$, $i,j=1,2,3$ are defined in \eqref{e:K} and \eqref{e:KK} by
\begin{align}
\K_j & = \frac{\partial^2}{\partial p_j^2} + \frac{d + 1 - 2 \Delta_j}{p_j} \frac{\partial}{\partial p_j}, \\
\K_{ij} & = \K_i - \K_j. \label{a:Kij}
\end{align}
By $\Delta_j$, $j=1,2,3$ we denote the conformal dimension of the $j$-th operator in a given 3-point function. For example in $\lla T^{\mu_1 \nu_1} J^{\mu_2} J^{\mu_3} \rra$ we have $\Delta_1 = d$ and $\Delta_2 = \Delta_3 = d - 1$.

The triple-$K$ integral \eqref{e:J} and its reduced version \eqref{e:Jred} are
\begin{align}
I_{\alpha \{ \beta_1 \beta_2 \beta_3 \}}(p_1, p_2, p_3) & = \int_0^\infty \D x \: x^\alpha \prod_{j=1}^3 p_j^{\beta_j} K_{\beta_j}(p_j x), \label{a:I} \\
J_{N \{ k_j \}} & = I_{\frac{d}{2} - 1 + N \{ \Delta_j - \frac{d}{2} + k_j \}}, \label{a:J}
\end{align}
where $K_\nu$ is the Bessel function $K$ (modified Bessel function of the second kind) and we use a shortened notation $\{k_j\} = \{k_1 k_2 k_3\}$.

Solutions to the primary CWIs, see section \ref{ch:sol_to_pri_CWIs}, are given as linear combinations of reduced triple-$K$ integrals multiplied by constants, denoted by $\alpha_j$ and called primary constants. If a primary constant is not restricted by means of the secondary CWIs, then it is a free parameter depending on the details of the theory.

If a triple-$K$ integral diverges it can be regularised by
\begin{equation} \label{a:scheme}
I_{\alpha \{\beta_1 \beta_2 \beta_3\}} \mapsto I_{\alpha + u \epsilon \{\beta_1 + v \epsilon, \beta_2 + v \epsilon, \beta_3 + v \epsilon\}},
\end{equation}
with $u = v = -1/2$ then substituted at the end of the calculation for any form factor.
If the regulator $\epsilon$ cannot be removed, then both triple-$K$ integrals and primary constants are power series in $\epsilon$ regularised in the dimensional regularisation scheme $d \mapsto d - \epsilon$, $\Delta_j \mapsto \Delta_j - \epsilon$.

The differential operators appearing in the secondary CWIs are defined by \eqref{e:L} and \eqref{e:R} and read
\begin{align}
\Lo_{N} & = p_1 (p_1^2 + p_2^2 - p_3^2) \frac{\partial}{\partial p_1} + 2 p_1^2 p_2 \frac{\partial}{\partial p_2} \nn\\
& \qquad + \: \left[ (2 d - \Delta_1 - 2 \Delta_2 + N) p_1^2 + (2 \Delta_1 - d) (p_3^2 - p_2^2) \right], \label{a:L} \\
\Ro & = p_1 \frac{\partial}{\partial p_1} - (2 \Delta_1 - d), \label{a:R} \\
\Lo'_{N} & = \Lo_{N} \text{ with } (p_1 \leftrightarrow p_2) \text{ and } (\Delta_1 \leftrightarrow \Delta_2), \label{a:L2} \\
\Ro' & = \Ro \text{ with } (p_1 \rightarrow p_2) \text{ and } (\Delta_1 \rightarrow \Delta_2). \label{a:R2}
\end{align}
The secondary CWIs denoted by an asterisk are redundant, \textit{i.e.}, they do not impose any additional constraints on primary constants, see section \ref{ch:tran_tjj}.

Finally, we use the constant $l_{\alpha \{ \beta_k \}}$ defined in \eqref{e:l},
\begin{equation} \label{a:l}
l_{\alpha \{ \beta_1 \beta_2 \beta_3 \}} = \frac{2^{\alpha - 3} \Gamma(\beta_3)}{\Gamma(\alpha - \beta_3 + 1)} \prod_{\epsilon_1, \epsilon_2 \in \{-1, 1\}} \Gamma \left( \frac{\alpha - \beta_3 + 1 + \epsilon_1 \beta_1 + \epsilon_2 \beta_2}{2} \right).
\end{equation}

\subsection*{Tensor decomposition}

In momentum space correlators may be expressed in terms of tensor structures constructed from momenta and the metric multiplied by scalar form factors. Due to momentum conservation not all momenta are independent and we use the convention to consider different momenta as being independent depending on their Lorentz indices are discussed (see section \ref{ch:Tensor_structure} and in particular \eqref{e:momenta_choice}):
\begin{equation} \label{a:momenta}
\bs{p}_1, \bs{p}_2 \text{ for } \mu_1, \nu_1; \ \bs{p}_2, \bs{p}_3 \text{ for } \mu_2, \nu_2 \text{  and  } \bs{p}_3, \bs{p}_1 \text{ for }\mu_3, \nu_3.
\end{equation}

The transverse and transverse-traceless projectors \eqref{e:pi} and \eqref{e:Pi} are
\begin{align}
\pi^{\mu}_{\alpha}(\bs{p}) & = \delta^{\mu}_{\alpha} - \frac{p^{\mu} p_{\alpha}}{p^2}, \\
\Pi^{\mu \nu}_{\alpha \beta}(\bs{p}) & = \frac{1}{2} \left( \pi^{\mu}_{\alpha}(\bs{p}) \pi^{\nu}_{\beta}(\bs{p}) + \pi^{\mu}_{\beta}(\bs{p}) \pi^{\nu}_{\alpha}(\bs{p}) \right) - \frac{1}{d - 1} \pi^{\mu \nu}(\bs{p}) \pi_{\alpha \beta}(\bs{p}).
\end{align}
The transverse(-traceless) and semi-local parts of the conserved current $J^\mu$ the stress-energy tensor $T^{\mu \nu}$ are given by \eqref{e:decompJ} and \eqref{e:decompT} and read
\begin{align}
j^{\mu} \equiv \pi^{\mu}_{\alpha} J^{\alpha}, & \qquad\qquad j_{\text{loc}}^{\mu} \equiv J^{\mu} - j^{\mu}, \\
t^{\mu \nu} \equiv \Pi^{\mu \nu}_{\alpha \beta} T^{\alpha \beta}, & \qquad\qquad t_{\text{loc}}^{\mu \nu} \equiv T^{\mu \nu} - t^{\mu \nu}.
\end{align}
The semi-local parts (denoted with the subscript `loc') can also be expressed as
\begin{equation}
j_{\text{loc}}^{\mu} = \frac{p^\mu}{p^2} r, \qquad t_{\text{loc}}^{\mu \nu} = \frac{p^\mu}{p^2} R^\nu + \frac{p^\nu}{p^2} R^\mu - \frac{p^\mu p^\nu}{p^4} R + \frac{1}{d-1} \pi^{\mu \nu} \left( T - \frac{R}{p^2} \right),
\end{equation}
where longitudinal and trace parts are
\begin{equation}
r = p_\mu J^\mu, \qquad R^\nu = p_\mu T^{\mu \nu}, \qquad R = p_\nu R^\nu, \qquad T = T^{\mu}_{\mu}.
\end{equation}
It will also be useful to define the operator $\mathscr{T}^{\mu\nu}_{\alpha}$ as in \eqref{e:curlyT}, namely
\begin{equation} \label{a:T}
\mathscr{T}^{\mu\nu}_{\alpha} (\bs{p}) = \frac{1}{p^2} \left[ 2 p^{(\mu} \delta^{\nu)}_\alpha - \frac{p_\alpha}{d-1} \left( \delta^{\mu\nu} + (d-2) \frac{p^\mu p^\nu}{p^2} \right) \right].
\end{equation}
We also denote $\mathscr{T}^{\mu\nu\alpha} = \delta^{\alpha \beta}\mathscr{T}^{\mu\nu}_{\beta}$.

\subsection*{Operators in the theory}

We assume the CFT contains the following data:
\begin{itemize}

\item A symmetry group $G$. The conserved current $J^{\mu a}$, $a = 1, \ldots, \dim G$, is then the Noether current associated with the symmetry and is sourced by a potential $A_\mu^a$.
Currents transform in the adjoint representation and we denote the structure constants as $f^{abc}$.  We assume the Killing form is diagonal, $\tr (T^a T^b) = \tfrac{1}{2} \delta^{ab}$, where $T^a$ are generators of the group.
\item Scalar primary operators $\mathcal{O}^I$ all of the same dimension $\Delta$. They are sourced by $\phi_0^I$ and transform in a representation $R$ of the symmetry group. The representation matrices are denoted by $(T_R^a)^{IJ}$.
\item A stress-energy tensor $T_{\mu \nu}$ sourced by a metric $g^{\mu \nu}$.
\end{itemize}

The relevant Ward identities in the CFT are discussed in section \ref{sec:CWIs}; in particular the transverse Ward identities are given in section \ref{ch:transverseWI}. Note that we define the 3-point function of the stress-energy tensor to be the correlator of three separate stress-energy tensor insertions (and this results in additional terms containing functional derivatives relative to other papers in the literature, see the discussion in section \ref{ch:transverseWI}).

The normalisation constants $c_{\mathcal{O}}$, $c_{J}$, $c_T$ for 2-point functions are
\begin{align}
& \lla \mathcal{O}^I(\bs{p}) \mathcal{O}^J(-\bs{p}) \rra = c_{\mathcal{O}} \delta^{IJ} \Gamma \left( \frac{d}{2} - \Delta - v \epsilon \right) p^{2 \Delta - d + 2 v \epsilon}, \label{a:OO} \\
& \lla J^{\mu a}(\bs{p}) J^{\nu b}(-\bs{p}) \rra = c_{J} \pi^{\mu
\nu}(\bs{p})\delta^{ab} \Gamma \left( 1 - \frac{d}{2} - v \epsilon \right) p^{d-2 + 2 v \epsilon}, \label{a:JJ} \\
& \lla T^{\mu \nu}(\bs{p}) T^{\rho \sigma}(-\bs{p}) \rra = c_{T} \Pi^{\mu \nu \rho \sigma}(\bs{p})\Gamma \left( - \frac{d}{2} - v \epsilon \right) p^{d + 2 v \epsilon}, \label{a:TT}
\end{align}
where we use the general regularisation scheme \eqref{e:genschemed} which takes $d \mapsto d + 2 u \epsilon$, $\Delta_j \mapsto \Delta_j + (u + v) \epsilon$. Notice that the parameter $u$ does not appear in the 2-point functions. Dimensional regularisation then corresponds to $u = v = -1/2$.

In the following, we will illustrate our general results with specific examples in $d=3$ and $5$ dimensions.  We consider for these purposes scalar operators both with dimensions $\Delta=d-2$ and with dimension $\Delta=d$.  The former may be constructed as $\mathcal{O}=\phi^2$ in a theory of free scalars, where $\phi$ is the fundamental field, while the latter presents an interesting case being marginal.

\section*{\texorpdfstring{$\< \mathcal{O} \mathcal{O} \mathcal{O} \>$}{<OOO>}}
\addcontentsline{toc}{subsection}{\numberline {2}\texorpdfstring{$\< \mathcal{O} \mathcal{O} \mathcal{O} \>$}{<OOO>}}
\setcounter{subsection}{2}
\setcounter{equation}{0}

\begin{equation}
\lla \mathcal{O}^{I_1}(\bs{p}_1) \mathcal{O}^{I_2}(\bs{p}_2) \mathcal{O}^{I_3}(\bs{p}_3) \rra = A_1^{I_1 I_2 I_3}(p_1, p_2, p_3),
\end{equation}
The primary CWIs are
\begin{equation} \label{e:cwiA1}
\K_{ij} A_1^{I_1 I_2 I_3} = 0, \qquad i,j = 1,2,3,
\end{equation}
The solution in terms of triple-$K$ integrals \eqref{a:J} is
\begin{equation}
A_1^{I_1 I_2 I_3} = \alpha_1^{I_1 I_2 I_3} J_{0 \{000\}},
\end{equation}
where $\alpha_1^{I_1 I_2 I_3}$ is an arbitrary constant.
(Note that primary constants inherit the group structure of the correlation function.)
For any permutation $\sigma$ of the set $\{1,2,3\}$ the $A_1$ form factor satisfies
\begin{equation}
A_1^{I_{\sigma(1)} I_{\sigma(2)} I_{\sigma(3)}}(p_{\sigma(1)}, p_{\sigma(2)}, p_{\sigma(3)}) = A_1^{I_1 I_2 I_3}(p_1, p_2, p_3).
\end{equation}

\section*{\texorpdfstring{$\< J^{\mu_1} \mathcal{O} \mathcal{O} \>$}{<JOO>}}
\addcontentsline{toc}{subsection}{\numberline {3}\texorpdfstring{$\< J^{\mu_1} \mathcal{O} \mathcal{O} \>$}{<JOO>}}
\setcounter{subsection}{3}
\setcounter{equation}{0}

\noindent \textbf{Ward identities.} The transverse Ward identity is
\begin{align}
& p_{1 \mu_1} \lla J^{\mu_1 a}(\bs{p}_1) \mathcal{O}^{I_2}(\bs{p}_2) \mathcal{O}^{I_3}(\bs{p}_3) \rra = \nn\\
& \qquad = - (T_R^a)^{K I_3} \lla \mathcal{O}^K(\bs{p}_2) \mathcal{O}^{I_2}(-\bs{p}_2) \rra - (T_R^a)^{K I_2} \lla \mathcal{O}^K(\bs{p}_3) \mathcal{O}^{I_3}(-\bs{p}_3) \rra.
\end{align}

\bigskip \noindent \textbf{Reconstruction formula.} The full 3-point function can be reconstructed from the transverse-traceless part as
\begin{align}
& \lla J^{\mu_1 a}(\bs{p}_1) \mathcal{O}^{I_2}(\bs{p}_2) \mathcal{O}^{I_3}(\bs{p}_3) \rra = \lla j^{\mu_1 a}(\bs{p}_1) \mathcal{O}^{I_2}(\bs{p}_2) \mathcal{O}^{I_3}(\bs{p}_3) \rra \nn\\
& \qquad - \: \frac{p_1^{\mu_1}}{p_1^2} \left[ (T_R^a)^{K I_3} \lla \mathcal{O}^K(\bs{p}_2) \mathcal{O}^{I_2}(-\bs{p}_2) \rra + (T_R^a)^{K I_2} \lla \mathcal{O}^K(\bs{p}_3) \mathcal{O}^{I_3}(-\bs{p}_3) \rra \right].
\end{align}

\bigskip \noindent \textbf{Decomposition of the 3-point function.} The tensor decomposition of the transverse-traceless part is
\begin{equation}
\lla j^{\mu_1 a}(\bs{p}_1) \mathcal{O}^{I_2}(\bs{p}_2) \mathcal{O}^{I_3}(\bs{p}_3) \rra = \pi^{\mu_1}_{\alpha_1}(\bs{p}_1) \cdot A_1^{a I_2 I_3} p_2^{\alpha_1},
\end{equation}
where the form factor $A_1$ depends on the momentum magnitudes. This form factor is symmetric under $(p_2, I_2) \leftrightarrow (p_3, I_3)$, \textit{i.e.},
\begin{equation}
A_1^{a I_3 I_2}(p_1, p_3, p_2) = A_1^{a I_2 I_3}(p_1, p_2, p_3).
\end{equation}
This form factor is given by
\begin{equation}
A_1^{a I_2 I_3} = \text{coefficient of } p_2^{\mu_1} \text{ in } \lla J^{\mu_1 a}(\bs{p}_1) \mathcal{O}^{I_2}(\bs{p}_2) \mathcal{O}^{I_3}(\bs{p}_3) \rra.
\end{equation}

\bigskip \noindent \textbf{Primary conformal Ward identities.} The primary CWIs are
\begin{equation}
\K_{ij} A_1^{a I_2 I_3} = 0, \qquad i,j = 1,2,3,
\end{equation}
The solution in terms of triple-$K$ integrals \eqref{a:J} is
\begin{equation} \label{a:JOO1}
A_1^{a I_2 I_3} = \alpha_1^{a I_2 I_3} J_{1 \{000\}},
\end{equation}
where $\a_1^{a I_2 I_3}$ is a constant. In particular $\a_1^{a I_3 I_2} = \a_1^{a I_2 I_3}$. If the integral diverges, the regularisation \eqref{a:scheme} should be used.

\bigskip \noindent \textbf{Secondary conformal Ward identities.} The independent secondary CWI is
\begin{equation}
\Lo_{1} A_1^{a I_2 I_3} = 2(d-2) \left[ p_{1 \mu_1} \lla J^{\mu_1 a}(\bs{p}_1) \mathcal{O}^{I_2}(\bs{p}_2) \mathcal{O}^{I_3}(\bs{p}_3) \rra \right],
\end{equation}
where $\Lo_{N}$ is given by \eqref{a:L}. Assuming the unitarity bound $\Delta_2 = \Delta_3 = \Delta \geq \frac{d}{2} - 1$ for the dimensions of the scalar operators we find
\begin{align}
& \alpha^{a I_2 I_3}_1 (-2 d + 2 \Delta - (u-v) \epsilon ) l_{\frac{d}{2} + u \epsilon, \{ \frac{d}{2} - 1 + v \epsilon, \Delta - \frac{d}{2} + v \epsilon, \Delta - \frac{d}{2} + v \epsilon \}} = \nn\\
& \qquad = -2 (-2 + d + 2 u \epsilon) \Gamma \left( \frac{d}{2} - \Delta - v \epsilon \right) (T_R^a)^{I_2 I_3} c_{\O},
\end{align}
where the constant $l_{\alpha\{\beta_j\}}$ is defined in \eqref{a:l}.
After the substitution of the solution of the secondary CWI to \eqref{a:JOO1}, the limit $u = v = -1/2$ should be taken. The form factor then represents the 3-point function regulated in the dimensional regularisation \eqref{e:dimreg}.

The 3-point function $\lla J^{\mu_1} \mathcal{O} \mathcal{O} \rra$ is therefore completely determined in terms of this normalisation.

\subsection*{Examples}

\bigskip \noindent \textbf{For $\bm{d=3}$ and $\bm{\Delta_2=\Delta_3 = 1}$} we find
\begin{equation}
A_1^{a I_2 I_3} = - \frac{2 \sqrt{\pi} (T_R^a)^{I_2 I_3} c_{\mathcal{O}}}{b_{23} a_{123}}.
\end{equation}

\bigskip \noindent \textbf{For $\bm{d=5}$ and $\bm{\Delta_2=\Delta_3 = 3}$} we find
\begin{equation}
A_1^{a I_2 I_3} = - 4 \sqrt{\pi} (T_R^a)^{I_2 I_3} c_{\mathcal{O}} \frac{p_1 + a_{123}}{a_{123}^2}.
\end{equation}

\section*{\texorpdfstring{$\< J^{\mu_1} J^{\mu_2} \mathcal{O} \>$}{<JJO>}}
\addcontentsline{toc}{subsection}{\numberline {4}\texorpdfstring{$\< J^{\mu_1} J^{\mu_2} \mathcal{O} \>$}{<JJO>}}
\setcounter{subsection}{4}
\setcounter{equation}{0}

\noindent \textbf{Ward identities.} The transverse Ward identity is
\begin{equation}
p_{1 \mu_1} \lla J^{\mu_1 a_1}(\bs{p}_1) J^{\mu_2 a_2}(\bs{p}_2) \mathcal{O}^I(\bs{p}_3) \rra = p_{1 \mu_1} \lla \frac{\delta J^{\mu_1 a_1}}{\delta A_{\mu_2}^{a_2}}(\bs{p}_1, \bs{p}_2) \mathcal{O}^I(\bs{p}_3) \rra.
\end{equation}

\bigskip \noindent \textbf{Reconstruction formula.} The full 3-point function can be reconstructed from the transverse-traceless part as
\begin{align}
& \lla J^{\mu_1 a_1}(\bs{p}_1) J^{\mu_2 a_2}(\bs{p}_2) \mathcal{O}^I(\bs{p}_3) \rra = \lla j^{\mu_1 a_1}(\bs{p}_1) j^{\mu_2 a_2}(\bs{p}_2) \mathcal{O}^I(\bs{p}_3) \rra \nn\\
& \qquad + \: \frac{p_1^{\mu_1} p_{1 \alpha}}{p_1^2} \lla \frac{\delta J^{\alpha a_1}}{\delta A_{\mu_2}^{a_2}}(\bs{p}_1, \bs{p}_2) \mathcal{O}^I(\bs{p}_3) \rra + \frac{p_2^{\mu_2} p_{2 \beta}}{p_2^2} \lla \frac{\delta J^{\beta a_2}}{\delta A_{\mu_1}^{a_1}}(\bs{p}_2, \bs{p}_1) \mathcal{O}^I(\bs{p}_3) \rra \nn\\
& \qquad - \: \frac{p_1^{\mu_1} p_2^{\mu_2} p_{1 \alpha} p_{2 \beta}}{p_1^2 p_2^2} \lla \frac{\delta J^{\alpha a_1}}{\delta A_{\beta}^{a_2}}(\bs{p}_1, \bs{p}_2) \mathcal{O}^I(\bs{p}_3) \rra.
\end{align}

\bigskip \noindent \textbf{Decomposition of the 3-point function.} The tensor decomposition of the transverse-traceless part is
\begin{equation}
\lla j^{\mu_1 a_1}(\bs{p}_1) j^{\mu_2 a_2}(\bs{p}_2) \mathcal{O}^I(\bs{p}_3) \rra = \pi^{\mu_1}_{\alpha_1}(\bs{p}_1) \pi^{\mu_2}_{\alpha_2}(\bs{p}_2) \left[ A_1^{a_1 a_2 I} p_2^{\alpha_1} p_3^{\alpha_2} + A_2^{a_1 a_2 I} \delta^{\alpha_1 \alpha_2} \right].
\end{equation}
The form factors $A_1$ and $A_2$ are functions of the momentum magnitudes. Both form factors are symmetric under $(p_1, a_1) \leftrightarrow (p_2, a_2)$, \textit{i.e.}, they satisfy
\begin{equation}
A_j^{a_2 a_1 I}(p_2, p_1, p_3) = A_j^{a_1 a_2 I}(p_1, p_2, p_3), \qquad j = 1,2.
\end{equation}
These form factors can be calculated as follows
\begin{align}
A_1^{a_1 a_2 I} & = \text{coefficient of } p_2^{\mu_1} p_3^{\mu_2}, \\
A_2^{a_1 a_2 I} & = \text{coefficient of } \delta^{\mu_1 \mu_2}
\end{align}
in $\lla J^{\mu_1 a_1}(\bs{p}_1) J^{\mu_2 a_2}(\bs{p}_2) \mathcal{O}^I(\bs{p}_3) \rra$.

\bigskip \noindent \textbf{Primary conformal Ward identities.} The primary CWIs are
\begin{equation}
\begin{array}{ll}
\K_{12} A_1^{a_1 a_2 I} = 0, & \qquad\qquad \K_{13} A_1^{a_1 a_2 I} = 0, \\
\K_{12} A_2^{a_1 a_2 I} = 0, & \qquad\qquad \K_{13} A_2^{a_1 a_2 I} = 2 A_1^{a_1 a_2 I},
\end{array}
\end{equation}
The solution in terms of triple-$K$ integrals \eqref{a:J} is
\begin{align}
A_1^{a_1 a_2 I} & = \alpha_1^{a_1 a_2 I} J_{2 \{000\}}, \label{a:JJO1} \\
A_2^{a_1 a_2 I} & = \alpha_1^{a_1 a_2 I} J_{1 \{001\}} + \alpha_2^{a_1 a_2 I} J_{0 \{000\}}, \label{a:JJOlast}
\end{align}
where $\a_j^{a I_2 I_3}$, $j=1,2$ are constants. In particular $\a_j^{a_2 a_1 I} = \a_j^{a_1 a_2 I}$ for $j=1,2$. If the integrals diverge, the regularisation \eqref{a:scheme} should be used.

\bigskip \noindent \textbf{Secondary conformal Ward identities.} The independent secondary CWI is
\begin{align}
& \Lo_{1} A_1^{a_1 a_2 I} + 2 \Ro A_2^{a_1 a_2 I} = \nn\\
& \qquad = 2 (d-2) \cdot \text{coefficient of } p^{\mu_2}_3 \text{ in } p_{1 \mu_1} \lla J^{\mu_1 a_1}(\bs{p}_1) J^{\mu_2 a_2}(\bs{p}_2) \mathcal{O}^I(\bs{p}_3) \rra, \label{sec_JJO}
\end{align}
where $\Lo$ and $\Ro$ operators are given by \eqref{a:L} and \eqref{a:R}. This leads to
\begin{align}
& l_{\frac{d}{2} + 1 + u \epsilon \{ \frac{d}{2} - 1 + v \epsilon, \frac{d}{2} - 1 + v \epsilon, \frac{d}{2} - \Delta - v \epsilon \}} \left[ - (2 + 2 \Delta + (u + 3v) \epsilon) \alpha_1^{a_1 a_2 I} \right.\nn\\
& \qquad\qquad\qquad \left. + \: \frac{4 (1 + \Delta + (u+v) \epsilon) \alpha_2^{a_1 a_2 I}}{(-2 + d - \Delta - (u-v)\epsilon)(\Delta + (u+v)\epsilon)} \right] = \nn\\
& \qquad = 2 \delta_{d, \Delta-2-2n} (-2 + d + 2 u \epsilon) \Gamma \left( \frac{d}{2} - \Delta - v \epsilon \right) c^{a_1 a_2 I} c_{\mathcal{O}},
\end{align}
where the constant $l_{\alpha\{\beta_j\}}$ is defined in \eqref{a:l} and $c^{a_1 a_2 K}$ is a constant defined by
\begin{align} \label{a:caaK}
& \left. p_{1 \mu_1} \lla \frac{\delta J^{\mu_1 a_1}}{\delta A^{a_2}_{\mu_2}}(\bs{p}_1, \bs{p}_2) \mathcal{O}^I(\bs{p}_3) \rra \right|_{p_1 = p_2 = p} = \nn\\
& \qquad\qquad = p_3^{\mu_2} \cdot c^{a_1 a_2 K} p^{d-\Delta_3-2} \lla \mathcal{O}^K(\bs{p}_3) \mathcal{O}^I(-\bs{p}_3) \rra + \ldots
\end{align}
\textit{i.e.}, we first write down the most general tensor decomposition for $p_{1 \mu_1} \lla \frac{\delta J^{\mu_1 a_1}}{\delta A^{a_2}_{\mu_2}}(\bs{p}_1, \bs{p}_2) \mathcal{O}^I(\bs{p}_3) \rra$ then extract the coefficient of $p_3^{\mu_2}$ and set $p_1 = p_2 = p$ in this coefficient\footnote{This correlator is non-vanishing only if $d - \Delta_3 - 2 = 2n$ for a non-negative integer $n$. Otherwise, the right-hand side of the secondary Ward identity \eqref{sec_JJO} vanishes.}.

After the substitution of the solution of the secondary CWI to \eqref{a:JJO1} - \eqref{a:JJOlast}, the limit $u = v = -1/2$ should be taken. The form factors then represent the 3-point function regulated in the dimensional regularisation \eqref{e:dimreg}.

In summary, the 3-point function $\lla J^{\mu_1} J^{\mu_2} \mathcal{O} \rra$ depends on the 2-point function normalisations $c_{\mathcal{O}}$ and $c^{a_1 a_2 K}$, and on one undetermined primary constant $\alpha_1^{a_1 a_2 I}$. Note that for these correlation functions to be non-zero the symmetric group $G$ must have an invariant tensor $r^{a_1 a_2 I}$ (which is a non-trivial condition). Then $\alpha_1^{a_1 a_2 I} = \alpha_1 r^{a_1 a_2 I}$.

\subsection*{Examples}

\bigskip \noindent \textbf{For $\bm{d=3}$ and $\bm{\Delta_3 = 1}$} we find
\begin{align}
A_1^{a_1 a_2 I} & = \frac{\alpha_1^{a_1 a_2 I}}{p_3 a_{123}^2}, \\
A_2^{a_1 a_2 I} & = \alpha_1^{a_1 a_2 I} \left( \frac{1}{a_{123}} - \frac{1}{2 p_3} \right) - \frac{\sqrt{\pi} c^{a_1 a_2 I} c_{\mathcal{O}}}{p_3}.
\end{align}

\bigskip \noindent \textbf{For $\bm{d=3}$ and $\bm{\Delta_3 = 3}$} we find
\begin{align}
A_1^{a_1 a_2 I} & = \alpha_1^{a_1 a_2 I} \frac{a_{123} + p_3}{a_{123}^2}, \\
A_2^{a_1 a_2 I} & = - \alpha_1^{a_1 a_2 I} \frac{-2 p_3^2 + p_3 a_{12} + a_{12}^2}{2 a_{123}}.
\end{align}

\bigskip \noindent \textbf{For $\bm{d=5}$ and $\bm{\Delta_3 = 3}$} we find
\begin{align}
A_1^{a_1 a_2 I} & = \alpha_1^{a_1 a_2 I} \frac{p_3^2 + 3 p_3 a_{12} + 2(a_{12}^2 + b_{12})}{a_{123}^3}, \\
A_2^{a_1 a_2 I} & = \alpha_1^{a_1 a_2 I} \frac{p_3^3 + 2 p_3^2 a_{12} + p_3 (-a_{12}^2 + 4 b_{12}) + 2 a_{13} (-a_{12}^2 + b_{12})}{2 a_{123}^2} \nn\\
& \qquad\qquad + 2 \sqrt{\pi} p_3 c^{a_1 a_2 I} c_{\mathcal{O}}.
\end{align}

\bigskip \noindent \textbf{For $\bm{d=5}$ and $\bm{\Delta_3 = 5}$} we find
\begin{align}
A_1^{a_1 a_2 I} & = \frac{\alpha_1^{a_1 a_2 I}}{a_{123}^3} \left[ -2 p_3^4 - 6 p_3^3 a_{12} - 2 p_3^2 (5 a_{12}^2 - 4 b_{12}) \right.\nn\\
& \qquad\qquad \left. - 3 a_{12}(a_{12} + 3 p_3) (a_{12}^2 - b_{12}) \right], \\
A_2^{a_1 a_2 I} & = \frac{\alpha_1^{a_1 a_2 I}}{2 a_{123}^2} \left[ -2 p_3^5 -4 p_3^4 a_{12} - 4 p_3^3 (a_{12}^2 - b_{12}) + p_3^2 a_{12} (a_{12}^2 - 7 b_{12}) \right.\nn\\
& \qquad\qquad \left. +\:3 a_{12}^2 (2 p_3 + a_{12}) (a_{12}^2 - 3 b_{12}) \right].
\end{align}

\section*{\texorpdfstring{$\< J^{\mu_1} J^{\mu_2} J^{\mu_3} \>$}{<JJJ>}}
\addcontentsline{toc}{subsection}{\numberline {5}\texorpdfstring{$\< J^{\mu_1} J^{\mu_2} J^{\mu_3} \>$}{<JJJ>}}
\setcounter{subsection}{5}
\setcounter{equation}{0}

\noindent \textbf{Ward identities.} The transverse Ward identity is
\begin{align}
& p_{1 \mu_1} \lla J^{\mu_1 a_1}(\bs{p}_1) J^{\mu_2 a_2}(\bs{p}_2) J^{\mu_3 a_3}(\bs{p}_3) \rra = \nn\\
& \qquad = \I f^{a_1 b a_3} \lla J^{\mu_3 b}(\bs{p}_2) J^{\mu_2 a_2}(-\bs{p}_2) \rra - \I f^{a_1 a_2 b} \lla J^{\mu_2 b}(\bs{p}_3) J^{\mu_3 a_3}(-\bs{p}_3) \rra,
\end{align}
where $f^{abc}$ are the structure constants of the symmetry group.

\bigskip \noindent \textbf{Reconstruction formula.} The full 3-point function can be reconstructed from the transverse-traceless part as
\begin{align}
& \lla J^{\mu_1 a_1}(\bs{p}_1) J^{\mu_2 a_2}(\bs{p}_2) J^{\mu_3 a_3}(\bs{p}_3) \rra = \lla j^{\mu_1 a_1}(\bs{p}_1) j^{\mu_2 a_2}(\bs{p}_2) j^{\mu_3 a_3}(\bs{p}_3) \rra \nn\\
& \qquad + \: \left[ \frac{p_1^{\mu_1}}{p_1^2} \left( \I f^{a_1 b a_3} \lla J^{\mu_3 b}(\bs{p}_2) J^{\mu_2 a_2}(-\bs{p}_2) \rra - \I f^{a_1 a_2 b} \lla J^{\mu_2 b}(\bs{p}_3) J^{\mu_3 a_3}(-\bs{p}_3) \rra \right) \right] \nn\\
& \qquad\qquad + \: [ (\mu_1, a_1, \bs{p}_1) \leftrightarrow (\mu_2, b_2, \bs{p}_2) ] + [ (\mu_1, a_1, \bs{p}_1) \leftrightarrow (\mu_3, a_3, \bs{p}_3) ] \nn\\
& \qquad + \: \left[ \frac{p_1^{\mu_1} p_2^{\mu_2}}{p_1^2 p_2^2} \I f^{a_1 a_2 b} p_{2 \alpha} \lla J^{\alpha b}(\bs{p}_3) J^{\mu_3 a_3}(-\bs{p}_3) \rra \right] \nn\\
& \qquad\qquad + \: [ (\mu_1, a_1, \bs{p}_1) \leftrightarrow (\mu_3, a_3, \bs{p}_3) ] + [ (\mu_2, a_2, \bs{p}_2) \leftrightarrow (\mu_3, a_3, \bs{p}_3) ].
\end{align}

\bigskip \noindent \textbf{Decomposition of the 3-point function.} The tensor decomposition of the transverse-traceless part is
\begin{align}
& \lla j^{\mu_1 a_1}(\bs{p}_1) j^{\mu_2 a_2}(\bs{p}_2) j^{\mu_3 a_3}(\bs{p}_3) \rra = \pi^{\mu_1}_{\alpha_1}(\bs{p}_1) \pi^{\mu_2}_{\alpha_2}(\bs{p}_2) \pi^{\mu_3}_{\alpha_3}(\bs{p}_3) \left[
A_1^{a_1 a_2 a_3} p_2^{\alpha_1} p_3^{\alpha_2} p_1^{\alpha_3} \right. \nonumber \\
& \left. \qquad \qquad + \: A_2^{a_1 a_2 a_3} \delta^{\alpha_1 \alpha_2} p_1^{\alpha_3} + A_2^{a_3 a_1 a_2}(p_3, p_1, p_2) \delta^{\alpha_1 \alpha_3} p_3^{\alpha_2} \right.\nn\\
& \qquad\qquad\qquad\qquad \left. + \: A_2^{a_2 a_3 a_1}(p_2, p_3, p_1) \delta^{\alpha_2 \alpha_3} p_2^{\alpha_1} \right].
\end{align}
The form factors $A_1$ and $A_2$ are functions of the momentum magnitudes. If no arguments are given, then we assume the standard ordering, $A_j = A_j(p_1, p_2, p_3)$.

The $A_1$ factor is completely antisymmetric, \textit{i.e.}, for any permutation $\sigma$ of the set $\{1,2,3\}$ it satisfies
\begin{equation}
A_1^{a_{\sigma(1)} a_{\sigma(2)} a_{\sigma(3)}}(p_{\sigma(1)}, p_{\sigma(2)}, p_{\sigma(3)}) = (-1)^\sigma A_1^{a_1 a_2 a_3}(p_1, p_2, p_3),
\end{equation}
where $(-1)^\sigma$ denotes the sign of the permutation $\sigma$.
Under a permutation of the momenta only, however, the form factor is completely symmetric,
\begin{equation}
A_1^{a_1 a_2 a_3}(p_{\sigma(1)}, p_{\sigma(2)}, p_{\sigma(3)}) = A_1^{a_1 a_2 a_3}(p_1, p_2, p_3).
\end{equation}
The form factor $A_2$ is antisymmetric under $(p_1, a_1) \leftrightarrow (p_2, a_2)$, \textit{i.e.},
\begin{equation}
A_2^{a_2 a_1 a_3}(p_2, p_1, p_3) = -A_2^{a_1 a_2 a_3}(p_1, p_2, p_3).
\end{equation}
Note that the group structure of the form factors requires the existence of tensors of the form $t^{a_1 a_2 a_3}$ with appropriate symmetry properties (fully antisymmetric for the one associated with $A_1$, and antisymmetric in its first two indices for the one associated with $A_2$\footnote{The secondary Ward identities then ensure this tensor is fully antisymmetric, as below.}). One such tensor 
is the structure constant $f^{a_1 a_2 a_3}$. As argued in \cite{Osborn:1993cr}, the correlation function vanishes if the symmetry group is Abelian.

The form factors can be calculated as follows
\begin{align}
A_1^{a_1 a_2 a_3} & = \text{coefficient of } p_2^{\mu_1} p_3^{\mu_2} p_1^{\mu_3}, \\
A_2^{a_1 a_2 a_3} & = \text{coefficient of } \delta^{\mu_1 \mu_2} p_1^{\mu_3}
\end{align}
in $\lla J^{\mu_1 a_1}(\bs{p}_1) J^{\mu_2 a_2}(\bs{p}_2) J^{\mu_3 a_3}(\bs{p}_3) \rra$.

\bigskip \noindent \textbf{Primary conformal Ward identities.} The primary CWIs are
\begin{equation}
\begin{array}{ll}
\K_{12} A_1^{a_1 a_2 a_3} = 0, & \qquad\qquad \K_{13} A_1^{a_1 a_2 a_3} = 0, \\
\K_{12} A_2^{a_1 a_2 a_3} = 0, & \qquad\qquad \K_{13} A_2^{a_1 a_2 a_3} = 2 A_1^{a_1 a_2 a_3}.
\end{array}
\end{equation}
The solution in terms of triple-$K$ integrals \eqref{a:J} is
\begin{align}
A_1^{a_1 a_2 a_3} & = \a_1^{a_1 a_2 a_3} J_{3 \{000\}}, \label{a:JJJ1} \\
A_2^{a_1 a_2 a_3} & = \a_1^{a_1 a_2 a_3} J_{2 \{001\}} + \a_2^{a_1 a_2 a_3} J_{1 \{000\}}, \label{a:JJJlast}
\end{align}
where $\a_j^{a_1 a_2 a_3}$, $j=1,2$ are constants. If the integrals diverge, the regularisation \eqref{a:scheme} should be used.

\bigskip \noindent \textbf{Secondary conformal Ward identities.} The independent secondary CWIs are
\begin{align}\label{eqn320}
& (*) \ \Lo_{3} A_1^{a_1 a_2 a_3} + 2 \Ro \left[ A_2^{a_1 a_2 a_3} - A_2^{a_3 a_1 a_2}(p_3, p_1, p_2) \right]  \nn\\
& \qquad = 2 (\Delta_1 - 1) \cdot \text{coefficient of } p_3^{\mu_2} p_1^{\mu_3} \text{ in } p_{1 \mu_1} \lla J^{\mu_1 a_1}(\bs{p}_1) J^{\mu_2 a_2}(\bs{p}_2) J^{\mu_3 a_3}(\bs{p}_3) \rra, \\[1ex]
& \Lo_{1} \left[ A_2^{a_2 a_3 a_1}(p_2, p_3, p_1) \right] + 2 p_1^2 \left[ A_2^{a_3 a_1 a_2}(p_3, p_1, p_2) - A_2^{a_1 a_2 a_3} \right]  \nn\\
& \qquad = 2 (\Delta_1 - 1) \cdot \text{coefficient of } \delta^{\mu_2 \mu_3} \text{ in } p_{1 \mu_1} \lla J^{\mu_1 a_1}(\bs{p}_1) J^{\mu_2 a_2}(\bs{p}_2) J^{\mu_3 a_3}(\bs{p}_3) \rra,
\end{align}
where $\Lo$ and $\Ro$ operators are given by \eqref{e:L} and \eqref{e:R}. Equation \eqref{eqn320} determines the symmetry properties of $\a_2^{a_1 a_2 a_3}$, enforcing $\a_2^{a_1 a_3 a_2} = -\a_2^{a_1 a_2 a_3}$ in any dimension. Given its antisymmetry in the first two indices, $\a_2^{a_1 a_2 a_3}$ is thus fully antisymmetric. Otherwise, this equation is redundant, with its right-hand side vanishing, and places no further constraints on the primary constants.

The secondary CWIs lead to
\begin{equation}
\a_2^{a_1 a_2 a_3} = (d-2+2 v \epsilon) \left[ - \a_1^{a_1 a_2 a_3} + \frac{2^{3 - \frac{d}{2} - v \epsilon} \cdot \I f^{a_1 a_2 a_3} c_{J}}{\Gamma^2 \left( \frac{d}{2} + v \epsilon \right)} \right],
\end{equation}
where $c_J$ is the 2-point function normalisation \eqref{a:JJ}.
After the substitution of the solution of the secondary CWI to \eqref{a:JJJ1} - \eqref{a:JJJlast}, the limit $u = v = -1/2$ should be taken. The form factors then represent the 3-point function regulated in the dimensional regularisation \eqref{e:dimreg}.

The 3-point function $\lla J^{\mu_1} J^{\mu_2} J^{\mu_3} \rra$ therefore depends on the 2-point function normalisation $c_J$ and on one undetermined primary constant $\a_1^{a_1 a_2 a_3}$.

\subsection*{Examples}

\bigskip \noindent \textbf{For $\bm{d=3}$} we find
\begin{align}
A_1^{a_1 a_2 a_3} & = \frac{2 \a_1^{a_1 a_2 a_3}}{a_{123}^3}, \\
A_2^{a_1 a_2 a_3} & = \a_1^{a_1 a_2 a_3} \frac{p_3}{a_{123}^2} + \frac{4 \sqrt{\pi} \I f^{a_1 a_2 a_3} c_{J}}{a_{123}}.
\end{align}

\bigskip \noindent \textbf{For $\bm{d=5}$} we find
\begin{align}
A_1^{a_1 a_2 a_3} & = \frac{2 \a_1^{a_1 a_2 a_3}}{a_{123}^4} \left[ a_{123}^3 + a_{123} b_{123} + 3 c_{123} \right], \\
A_2^{a_1 a_2 a_3} & = \frac{\a_1^{a_1 a_2 a_3} p_3^2}{a_{123}^3} \left[ p_3^2 + 3 p_3 a_{12} + 2 (a_{12}^2 + b_{12}) \right] \nn\\
& \qquad\qquad - \: \frac{8 \sqrt{\pi} \I f^{a_1 a_2 a_3} c_{J}}{3 a_{123}^2} \left[ a_{123}^3 - a_{123} b_{123} - c_{123} \right].
\end{align}

\section*{\texorpdfstring{$\< T^{\mu_1 \nu_1} \mathcal{O} \mathcal{O} \>$}{<TOO>}}
\addcontentsline{toc}{subsection}{\numberline {6}\texorpdfstring{$\< T^{\mu_1 \nu_1} \mathcal{O} \mathcal{O} \>$}{<TOO>}}
\setcounter{subsection}{6}
\setcounter{equation}{0}

\noindent \textbf{Ward identities.} The transverse and trace Ward identities are
\begin{align}
& p_1^{\nu_1} \lla T_{\mu_1 \nu_1}(\bs{p}_1) \mathcal{O}^{I_2}(\bs{p}_2) \mathcal{O}^{I_3}(\bs{p}_3) \rra = \nn\\
& \qquad = p_{3 \mu_1} \lla \mathcal{O}^{I_2}(\bs{p}_3) \mathcal{O}^{I_3}(-\bs{p}_3) \rra + p_{2 \mu_1} \lla \mathcal{O}^{I_2}(\bs{p}_2) \mathcal{O}^{I_3}(-\bs{p}_2) \rra, \\
& \lla T(\bs{p}_1) \mathcal{O}^{I_2}(\bs{p}_2) \mathcal{O}^{I_3}(\bs{p}_3) \rra = - \Delta_3 \left[ \lla \mathcal{O}^{I_2}(\bs{p}_3) \mathcal{O}^{I_3}(-\bs{p}_3) \rra + \lla \mathcal{O}^{I_2}(\bs{p}_2) \mathcal{O}^{I_3}(-\bs{p}_2) \rra \right].
\end{align}

\bigskip \noindent \textbf{Reconstruction formula.} The full 3-point function can be reconstructed from the transverse-traceless part as
\begin{align}
& \lla T^{\mu_1 \nu_1}(\bs{p}_1) \mathcal{O}^{I_2}(\bs{p}_2) \mathcal{O}^{I_3}(\bs{p}_3) \rra = \lla t^{\mu_1 \nu_1}(\bs{p}_1) \mathcal{O}^{I_2}(\bs{p}_2) \mathcal{O}^{I_3}(\bs{p}_3) \rra \nn\\
& \qquad + \left[ p_2^\alpha \mathscr{T}^{\mu_1 \nu_1}_{\alpha}(\bs{p}_1) - \frac{\Delta_3}{d-1} \pi^{\mu_1 \nu_1}(\bs{p}_1) \right] \lla \mathcal{O}^{I_2}(\bs{p}_2) \mathcal{O}^{I_3}(-\bs{p}_2) \rra + [ \bs{p}_2 \leftrightarrow \bs{p}_3 ],
\end{align}
where $\mathscr{T}^{\mu_1 \nu_1}_{\alpha}$ is defined in \eqref{a:T}.

\bigskip \noindent \textbf{Decomposition of the 3-point function.} The tensor decomposition of the transverse-traceless part is
\begin{equation}
\lla t^{\mu_1 \nu_1}(\bs{p}_1) \mathcal{O}^{I_2}(\bs{p}_2) \mathcal{O}^{I_3}(\bs{p}_3) \rra = \Pi^{\mu_1 \nu_1}_{\alpha_1 \beta_1}(\bs{p}_1) \cdot A_1^{I_2 I_3} p_2^{\alpha_1} p_2^{\beta_1},
\end{equation}
where $A_1$ is a form factor depending on the momentum magnitudes. This form factor is symmetric under $(p_2, I_2) \leftrightarrow (p_3, I_3)$, \textit{i.e.},
\begin{equation}
A_1^{I_3 I_2}(p_1, p_3, p_2) = A_1^{I_2 I_3}(p_1, p_2, p_3)
\end{equation}
and may be calculated as
\begin{equation}
A_1^{I_2 I_3} = \text{coefficient of } p_2^{\mu_1} p_2^{\nu_1} \text{ in } \lla T^{\mu_1 \nu_1}(\bs{p}_1) \mathcal{O}^{I_2}(\bs{p}_2) \mathcal{O}^{I_3}(\bs{p}_3) \rra.
\end{equation}

\bigskip \noindent \textbf{Primary conformal Ward identities.} The primary CWIs are
\begin{equation}
\K_{ij} A_1^{I_2 I_3} = 0, \qquad i,j = 1,2,3.
\end{equation}
The solution in terms of triple-$K$ integrals \eqref{a:J} is
\begin{equation} \label{a:TOO1}
A_1^{I_2 I_3} = \a_1^{I_2 I_3} J_{2 \{000\}},
\end{equation}
where $\a_1^{I_2 I_3}$ is a constant (note $\a_1^{I_3 I_2} = \a_1^{I_2 I_3}$). If the integral diverges, the regularisation \eqref{a:scheme} should be used.

\bigskip \noindent \textbf{Secondary conformal Ward identities.} The independent secondary CWI is
\begin{equation}
\Lo_{2} A_1^{I_2 I_3} = 2 d \cdot \text{coefficient of } p_2^{\mu_1} \text{ in } p_{1 \nu_1} \lla T^{\mu_1 \nu_1}(\bs{p}_1) \mathcal{O}^{I_2}(\bs{p}_2) \mathcal{O}^{I_3}(\bs{p}_3) \rra,
\end{equation}
where $\Lo_{N}$ is defined in \eqref{a:L}. Assuming the unitarity bound for the conformal dimension of the scalar operator $\Delta_2 = \Delta_3 = \Delta \geq \frac{d}{2} - 1$ we find
\begin{align}
& \a_1^{I_2 I_3} (-2 - 2d + 2\Delta - (u-v) \epsilon) l_{\frac{d}{2} + 1 + u \epsilon \{ \frac{d}{2} + v \epsilon, \Delta - \frac{d}{2} + v \epsilon, \Delta - \frac{d}{2} + v \epsilon \}} = \nn\\
& \qquad = - 2 (d + 2 u \epsilon) \Gamma \left( \frac{d}{2} - \Delta - v \epsilon \right) \delta^{I_2 I_3} c_{\O},
\end{align}
where the constant $l_{\alpha\{\beta_j\}}$ is defined in \eqref{a:l}.
After the substitution of the solution of the secondary CWI to \eqref{a:TOO1}, the limit $u = v = -1/2$ should be taken. The form factor then represents the 3-point function regulated in the dimensional regularisation \eqref{e:dimreg}.

The 3-point function $\lla T^{\mu_1 \nu_1} \mathcal{O} \mathcal{O} \rra$ is thus uniquely determined in terms of the 2-point function normalisation $c_{\mathcal{O}}$.

\subsection*{Examples}

\bigskip \noindent \textbf{For $\bm{d=3}$ and $\bm{\Delta_2=\Delta_3 = 1}$} we find
\begin{equation}
A_1^{I_2 I_3} = - 2 \sqrt{\pi} c_{\mathcal{O}} \delta^{I_2 I_3} \frac{p_1 + a_{123}}{b_{23} a_{123}^2}.
\end{equation}

\bigskip \noindent \textbf{For $\bm{d=3}$ and $\bm{\Delta_2=\Delta_3 = 3}$} we find
\begin{equation} \label{res:TOOd3}
A_1^{I_2 I_3} =  \tfrac{8}{3} \sqrt{\pi} c_{\mathcal{O}} \delta^{I_2 I_3} \frac{a_{123}^3 - a_{123} b_{123} - c_{123}}{a_{123}^2}.
\end{equation}

\bigskip \noindent \textbf{For $\bm{d=5}$ and $\bm{\Delta_2=\Delta_3 = 3}$} we find
\begin{equation}
A_1^{I_2 I_3} = - \tfrac{4}{3} \sqrt{\pi} c_{\mathcal{O}} \delta^{I_2 I_3} \frac{8 p_1^2 + 9 p_1 a_{23} + 3 a_{23}^2}{a_{123}^3}.
\end{equation}

\bigskip \noindent \textbf{For $\bm{d=5}$ and $\bm{\Delta_2=\Delta_3 = 5}$} we find
\begin{align}
A_1^{I_2 I_3} & = - \frac{16 \sqrt{\pi} c_{\mathcal{O}} \delta^{I_2 I_3}}{45 a_{123}^3} \left[ 3 a_{123}^6 - 9 a_{123}^4 b_{123} + 3 a_{123}^2 b_{123}^2 \right.\nn\\
& \qquad\qquad \left. +3 a_{123}^3 c_{123} + 3 a_{123} b_{123} c_{123} + 2 c_{123}^2 \right].
\end{align}

\section*{\texorpdfstring{$\< T^{\mu_1 \nu_1} J^{\mu_2} \mathcal{O} \>$}{<TJO>}}
\addcontentsline{toc}{subsection}{\numberline {7}\texorpdfstring{$\< T^{\mu_1 \nu_1} J^{\mu_2} \mathcal{O} \>$}{<TJO>}}
\setcounter{subsection}{7}
\setcounter{equation}{0}

The transverse-traceless part of $\lla T^{\mu_1 \nu_1} J^{\mu_2} \mathcal{O} \rra$ vanishes. We present the detailed analysis of this case in appendix \ref{ch:vanish}.

\bigskip \noindent \textbf{Ward identities.} The transverse and trace Ward identities are
\begin{align}
& p_1^{\nu_1} \lla T_{\mu_1 \nu_1}(\bs{p}_1) J^{\mu_2 a}(\bs{p}_2) \mathcal{O}^I(\bs{p}_3) \rra = p_1^{\nu_1} \lla \frac{\delta T_{\mu_1 \nu_1}}{\delta A_{\mu_2}^a}(\bs{p}_1, \bs{p}_2) \mathcal{O}^I(\bs{p}_3) \rra, \\
& p_{2 \mu_2} \lla T_{\mu_1 \nu_1}(\bs{p}_1) J^{\mu_2 a}(\bs{p}_2) \mathcal{O}^I(\bs{p}_3) \rra = 2 p_{2 \mu_2} \lla \frac{\delta J^{\mu_2 a}}{\delta g^{\mu_1 \nu_1}}(\bs{p}_2, \bs{p}_1) \mathcal{O}^I(\bs{p}_3) \rra, \\
& \lla T(\bs{p}_1) J^{\mu_2 a}(\bs{p}_2) \mathcal{O}^I(\bs{p}_3) \rra = 0.
\end{align}

\bigskip \noindent \textbf{Reconstruction formula.} The full 3-point function can be reconstructed from the transverse-traceless part as
\begin{align}
& \lla T^{\mu_1 \nu_1}(\bs{p}_1) J^{\mu_2 a}(\bs{p}_2) \mathcal{O}^I(\bs{p}_3) \rra = \lla t^{\mu_1 \nu_1}(\bs{p}_1) j^{\mu_2 a}(\bs{p}_2) \mathcal{O}^I(\bs{p}_3) \rra \nn\\
& \qquad + \frac{\pi^{\mu_2}_{\alpha_2}(\bs{p}_2) p_1^{\beta_1}}{p_1^2} \mathscr{T}^{\mu_1 \nu_1 \alpha_1}(\bs{p}_1) \lla \frac{\delta T_{\alpha_1 \beta_1}}{\delta A^a_{\alpha_2}}(\bs{p}_1, \bs{p}_2) \mathcal{O}^I(\bs{p}_3) \rra \nn\\
& \qquad\qquad + \: \frac{2 p_2^{\mu_2} p_{2 \alpha_2}}{p_2^2} \delta^{\mu_1 \alpha_1} \delta^{\nu_1 \beta_1} \lla \frac{\delta J^{\alpha_2 a}}{\delta g^{\alpha_1 \beta_1}}(\bs{p}_2, \bs{p}_1) \mathcal{O}^I(\bs{p}_3) \rra,
\end{align}
where $\mathscr{T}^{\mu_1 \nu_1}_{\alpha}$ is defined in \eqref{a:T}.

\bigskip \noindent \textbf{Decomposition of the 3-point function.} The tensor decomposition of the transverse-traceless part is
\begin{equation}
\lla t^{\mu_1 \nu_1}(\bs{p}_1) j^{\mu_2 a}(\bs{p}_2) \mathcal{O}^I(\bs{p}_3) \rra = \Pi^{\mu_1 \nu_1}_{\alpha_1 \beta_1}(\bs{p}_1) \pi^{\mu_2}_{\alpha_2}(\bs{p}_2) \left[ A_1^{aI} p_2^{\alpha_1} p_2^{\beta_1} p_3^{\alpha_2} + A_2^{aI} \delta^{\alpha_1 \alpha_2} p_2^{\beta_1} \right].
\end{equation}
The form factors $A_1$ and $A_2$ depend on the momentum magnitudes, and may be calculated as follows
\begin{align}
A_1^{aI} & = \text{coefficient of } p_2^{\mu_1} p_2^{\nu_1} p_3^{\mu_2}, \\
A_2^{aI} & = 2 \cdot \text{coefficient of } \delta^{\mu_1 \mu_2} p_2^{\nu_1}
\end{align}
in $\lla T^{\mu_1 \nu_1}(\bs{p}_1) J^{\mu_2 a}(\bs{p}_2) \mathcal{O}^I(\bs{p}_3) \rra$. These form factors do not exhibit any symmetry properties.

\bigskip \noindent \textbf{Primary conformal Ward identities.} The primary CWIs are
\begin{equation}
\begin{array}{ll}
\K_{12} A_1^{aI} = 0, & \qquad\qquad \K_{13} A_1^{aI} = 0, \\
\K_{12} A_2^{aI} = 0, & \qquad\qquad \K_{13} A_2^{aI} = 4 A_1^{aI}.
\end{array}
\end{equation}
The solution in terms of triple-$K$ integrals \eqref{a:J} is
\begin{align}
A_1^{aI} & = \a_1^{aI} J_{3 \{000\}}, \\
A_2^{aI} & = 2 \a_1^{aI} J_{2 \{001\}} + \a_2^{aI} J_{1 \{000\}},
\end{align}
where $\a_j^{aI}$, $j=1,2$ are constants. If the integrals diverge, the regularisation \eqref{a:scheme} should be used.

\bigskip \noindent \textbf{Secondary conformal Ward identities.} The independent secondary CWIs are
\begin{align}
& \Lo_{2} A_1^{aI} + \Ro A_2^{aI} = 2 d \cdot \text{coefficient of } p_2^{\mu_1} p_3^{\mu_2} \text{ in } p_{1 \nu_1} \lla T^{\mu_1 \nu_1}(\bs{p}_1) J^{\mu_2 a}(\bs{p}_2) \mathcal{O}^I(\bs{p}_3) \rra, \\
& \Lo'_{1} A_1^{aI} + 2 \Ro' A_2^{aI} = \nn\\
& \qquad = - 2 (d - 2) \cdot \text{coefficient of } p_2^{\mu_1} p_2^{\nu_1} \text{ in } p_{2 \mu_2} \lla T^{\mu_1 \nu_1}(\bs{p}_1) J^{\mu_2 a}(\bs{p}_2) \mathcal{O}^I(\bs{p}_3) \rra, \\
& \Lo_{2} A_2^{aI} = 4 d \cdot \text{coefficient of } \delta^{\mu_1 \mu_2} \text{ in } p_{1 \nu_1} \lla T^{\mu_1 \nu_1}(\bs{p}_1) J^{\mu_2 a}(\bs{p}_2) \mathcal{O}^I(\bs{p}_3) \rra,
\end{align}
where the $\Lo$ and $\Ro$ operators are given by \eqref{a:L} and \eqref{a:R}. This leads to the trivial solution
\begin{equation}
\a_1^{aI} = \a_2^{aI} = 0.
\end{equation}

\section*{\texorpdfstring{$\< T^{\mu_1 \nu_1} J^{\mu_2} J^{\mu_3} \>$}{<TJJ>}}
\addcontentsline{toc}{subsection}{\numberline {8}\texorpdfstring{$\< T^{\mu_1 \nu_1} J^{\mu_2} J^{\mu_3} \>$}{<TJJ>}}
\setcounter{subsection}{8}
\setcounter{equation}{0}

\noindent \textbf{Ward identities.} The transverse and trace Ward identities are
\begin{align}
& p_1^{\nu_1} \lla T_{\mu_1 \nu_1}(\bs{p}_1) J^{\mu_2 a_2}(\bs{p}_2) J^{\mu_3 a_3}(\bs{p}_3) \rra = \nn\\
& \qquad = p_1^{\nu_1} \lla \frac{\delta T_{\mu_1 \nu_1}}{\delta A_{\mu_3}^{a_3}}(\bs{p}_1, \bs{p}_3) J^{\mu_2 a_2}(\bs{p}_2) \rra + p_1^{\nu_1} \lla \frac{\delta T_{\mu_1 \nu_1}}{\delta A_{\mu_2}^{a_2}}(\bs{p}_1, \bs{p}_2) J^{\mu_3 a_3}(\bs{p}_3) \rra \nn\\
& \qquad\qquad + \: 2 \delta^{\mu_3 [\mu_1} p_3^{\alpha]} \lla J^{\mu_2 a_2}(\bs{p}_2) J_{\alpha}^{a_3}(-\bs{p}_2) \rra + 2 \delta^{\mu_2 [\mu_1} p_2^{\alpha]} \lla J_{\alpha}^{a_2}(\bs{p}_3) J^{\mu_3 a_3}(-\bs{p}_3) \rra, \\
& p_{2 \mu_2} \lla T_{\mu_1 \nu_1}(\bs{p}_1) J^{\mu_2 a_2}(\bs{p}_2) J^{\mu_3 a_3}(\bs{p}_3) \rra = \nn\\
& \qquad = 2 p_{2 \mu_2} \lla \frac{\delta J^{\mu_2 a_2}}{\delta g^{\mu_1 \nu_1}}(\bs{p}_2, \bs{p}_1) J^{\mu_3 a_3}(\bs{p}_3) \rra + \delta_{\mu_1 \nu_1} p_{1 \alpha} \lla J^{\alpha a_2}(\bs{p}_3) J^{\mu_3 a_3}(-\bs{p}_3) \rra, \\
& \lla T(\bs{p}_1) J^{\mu_2 a_2}(\bs{p}_2) J^{\mu_3 a_3}(\bs{p}_3) \rra = \nn\\
& \qquad = \lla \frac{\delta T}{\delta A^{a_2}_{\mu_2}}(\bs{p}_1, \bs{p}_2) J^{\mu_3 a_3}(\bs{p}_3) \rra + \lla \frac{\delta T}{\delta A^{a_3}_{\mu_3}}(\bs{p}_1, \bs{p}_3) J^{\mu_2 a_2}(\bs{p}_2) \rra.
\end{align}

\bigskip \noindent \textbf{Reconstruction formula.} The full 3-point function can be reconstructed from the transverse-traceless part as
\begin{align}
& \lla T^{\mu_1 \nu_1}(\bs{p}_1) J^{\mu_2 a_2}(\bs{p}_2) J^{\mu_3 a_3}(\bs{p}_3) \rra = \lla t^{\mu_1 \nu_1}(\bs{p}_1) j^{\mu_2 a_2}(\bs{p}_2) j^{\mu_3 a_3}(\bs{p}_3) \rra \nn\\
& \qquad + \left[ 2 \mathscr{T}^{\mu_1 \nu_1}_{\alpha}(\bs{p}_1) \pi^{\mu_3 [\alpha}(\bs{p}_3) p_3^{\beta]} + \frac{p_3^{\mu_3}}{p_3^2} \delta^{\mu_1 \nu_1} p_{1 \beta} \right] \lla J^{\mu_2 a_2}(\bs{p}_2) J_{\beta}^{a_3}(-\bs{p}_2) \rra \nn\\
& \qquad + \: \pi^{\mu_3}_{\alpha_3}(\bs{p}_3) \left[ \mathscr{T}^{\mu_1 \nu_1 \alpha_1}(\bs{p}_1) p_1^{\beta_1} + \frac{\pi^{\mu_1 \nu_1}(\bs{p}_1)}{d-1} \delta^{\alpha_1 \beta_1} \right] \lla \frac{\delta T_{\alpha_1 \beta_1}}{\delta A_{\alpha_3}^{a_3}}(\bs{p}_1, \bs{p}_3) J^{\mu_2 a_2}(\bs{p}_2) \rra \nn\\
& \qquad + \frac{2 p_3^{\mu_3} p_{3 \alpha_3}}{p_3^2} \delta^{\mu_1 \alpha_1} \delta^{\nu_1 \beta_1} \lla \frac{\delta J^{\alpha_3 a_3}}{\delta g^{\alpha_1 \beta_1}}(\bs{p}_3, \bs{p}_1) J^{\mu_2 a_2}(\bs{p}_2) \rra \nn\\
& \qquad + \: \text{everything with } (\bs{p}_2, a_2, \mu_2) \leftrightarrow (\bs{p}_3, a_3, \mu_3), \label{a:tjjrecon}
\end{align}
where $\mathscr{T}^{\mu_1 \nu_1}_{\alpha}$ is defined in \eqref{a:T}.

\bigskip \noindent \textbf{Decomposition of the 3-point function.} The tensor decomposition of the transverse-traceless part is
\begin{align}
& \lla t^{\mu_1 \nu_1}(\bs{p}_1) j^{\mu_2 a_2}(\bs{p}_2) j^{\mu_3 a_3}(\bs{p}_3) \rra \nn\\
& \qquad = \Pi^{\mu_1 \nu_1}_{\alpha_1 \beta_1}(\bs{p}_1) \pi^{\mu_2}_{\alpha_2}(\bs{p}_2) \pi^{\mu_3}_{\alpha_3}(\bs{p}_3) \left[ A_1^{a_2 a_3} p_2^{\alpha_1} p_2^{\beta_1} p_3^{\alpha_2} p_1^{\alpha_3} + A_2^{a_2 a_3} \delta^{\alpha_2 \alpha_3} p_2^{\alpha_1} p_2^{\beta_1} \right.\nn\\
& \qquad \qquad + \: A_3^{a_2 a_3} \delta^{\alpha_1 \alpha_2} p_2^{\beta_1} p_1^{\alpha_3} + A_3^{a_3 a_2}(p_2 \leftrightarrow p_3) \delta^{\alpha_1 \alpha_3} p_2^{\beta_1} p_3^{\alpha_2} \nonumber \\
& \left. \qquad \qquad + \: A_4^{a_2 a_3} \delta^{\alpha_1 \alpha_3} \delta^{\alpha_2 \beta_1} \right].
\end{align}
The form factors $A_j$, $j=1,2,3,4$ are functions of the momentum magnitudes. If no arguments are specified then the standard ordering is assumed, $A_j = A_j(p_1, p_2, p_3)$, while by $p_i \leftrightarrow p_j$ we denote the exchange of the two momenta, \textit{e.g.}, $A_3(p_2 \leftrightarrow p_3) = A_3(p_1, p_3, p_2)$.

The form factors $A_1$, $A_2$ and $A_4$ are symmetric under $(p_2, a_2) \leftrightarrow (p_3, a_3)$, \textit{i.e.}, they satisfy,
\begin{equation}
A_j^{a_3 a_2}(p_1, p_3, p_2) = A_j^{a_2 a_3}(p_1, p_2, p_3), \qquad j \in \{1,2,4\},
\end{equation}
while the form factor $A_3$ does not exhibit any symmetry properties.

The form factors may be determined as follows
\begin{align}
A_1^{a_2 a_3} & = \text{coefficient of } p_2^{\mu_1} p_2^{\nu_1} p_3^{\mu_2} p_1^{\mu_3}, \\
A_2^{a_2 a_3} & = \text{coefficient of } \delta^{\mu_2 \mu_3} p_2^{\mu_1} p_2^{\nu_1}, \\
A_3^{a_2 a_3} & = 2 \cdot \text{coefficient of } \delta^{\mu_1 \mu_2} p_2^{\nu_1} p_1^{\mu_3}, \\
A_4^{a_2 a_3} & = 2 \cdot \text{coefficient of } \delta^{\mu_1 \mu_2} \delta^{\mu_3 \nu_1},
\end{align}
in $\lla T^{\mu_1 \nu_1}(\bs{p}_1) J^{\mu_2 a_2}(\bs{p}_2) J^{\mu_3 a_3}(\bs{p}_3) \rra$.

\bigskip \noindent \textbf{Primary conformal Ward identities.} The primary CWIs are
\begin{equation}
\begin{array}{ll}
\K_{12} A_1^{a_2 a_3} = 0, & \qquad\qquad \K_{13} A_1^{a_2 a_3} = 0, \\
\K_{12} A_2^{a_2 a_3} = -2 A_1^{a_2 a_3}, & \qquad\qquad \K_{13} A_2^{a_2 a_3} = - 2 A_1^{a_2 a_3}, \\
\K_{12} A_3^{a_2 a_3} = 0, & \qquad\qquad \K_{13} A_3^{a_2 a_3} = 4 A_1^{a_2 a_3}, \\
\K_{12} A_4^{a_2 a_3} = 2 A_3^{a_2 a_3}, & \qquad\qquad \K_{13} A_4^{a_2 a_3} = 2 A_3^{a_3 a_2}(p_2 \leftrightarrow p_3),
\end{array}
\end{equation}
The solution in terms of triple-$K$ integrals \eqref{a:J} is
\begin{align}
A_1^{a_2 a_3} & = \a_1^{a_2 a_3} J_{4 \{000\}}, \label{e:restjj1} \\
A_2^{a_2 a_3} & = \a_1^{a_2 a_3} J_{3 \{100\}} + \a_2^{a_2 a_3} J_{2 \{000\}}, \\
A_3^{a_2 a_3} & = 2 \a_1^{a_2 a_3} J_{3 \{001\}} + \a_3^{a_2 a_3} J_{2 \{000\}}, \\
A_4^{a_2 a_3} & = 2 \a_1^{a_2 a_3} J_{2 \{011\}} +  \a_3^{a_2 a_3} \left( J_{1 \{010\}} + J_{1 \{001\}} \right) + \a_4^{a_2 a_3} J_{0 \{000\}}, \label{e:restjj4}
\end{align}
where $\a_j^{a_2 a_3}$, $j=1,2,3,4$ are constants. In particular all constants are symmetric in the group indices, $\a_j^{a_3 a_2} = \a_j^{a_2 a_3}$, $j=1,2,3,4$. If the integrals diverge, the regularisation \eqref{a:scheme} should be used.

\bigskip \noindent \textbf{Secondary conformal Ward identities.} The independent secondary CWIs are
\begin{align}
& (*) \ \Lo_{4} A_1^{a_2 a_3} + \Ro \left[ A_3^{a_2 a_3} - A_3^{a_2 a_3}(p_2 \leftrightarrow p_3) \right] = \nn\\
& \qquad = 2 d \cdot \text{coefficient of } p_2^{\mu_1} p_3^{\mu_2} p_1^{\mu_3} \text{ in } p_{1 \nu_1} \lla T^{\mu_1 \nu_1}(\bs{p}_1) J^{\mu_2 a_2}(\bs{p}_2) J^{\mu_3 a_3}(\bs{p}_3) \rra, \label{e:tjjseccwi1} \\
& \Lo'_{3} A_1^{a_2 a_3} + 2 \Ro' \left[ A_3^{a_2 a_3} - A_2^{a_2 a_3} \right] = \nn\\
& \qquad = 2 d \cdot \text{coefficient of } p_2^{\mu_1} p_2^{\nu_1} p_1^{\mu_3} \text{ in } p_{2 \mu_2} \lla T^{\mu_1 \nu_1}(\bs{p}_1) J^{\mu_2 a_2}(\bs{p}_2) J^{\mu_3 a_3}(\bs{p}_3) \rra, \label{e:tjjseccwi1a} \\
& \Lo_{2} A_2^{a_2 a_3} - p_1^2 \left[ A_3^{a_2 a_3} - A_3^{a_2 a_3}(p_2 \leftrightarrow p_3) \right] = \nn\\
& \qquad = 2 d \cdot \text{coefficient of } \delta^{\mu_2 \mu_3} p_2^{\mu_1} \text{ in } p_{1 \nu_1} \lla T^{\mu_1 \nu_1}(\bs{p}_1) J^{\mu_2 a_2}(\bs{p}_2) J^{\mu_3 a_3}(\bs{p}_3) \rra, \\
& \Lo_{4} A_3^{a_2 a_3} - 2 \Ro A_4^{a_2 a_3} = \nn\\
& \qquad = 4 d \cdot \text{coefficient of } \delta^{\mu_1 \mu_2} p_1^{\mu_3} \text{ in } p_{1 \nu_1} \lla T^{\mu_1 \nu_1}(\bs{p}_1) J^{\mu_2 a_2}(\bs{p}_2) J^{\mu_3 a_3}(\bs{p}_3)) \rra, \label{e:tjjseccwi2}
\end{align}
where $\Lo$ and $\Ro$ are given by \eqref{a:L} and \eqref{a:R}. The identity denoted by the asterisk is redundant, \textit{i.e.}, it is trivially satisfied in all cases and does not impose any additional conditions on the primary constants. Furthermore, the transverse Ward identities imply that the right-hand sides of \eqref{e:tjjseccwi1} and \eqref{e:tjjseccwi1a} vanish. The secondary CWIs lead to the following relations,
\begin{align}
\alpha_2^{a_2 a_3} & = - (d + 2 v \epsilon) \alpha_1^{a_2 a_3} + \frac{2^{3 - \frac{d}{2} - v \epsilon} \delta^{a_2 a_3} c_J}{\Gamma^2 \left( \frac{d}{2} + v \epsilon \right)},  \\
\alpha_3^{a_2 a_3} & = \alpha_2^{a_2 a_3}, \\
\alpha_4^{a_2 a_3} & = - (d - 2 + 2 v \epsilon) \alpha_2^{a_2 a_3} - \frac{1}{2} \epsilon (u - v) \left[ (d + 2 v \epsilon) \left( 2( d - 2 + 2 v \epsilon) \alpha_1^{a_2 a_3} + \alpha_2^{a_2 a_3} \right) \right.\nn\\
& \qquad\qquad\qquad \left. - \: \frac{2^{5 - \frac{d}{2} - v \epsilon} c_J c_3^{a_2 a_3}}{\Gamma \left( \frac{d}{2} - 1 + v \epsilon \right) \Gamma \left( \frac{d}{2} + v \epsilon \right)} \right] + O((u-v)^2 \epsilon^2),
\end{align}
where $c_J$ is the normalisation of the 2-point function \eqref{a:JJ}, while the constant $c^{ab}$ is defined as
\begin{equation} \label{a:caa}
\lla \frac{\delta T_{\mu_1 \nu_1}}{\delta A_{\mu_2}^a}(\bs{p}_1, \bs{p}_2) J^{\mu_3 b}(\bs{p}_3) \rra = 2 c^{ab} \delta^{\mu_2}_{(\mu_1} \lla J^a_{\nu_1)}(\bs{p}_3) J^{\mu_3 b}(- \bs{p}_3) \rra.
\end{equation}
After the substitution of the solution of the secondary CWIs to \eqref{e:restjj1} - \eqref{e:restjj4}, the limit $u = v = -1/2$ should be taken. The form factors then represent the 3-point function regulated in the dimensional regularisation \eqref{e:dimreg}.

The 3-point function $\lla T^{\mu_1 \nu_1} J^{\mu_2} J^{\mu_3} \rra$ thus depends on the 2-point function normalisations $c_J$ and $c^{ab}$ and one undetermined primary constant $\a_1^{a_2 a_3}$. The dependence of this correlator on two 2-point function normalisations rather than the one found in \cite{Osborn:1993cr} is related to
our definition of this correlator, as discussed above \eqref{e:defTJJ}.

\subsection*{Examples}

\bigskip \noindent \textbf{For $\bm{d=3}$} we find
\begin{align}
A^{a_2 a_3}_1 & = \a_1^{a_2 a_3} \frac{2(4 p_1 + a_{23})}{a_{123}^4}, \\
A^{a_2 a_3}_2 & = \a_1^{a_2 a_3} \frac{2 p_1^2}{a_{123}^3} + \frac{4 \sqrt{\pi} (2 p_1 + a_{23})}{a_{123}^2} c_{J} \delta^{a_2 a_3}, \\
A^{a_2 a_3}_3 & = \frac{\a_1^{a_2 a_3}}{a_{123}^3} \left[ - 2 p_1^2 - p_2^2+p_3^2 - 3 p_1 p_2 + 3 p_1 p_3 \right] + \frac{4 \sqrt{\pi} (2 p_1 + a_{23})}{a_{123}^2} c_J \delta^{a_2 a_3}, \\
A^{a_2 a_3}_4 & = \a_1^{a_2 a_3} \frac{(2 p_1 + a_{23}) (p_1^2 - a_{23}^2 + 4 b_{23})}{2 a_{123}^2} \nn\\
& \qquad \qquad - \: 2 \sqrt{\pi} \left( \frac{2 p_1^2}{a_{123}} - a_{23} \right) c_J \delta^{a_2 a_3} - 4 \sqrt{\pi} a_{23} c^{a_3 a_2} c_{J}.
\end{align}

\bigskip \noindent \textbf{For $\bm{d=5}$} we find
\begin{align}
A_1^{a_2 a_3} & = \frac{2 \a_1^{a_2 a_3}}{a_{123}^5} \left[ 4 p_1^4 + 20 p_1^3 a_{23} + 4 p_1^2 (7 a_{23}^2 + 6 b_{23}) \right.\nn\\
& \qquad\qquad \left. + \: 15 p_1 a_{23} (a_{23}^2 + b_{23}) + 3 a_{23}^2 (a_{23}^2 + b_{23}) \right],
\end{align}
\begin{align}
A_2^{a_2 a_3} &= \frac{2 \a_1^{a_2 a_3} p_1^2}{a_{123}^4} \left[ a_{123}^3 + a_{123} b_{123} + 3 c_{123} \right] \nn\\
& \qquad - \: \frac{8 \sqrt{\pi} c_{J} \delta^{a_2 a_3}}{9 a_{123}^3} \left[ 2 p_1^4 + 6 p_1^3 a_{23} + 2 p_1^2 (5 a_{23}^2 - 4 b_{23}) \right.\nn\\
& \qquad\qquad \left. + \: 9 p_1 (a_{23}^3 - a_{23} b_{23}) + 3 a_{23}^2 (a_{23}^2 - b_{23}) \right],
\end{align}
\begin{align}
A_3^{a_2 a_3} &= \frac{\a_1^{a_2 a_3}}{a_{123}^4} \left[ -2 p_1^5 - 8 p_1^4 (p_2 + p_3) - 8 p_1^3 p_2 (2 p_2 + 3 p_3) \right.\nn\\
& \qquad\qquad + \: p_1^2 (-19 p_2^3 - 40 p_2^2 p_3 + 24 p_2 p_3^2 + 15 p_3^3) \nn\\
& \qquad\qquad \left. - \: 3 (4 p_1 + p_2 + p_3) (p_2^2 - p_3^2) (p_2^2 + 3 p_2 p_3 + p_3^2) \right] \nn\\
& \qquad - \: \frac{8 \sqrt{\pi} c_{J} \delta^{a_2 a_3}}{9 a_{123}^3} \left[ 2 p_1^4 + 6 p_1^3 a_{23} + 2 p_1^2 (5 a_{23}^2 - 4 b_{23}) \right.\nn\\
& \qquad\qquad \left. + \: 9 p_1 a_{23} (a_{23}^2 - b_{23}) + 3 a_{23}^2 (a_{23}^2 - b_{23}) \right],
\end{align}
\begin{align}
A_4^{a_2 a_3} &= \frac{\a_1^{a_2 a_3}}{2 a_{123}^3} \left[ 2 p_1^6 + 6 p_1^5 a_{23} + 4 p_1^4 (2 a_{23}^2 - b_{23}) + p_1^2 (a_{23}^2 + b_{23}) (3 p_1 a_{23} - 7 a_{23}^2 + 32 b_{23}) \right.\nn\\
&\qquad\qquad \left. - \: 3 a_{23} (3 p_1 + a_{23}) (a_{23}^2 - 4 b_{23}) (a_{23}^2 + b_{23}) \right] \nn\\
& \qquad + \: \frac{4 \sqrt{\pi} c_{J} \delta^{a_2 a_3}}{9 a_{123}^2} \left[ 2 p_1^5 + 4 p_1^4 a_{23} + 4 p_1^3 (a_{23}^2 - b_{23}) - p_1^2 a_{23} (a_{23}^2 - 7 b_{23}) \right.\nn\\
&\qquad\qquad \left. - \: 6 p_1 a_{23}^2 (a_{23}^2 - 3 b_{23}) - 3 a_{23}^3 (a_{23}^2 - 3 b_{23}) \right] \nn\\
&\qquad + \: 8 \sqrt{\pi} (p_2^3 + p_3^3) c^{a_3 a_2} c_{J}.
\end{align}

\section*{\texorpdfstring{$\< T^{\mu_1 \nu_1} T^{\mu_2 \nu_2} \mathcal{O} \>$}{<TTO>}}
\addcontentsline{toc}{subsection}{\numberline {9}\texorpdfstring{$\< T^{\mu_1 \nu_1} T^{\mu_2 \nu_2} \mathcal{O} \>$}{<TTO>}}
\setcounter{subsection}{9}
\setcounter{equation}{0}

\noindent \textbf{Ward identities.} The transverse and trace Ward identities are
\begin{align}
& p_1^{\nu_1} \lla T_{\mu_1 \nu_1}(\bs{p}_1) T_{\mu_2 \nu_2}(\bs{p}_2) \mathcal{O}^I(\bs{p}_3) \rra = 2 p_1^{\nu_1} \lla \frac{\delta T_{\mu_1 \nu_1}}{\delta g^{\mu_2 \nu_2}}(\bs{p}_1, \bs{p}_2) \mathcal{O}^I(\bs{p}_3) \rra, \\
& \lla T(\bs{p}_1) T_{\mu_2 \nu_2}(\bs{p}_2) \mathcal{O}^I(\bs{p}_3) \rra = 2 \lla \frac{\delta T}{\delta g^{\mu_2 \nu_2}}(\bs{p}_1, \bs{p}_2) \mathcal{O}^I(\bs{p}_3) \rra.
\end{align}

\bigskip \noindent \textbf{Reconstruction formula.} The full 3-point function can be reconstructed from the transverse-traceless part as
\begin{align}
& \lla T^{\mu_1 \nu_1}(\bs{p}_1) T^{\mu_2 \nu_2}(\bs{p}_2) \mathcal{O}^I(\bs{p}_3) \rra = \lla t^{\mu_1 \nu_1}(\bs{p}_1) t^{\mu_2 \nu_2}(\bs{p}_2) \mathcal{O}^I(\bs{p}_3) \rra \nn\\
& \qquad + 2 \left[ \mathscr{T}^{\mu_1 \nu_1 \alpha_1}(\bs{p}_1) p_1^{\beta_1} + \frac{\pi^{\mu_1 \nu_1}(\bs{p}_1)}{d-1} \delta^{\alpha_1 \beta_1} \right] \delta^{\mu_2 \alpha_2} \delta^{\nu_2 \beta_2} \lla \frac{\delta T_{\alpha_1 \beta_1}}{\delta g^{\alpha_2 \beta_2}}(\bs{p}_1, \bs{p}_2) \mathcal{O}^I(\bs{p}_3) \rra \nn\\
& \qquad + 2 [ (\mu_1, \nu_1, \bs{p}_1) \leftrightarrow (\mu_2, \nu_2, \bs{p}_2) ] \nn\\
& \qquad - 4 \left[ \mathscr{T}^{\mu_1 \nu_1 \alpha_1}(\bs{p}_1) p_1^{\beta_1} + \frac{\pi^{\mu_1 \nu_1}(\bs{p}_1)}{d-1} \delta^{\alpha_1 \beta_1} \right] \left[ \mathscr{T}^{\mu_2 \nu_2 \alpha_2}(\bs{p}_2) p_2^{\beta_2} + \frac{\pi^{\mu_2 \nu_2}(\bs{p}_2)}{d-1} \delta^{\alpha_2 \beta_2} \right] \times \nn\\
& \qquad\qquad\qquad\qquad \times \: \lla \frac{\delta T_{\alpha_1 \beta_1}}{\delta g^{\alpha_2 \beta_2}}(\bs{p}_1, \bs{p}_2) \mathcal{O}^I(\bs{p}_3) \rra,
\end{align}
where $\mathscr{T}^{\mu \nu}_{\alpha}$ is defined in \eqref{a:T}.

\bigskip \noindent \textbf{Decomposition of the 3-point function.} The tensor decomposition of the transverse-traceless part is
\begin{align}
& \lla t^{\mu_1 \nu_1}(\bs{p}_1) t^{\mu_2 \nu_2}(\bs{p}_2) \mathcal{O}^I(p_3) \rra = \Pi^{\mu_1 \nu_1}_{\alpha_1 \beta_1}(\bs{p}_1) \Pi^{\mu_2 \nu_2}_{\alpha_2 \beta_2}(\bs{p}_2) \left[
A_1^I p_2^{\alpha_1} p_2^{\beta_1} p_3^{\alpha_2} p_3^{\beta_2} \right. \nonumber \\
& \qquad \qquad \left. + \: A_2^I \delta^{\alpha_1 \alpha_2} p_2^{\beta_1} p_3^{\beta_2} + A_3^I \delta^{\alpha_1 \alpha_2} \delta^{\beta_1 \beta_2} \right].
\end{align}
The form factors $A_j$, $j=1,2,3$ are functions of the momentum magnitudes. All form factors are symmetric under $p_1 \leftrightarrow p_2$, \textit{i.e.}, they satisfy
\begin{equation}
A_j^I(p_2, p_1, p_3) = A_j^I(p_1, p_2, p_3), \qquad j=1,2,3.
\end{equation}
These form factors may be calculated using
\begin{align}
A_1^I & = \text{coefficient of } p_2^{\mu_1} p_2^{\nu_1} p_3^{\mu_2} p_3^{\nu_2}, \\
A_2^I & = 4 \cdot \text{coefficient of } \delta^{\mu_1 \mu_2} p_2^{\nu_1} p_3^{\nu_2}, \\
A_3^I & = 2 \cdot \text{coefficient of } \delta^{\mu_1 \mu_2} \delta^{\nu_1 \nu_2},
\end{align}
in $\lla T^{\mu_1 \nu_1}(\bs{p}_1) T^{\mu_2 \nu_2}(\bs{p}_2) \mathcal{O}^I(\bs{p}_3) \rra$.

\bigskip \noindent \textbf{Primary conformal Ward identities.} The primary CWIs are
\begin{equation}
\begin{array}{ll}
\K_{12} A_1^I = 0, &\qquad\qquad \K_{13} A_1^I = 0, \\
\K_{12} A_2^I = 0, &\qquad\qquad \K_{13} A_2^I = 8 A_1^I, \\
\K_{12} A_3^I = 0, &\qquad\qquad \K_{13} A_3^I = 2 A_2^I,
\end{array}
\end{equation}
The solution in terms of triple-$K$ integrals \eqref{a:J} is
\begin{align}
A_1^I & = \a_1^I J_{4 \{000\}}, \label{a:TTO1} \\
A_2^I & = 4 \a_1^I J_{3 \{001\}} + \a_2^I J_{2 \{000\}}, \\
A_3^I & = 2 \a_1^I J_{2 \{002\}} + \a_2^I J_{1 \{001\}} + \a_3^I J_{0 \{000\}}, \label{a:TTO3}
\end{align}
where $\a_j^I$, $j=1,2,3$ are constants. If the integrals diverge, the regularisation \eqref{a:scheme} should be used.

\bigskip \noindent \textbf{Secondary conformal Ward identities.} The independent secondary CWIs are
\begin{align}
& \Lo_{2} A_1^I + \Ro A_2^I = \nn\\
& \qquad = 2 d \cdot \text{coefficient of } p_2^{\mu_1} p_3^{\mu_2} p_3^{\nu_2} \text{ in } p_{1 \nu_1} \lla T^{\mu_1 \nu_1}(\bs{p}_1) T^{\mu_2 \nu_2}(\bs{p}_2) \mathcal{O}^I(\bs{p}_3) \rra, \\
& \Lo_{2} A_2^I + 4 \Ro A_3^I = \nn\\
& \qquad = 8 d \cdot \text{coefficient of } \delta^{\mu_1 \mu_2} p_3^{\mu_2} \text{ in } p_{1 \nu_1} \lla T^{\mu_1 \nu_1}(\bs{p}_1) T^{\mu_2 \nu_2}(\bs{p}_2) \mathcal{O}^I(\bs{p}_3) \rra.
\end{align}
where $\Lo$ and $\Ro$ are defined in \eqref{a:L} and \eqref{a:R}. They lead to the following relations
\begin{align}
& l_{\frac{d}{2} + 3 + u \epsilon, \{ \frac{d}{2} + v \epsilon, \frac{d}{2} + v \epsilon, \frac{d}{2} - \Delta - v \epsilon \}} \left[ -(6 + 2 \Delta + (u+3v) \epsilon) \alpha_1^I \right.\nn\\
& \qquad\qquad\qquad \left. + \: \frac{2(3+\Delta + (u+v)\epsilon)}{(-2+d-\Delta-(u-v)\epsilon)(2+\Delta+(u+v)\epsilon)} \alpha_2^I \right] = \nn\\
& \qquad = 4 \delta_{d, \Delta-2-2n} (d + 2 u\epsilon) \Gamma \left( \frac{d}{2} - \Delta - v \epsilon \right) c_1^I c_{\O}, \\
& l_{\frac{d}{2} + 1 + u \epsilon, \{ \frac{d}{2} + v \epsilon, \frac{d}{2} + v \epsilon, \frac{d}{2} - \Delta - v \epsilon \}} \left[ -(2 + 2 \Delta + (u+3v) \epsilon) \alpha_2^I \right.\nn\\
& \qquad\qquad\qquad \left. + \: \frac{8(1+\Delta + (u+v)\epsilon)}{(d-\Delta-(u-v)\epsilon)(\Delta+(u+v)\epsilon)} \alpha_3^I \right] = \nn\\
& \qquad = 16 \delta_{d, \Delta-2n} (d + 2 u\epsilon) \Gamma \left( \frac{d}{2} - \Delta - v \epsilon \right) c_2^I c_{\O},
\end{align}
where the constant $l_{\alpha\{\beta_j\}}$ is defined in \eqref{a:l} and the constants $c_1^K$ and $c_2^K$ are defined as
\begin{align}
& \left. p_1^{\nu_1} \lla \frac{\delta T_{\mu_1 \nu_1}}{\delta g^{\mu_2 \nu_2}}(\bs{p}_1, \bs{p}_2) \mathcal{O}^K(\bs{p})_3 \rra \right|_{p_1 = p_2 = p} = c^K_1 p_2^{\mu_1} p_3^{\mu_2} p_3^{\nu_2} p^{d - \Delta_3 - 2} \lla \mathcal{O}^I(\bs{p}_3) \mathcal{O}^K(-\bs{p}_3) \rra \nn\\
& \qquad\qquad + \: c^K_2 \delta^{\mu_1 \mu_2} p_3^{\nu_2} p^{d - \Delta_3} \lla \mathcal{O}^I(\bs{p}_3) \mathcal{O}^K(-\bs{p}_3) \rra + \ldots \label{a:c12K}
\end{align}
{\it i.e.,}  we first write down the most general tensor decomposition for $p_1^{\nu_1} \lla \frac{\delta T_{\mu_1 \nu_1}}{\delta g^{\mu_2 \nu_2}}(\bs{p}_1, \bs{p}_2) \mathcal{O}^K(\bs{p})_3 \rra$, then extract the coefficient of $p_2^{\mu_1} p_3^{\mu_2} p_3^{\nu_2}$ and set $p_1 = p_2 = p$ in this expression\footnote{The coefficient $c_1^K$ can be non-vanishing only if $d - \Delta_3 - 2 = 2n$ for a non-negative integer $n$. Similarly, for $c_2^K$ to be non-zero one needs $d - \Delta_3 = 2n'$ for a non-negative integer $n'$. Otherwise, the right-hand sides of the secondary Ward identities vanish accordingly.}.

After the substitution of the solution of the secondary CWIs to \eqref{a:TTO1} - \eqref{a:TTO3}, the limit $u = v = -1/2$ should be taken. The form factors then represent the 3-point function regulated in the dimensional regularisation \eqref{e:dimreg}.

The 3-point function $\lla T^{\mu_1 \nu_1} T^{\mu_2 \nu_2} \mathcal{O} \rra$ thus depends on the 2-point function normalisations $c_{\mathcal{O}}$, $c_1^K$ and $c_2^K$ and one undetermined primary constant $\a_1^I$.

\subsection*{Examples}

\bigskip \noindent \textbf{For $\bm{d=3}$ and $\bm{\Delta_3 = 1}$} we find
\begin{align}
A_1^I &= \frac{\a_1^I}{p_3 a_{123}^4} \left[ p_3^2 + 4 p_3 a_{12} + 3 (a_{12}^2 + 2 b_{12}) \right], \\
A_2^I &= \frac{\a_1^I}{p_3 a_{123}^3} \left[ p_3^3 + 3 p_2^2 a_{12} + p_3 (-a_{12}^2 + 8 b_{12}) - 3 a_{12}^3 \right] - \frac{4 \sqrt{\pi} c_1^I c_{\mathcal{O}}}{p_3}, \\
A_3^I &= \frac{\a_1^I (a_{12} - p_3)}{4 p_3 a_{123}^2} \left[ - p_3^3 - 3 p_3^2 a_{12} + p_3 (a_{12}^2 - 10 b_{12}) + 3 a_{12} ( a_{12}^2 - 2 b_{12}) \right] \nn\\
& \qquad + \frac{\sqrt{\pi} c_{\mathcal{O}}}{p_3} \left[ (c_1^I - 3 c_2^I) (p_1^2 + p_2^2) + 3 (c_1^I + c_2^I) p_3^2 \right].
\end{align}

\bigskip \noindent \textbf{For $\bm{d=3}$ and $\bm{\Delta_3 = 3}$} we find
\begin{align}
A_1^I &= \frac{2 \a_1^I}{a_{123}^4} \left[ a_{123}^3 + a_{123} b_{123} + 3 c_{123} \right], \\
A_2^I &= \frac{2 \a_1^I}{a_{123}^3} \left[ p_3^4 + 3 p_3^3 a_{12} + 6 p_3^2 b_{12} + 3 p_3 a_{12} (- a_{12}^2 + b_{12}) + a_{12}^2 (b_{12} - a_{12}^2) \right], \\
A_3^I &= \frac{\a_1^I}{2 a_{123}^2} \left[ p_3^5 + 2 p_3^4 a_{12} + p_3^3 (-3 a_{12}^2 + 8 b_{12}) - p_3^2 (a_{12}^3 + 5 a_{12} b_{12}) \right.\nn\\
&\qquad\qquad \left. + \: 6 p_3 (a_{12}^4 - 3 a_{12}^2 b_{12}) + 3 a_{12}^3 (a_{12}^2 - 3 b_{12}) \right] \nn\\
&\qquad - \: \tfrac{16}{3} \sqrt{\pi} c_2^I c_{\mathcal{O}} (p_1^3 + p_2^3 + p_3^3).
\end{align}

\bigskip \noindent \textbf{For $\bm{d=5}$ and $\bm{\Delta_3 = 3}$} we find
\begin{align}
A_1^I &= \frac{3 \a_1^I}{a_{123}^5} \left[ 3 p_3^4 + 15 p_3^3 a_{12} + p_3^2 (29 a_{12}^2 + 2 b_{12}) + 5 p_3 a_{12} (5 a_{12}^2 + 2 b_{12}) \right.\nn\\
& \qquad\qquad \left. + \: 8 (a_{12}^4 + a_{12}^2 b_{12} + b_{12}^2) \right], \\
A_2^I &= \frac{\a_1^I}{a_{123}^4} \left[ 9 p_3^5 + 36 p_3^4 a_{12} + 6 p_3^3 (7 a_{12}^2 + 4 b_{12}) - 12 p_3^2 a_{12} (a_{12}^2 - 8 b_{12}) \right.\nn\\
&\qquad\qquad \left. + \: p_3 (- 51 a_{12}^4 + 96 a_{12}^2 b_{12} + 32 b_{12}^2) + 8 a_{12} (- 3 a_{12}^4 + 3 a_{12}^2 b_{12} + b_{12}^2) \right] \nn\\
&\qquad + \: 8 \sqrt{\pi} p_3 c_1^I c_{\mathcal{O}}, \\
A_3^I &= \frac{\a_1^I}{4 a_{123}^3} \left[ 9 p_3^6 + 27 p_3^5 a_{12} + 6 p_3^4 (a_{12}^2 + 7 b_{12}) + 18 p_3^3 a_{12} (-3 a_{12}^2 + 7 b_{12}) \right.\nn\\
&\qquad\qquad + \: p_3^2 (-39 a_{12}^4 + 30 a_{12}^2 b_{12} + 64 b_{12}^2) + 9 p_3 a_{12} (a_{12}^2 - 4 b_{12}) (3 a_{12}^2 - 2 b_{12}) \nn\\
&\qquad\qquad \left. + \: 24 a_{12}^2 (a_{12}^4 - 3 a_{12}^2 b_{12} + b_{12}^2) \right] \nn\\
&\qquad - \: \tfrac{1}{3} \sqrt{\pi} c_{\mathcal{O}} p_3 \left[ 3 ( 2 c_1^I - 5 c_2^I ) (p_1^2 + p_2^2) + (2 c_1^I + 15 c_2^I ) p_3^2 \right].
\end{align}

\bigskip \noindent \textbf{For $\bm{d=5}$ and $\bm{\Delta_3 = 5}$} we find
\begin{align}
A_1^I &= \frac{6 \a_1^I}{a_{123}^5} \left[ -3 a_{123}^6 + 3 a_{123}^4 b_{123} + a_{123}^2 b_{123}^2 + a_{123}^3 c_{123} + 3 a_{123} b_{123} c_{123} + 4 c_{123}^2 \right], \\
A_2^I &= \frac{2 \a_1^I}{a_{123}^4} \left[ -9 p_3^7 - 36 p_3^6 a_{12} - 3 p_3^5 (17 a_{12}^2 + 2 b_{12}) - 24 p_3^4 a_{12} (a_{12}^2 + b_{12}) \right.\nn\\
&\qquad\qquad + \: 8 p_3^3 (3 a_{12}^4 - 12 a_{12}^2 b_{12} + 5 b_{12}^2) + p_3^2 (51 a_{12}^5 - 159 a_{12}^3 b_{12} + 55 a_{12} b_{12}^2) \nn\\
&\qquad\qquad \left. + \: 9 a_{12}^2 (a_{12} + 4 p_3) (a_{12}^4 - 3 a_{12}^2 b_{12} + b_{12}^2) \right], \\
A_3^I &= \frac{\a_1^I}{2 a_{123}^3} \left[ -9 p_3^8 - 27 p_3^7 a_{12} - 3 p_3^6 (5 a_{12}^2 + 8 b_{12}) + 9 p_3^5 (3 a_{12}^3 - 8 a_{12} b_{12}) \right.\nn\\
&\qquad\qquad + \: 8 p_3^4 (6 a_{12}^4 - 15 a_{12}^2 b_{12} + 4 b_{12}^2) + 9 p_3^3 (a_{12}^5 + 3 a_{12}^3 b_{12} - 11 a_{12} b_{12}^2) \nn\\
&\qquad\qquad \left. + \: p_3^2 (-69 a_{12}^6 + 369 a_{12}^4 b_{12} - 393 a_{12}^2 b_{12}^2) - 27 a_{12}^3 (a_{12} + 3 p_3) (a_{12}^4 - 5 a_{12}^2 b_{12} + 5 b_{12}^2) \right] \nn\\
&\qquad + \: \tfrac{32}{15} \sqrt{\pi} c_2^I c_{\mathcal{O}} (p_1^5 + p_2^5 + p_3^5).
\end{align}

\section*{\texorpdfstring{$\< T^{\mu_1 \nu_1} T^{\mu_2 \nu_2} J^{\mu_3} \>$}{<TTJ>}}
\addcontentsline{toc}{subsection}{\numberline {10}\texorpdfstring{$\< T^{\mu_1 \nu_1} T^{\mu_2 \nu_2} J^{\mu_3} \>$}{<TTJ>}}
\setcounter{subsection}{10}
\setcounter{equation}{0}

This correlation function is at most semi-local, as was proved in \cite{Costa:2011a} through a position space analysis.
Our result confirms the triviality of this correlator through independent calculations in momentum space. In appendix \ref{ch:vanish} we discuss the triviality of $\lla T^{\mu_1 \nu_1} J^{\mu_2} \mathcal{O} \rra$, which is very similar to $\lla T^{\mu_1 \nu_1} T^{\mu_2 \nu_2} J^{\mu_3} \rra$.

\bigskip \noindent \textbf{Ward identities.} The transverse and trace Ward identities are
\begin{align}
& p_1^{\nu_1} \lla T_{\mu_1 \nu_1}(\bs{p}_1) T_{\mu_2 \nu_2}(\bs{p}_2) J^{\mu_3 a}(\bs{p}_3) \rra = 2 p_1^{\nu_1} \lla \frac{\delta T_{\mu_1 \nu_1}}{\delta g^{\mu_2 \nu_2}}(\bs{p}_1, \bs{p}_2) J^{\mu_3 a}(\bs{p}_3) \rra, \\
& p_{3 \mu_3} \lla T_{\mu_1 \nu_1}(\bs{p}_1) T_{\mu_2 \nu_2}(\bs{p}_2) J^{\mu_3 a}(\bs{p}_3) \rra = 0, \\
& \lla T(\bs{p}_1) T_{\mu_2 \nu_2}(\bs{p}_2) J^{\mu_3 a}(\bs{p}_3) \rra = 2 \lla \frac{\delta T}{\delta g^{\mu_2 \nu_2}}(\bs{p}_1, \bs{p}_2) J^{\mu_3 a}(\bs{p}_3) \rra.
\end{align}

\bigskip \noindent \textbf{Reconstruction formula.} The full 3-point function can be reconstructed from the transverse-traceless part as
\begin{align}
& \lla T^{\mu_1 \nu_1}(\bs{p}_1) T^{\mu_2 \nu_2}(\bs{p}_2) J^{\mu_3 a}(\bs{p}_3) \rra = \lla t^{\mu_1 \nu_1}(\bs{p}_1) t^{\mu_2 \nu_2}(\bs{p}_2) j^{\mu_3 a}(\bs{p}_3) \rra \nn\\
& \qquad + \: 2 \left[ \mathscr{T}^{\mu_1 \nu_1 \alpha_1}(\bs{p}_1) p_1^{\beta_1} + \frac{\pi^{\mu_1 \nu_1}(\bs{p}_1)}{d-1} \delta_{\alpha_1 \beta_1} \right] \delta^{\mu_2 \alpha_2} \delta^{\nu_2 \beta_2} \lla \frac{\delta T_{\alpha_1 \beta_1}}{\delta g^{\alpha_2 \beta_2}}(\bs{p}_1, \bs{p}_2) J^{\mu_3 a}(\bs{p}_3) \rra \nn\\
& \qquad + \: 2 [ (\mu_1, \nu_1, \bs{p}_1) \leftrightarrow (\mu_2, \nu_2, \bs{p}_2) ] \nn\\
& \qquad - \: 4 \left[ \mathscr{T}^{\mu_1 \nu_1 \alpha_1}(\bs{p}_1) p_1^{\beta_1} + \frac{\pi^{\mu_1 \nu_2}(\bs{p}_1)}{d-1} \delta_{\alpha_1 \beta_1} \right] \left[ \mathscr{T}^{\mu_2 \nu_2 \alpha_2}(\bs{p}_2) p_2^{\beta_2} + \frac{\pi^{\mu_2 \nu_2}(\bs{p}_2)}{d-1} \delta_{\alpha_2 \beta_2} \right] \times \nn\\
& \qquad\qquad\qquad\qquad \times \: \lla \frac{\delta T_{\alpha_1 \beta_1}}{\delta g^{\alpha_2 \beta_2}}(\bs{p}_1, \bs{p}_2) J^{\mu_3 a}(\bs{p}_3) \rra,
\end{align}
where $\mathscr{T}^{\mu \nu \alpha}$ is defined in \eqref{a:T}.

\bigskip \noindent \textbf{Decomposition of the 3-point function.} The tensor decomposition of the transverse-traceless part is
\begin{align}
& \lla t^{\mu_1 \nu_1}(\bs{p}_1) t^{\mu_2 \nu_2}(\bs{p}_2) j^{\mu_3 a}(\bs{p}_3) \rra = \Pi^{\mu_1 \nu_1}_{\alpha_1 \beta_1}(\bs{p}_1) \Pi^{\mu_2 \nu_2}_{\alpha_2 \beta_2}(\bs{p}_2) \pi^{\mu_3}_{\alpha_3}(\bs{p}_3) \left[
A_1^a p_2^{\alpha_1} p_2^{\beta_1} p_3^{\alpha_2} p_3^{\beta_2} p_1^{\alpha_3} \right. \nonumber \\
& \qquad \qquad + \: A_2^a \delta^{\beta_1 \beta_2} p_2^{\alpha_1} p_3^{\alpha_2} p_1^{\alpha_3} \nonumber \\
& \qquad \qquad + \: A_3^a \delta^{\alpha_1 \alpha_3} p_2^{\beta_1} p_3^{\alpha_2} p_3^{\beta_2} - A_3^a(p_1 \leftrightarrow p_2) \delta^{\alpha_2 \alpha_3} p_2^{\alpha_1} p_2^{\beta_1} p_3^{\beta_2} \nonumber \\
& \qquad \qquad + \: A_4^a \delta^{\alpha_1 \alpha_2} \delta^{\beta_1 \beta_2} p_1^{\alpha_3} \nonumber \\
& \left. \qquad \qquad + \: A_5^a \delta^{\alpha_1 \alpha_2} \delta^{\alpha_3 \beta_2} p_2^{\beta_1} - A_5(p_1 \leftrightarrow p_2) \delta^{\alpha_1 \alpha_2} \delta^{\alpha_3 \beta_1} p_3^{\beta_2} \right].
\end{align}
The form factors $A_j$, $j=1, \ldots, 5$ are functions of the momentum magnitudes. If no arguments are specified then the standard ordering is assumed, $A_j = A_j(p_1, p_2, p_3)$, while by $p_i \leftrightarrow p_j$ we denote the exchange of the two momenta, \textit{e.g.}, $A_3(p_1 \leftrightarrow p_2) = A_3(p_2, p_1, p_3)$.

The form factors $A_1$, $A_2$ and $A_4$ are antisymmetric under $p_1 \leftrightarrow p_2$, \textit{i.e.}, they satisfy
\begin{equation}
A_j^a(p_2, p_1, p_3) = - A_j^a(p_1, p_2, p_3), \qquad j \in \{1,2,4\}.
\end{equation}
The remaining form factors $A_3$ and $A_5$ do not exhibit any symmetry properties.

The form factors can be calculated as follows
\begin{align}
A_1^a & = \text{coefficient of } p_2^{\mu_1} p_2^{\nu_1} p_3^{\mu_2} p_3^{\nu_2} p_1^{\mu_3}, \\
A_2^a & = 4 \cdot \text{coefficient of } \delta^{\nu_1 \nu_2} p_2^{\mu_1} p_3^{\mu_2} p_1^{\mu_3}, \\
A_3^a & = 2 \cdot \text{coefficient of } \delta^{\mu_1 \mu_3} p_2^{\nu_1} p_3^{\mu_2} p_3^{\nu_2}, \\
A_4^a & = 2 \cdot \text{coefficient of } \delta^{\mu_1 \mu_2} \delta^{\nu_1 \nu_2} p_1^{\mu_3}, \\
A_5^a & = 4 \cdot \text{coefficient of } \delta^{\mu_1 \mu_2} \delta^{\mu_3 \nu_2} p_2^{\nu_1}
\end{align}
in $\lla T^{\mu_1 \nu_1}(\bs{p}_1) T^{\mu_2 \nu_2}(\bs{p}_2) J^{\mu_3 a}(\bs{p}_3) \rra$.

\bigskip \noindent \textbf{Primary conformal Ward identities.} The primary CWIs are
\begin{equation}
\begin{array}{ll}
\K_{12} A_1^a = 0, &\qquad\qquad \K_{13} A_1^a = 0, \\
\K_{12} A_2^a = 0, &\qquad\qquad \K_{13} A_2^a = 2 A_1^a, \\
\K_{12} A_3^a = 2 A_1^a, &\qquad\qquad \K_{13} A_3^a = 0, \\
\K_{12} A_4^a = 0, &\qquad\qquad \K_{13} A_4^a = 4 A_2^a, \\
\K_{12} A_5^a = -2 A_2^a, &\qquad\qquad \K_{13} A_5^a = - 2 \left[ A_2^a + A_3^a(p_1 \leftrightarrow p_2) \right].
\end{array}
\end{equation}
The solution in terms of triple-$K$ integrals \eqref{a:J} is
\begin{align}
A_1^a & = \a_1^a J_{5 \{000\}}, \\
A_2^a & = \a_1^a J_{4 \{001\}} + \a_2^a J_{3 \{000\}}, \\
A_3^a & = \a_1^a J_{4 \{010\}} + \a_3^a J_{3 \{000\}}, \\
A_4^a & = \a_1^a J_{3 \{002\}} + 2 \a_2^a J_{2 \{001\}} + \a_4^a I_{1 \{000\}}, \\
A_5^a & = \a_1^a J_{3 \{101\}} + \a_2^a J_{2 \{100\}} + \a_3^a I_{2 \{001\}} + \a_5^a J_{1 \{000\}},
\end{align}
where $\a_j^a$, $j=1, \ldots, 5$ are constants. If the integrals diverge, the regularisation \eqref{a:scheme} should be used.

\bigskip \noindent \textbf{Secondary conformal Ward identities.} The independent secondary CWIs are
\begin{align}
& \Lo_{4} A_1^a + \Ro \left[ A_2^a - A_3^a \right] = \nn\\
& \qquad = 2 d \cdot \text{coefficient of } p_2^{\mu_1} p_3^{\mu_2} p_3^{\nu_2} p_1^{\mu_3} \text{ in } p_{1 \nu_1} \lla T^{\mu_1 \nu_1}(\bs{p}_1) T^{\mu_2 \nu_2}(\bs{p}_2) J^{\mu_3 a}(\bs{p}_3) \rra,  \\
& \Lo_{4} A_2^a + 2 \Ro \left[ 2 A_4^a + A_5^a (p_1 \leftrightarrow p_2) \right] = \nn\\
& \qquad = 4 d \cdot \text{coefficient of } \delta^{\mu_1 \mu_3} p_3^{\mu_2} p_3^{\nu_2} \text{ in } p_{1 \nu_1} \lla T^{\mu_1 \nu_1}(\bs{p}_1) T^{\mu_2 \nu_2}(\bs{p}_2) J^{\mu_3 a}(\bs{p}_3) \rra,  \\
& \Lo_{0} A_3^a - 2 \Ro \left[ A_5^a(p_1 \leftrightarrow p_2) \right] = \nn\\
& \qquad = 8 d \cdot \text{coefficient of } \delta^{\mu_1 \mu_2} p_3^{\nu_2} p_1^{\mu_3} \text{ in } p_{1 \nu_1} \lla T^{\mu_1 \nu_1}(\bs{p}_1) T^{\mu_2 \nu_2}(\bs{p}_2) J^{\mu_3 a}(\bs{p}_3) \rra,  \\
& \Lo_{2} A_5^a - 2 p_1^2 \left[ 2 A_4^a + A_5^a(p_1 \leftrightarrow p_2) \right] = \nn\\
& \qquad = 8 d \cdot \text{coefficient of } \delta^{\mu_2 \mu_3} \delta^{\mu_1 \nu_2} \text{ in } p_{1 \nu_1} \lla T^{\mu_1 \nu_1}(\bs{p}_1) T^{\mu_2 \nu_2}(\bs{p}_2) J^{\mu_3 a}(\bs{p}_3) \rra,
\end{align}
where $\Lo$ and $\Ro$ are given by \eqref{a:R} and \eqref{a:R}. They lead to
\begin{align}
& \a_1^a = \a_2^a = \a_3^a = \a_4^a = 0, \\
& \a_5^a = O(\epsilon).
\end{align}

\section*{\texorpdfstring{$\< T^{\mu_1 \nu_1} T^{\mu_2 \nu_2} T^{\mu_3 \nu_3} \>$}{<TTT>}}
\addcontentsline{toc}{subsection}{\numberline {11}\texorpdfstring{$\< T^{\mu_1 \nu_1} T^{\mu_2 \nu_2} T^{\mu_3 \nu_3} \>$}{<TTT>}}
\setcounter{subsection}{11}
\setcounter{equation}{0}

\noindent \textbf{Ward identities.} The transverse and trace Ward identities are
\begin{align}
& p_1^{\nu_1} \lla T_{\mu_1 \nu_1}(\bs{p}_1) T_{\mu_2 \nu_2}(\bs{p}_2) T_{\mu_3 \nu_3}(\bs{p}_3) \rra = \nn\\
& \qquad = 2 p_1^{\nu_1} \lla \frac{\delta T_{\mu_1 \nu_1}}{\delta g^{\mu_3 \nu_3}}(\bs{p}_1, \bs{p}_3) T_{\mu_2 \nu_2}(\bs{p}_2) \rra
+ 2 p_1^{\nu_1} \lla \frac{\delta T_{\mu_1 \nu_1}}{\delta g^{\mu_2 \nu_2}}(\bs{p}_1, \bs{p}_2) T_{\mu_3 \nu_3}(\bs{p}_3) \rra \nn\\
& \qquad + \: 2 p_{1 (\mu_3} \lla T_{\nu_3) \mu_1}(\bs{p}_2) T_{\mu_2 \nu_2}(-\bs{p}_2) \rra
+ 2 p_{1 (\mu_2} \lla T_{\nu_2) \mu_1}(\bs{p}_3) T_{\mu_3 \nu_3}(-\bs{p}_3) \rra \nn\\
& \qquad + \: \delta_{\mu_3 \nu_3} p_3^\alpha \lla T_{\alpha \mu_1}(\bs{p}_2) T_{\mu_2 \nu_2}(-\bs{p}_2) \rra
+ \delta_{\mu_2 \nu_2} p_2^\alpha \lla T_{\alpha \mu_1}(\bs{p}_3) T_{\mu_3 \nu_3}(-\bs{p}_3) \rra \nn\\
& \qquad - \: p_{3 \mu_1} \lla T_{\mu_2 \nu_2}(\bs{p}_2) T_{\mu_3 \nu_3}(-\bs{p}_2) \rra
- p_{2 \mu_1} \lla T_{\mu_2 \nu_2}(\bs{p}_3) T_{\mu_3 \nu_3}(-\bs{p}_3) \rra, \label{e:WardTTT} \\
& \lla T(\bs{p}_1) T_{\mu_2 \nu_2}(\bs{p}_2) T_{\mu_3 \nu_3}(\bs{p}_3) \rra = \nn\\
& \qquad = 2 \lla \frac{\delta T}{\delta g^{\mu_2 \nu_2}}(\bs{p}_1, \bs{p}_2) T_{\mu_3 \nu_3}(\bs{p}_3) \rra + 2 \lla \frac{\delta T}{\delta g^{\mu_3 \nu_3}}(\bs{p}_1, \bs{p}_3) T_{\mu_2 \nu_2}(\bs{p}_2) \rra.
\end{align}

\bigskip \noindent \textbf{Reconstruction formula.} Define
\begin{align}
& \mathcal{L}^{\mu_1 \nu_1 \mu_2 \nu_2 \mu_3 \nu_3}(\bs{p}_1, \bs{p}_2, \bs{p}_3) = \nn\\
& \qquad = 2 \left[ \mathscr{T}^{\mu_1 \nu_1 \alpha_1}(\bs{p}_1) p_1^{\beta_1} + \frac{\pi^{\mu_1 \nu_1}(\bs{p}_1)}{d-1} \delta^{\alpha_1 \beta_1} \right] \delta^{\mu_3 \alpha_3} \delta^{\nu_3 \beta_3} \lla \frac{\delta T_{\alpha_1 \beta_1}}{\delta g^{\alpha_3 \beta_3}}(\bs{p}_1, \bs{p}_3) T^{\mu_2 \nu_2}(\bs{p}_2) \rra \nn\\
& \qquad\qquad + \left[ \mathscr{T}^{\mu_1 \nu_1 \beta_3}(\bs{p}_1) ( 2 p_1^{(\mu_3} \delta^{\nu_3) \alpha_3} + p_3^{\alpha_3} \delta^{\mu_3 \nu_3} ) - p_3^{\alpha} \mathscr{T}^{\mu_1 \nu_1}_{\alpha}(\bs{p}_1) \delta^{\mu_3 \alpha_3} \delta^{\nu_3 \beta_3} \right. \nn\\
& \qquad\qquad\qquad\qquad \left. + \: \frac{2 \pi^{\mu_1 \nu_1}(\bs{p}_1)}{d-1} \delta^{\mu_3 \alpha_3} \delta^{\nu_3 \beta_3} \right] \lla T_{\alpha_3 \beta_3}(\bs{p}_2) T^{\mu_2 \nu_2}(- \bs{p}_2) \rra, \label{e:L_TTT}
\end{align}
where $\mathscr{T}^{\mu \nu \alpha}$ is defined in \eqref{a:T}. The full 3-point function can be reconstructed from the transverse-traceless part as
\begin{align}
& \lla T^{\mu_1 \nu_1}(\bs{p}_1) T^{\mu_2 \nu_2}(\bs{p}_2) T^{\mu_3 \nu_3}(\bs{p}_3) \rra = \lla t^{\mu_1 \nu_1}(\bs{p}_1) t^{\mu_2 \nu_2}(\bs{p}_2) t^{\mu_3 \nu_3}(\bs{p}_3) \rra \nn\\
& \qquad + \: \sum_{\sigma} \mathcal{L}^{\mu_{\sigma(1)} \nu_{\sigma(1)} \mu_{\sigma(2)} \nu_{\sigma(2)} \mu_{\sigma(3)} \nu_{\sigma(3)}}(\bs{p}_{\sigma(1)}, \bs{p}_{\sigma(2)}, \bs{p}_{\sigma(3)}) \nn\\
& \qquad - \left[ \mathscr{T}^{\mu_3 \nu_3}_{\alpha_3}(\bs{p}_3) p_{3 \beta_3} + \frac{\pi^{\mu_3 \nu_3}(\bs{p}_3)}{d-1} \delta_{\alpha_3 \beta_3} \right] \mathcal{L}^{\mu_1 \nu_1 \mu_2 \nu_2 \alpha_3 \beta_3}(\bs{p}_1, \bs{p}_2, \bs{p}_3) \nn\\
& \qquad - [ (\mu_1, \nu_1, \bs{p}_1) \mapsto (\mu_2, \nu_2, \bs{p}_2) \mapsto (\mu_3, \nu_3, \bs{p}_3) \mapsto (\mu_1, \nu_1, \bs{p}_1) ] \nn\\
& \qquad - [ (\mu_1, \nu_1, \bs{p}_1) \mapsto (\mu_3, \nu_3, \bs{p}_3) \mapsto (\mu_2, \nu_2, \bs{p}_2) \mapsto (\mu_1, \nu_1, \bs{p}_1) ], \label{re_TTT}
\end{align}
where the sum is taken over all six permutations $\sigma$ of the set $\{1,2,3\}$.
\newpage
\bigskip \noindent \textbf{Decomposition of the 3-point function.} The tensor decomposition of the transverse-traceless part is
\begin{align}
& \lla t^{\mu_1 \nu_1}(\bs{p}_1) t^{\mu_2 \nu_2}(\bs{p}_2) t^{\mu_3 \nu_3}(\bs{p}_3) \rra \nonumber\\
& \qquad = \Pi^{\mu_1 \nu_1}_{\alpha_1 \beta_1}(\bs{p}_1) \Pi^{\mu_2 \nu_2}_{\alpha_2 \beta_2}(\bs{p}_2) \Pi^{\mu_3 \nu_3}_{\alpha_3 \beta_3}(\bs{p}_3) \left[
A_1 p_2^{\alpha_1} p_2^{\beta_1} p_3^{\alpha_2} p_3^{\beta_2} p_1^{\alpha_3} p_1^{\beta_3} \right. \nonumber \\
& \qquad \qquad + \: A_2 \delta^{\beta_1 \beta_2} p_2^{\alpha_1} p_3^{\alpha_2} p_1^{\alpha_3} p_1^{\beta_3} + A_2(p_1 \leftrightarrow p_3) \delta^{\beta_2 \beta_3} p_2^{\alpha_1} p_2^{\beta_1} p_3^{\alpha_2} p_1^{\alpha_3} \nonumber \\
& \qquad \qquad \qquad \qquad + \: A_2(p_2 \leftrightarrow p_3) \delta^{\beta_1 \beta_3} p_2^{\alpha_1} p_3^{\alpha_2} p_3^{\beta_2} p_1^{\alpha_3} \nonumber \\
& \qquad \qquad + \: A_3 \delta^{\alpha_1 \alpha_2} \delta^{\beta_1 \beta_2} p_1^{\alpha_3} p_1^{\beta_3} + A_3(p_1 \leftrightarrow p_3) \delta^{\alpha_2 \alpha_3} \delta^{\beta_2 \beta_3} p_2^{\alpha_1} p_2^{\beta_1} \nonumber \\
& \qquad \qquad \qquad \qquad + \: A_3(p_2 \leftrightarrow p_3) \delta^{\alpha_1 \alpha_3} \delta^{\beta_1 \beta_3} p_3^{\alpha_2} p_3^{\beta_2} \nonumber \\
& \qquad \qquad + \: A_4 \delta^{\alpha_1 \alpha_3} \delta^{\alpha_2 \beta_3} p_2^{\beta_1} p_3^{\beta_2} + A_4(p_1 \leftrightarrow p_3) \delta^{\alpha_1 \alpha_3} \delta^{\alpha_2 \beta_1} p_3^{\beta_2} p_1^{\beta_3} \nonumber \\
 & \qquad \qquad \qquad \qquad + \: A_4(p_2 \leftrightarrow p_3) \delta^{\alpha_1 \alpha_2} \delta^{\alpha_3 \beta_2} p_2^{\beta_1} p_1^{\beta_3} \nonumber \\
& \left. \qquad \qquad + \: A_5 \delta^{\alpha_1 \beta_2} \delta^{\alpha_2 \beta_3} \delta^{\alpha_3 \beta_1} \right]. \label{e:TTTdecomp}
\end{align}
The form factors $A_j$, $j=1, \ldots, 5$ are functions of the momentum magnitudes. If no arguments are specified then the standard ordering is assumed, $A_j = A_j(p_1, p_2, p_3)$, while by $p_i \leftrightarrow p_j$ we denote the exchange of the two momenta, \textit{e.g.}, $A_1(p_1 \leftrightarrow p_3) = A_2(p_3, p_2, p_1)$.

The form factors $A_1$ and $A_5$ are symmetric under any permutation of momenta, \textit{i.e.}, for any permutation $\sigma$ of the set $\{1,2,3\}$,
\begin{equation}
A_j(p_{\sigma(1)}, p_{\sigma(2)}, p_{\sigma(3)}) = A_j(p_1, p_2, p_3), \qquad j \in \{1,5\}.
\end{equation}
The remaining form factors are symmetric under $p_1 \leftrightarrow p_2$, \textit{i.e.}, they satisfy
\begin{equation}
A_j(p_2, p_1, p_3) = A_j(p_1, p_2, p_3), \qquad j \in \{2,3,4\}.
\end{equation}

The form factors can be calculated as
\begin{align}
A_1 & = \text{coefficient of } p_2^{\mu_1} p_2^{\nu_1} p_3^{\mu_2} p_3^{\nu_2} p_1^{\mu_3} p_1^{\nu_3}, \\
A_2 & = 4 \cdot \text{coefficient of } \delta^{\nu_1 \nu_2} p_2^{\mu_1} p_3^{\mu_2} p_1^{\mu_3} p_1^{\nu_3}, \\
A_3 & = 2 \cdot \text{coefficient of } \delta^{\mu_1 \mu_2} \delta^{\nu_1 \nu_2} p_1^{\mu_3} p_1^{\nu_3}, \\
A_4 & = 8 \cdot \text{coefficient of } \delta^{\mu_1 \mu_3} \delta^{\mu_2 \nu_3} p_2^{\nu_1} p_3^{\nu_2}, \\
A_5 & = 8 \cdot \text{coefficient of } \delta^{\mu_1 \nu_2} \delta^{\mu_2 \nu_3} \delta^{\mu_3 \nu_1},
\end{align}
in $\lla T^{\mu_1 \nu_1}(\bs{p}_1) T^{\mu_2 \nu_2}(\bs{p}_2) T^{\mu_3 \nu_3}(\bs{p}_3) \rra$.

\bigskip \noindent \textbf{Primary conformal Ward identities.} The primary CWIs are
\begin{equation}
\begin{array}{ll}
\K_{12} A_1 = 0, &\qquad \K_{13} A_1 = 0, \\
\K_{12} A_2 = 0, &\qquad \K_{13} A_2 = 8 A_1, \\
\K_{12} A_3 = 0, &\qquad \K_{13} A_3 = 2 A_2, \\
\K_{12} A_4 = 4 \left[ A_2(p_1 \leftrightarrow p_3) - A_2(p_2 \leftrightarrow p_3) \right], &\qquad \K_{13} A_4 = - 4 A_2(p_2 \leftrightarrow p_3), \\
\K_{12} A_5 = 2 \left[ A_4(p_2 \leftrightarrow p_3) - A_4(p_1 \leftrightarrow p_3) \right], &\qquad \K_{13} A_5 = 2 \left[ A_4 - A_4(p_1 \leftrightarrow p_3) \right].
\end{array}
\end{equation}
The solution in terms of triple-$K$ integrals \eqref{a:J} is
\begin{align}
A_1 & = \a_1 J_{6 \{000\}}, \label{a:TTT1} \\
A_2 & = 4 \a_1 J_{5 \{001\}} + \a_2 J_{4 \{000\}}, \\
A_3 & = 2 \a_1 J_{4 \{002\}} + \a_2 J_{3 \{001\}} + \a_3 J_{2 \{000\}}, \\
A_4 & = 8 \a_1 J_{4 \{110\}} - 2 \a_2 J_{3 \{001\}} + \a_4 J_{2 \{000\}}, \\
A_5 & = 8 \a_1 J_{3 \{111\}} + 2 \a_2 \left( J_{2 \{110\}} + J_{2 \{101\}} + J_{2 \{011\}} \right) + \a_5 J_{0 \{000\}}, \label{a:TTTlast}
\end{align}
where $\a_j$, $j=1,\ldots,5$ are constants. If the integrals diverge, the regularisation \eqref{a:scheme} should be used.

\bigskip \noindent \textbf{Secondary conformal Ward identities.} The independent secondary CWIs are
\begin{align}
& (*) \ \Lo_{6} A_1 + \Ro \left[ A_2 - A_2(p_2 \leftrightarrow p_3) \right] = \label{e:tttseccwi1}\\
& \qquad = 2 d \cdot \text{coeff. of } p_2^{\mu_1} p_3^{\mu_2} p_3^{\nu_2} p_1^{\mu_3} p_1^{\nu_3} \text{ in } p_{1 \nu_1} \lla T^{\mu_1 \nu_1}(\bs{p}_1) T^{\mu_2 \nu_2}(\bs{p}_2) T^{\mu_3 \nu_3}(\bs{p}_3) \rra,  \nn \\
& \Lo_{6} A_2 + 2 \Ro \left[ 2 A_3 - A_4(p_1 \leftrightarrow p_3) \right] = \\
& \qquad = 8 d \cdot \text{coefficient of } \delta^{\mu_1 \mu_2} p_3^{\nu_2} p_1^{\mu_3} p_1^{\nu_3} \text{ in } p_{1 \nu_1} \lla T^{\mu_1 \nu_1}(\bs{p}_1) T^{\mu_2 \nu_2}(\bs{p}_2) T^{\mu_3 \nu_3}(\bs{p}_3) \rra\nn,  \\
& (*) \ \Lo_{4} \left[ A_2(p_1 \leftrightarrow p_3) \right] + \Ro  \left[ A_4(p_2 \leftrightarrow p_3) - A_4 \right] + 2 p_1^2 \left[ A_2(p_2 \leftrightarrow p_3) - A_2 \right] = \label{e:tttseccwi3}\\
& \qquad = 8 d \cdot \text{coefficient of } \delta^{\mu_2 \mu_3} p_2^{\mu_1} p_3^{\nu_2} p_1^{\nu_3} \text{ in } p_{1 \nu_1} \lla T^{\mu_1 \nu_1}(\bs{p}_1) T^{\mu_2 \nu_2}(\bs{p}_2) T^{\mu_3 \nu_3}(\bs{p}_3) \rra\nn,  \\
& \Lo_{4} \left[ A_4(p_2 \leftrightarrow p_3) \right] - 2 \Ro A_5 + 2 p_1^2 \left[ A_4(p_1 \leftrightarrow p_3) - 4 A_3 \right] = \\
& \qquad = 16 d \cdot \text{coefficient of } \delta^{\mu_1 \mu_2} \delta^{\mu_3 \nu_2} p_1^{\nu_3} \text{ in } p_{1 \nu_1} \lla T^{\mu_1 \nu_1}(\bs{p}_1) T^{\mu_2 \nu_2}(\bs{p}_2) T^{\mu_3 \nu_3}(\bs{p}_3) \rra, \nn\\
& \Lo_{2} \left[ A_3(p_1 \leftrightarrow p_3) \right] + p_1^2 \left[ A_4 - A_4(p_2 \leftrightarrow p_3) \right] = \\
& \qquad = 4 d \cdot \text{coefficient of } \delta^{\mu_2 \mu_3} \delta^{\nu_2 \nu_3} p_2^{\mu_1} \text{ in } p_{1 \nu_1} \lla T^{\mu_1 \nu_1}(\bs{p}_1) T^{\mu_2 \nu_2}(\bs{p}_2) T^{\mu_3 \nu_3}(\bs{p}_3) \rra, \nn
\end{align}
where the operators $\Lo$ and $\Ro$ are defined in \eqref{a:L} and \eqref{a:R}. The identities denoted by asterisks are redundant, \textit{i.e.}, they are trivially satisfied in all cases and do not impose any additional conditions on the primary constants. Furthermore, the transverse Ward identities imply that the right-hand sides of \eqref{e:tttseccwi1} -- \eqref{e:tttseccwi3} vanish. The secondary CWIs lead to
\begin{align}
\alpha_3 & = - (d + 2 v \epsilon) \left( 2 (2 + d + 2 v \epsilon) \alpha_1 + \alpha_2 \right) + \frac{2^{3-\frac{d}{2} - v \epsilon} c_T}{\Gamma \left( \frac{d}{2} + v \epsilon \right) \Gamma \left(1 + \frac{d}{2} + v \epsilon \right)}, \\
\alpha_4 & = (2 + 3d + 6 v \epsilon)\alpha_2 + 2 \alpha_3, \\
\alpha_5 & = - 2(d + 2 v \epsilon)^2 \alpha_2 + \frac{2^{5 - \frac{d}{2} - v \epsilon}(1 + 2 c_g) c_T (u - v) \epsilon}{\Gamma^2 \left( \frac{d}{2} + v \epsilon \right)} \nn\\
& \qquad - \: \frac{1}{2}(u - v)\epsilon(d+2 v \epsilon) \left[ (2+d+2 v \epsilon) \left( 8(d+2 v \epsilon) \alpha_1 + 3 \alpha_2 \right) + 2 \alpha_3 \right].
\end{align}
The constant $c_T$ is the 2-point function normalisation  \eqref{a:TT}, while the constant $c_g$ is defined as
\begin{equation} \label{a:cg}
\lla \frac{\delta T_{\mu_1 \nu_1}}{\delta g^{\mu_2 \nu_2}}(\bs{p}_1, \bs{p}_2) T_{\mu_3 \nu_3}(\bs{p}_3) \rra = 4 c_g \delta_{(\mu_1(\mu_2} \lla T_{\nu_1)\nu_2)}(\bs{p}_3) T_{\mu_3 \nu_3}(- \bs{p}_3) \rra + \ldots
\end{equation}
The omitted terms do not contain the tensor structures listed explicitly.
After the substitution of the solution of the secondary CWIs to \eqref{a:TTT1} - \eqref{a:TTTlast}, the limit $u = v = -1/2$ should be taken. The form factors then represent the 3-point function regulated in the dimensional regularisation \eqref{e:dimreg}.

The 3-point function $\lla T^{\mu_1 \nu_1} T^{\mu_2 \nu_2} T^{\mu_3 \nu_3} \rra$ therefore depends on the 2-point function normalisations $c_T$ and $c_g$ and two undetermined primary constants $\a_1$ and $\a_2$. The dependence of this correlator on two 2-point function normalisations rather than only one as found in \cite{Osborn:1993cr} is related to the definition \eqref{e:defTTT} we adopt for this correlator.  Our definition differs by the semi-local terms on the right-hand side of \eqref{e:defTTT}, and it is these terms that produce the dependence of our solution on $c_g$ through \eqref{a:cg}.
(Similar considerations also apply for $\lla T_{\mu_1 \nu_1} J^{\mu_2} J^{\mu_3} \rra$ as discussed above \eqref{e:defTJJ}.)

An additional effect in dimension $d=3$ is that the tensor decomposition becomes degenerate meaning there are only two instead of the usual five form factors.  In consequence, the stress-energy tensor 3-point function in $d=3$ only depends on the primary constant $\alpha_1$, rather than on both $\alpha_1$ and $\alpha_2$.  Along with the two 2-point function normalisations, this makes three parameters in total (or two using the definition of the 3-point function in \cite{Osborn:1993cr}).
We present a discussion of this degeneracy in appendix \ref{ch:degeneracy}, the results of which we make use of below.

\subsection*{Examples}

\bigskip \noindent \textbf{For $\bm{d=3}$} we find
\begin{equation}
A_1 = \frac{8 \a_1}{a_{123}^6} \left[ a_{123}^3 + 3 a_{123} b_{123} + 15 c_{123} \right], \label{e:TTTA1}
\end{equation}
\begin{align}
A_2 & = \frac{8 \a_1}{a_{123}^5} \left[ 4 p_3^4 + 20 p_3^3 a_{12} + 4 p_3^2 (7 a_{12}^2 + 6 b_{12}) + 15 p_3 a_{12} (a_{12}^2 + b_{12}) + 3 a_{12}^2 (a_{12}^2 + b_{12}) \right] \nn\\
& \qquad + \: \frac{2 \a_2}{a_{123}^4} \left[ a_{123}^3 + a_{123} b_{123} + 3 c_{123} \right], \label{e:TTTA2}
\end{align}
\begin{align}
A_3 & = \frac{2 \a_1 p_3^2}{a_{123}^4} \left[ 7 p_3^3 + 28 p_3^2 a_{12} + 3 p_3 (11 a_{12}^2 + 6 b_{12}) + 12 a_{12} ( a_{12}^2 + b_{12} ) \right] \nn\\
& \qquad + \: \frac{\a_2 p_3^2}{a_{123}^3} \left[ p_3^2 + 3 p_3 a_{12} + 2 (a_{12}^2 + b_{12}) \right] - \frac{8 \sqrt{\pi} c_{T}}{3 a_{123}^2} \left[ a_{123}^3 - a_{123} b_{123} - c_{123} \right], \label{e:TTTA3}
\end{align}
\begin{align}
A_4 & = \frac{4 \a_1}{a_{123}^4} \left[ -3 p_3^5 - 12 p_3^4 a_{12} - 9 p_3^3 (a_{12}^2 + 2 b_{12}) + 9 p_3^2 a_{12} (a_{12}^2 - 3 b_{12}) \right.\nn\\
&\qquad\qquad \left. + \: (4 p_3 + a_{12}) (3 a_{12}^4 - 3 a_{12}^2 b_{12} + 4 b_{12}^2) \right] \nn\\
& \qquad + \: \frac{\a_2}{a_{123}^3} \left[ -p_3^4 - 3 p_3^3 a_{12} - 6 p_3^2 b_{12} + a_{12} (a_{12}^2 - b_{12}) (3 p_3 + a_{12}) \right] \nn\\
& \qquad - \: \frac{16 \sqrt{\pi} c_{T}}{3 a_{123}^2} \left[ a_{123}^3 - a_{123} b_{123} - c_{123} \right], \label{e:TTTA4}
\end{align}
\begin{align}
A_5 & = \frac{2 \a_1}{a_{123}^3} \left[ -3 a_{123}^6 + 9 a_{123}^4 b_{123} + 12 a_{123}^2 b_{123}^2 - 33 a_{123}^3 c_{123} + 12 a_{123} b_{123} c_{123} + 8 c_{123}^2 \right] \nn\\
& \qquad + \: \frac{\a_2}{2 a_{123}^2} \left[ -a_{123}^5 + 3 a_{123}^3 b_{123} + 4 a_{123} b_{123}^2 - 11 a_{123}^2 c_{123} + 4 b_{123} c_{123} \right] \nn\\
& \qquad + \: \tfrac{8}{3} \sqrt{\pi} ( c_{T} + 4 c_{g} c_{T} ) (p_1^3 + p_2^3 + p_3^3). \label{e:TTTA5}
\end{align}

\bigskip \noindent \textbf{For $\bm{d=5}$} we find
\begin{equation}
A_1 = \frac{72 \a_1}{a_{123}^7} \left[ a_{123}^2 (a_{123}^4 + a_{123}^2 b_{123} + b_{123}^2) + a_{123} (a_{123}^2 + 5 b_{123}) c_{123} + 10 c_{123}^2 \right],
\end{equation}
\begin{align}
A_2 & = \frac{24 \a_1}{a_{123}^6} \left[ - 12 p_3^7 - 72 p_3^6 a_{12} + 24 p_3^5 (-8 a_{12}^2 + b_{12}) + 24 p_3^4 a_{12} (-13 a_{12}^2 + 6 b_{12}) \right.\nn\\
&\qquad\qquad + \: 8 p_3^3 (-42 a_{12}^4 + 33 a_{12}^2 b_{12} + 8 b_{12}^2) + 3 p_3^2 a_{12} (-77 a_{12}^4 + 73 a_{12}^2 b_{12} + 23 b_{12}^2) \nn\\
&\qquad\qquad \left. + \: 30 p_3 a_{12}^2 (-3 a_{12}^4 + 3 a_{12}^2 b_{12} + b_{12}^2) + 5 a_{12}^3 (-3 a_{12}^4 + 3 a_{12}^2 b_{12} + b_{12}^2) \right] \nn\\
& \qquad + \: \frac{6 \a_2}{a_{123}^5} \left[ -3 a_{123}^6 + 3 a_{123}^4 b_{123} + a_{123}^2 b_{123}^2 + a_{123}^3 c_{123} + 3 a_{123} b_{123} c_{123} + 4 c_{123}^2 \right],
\end{align}
\begin{align}
A_3 & = \frac{2 \a_1 p_3^2}{a_{123}^5} \left[ -81 p_3^6 - 405 p_3^5 a_{12} - 3 p_3^4 (281 a_{12}^2 - 22 b_{12}) - 15 p_3^3 a_{12} (65 a_{12}^2 - 22 b_{12}) \right.\nn\\
&\qquad\qquad - \: 8 p_3^2 (87 a_{12}^4 - 63 a_{12}^2 b_{12} - 13 b_{12}^2) - 100 p_3 a_{12} (3 a_{12}^4 - 3 a_{12}^2 b_{12} - b_{12}^2) \nn\\
&\qquad\qquad \left. - \: 20 a_{12}^2 (3 a_{12}^4 - 3 a_{12}^2 b_{12} - b_{12}^2) \right] \nn\\
& \qquad + \: \frac{a_2 p_3^2}{a_{123}^4} \left[ -9 p_3^5 - 36 p_3^4 a_{12} - 3 p_3^3 (19 a_{12}^2 - 2 b_{12}) - 24 p_3^2 a_{12} (2 a_{12}^2 - b_{12}) \right.\nn\\
&\qquad\qquad \left. - \: 8 p_3 (3 a_{12}^4 - 3 a_{12}^2 b_{12} - b_{12}^2) - 2 a_{12} (3 a_{12}^4 - 3 a_{12}^2 b_{12} - b_{12}^2) \right] \nn\\
& \qquad + \: \frac{16 \sqrt{\pi} c_{T}}{45 a_{123}^3} \left[ 3 a_{123}^2 (a_{123}^4 - 3 a_{123}^2 b_{123} + b_{123}^2) + 3 a_{123} (a_{123}^2 + b_{123}) c_{123} + 2 c_{123}^2 \right],
\end{align}
\begin{align}
A_4 & = \frac{4 \a_1}{a_{123}^5} \left[ 45 p_3^8 + 225 p_3^7 a_{12} + 15 p_3^6 (29 a_{12}^2 + 2 b_{12}) + 75 p_3^5 a_{12} (5 a_{12}^2 + 2 b_{12}) \right.\nn\\
&\qquad\qquad + \: 8 p_3^4 (75 a_{12}^2 - 23 b_{12}) b_{12} - 5 p_3^3 a_{12} (75 a_{12}^4 - 255 a_{12}^2 b_{12} + 79 b_{12}^2) \nn\\
&\qquad\qquad - \: p_3^2 (435 a_{12}^6 - 1335 a_{12}^4 b_{12} + 343 a_{12}^2 b_{12}^2 - 96 b_{12}^3) \nn\\
&\qquad\qquad \left. - \: 3 a_{12} (5 p_3 + a_{12}) (15 a_{12}^6 - 45 a_{12}^4 b_{12} + 11 a_{12}^2 b_{12}^2 - 4 b_{12}^3) \right] \nn\\
& \qquad + \: \frac{\a_2}{a_{123}^4} \left[9 p_3^7 + 36 p_3^6 a_{12} + 3 p_3^5 (17 a_{12}^2 + 2 b_{12}) + 24 p_3^4 a_{12} (a_{12}^2 + b_{12}) \right.\nn\\
&\qquad\qquad - \: 8 p_3^3 (3 a_{12}^4 - 12 a_{12}^2 b_{12} + 5 b_{12}^2) - p_3^2 a_{12} (51 a_{12}^4 - 159 a_{12}^2 b_{12} + 55 b_{12}^2) \nn\\
&\qquad\qquad \left. - \: 9 a_{12}^2 (4 p_3 + a_{12}) (a_{12}^4 - 3 a_{12}^2 b_{12} + b_{12}^2) \right] \nn\\
& \qquad + \: \frac{32 \sqrt{\pi} c_{T}}{45 a_{123}^3} \left[ 3 a_{123}^2 (a_{123}^4 - 3 a_{123}^2 b_{123} + b_{123}^2) + 3 a_{123} (a_{123}^2 + b_{123}) c_{123} + 2 c_{123}^2 \right],
\end{align}
\begin{align}
A_5 & = \frac{6 \a_1}{(p_1 + p_2 + p_3)^4} \left[ 5 a_{123}^3 (a_{123}^2 - 4 b_{123}) (3 a_{123}^4 - 3 a_{123}^2 b_{123} - b_{123}^2) \right. \nn\\
& \qquad\qquad \left. + \: 5 a_{123}^2 (23 a_{123}^4 - 23 a_{123}^2 b_{123} + 4 b_{123}^2) c_{123} - 4 a_{123} (a_{123}^2 - 4 b_{123}) c_{123}^2 + 8 c_{123}^3 \right] \nn\\
& \qquad + \: \frac{\a_2}{2 (p_1 + p_2 + p_3)^3} \left[ 3 a_{123}^2 (a_{123}^2 - 4 b_{123}) (3 a_{123}^4 - 3 a_{123}^2 b_{123} - b_{123}^2) \right.\nn\\
& \qquad\qquad \left. + \: 3 a_{123} (23 a_{123}^4 - 23 a_{123}^2 b_{123} + 4 b_{123}^2) c_{123} - 4 (a_{123}^2 - 2 b_{123}) c_{123}^2 \right] \nn\\
& \qquad - \: \tfrac{16}{15} \sqrt{\pi} ( c_{T} + 4 c_{g} c_{T} ) (p_1^5 + p_2^5 + p_3^5).
\end{align}


\appendix

\newpage
\part*{Appendix}
\addcontentsline{toc}{section}{Appendix}
\setcounter{section}{1}
\setcounter{subsection}{0}
\setcounter{equation}{0}

\subsection{\texorpdfstring{Decomposition of $\< T^{\mu_1 \nu_1} T^{\mu_2 \nu_2} T^{\mu_3 \nu_3} \>$ in non-conformal case}{Decomposition of <TTT> in non-conformal case}} \label{ch:decTTT}

In this section we present the decomposition of the stress-energy tensor 3-point function for a general quantum field theory. 
As the stress-energy tensor in a general theory is no longer traceless, our arguments in the main text need some minor modifications.  First, we discuss how to reconstruct the full correlation function from the purely transverse part, making use of the transverse Ward identities in a similar fashion to section \ref{ch:finding}. We then proceed to construct the general tensor decomposition of this transverse part in terms of ten independent form factors.

As in the main text, we will denote the transverse-traceless part of the stress-energy tensor by $t^{\mu \nu} = \Pi^{\mu \nu}_{\alpha \beta} T^{\alpha \beta}$.  Here, we will also make use of the purely transverse part, $t_T^{\mu \nu} = \pi^\mu_\alpha \pi^\nu_\beta T^{\alpha \beta}$,
which includes a nonvanishing trace part $(t_T)^\mu_\mu$.
The difference between the stress-energy tensor and its transverse part can then be written $\tilde{t}_{\text{loc}}^{\mu \nu} = T^{\mu \nu} - t^{\mu \nu}_T$, \textit{i.e.},
\begin{equation}
\tilde{t}^{\mu \nu}_{\text{loc}} = \left( \frac{p^\mu}{p^2} \delta^{\nu}_\alpha + \frac{p^\nu}{p^2} \delta^{\mu}_\alpha - \frac{p^\mu p^\nu p_\alpha}{p^4} \right) p_\beta T^{\alpha \beta}.
\end{equation}

To obtain the reconstruction formula, we use the Ward identity \eqref{e:WardTTT} to re-express $p_\beta T^{\alpha \beta}$ in terms of 2-point functions when the expectation value of $\tilde{t}^{\mu \nu}_{\text{loc}}$ with other operators is taken.
Defining the operator
\begin{align}
& \tilde{\mathcal{L}}^{\mu_1 \nu_1 \mu_2 \nu_2 \mu_3 \nu_3}(\bs{p}_1, \bs{p}_2, \bs{p}_3) = \frac{1}{p_1^2} \left( 2 p_1^{(\mu_1} \delta^{\nu_1)}_{\alpha_1} - \frac{p_1^{\mu_1} p_1^{\nu_1} p_{1 \alpha_1}}{p_1^2} \right) \times \nn\\
& \qquad \times \left[ 2 \delta^{\mu_3 \alpha_3} \delta^{\nu_3 \alpha_3} p_1^{\beta_1} \lla \frac{\delta T_{\alpha_1 \beta_1}}{\delta g^{\alpha_3 \beta_3}}(\bs{p}_1, \bs{p}_3) T^{\mu_2 \nu_2}(\bs{p}_2) \rra \right.\nn\\
& \qquad\qquad \left. + \left( \delta^{\beta_3 \alpha_1} (2 p_1^{(\mu_3} \delta^{\nu_3) \alpha_3}  + p_3^{\alpha_3} \delta^{\mu_3 \nu_3}) - p_3^{\alpha_1} \delta^{\alpha_3 \mu_3} \delta^{\beta_3 \nu_3} \right) \lla T_{\alpha_3 \beta_3}(\bs{p}_2) T^{\mu_2 \nu_2}(-\bs{p}_2) \rra \right],
\end{align}
the reconstruction formula takes the form
\begin{align}
& \lla T^{\mu_1 \nu_1}(\bs{p}_1) T^{\mu_2 \nu_2}(\bs{p}_2) T^{\mu_3 \nu_3}(\bs{p}_3) \rra = \lla t_T^{\mu_1 \nu_1}(\bs{p}_1) t_T^{\mu_2 \nu_2}(\bs{p}_2) t_T^{\mu_3 \nu_3}(\bs{p}_3) \rra \nn\\
& \qquad + \: \sum_{\sigma} \tilde{\mathcal{L}}^{\mu_{\sigma(1)} \nu_{\sigma(1)} \mu_{\sigma(2)} \nu_{\sigma(2)} \mu_{\sigma(3)} \nu_{\sigma(3)}}(\bs{p}_{\sigma(1)}, \bs{p}_{\sigma(2)}, \bs{p}_{\sigma(3)}) \nn\\
& \qquad - \frac{1}{p_3^2} \left( 2 p_3^{(\mu_3} \delta^{\nu_3)}_{\alpha_3} - \frac{p_3^{\mu_3} p_3^{\nu_3} p_{3 \alpha_3}}{p_3^2} \right) p_{3 \beta_3} \tilde{\mathcal{L}}^{\mu_1 \nu_1 \mu_2 \nu_2 \alpha_3 \beta_3}(\bs{p}_1, \bs{p}_2, \bs{p}_3) \nn\\
& \qquad - [ (\mu_1, \nu_1, \bs{p}_1) \mapsto (\mu_2, \nu_2, \bs{p}_2) \mapsto (\mu_3, \nu_3, \bs{p}_3) \mapsto (\mu_1, \nu_1, \bs{p}_1) ] \nn\\
& \qquad - [ (\mu_1, \nu_1, \bs{p}_1) \mapsto (\mu_3, \nu_3, \bs{p}_3) \mapsto (\mu_2, \nu_2, \bs{p}_2) \mapsto (\mu_1, \nu_1, \bs{p}_1) ],
\end{align}
where the sum is taken over all six permutations $\sigma$ of the set $\{1,2,3\}$. Note the similarity between these expression and (\ref{e:L_TTT}, \ref{re_TTT}).

We turn now to the tensor decomposition of the purely transverse part of the 3-point function.
The most general form of this is
\begin{align}
& \lla t_T^{\mu_1 \nu_1}(\bs{p}_1) t_T^{\mu_2 \nu_2}(\bs{p}_2) t_T^{\mu_3 \nu_3}(\bs{p}_3) \rra = \nn\\
& \qquad = \pi^{\mu_1}_{(\alpha_1}(\bs{p}_1) \pi^{\nu_1}_{\beta_1)}(\bs{p}_1) \pi^{\mu_2}_{(\alpha_2}(\bs{p}_2) \pi^{\nu_2}_{\beta_2)}(\bs{p}_2) \pi^{\mu_3}_{(\alpha_3}(\bs{p}_3) \pi^{\nu_3}_{\beta_3)}(\bs{p}_3) X^{\alpha_1 \beta_1 \alpha_2 \beta_2 \alpha_3 \beta_3},
\end{align}
where $X^{\alpha_1 \beta_1 \alpha_2 \beta_2 \alpha_3 \beta_3}$ is a general tensor built from the metric $\delta^{\mu \nu}$ and two independent momenta, with a kinematic dependence on the momentum magnitudes $p_1$, $p_2$ and $p_3$. Note, however, that if $X^{\alpha_1 \beta_1 \alpha_2 \beta_2 \alpha_3 \beta_3}$ contains $p_j^{\alpha_j}$ or $p_j^{\beta_j}$ for $j \in \{1,2,3\}$ then the contractions with the corresponding transverse projectors vanish. We will assume that $X^{\alpha_1 \beta_1 \alpha_2 \beta_2 \alpha_3 \beta_3}$ is symmetric under $\alpha_j\leftrightarrow\beta_j$ and we use the convention \eqref{e:momenta_choice} (explained in detail in section \ref{ch:Tensor_structure}) for the momenta appearing under the various Lorentz indices:
\begin{equation}
\bs{p}_1, \bs{p}_2 \text{ for } \mu_1, \nu_1; \ \bs{p}_2, \bs{p}_3 \text{ for } \mu_2, \nu_2 \text{  and  } \bs{p}_3, \bs{p}_1 \text{ for }\mu_3, \nu_3.
\end{equation}

The following table lists all $24$ simple tensors from which $X^{\alpha_1 \beta_1 \alpha_2 \beta_2 \alpha_3 \beta_3}$ may be built.
\begin{table}[ht] 
\begin{center}
\begin{tabular}{|c|c|c|}
\hline
& $p_2^{\alpha_1} p_2^{\beta_1} p_3^{\alpha_2} p_3^{\beta_2} p_1^{\alpha_3} p_1^{\beta_3}$ & \\ \hline

$\delta^{\beta_1 \beta_2} p_2^{\alpha_1} p_3^{\alpha_2} p_1^{\alpha_3} p_1^{\beta_3}$ &
$\delta^{\beta_2 \beta_3} p_2^{\alpha_1} p_2^{\beta_1} p_3^{\alpha_2} p_1^{\alpha_3}$ &
$\delta^{\beta_1 \beta_3} p_2^{\alpha_1} p_3^{\alpha_2} p_3^{\beta_2} p_1^{\alpha_3}$ \\ \hline

$\delta^{\alpha_1 \alpha_2} \delta^{\beta_1 \beta_2} p_1^{\alpha_3} p_1^{\beta_3}$ &
$\delta^{\alpha_2 \alpha_3} \delta^{\beta_2 \beta_3} p_2^{\alpha_1} p_2^{\beta_1}$ &
$\delta^{\alpha_1 \alpha_3} \delta^{\beta_1 \beta_3} p_3^{\alpha_2} p_3^{\beta_2}$ \\ \hline

$\delta^{\alpha_1 \alpha_3} \delta^{\alpha_2 \beta_3} p_2^{\beta_1} p_3^{\beta_2}$ &
$\delta^{\alpha_1 \alpha_3} \delta^{\alpha_2 \beta_1} p_3^{\beta_2} p_1^{\beta_3}$ &
$\delta^{\alpha_1 \alpha_2} \delta^{\alpha_3 \beta_2} p_2^{\beta_1} p_1^{\beta_3}$ \\ \hline

& $\delta^{\alpha_1 \beta_2} \delta^{\alpha_2 \beta_3} \delta^{\alpha_3 \beta_1}$ & \\ \hline

$\delta^{\alpha_3 \beta_3} p_3^{\alpha_1} p_3^{\beta_1} p_3^{\alpha_2} p_3^{\beta_2}$ &
$\delta^{\alpha_1 \beta_1} p_3^{\alpha_2} p_3^{\beta_2} p_1^{\alpha_3} p_1^{\beta_3}$ &
$\delta^{\alpha_2 \beta_2} p_3^{\alpha_1} p_3^{\beta_1} p_1^{\alpha_3} p_1^{\beta_3}$ \\ \hline

$\delta^{\alpha_3 \beta_3} \delta^{\beta_1 \beta_2} p_2^{\alpha_1} p_3^{\alpha_2}$ &
$\delta^{\alpha_1 \beta_1} \delta^{\beta_2 \beta_3} p_3^{\alpha_2} p_1^{\alpha_3}$ &
$\delta^{\alpha_2 \beta_2} \delta^{\beta_1 \beta_3} p_2^{\alpha_1} p_1^{\alpha_3}$ \\ \hline

$\delta^{\alpha_3 \beta_3} \delta^{\alpha_1 \alpha_2} \delta^{\beta_1 \beta_2}$ &
$\delta^{\alpha_1 \beta_1} \delta^{\alpha_2 \alpha_3} \delta^{\beta_2 \beta_3}$ &
$\delta^{\alpha_2 \beta_2} \delta^{\alpha_1 \alpha_3} \delta^{\beta_1 \beta_3}$ \\ \hline

$\delta^{\alpha_1 \beta_1} \delta^{\alpha_2 \beta_2} p_1^{\alpha_3} p_1^{\beta_3}$ &
$\delta^{\alpha_2 \beta_2} \delta^{\alpha_3 \beta_3} p_2^{\alpha_1} p_2^{\beta_1}$ &
$\delta^{\alpha_1 \beta_1} \delta^{\alpha_3 \beta_3} p_3^{\alpha_2} p_3^{\beta_2}$ \\ \hline

& $\delta^{\alpha_1 \beta_1} \delta^{\alpha_2 \beta_2} \delta^{\alpha_3 \beta_3}$ & \\ \hline

\end{tabular}
\end{center}
\label{Table24}
\caption{\textit{When contracted with the transverse projectors, this table presents all $24$ tensor structures in the decomposition of the transverse part of $\lla T^{\mu_1 \nu_1} T^{\mu_2 \nu_2} T^{\mu_3 \nu_3} \rra$. Tensors are divided into $10$ orbits of the action of the symmetry group $S_3$, after the contractions with the transverse projectors are taken.}}
\end{table}

Contracting each tensor in the table with the transverse projectors we obtain $24$ transverse tensors denoted by $P_a$, $a = 1,2,\ldots, 24$. Each tensor $P_a$ can then be multiplied by a form factor $B_a$ to obtain the decomposition
\begin{equation}
\lla t_T^{\mu_1 \nu_1}(\bs{p}_1) t_T^{\mu_2 \nu_2}(\bs{p}_2) t_T^{\mu_3 \nu_3}(\bs{p}_3) \rra = \sum_{a=1}^{24} B_a(p_1, p_2, p_3) P_a^{\mu_1 \nu_1 \mu_2 \nu_2 \mu_3 \nu_3}.
\end{equation}
However, the number of independent form factors may be reduced by looking at the symmetry properties. If  we denote the permutation group of the set $\{1,2,3\}$ by $S_3$, then the 3-point function is $S_3$-invariant, \textit{i.e.}, for any $\sigma \in S_3$,
\begin{equation}
\lla t_T^{\mu_1 \nu_1}(\bs{p}_1) t_T^{\mu_2 \nu_2}(\bs{p}_2) t_T^{\mu_3 \nu_3}(\bs{p}_3) \rra = \lla t_T^{\mu_{\sigma(1)} \nu_{\sigma(1)}}(\bs{p}_{\sigma(1)}) t_T^{\mu_{\sigma(2)} \nu_{\sigma(2)}}(\bs{p}_{\sigma(2)}) t_T^{\mu_{\sigma(3)} \nu_{\sigma(3)}}(\bs{p}_{\sigma(3)}) \rra.
\end{equation}
When contracted with the transverse projectors, the tensors at the first, fifth and the last row of the table lead to the $S_3$-invariant tensors. Therefore, corresponding form factors are invariant under any permutation of their arguments, for example
\begin{equation}
B_1(p_1, p_2, p_3) = B_1(p_{\sigma(1)}, p_{\sigma(2)}, p_{\sigma(3)})
\end{equation}
for any $\sigma \in S_3$. The remaining tensors transform non-trivially under the action of $S_3$. For concreteness, consider the second line of the table, \textit{i.e.}, the part of the decomposition
\begin{equation} \label{e:exsym1}
B_2(p_1, p_2, p_3) P_2^{\mu_1 \nu_1 \mu_2 \nu_2 \mu_3 \nu_3} + B_3(p_1, p_2, p_3) P_3^{\mu_1 \nu_1 \mu_2 \nu_2 \mu_3 \nu_3} + B_4(p_1, p_2, p_3) P_4^{\mu_1 \nu_1 \mu_2 \nu_2 \mu_3 \nu_3}.
\end{equation}
Under the action of the symmetry group the tensors $P_2, P_3, P_4$ shuffle among each other. For example, under the action of the transposition $(\bs{p}_1, \mu_1, \nu_1) \leftrightarrow (\bs{p}_3, \mu_3, \nu_3)$ we obtain
\begin{equation} \label{e:exsym2}
B_2(p_3, p_2, p_1) P_3^{\mu_1 \nu_1 \mu_2 \nu_2 \mu_3 \nu_3} + B_3(p_3, p_2, p_1) P_2^{\mu_1 \nu_1 \mu_2 \nu_2 \mu_3 \nu_3} + B_4(p_3, p_2, p_1) P_4^{\mu_1 \nu_1 \mu_2 \nu_2 \mu_3 \nu_3}.
\end{equation}
Since the entire 3-point function is $S_3$-invariant, this implies that \eqref{e:exsym1} and \eqref{e:exsym2} are equal. Since all tensor structures $P_a$ are independent, we find
\begin{equation}
B_3(p_1, p_2, p_3) = B_2(p_3, p_2, p_1), \qquad B_4(p_1, p_2, p_3) = B_4(p_3, p_2, p_1).
\end{equation}
By analysing other symmetries we find that \eqref{e:exsym1} depends on one form factor only, say $B_2$,
\begin{equation}
B_2(p_1, p_2, p_3) P_2^{\mu_1 \nu_1 \mu_2 \nu_2 \mu_3 \nu_3} + B_2(p_1 \leftrightarrow p_3) P_3^{\mu_1 \nu_1 \mu_2 \nu_2 \mu_3 \nu_3} + B_2(p_2 \leftrightarrow p_3) P_4^{\mu_1 \nu_1 \mu_2 \nu_2 \mu_3 \nu_3}.
\end{equation}
Moreover, $B_2(p_1, p_2, p_3) = B_2(p_1 \leftrightarrow p_2)$.

The described procedure reduces the number of independent form factors from $24$ down to $10$. The same procedure applied to the transverse-traceless part of the 3-point function reduces the number of independent tensors from $11$ down to $5$. In this case the decomposition is given by \eqref{e:decompTTT}.

\subsection{\texorpdfstring{Degeneracy of the tensor structure of $\< T^{\mu_1 \nu_1} T^{\mu_2 \nu_2} T^{\mu_3 \nu_3} \>$ in $d=3$}{Degeneracy of the tensor structure of <TTT> in d=3}} \label{ch:degeneracy}

In dimension $d=3$, a special degeneracy occurs which allows the transverse-traceless part of $\lla T^{\mu_1 \nu_1} T^{\mu_2 \nu_2} T^{\mu_3 \nu_3} \rra$ to be decomposed in terms of only two form factors rather than five.

To see this, we first define the cross-product
\begin{equation}
\bs{n} = \bs{p}_1 \times \bs{p}_2 = \bs{p}_2 \times \bs{p}_3 = \bs{p}_3 \times \bs{p}_1
\end{equation}
and note that $n^2 = J^2/4$, where $J^2$ is defined in \eqref{e:lambda}. Using \eqref{e:metric_as_momenta} we find
\begin{equation} \label{e:delta3}
\delta^{\mu \nu} = \frac{4}{J^2} \left[ p_i^2 p_j^\mu p_j^\nu + p_j^2 p_i^\mu p_i^\nu - \bs{p}_i \cdot \bs{p}_j ( p_i^\mu p_j^\nu + p_j^\mu p_i^\nu) + n^\mu n^\nu \right]
\end{equation}
for any $i,j=1,2,3$ and $i \neq j$. From the fact that $\delta^{\alpha \beta} \Pi^{\mu \nu}_{\alpha \beta}(\bs{p}_j) = 0$, we find
\begin{equation} \label{e:Pinn}
\Pi^{\mu \nu}_{\alpha \beta}(\bs{p}_j) n^\alpha n^\beta = - p_j^2 \Pi^{\mu \nu}_{\alpha \beta}(\bs{p}_j) \: p_{(j+1)\:\text{mod}\:3}^\alpha \ p_{(j+1)\:\text{mod}\:3}^\beta, \qquad j=1,2,3.
\end{equation}
We can now go back to the decomposition of the transverse-traceless part of $\lla T^{\mu_1 \nu_1} T^{\mu_2 \nu_2} T^{\mu_3 \nu_3} \rra$, equation \eqref{e:TTTdecomp}, and exchange all $\delta^{\alpha \beta}$ for \eqref{e:delta3}. However, if one transverse-traceless projector is contracted with two vectors $\bs{n}$, then, according to \eqref{e:Pinn}, we can replace such a contraction with a contraction of two momenta with appropriate prefactors. Therefore, the only terms surviving in \eqref{e:TTTdecomp} are terms with either zero or two vectors $\bs{n}$. Hence we find only two tensor structures in the decomposition of $\lla t^{\mu_1 \nu_1} t^{\mu_2 \nu_2} t^{\mu_3 \nu_3} \rra$,
\begin{align}
& \lla t^{\mu_1 \nu_1}(\bs{p}_1) t^{\mu_2 \nu_2}(\bs{p}_2) t^{\mu_3 \nu_3}(\bs{p}_3) \rra \nonumber\\
& \qquad = \Pi^{\mu_1 \nu_1}_{\alpha_1 \beta_1}(\bs{p}_1) \Pi^{\mu_2 \nu_2}_{\alpha_2 \beta_2}(\bs{p}_2) \Pi^{\mu_3 \nu_3}_{\alpha_3 \beta_3}(\bs{p}_3) \left[
B_1 p_2^{\alpha_1} p_2^{\beta_1} p_3^{\alpha_2} p_3^{\beta_2} p_1^{\alpha_3} p_1^{\beta_3} \right. \nonumber \\
& \qquad \qquad + \: B_2 n^{\beta_1} n^{\beta_2} p_2^{\alpha_1} p_3^{\alpha_2} p_1^{\alpha_3} p_1^{\beta_3} + B_2(p_1 \leftrightarrow p_3) n^{\beta_2} n^{\beta_3} p_2^{\alpha_1} p_2^{\beta_1} p_3^{\alpha_2} p_1^{\alpha_3} \nonumber \\
& \left. \qquad \qquad \qquad \qquad + \: B_2(p_2 \leftrightarrow p_3) n^{\beta_1} n^{\beta_3} p_2^{\alpha_1} p_3^{\alpha_2} p_3^{\beta_2} p_1^{\alpha_3} \right].
\end{align}
The new form factors $B_j$ are functions of the momentum magnitudes. As usual, if no arguments are specified then the standard ordering is assumed, $B_j = B_j(p_1, p_2, p_3)$, while by $p_i \leftrightarrow p_j$ we denote the exchange of the two momenta, \textit{e.g.}, $B_2(p_1 \leftrightarrow p_3) = B_2(p_3, p_2, p_1)$.

We can now express the new form factors $B_j$ in terms of the old ones, $A_j$, defined in \eqref{e:TTTdecomp}. Using equation \eqref{e:Pinn}, we write the explicit form of the contraction of two transverse-traceless projectors with a metric as
\begin{equation}
\Pi^{\mu_1 \nu_1}_{\alpha_1 \beta_1}(\bs{p}_1) \Pi^{\mu_2 \nu_2}_{\alpha_2 \beta_2}(\bs{p}_2) \delta^{\beta_1 \beta_2} = \frac{4}{J^2} \Pi^{\mu_1 \nu_1}_{\alpha_1 \beta_1}(\bs{p}_1) \Pi^{\mu_2 \nu_2}_{\alpha_2 \beta_2}(\bs{p}_2) \left[ n^{\beta_1} n^{\beta_2} + \frac{1}{2} ( p_3^2 - p_1^2 - p_2^2 ) p_2^{\beta_1} p_3^{\beta_2} \right],
\end{equation}
from which we find
\begin{align}
B_1 & = A_1 + \frac{2}{J^2} \left[ ( p_3^2 - p_1^2 - p_2^2) A_2(p_1, p_2, p_3) + ( p_1 \leftrightarrow p_3 ) + ( p_2 \leftrightarrow p_3 ) \right] \nn\\
& \qquad + \: \frac{4}{J^4} \left[  \left( ( 8 p_1^2 p_2^2 - J^2 ) A_3 + ( p_3^4 - (p_1^2 - p_2^2)^2 ) A_4 \right) + ( p_1 \leftrightarrow p_3 ) + ( p_2 \leftrightarrow p_3 ) \right] \nn\\
& \qquad - \: \frac{8}{J^4} (p_1^2 + p_2^2 + p_3^2) A_5, \\
B_2 & = \frac{4}{J^2} A_2 + \frac{16}{J^4} \left[ (p_3^2 - p_1^2 - p_2^2) A_3 - p_3^2 A_4 + \right.\nn\\
& \qquad\qquad \left. + \: \frac{1}{2} ( p_2^2 - p_1^2 - p_3^2 ) A_4(p_1 \leftrightarrow p_3) + \frac{1}{2} ( p_1^2 - p_2^2 - p_3^2 ) A_4(p_2 \leftrightarrow p_3) \right] \nn\\
& \qquad + \: \frac{16}{J^4} A_5.
\end{align}
Using the general expressions \eqref{e:TTTA1} - \eqref{e:TTTA5} for the form factors in $d=3$, we arrive at the final result
\begin{align}
B_1 & = 1920 \alpha_1 \frac{c_{123}^3}{J^4 a_{123}^4} - \frac{32 \sqrt{\pi} c_T}{3 J^4 a_{123}^2} \left[ (3 + 8 c_g) a_{123}^5 ( a_{123}^2 - 5 b_{123} ) + 24(1 + 2 c_g) a_{123}^3 b_{123}^2 \right.\nn\\
& \qquad\qquad \left. - \: 8 a_{123} b_{123}^3 + \left( 3(8 c_g - 1) a_{123}^4 - 48 c_g a_{123}^2 b_{123} - 8 b_{123}^2 \right) c_{123} + 8 a_{123} c_{123}^2 \right], \\
B_2 & = - 1920 \alpha_1 \frac{c_{123}^2 p_3}{J^4 a_{123}^4} + \frac{256 \sqrt{\pi} c_T}{3 J^4 a_{123}^2} \left[ 2 (1 + c_g) p_3^4 (p_3 + 2 a_{12}) \right.\nn\\
& \qquad\qquad + \: p_3^3 \left(2 (2 + c_g) a_{12}^2 - 4 b_{12} \right) + p_3^2 a_{12} \left( (3 + 2 c_g) a_{12}^2 - (5 + 6 c_g) b_{12} \right) \nn\\
& \qquad\qquad + \: 2 p_3 \left( (1 + 2 c_g) a_{12}^2 (a_{12}^2 - 3 b_{12} ) + b_{12}^2 \right) \nn\\
& \qquad\qquad \left. - \: 3 (1 + 2 c_g) a_{12}^3 b_{12} + a_{12} b_{12}^2 + (1 + 2 c_g) a_{12}^5 \right].
\end{align}
The variables used in this expression are symmetric polynomials of the momentum magnitudes as defined in \eqref{e:variables}. Note that this expression has no dependence on the primary constant $\alpha_2$. Therefore, in $d=3$, the most general form of the correlation function $\lla T^{\mu_1 \nu_1} T^{\mu_2 \nu_2} T^{\mu_3 \nu_3} \rra$ depends on only one undetermined primary constant and on two 2-point function normalisations $c_T$ and $c_g$. This is in agreement with \cite{Osborn:1993cr}, noting that the normalisation constant $c_g$ arises through our definition  of the 3-point function in \eqref{e:defTTT}.

Finally, while similar considerations hold for other 3-point correlators in $d=3$ involving the stress-energy tensor, in these cases it turns out that the use of equation \eqref{e:delta3} does not reduce the number of independent primary constants in the final result.

\subsection{\texorpdfstring{Triple-$K$ integrals by Fourier transform}{Triple-K integrals by Fourier transform}} \label{ch:toKKK}

Here we present the computation leading from the momentum space integral in \eqref{e:Otriple-$K$} to the triple-$K$ representation. The method generalises the results of \cite{Barnes:2010}, as we consider a general integral
\begin{equation} \label{e:kint}
\int \frac{\D^d \bs{k}}{(2 \pi)^d} \frac{k^{\mu_1} \ldots k^{\mu_r}}{|\bs{k}|^{2 \delta_{3}} |\bs{k} - \bs{p}_1|^{2 \delta_{2}} |\bs{k} + \bs{p}_2|^{2 \delta_{1}}}.
\end{equation}
Using Schwinger parameters
\begin{equation}
\frac{1}{A^\alpha} = \frac{1}{\Gamma(\alpha)} \int_0^\infty \D s \: s^{\alpha-1} e^{-sA} \D s, \qquad \alpha > 0.
\end{equation}
we can write
\begin{align}
& \int \frac{\D^d \bs{k}}{(2 \pi)^d} \frac{k^{\mu_1} \ldots k^{\mu_r}}{\bs{k}^{2 \delta_{3}} |\bs{k} - \bs{p}_1|^{2 \delta_{2}} |\bs{k} + \bs{p}_2|^{2 \delta_{1}}} \nn\\
& = \Gamma^{-3} \int \frac{\D^d \bs{k}}{(2 \pi)^d} k^{\mu_1} \ldots k^{\mu_r} \int_{\R_+^3} \D \vec{s} \: s_1^{\delta_{1} - 1} s_2^{\delta_{2} - 1} s_3^{\delta_{3} - 1} \times \nn\\
& \qquad \times \exp \left[ - ( s_3 \bs{k}^2 + s_2 |\bs{k} - \bs{p}_1|^2 + s_1 |\bs{k} + \bs{p}_2|^2 ) \right],
\end{align}
where we use the following abbreviations
\begin{equation}
\Gamma^3 = \Gamma(\delta_1) \Gamma(\delta_2) \Gamma(\delta_3), \qquad \qquad \D \vec{s} = \D s_1 \D s_2 \D s_3.
\end{equation}
Denoting $s_t = s_1 + s_2 + s_3$, we rewrite the expression in the exponent as
\begin{equation}
s_3 \bs{k}^2 + s_2 |\bs{k} - \bs{p}_1|^2 + s_1 |\bs{k} + \bs{p}_2|^2 = s_t l^2 + \Delta,
\end{equation}
where
\begin{equation}
\bs{l} = \bs{k} + \frac{s_1 \bs{p}_2 - s_2 \bs{p}_1}{s_t}, \qquad \qquad \Delta = \frac{s_1 s_2 p_3^2 + s_1 s_3 p_2^2 + s_2 s_3 p_1^2}{s_t}.
\end{equation}
We can now re-express the integral \eqref{e:kint} as
\begin{equation} \label{e:kint2}
\Gamma^{-3} \int_{\R_+^3} \D \vec{s} \: s_1^{\delta_{1} - 1} s_2^{\delta_{2} - 1} s_3^{\delta_{3} - 1} e^{- \Delta} \int \frac{\D^d \bs{l}}{(2 \pi)^d} e^{-s_t l^2} \prod_{j=1}^r \left( l^{\mu_j} + \frac{s_2 p_1^{\mu_j} - s_1 p_2^{\mu_j}}{s_t} \right).
\end{equation}
This expression can be expanded and split up into a sum of integrals. The integral over $\bs{l}$ gives some moment of a Gaussian random variable. For any $a$ such that $\re a > 0$ we can find
\begin{align}
\int \frac{\D^d \bs{l}}{(2 \pi)^d} l^{2m} e^{-a l^2} & = \frac{\Gamma \left( \frac{d}{2} + m \right)}{(4 \pi)^{\frac{d}{2}} \Gamma \left( \frac{d}{2} \right)} \cdot \frac{1}{a^{\frac{d}{2} + m}}, \\
\int \frac{\D^d \bs{l}}{(2 \pi)^d} l^{\mu_1} \ldots l^{\mu_{2m}} e^{-a l^2} & = \frac{S^{\mu_1 \ldots \mu_{2m}}}{(4 \pi)^{\frac{d}{2}} 2^m a^{\frac{d}{2} + m}},
\end{align}
and integrals with an odd number of $l$ vanish. $S^{\mu_1 \ldots \mu_{2m}}$ is a completely symmetric tensor built from metrics only, with each coefficient equal to $1$, \textit{e.g.},
\begin{equation}
S^{\mu_1 \mu_2 \mu_3 \mu_4} = \delta^{\mu_1 \mu_2} \delta^{\mu_3 \mu_4} + \delta^{\mu_1 \mu_3} \delta^{\mu_2 \mu_4} + \delta^{\mu_1 \mu_4} \delta^{\mu_2 \mu_3}.
\end{equation}
The calculations of the integral \eqref{e:kint} therefore boil down to the calculation of several integrals of the form
\begin{equation} \label{e:idmj}
i_{d,m,\{\delta_j\}} = \frac{1}{(4 \pi)^{\frac{d}{2}} 2^m \Gamma^3} \int_{\R_+^3} \D \vec{s} \: s_t^{- \frac{d}{2} - m} s_1^{\delta_{1} - 1} s_2^{\delta_{2} - 1} s_3^{\delta_{3} - 1} e^{- \Delta}.
\end{equation}
This expression gives the coefficient of the completely symmetric tensor $S^{\mu_1 \ldots \mu_{2m}}$ when we evaluate \eqref{e:kint2}.  In fact, the coefficient of all the tensors in \eqref{e:kint2} can similarly be expressed in terms of $i_{d,m,\{\delta_j\}}$ for some values of $m$ and $\delta_j$.  (In this case, however, the $\delta_j$ parameters are no longer equal to those in  \eqref{e:kint}, since each momentum in \eqref{e:kint2} is accompanied by a Schwinger parameter.)

Let us now express the integral \eqref{e:idmj} in terms of the triple-$K$ integral \eqref{e:J}. Defining $\delta_t = \delta_1 + \delta_2 + \delta_3$, we make the following substitution in \eqref{e:idmj}
\begin{equation}
s_j = \frac{v_1 v_2 + v_1 v_3 + v_2 v_3}{2 v_j} = \frac{V}{2 v_j}, \qquad j=1,2,3,
\end{equation}
giving
\begin{equation}
i_{d,m,\{\delta_j\}} =  \frac{2^{\frac{d}{2} - \delta_t}}{(4 \pi)^{\frac{d}{2}} \Gamma^3} \int_{\R_+^3} \D \vec{v} \: V^{\delta_t - d - 2m} \prod_{j=1}^3 v_j^{\frac{d}{2} + m - \delta_j - 1} e^{-\frac{v_j p_j^2}{2}}.
\end{equation}
Observing that
\begin{equation}
V = v_1 v_2 v_3 \left( v_1^{-1} + v_2^{-1} + v_3^{-1} \right)
\end{equation}
and introducing a new Schwinger parameter $t$ to exponentiate the term in brackets, we find
\begin{align}
i_{d,m,\{\delta_j\}} & = \frac{2^{\frac{d}{2} - \delta_t}}{(4 \pi)^{\frac{d}{2}} \Gamma^3 \Gamma(d + 2m - \delta_t)} \int_0^\infty \D t \: t^{d + 2m - \delta_t - 1} \times \nn\\
& \qquad \times \: \int_{\R_+^3} \D \vec{v} \: \prod_{j=1}^3 v_j^{-\frac{d}{2} - m + \delta_t - \delta_j - 1} e^{-\frac{v_j p_j^2}{2} - \frac{t}{v_j}} \nn\\
& = \frac{2^{-d - 3m + \delta_t}}{(4 \pi)^{\frac{d}{2}} \Gamma^3 \Gamma(d + 2m - \delta_t)} \int_0^\infty \D t \: t^{d + 2m - \delta_t - 1} \times \nn\\
& \qquad \times \: \int_{\R_+^3} \D \vec{u} \: \prod_{j=1}^3 p_j^{d+2m - 2\delta_t + 2\delta_j} u_j^{-\frac{d}{2} - m + \delta_t - \delta_j - 1} e^{-u_j - \frac{t p_j^2}{2 u_j}}.
\end{align}
Using the standard formula \cite{Abramowitz}
\begin{equation}
K_\nu(z) = \frac{1}{2} \left( \frac{z}{2} \right)^\nu \int_0^\infty e^{-u-\frac{z^2}{4u}} u^{-\nu-1} \D u, \qquad | \arg z | < \frac{\pi}{4}
\end{equation}
we now obtain our final result
\begin{align}
i_{d,m,\{\delta_j\}} & = \frac{2^{-\frac{d}{2}-2m+4}}{(4 \pi)^{\frac{d}{2}} \Gamma^3 \Gamma(d + 2m - \delta_t)} \int_0^\infty \D t ( \sqrt{2 t} )^{\frac{d}{2} + m - 2} \prod_{j=1}^3 p_j^{\frac{d}{2}+m-\delta_t+\delta_j} K_{\frac{d}{2}+m-\delta_t+\delta_j}(\sqrt{2t} p_j) \nn\\
& = \frac{2^{-\frac{d}{2}-2m+4}}{(4 \pi)^{\frac{d}{2}} \Gamma^3 \Gamma(d + 2m - \delta_t)} \int_0^\infty \D x \: x^{\frac{d}{2} + m - 1} \prod_{j=1}^3 p_j^{\frac{d}{2}+m-\delta_t+\delta_j} K_{\frac{d}{2}+m-\delta_t+\delta_j}(p_j x) \nn\\
& = \frac{2^{-\frac{d}{2}-2m+4}}{(4 \pi)^{\frac{d}{2}} \Gamma^3 \Gamma(d + 2m - \delta_t)} I_{\frac{d}{2} + m - 1 \{ \frac{d}{2} + m - \delta_t + \delta_j \}}, \label{e:Idm}
\end{align}
where $\Gamma^3 = \Gamma(\delta_1) \Gamma(\delta_2) \Gamma(\delta_3)$ and $I$ stands for the triple-$K$ integral \eqref{e:J}.

\subsubsection{\texorpdfstring{Triple-$K$ integrals via Feynman parametrisation}{Triple-K integrals via Feynman parametrisation}}

Instead of Schwinger parameters, one can use the more familiar Feynman parametrisation in order to evaluate \eqref{e:kint}. Using standard results \cite{Peskin}, we can write \eqref{e:kint} as
\begin{equation}
\frac{\Gamma(\delta_t)}{\Gamma^3} \int_{[0,1]^3} \D X \: \prod_{j=1}^3 x_j^{\delta_j - 1} \int \frac{\D^d \bs{l}}{(2 \pi)^d} \frac{1}{(l^2 + D)^{\delta_t}} \prod_{j=1}^r \left( l^{\mu_j} + x_2 p_1^{\mu_j} - x_1 p_2^{\mu_j} \right),
\end{equation}
where
\begin{align}
\D X &= \D x_1 \D x_2 \D x_3 \: \delta(x_1 + x_2 + x_3 - 1), \\
\bs{l} &= \bs{k} - x_2 \bs{p}_1 + x_1 \bs{p}_2, \\
D &= p_1^2 x_2 x_3 + p_2^2 x_1 x_3 + p_3^2 x_1 x_2.
\end{align}
Looking at the coefficient of the tensor $S^{\mu_1 \ldots \mu_{2m}}$ defined in the previous section, we obtain
\begin{equation}
i_{d,m,\{\delta_j\}} = \frac{\Gamma \left( \delta_t - m - \frac{d}{2} \right)}{(4 \pi)^{\frac{d}{2}} 2^m \Gamma^3} \int_{[0,1]^3} \D X x_1^{\delta_1 - 1} x_2^{\delta_2 - 1} x_3^{\delta_3 - 1} D^{\frac{d}{2} + m - \delta_t}.
\end{equation}
Comparing with \eqref{e:Idm}, we then find
\begin{align}
I_{\alpha \{ \beta_1 \beta_2 \beta_3 \}} & = \int_0^\infty \D x \: x^\alpha \prod_{j=1}^3 p_j^{\beta_j} K_{\beta_j}(p_j x) \nn\\
& = 2^{\alpha - 3} \Gamma \left( \frac{\alpha - \beta_t + 1}{2} \right) \Gamma \left( \frac{\alpha + \beta_t + 1}{2} \right) \times \nn\\
& \qquad \times \int_{[0,1]^3} \D X \: D^{\frac{1}{2} ( \beta_t - \alpha - 1)} \prod_{j=1}^3 x_j^{\frac{1}{2} ( \alpha - 1 - \beta_t ) + \beta_j}, \label{e:FeynToJ}
\end{align}
where $\beta_t = \beta_1 + \beta_2 + \beta_3$ and $I$ is the triple-$K$ integral \eqref{e:J}.

\subsection{\texorpdfstring{Properties of triple-$K$ integrals}{Properties of triple-K integrals}} \label{ch:prop}

In this appendix we list some properties of modified Bessel functions used in the main text.  For further references, see \textit{e.g.}, \cite{Abramowitz}.

The Bessel function $I$ (modified Bessel function of the first kind) is given by the series
\begin{equation} \label{e:serI}
I_\nu(x) = \sum_{j=0}^\infty \frac{1}{j! \Gamma(\nu + j + 1)} \left( \frac{x}{2} \right)^{\nu + 2j}, \quad \nu \neq -1, -2, -3, \ldots
\end{equation}
The Bessel function $K$ (modified Bessel function of the second kind) is defined by
\begin{align}
K_\nu(x) & = \frac{\pi}{2 \sin (\nu \pi)} \left[ I_{-\nu}(x) - I_{\nu}(x) \right], \quad \nu \notin \Z, \label{e:defK} \\
K_n(x) & = \lim_{\epsilon \rightarrow 0} K_{n + \epsilon}(x), \quad n \in \Z.
\end{align}
The finite pointwise limit for $x > 0$ exists for any integer $n$. $K_\nu$ is an even function of $\nu$, \textit{i.e.}, $K_{-\nu}(x) = K_\nu(x)$ for any $\nu \in \R$. If $\nu = \frac{1}{2} + n$, for an integer $n$, the Bessel function reduces to elementary functions
\begin{equation} \label{e:Khalf}
K_\nu(x) = \sqrt{\frac{\pi}{2}} \frac{e^{-x}}{\sqrt{x}} \sum_{j=0}^{\left\lfloor |\nu| - \frac{1}{2} \right\rfloor} \frac{ \left( |\nu| - \frac{1}{2} + j \right)!}{j! \left(|\nu| - \frac{1}{2} - j \right)!} \frac{1}{(2 x)^j}, \quad \nu + \frac{1}{2} \in \Z,
\end{equation}
and in particular
\begin{equation} \label{e:Khalf1}
\begin{array}{lll}
K_{\frac{1}{2}}(x) = \sqrt{\dfrac{\pi}{2}} \dfrac{e^{-x}}{x^{\frac{1}{2}}}, && K_{\frac{3}{2}}(x) = \sqrt{\dfrac{\pi}{2}} \dfrac{e^{-x}}{x^{\frac{3}{2}}} (1 + x), \\
K_{\frac{5}{2}}(x) = \sqrt{\dfrac{\pi}{2}} \dfrac{e^{-x}}{x^{\frac{5}{2}}} (x^2 + 3 x + 3), && K_{\frac{7}{2}}(x) = \sqrt{\dfrac{\pi}{2}} \dfrac{e^{-x}}{x^{\frac{7}{2}}} (x^3 + 6 x^2 + 15 x + 5).
\end{array}
\end{equation}

The series expansion of the Bessel function $K_\nu$ for $\nu \notin \Z$ is given directly in terms of the expansion \eqref{e:serI} via the definition \eqref{e:defK}. In particular
\begin{equation}
K_{\nu}(x) = \left[ \Gamma(-\nu) 2^{-\nu - 1} x^\nu + O(x^{2 - \nu}) \right] + \left[ \frac{\Gamma(\nu) 2^{\nu - 1}}{x^\nu} + O(x^{2 + \nu}) \right], \quad \nu \notin \Z.
\end{equation}
For a non-negative integer index $n$, the expansion reads
\begin{align} \label{e:expKn}
K_n(x) & = \frac{1}{2} \left( \frac{x}{2} \right)^{-n} \sum_{j=0}^{n-1} \frac{(n-j-1)!}{j!} (-1)^j \left( \frac{x}{2} \right)^{2j} \nn\\
& \qquad + \: (-1)^{n+1} \log \left( \frac{x}{2} \right) I_n(x) \nn\\
& \qquad + \: (-1)^n \frac{1}{2} \left( \frac{x}{2} \right)^n \sum_{j=0}^\infty \frac{\psi(j+1) + \psi(n+j+1)}{j! (n+j)!} \left( \frac{x}{2} \right)^{2j},
\end{align}
where $\psi$ is the digamma function. At large $x$, the Bessel functions have the asymptotic expansions
\begin{equation} \label{e:asymIK}
I_\nu(x) = \frac{1}{\sqrt{2 \pi}} \frac{e^x}{\sqrt{x}} + \ldots, \qquad K_\nu(x) = \sqrt{\frac{\pi}{2}} \frac{e^{-x}}{\sqrt{x}} + \ldots, \quad \nu \in \R.
\end{equation}

For any index $\nu \in \R$, the Bessel function $K$ satisfies the following identities
\begin{align}
\frac{\partial}{\partial a} \left[ a^\nu K_\nu(a x) \right] & = - x a^\nu K_{\nu - 1}(a x), \\
K_{\nu-1}(x) + \frac{2 \nu}{x} K_{\nu}(x) & = K_{\nu + 1}(x), \\
K_{-\nu}(x) & = K_{\nu}(x).
\end{align}

\subsection{\texorpdfstring{Appell's $F_4$ function}{Appell's F4 function}} \label{ch:F4}

Appell's $F_4$ function can be defined by the following double series \cite{Appell,Erdelyi}
\begin{equation}
F_4(\alpha, \beta; \gamma, \gamma'; \xi, \eta) = \sum_{i,j = 0}^\infty \frac{(\alpha)_{i+j} (\beta)_{i+j}}{(\gamma)_i (\gamma')_j i! j!} \xi^i \eta^j, \quad \sqrt{|\xi|} + \sqrt{|\eta|} < 1,
\end{equation}
where $(\alpha)_i$ is a Pochhammer symbol. Notice that
\begin{equation}
F_4(\alpha, \beta; \gamma, \gamma'; \xi, \eta) = F_4(\beta, \alpha; \gamma, \gamma'; \xi, \eta) = F_4(\alpha, \beta; \gamma', \gamma; \eta, \xi).
\end{equation}
The series representation, however, is not very useful as in our case
\begin{equation}
\xi = \frac{p_1^2}{p_3^2}, \qquad \qquad \eta = \frac{p_2^2}{p_3^2}
\end{equation}
and the series converges when $p_3 > p_1 + p_2$, which is opposite to the triangle inequality.

As in the case of ordinary hypergeometric functions, the $F_4$ function satisfies certain differential equations. Let $\alpha, \beta, \gamma, \gamma'$ be fixed numbers. The following system of equations
\begin{align}
0 & = \left[ \xi (1 - \xi) \frac{\partial^2}{\partial \xi^2} - \eta^2 \frac{\partial^2}{\partial \eta^2} - 2 \xi \eta \frac{\partial^2}{\partial \xi \partial \eta} \right. \nn\\
& \qquad \left. + \: \left( \gamma - ( \alpha + \beta + 1) \xi \right) \frac{\partial}{\partial \xi} - (\alpha + \beta + 1) \eta \frac{\partial}{\partial \eta} - \alpha \beta \right] F(\xi, \eta), \\
0 & = \left[ \eta (1 - \eta) \frac{\partial^2}{\partial \eta^2} - \xi^2 \frac{\partial^2}{\partial \xi^2} - 2 \xi \eta \frac{\partial^2}{\partial \xi \partial \eta} \right. \nn\\
& \qquad \left. + \: \left( \gamma' - ( \alpha + \beta + 1) \eta \right) \frac{\partial}{\partial \eta} - (\alpha + \beta + 1) \xi \frac{\partial}{\partial \xi} - \alpha \beta \right] F(\xi, \eta),
\end{align}
has exactly four solutions given by \cite{Erdelyi, Exton:1995}
\begin{align}
& F_4(\alpha, \beta; \gamma, \gamma'; \xi, \eta), \\
& \xi^{1 - \gamma} F_4(\alpha + 1- \gamma, \beta + 1 - \gamma; 2 - \gamma, \gamma'; \xi, \eta), \\
& \eta^{1 - \gamma'} F_4(\alpha + 1- \gamma', \beta + 1 - \gamma'; \gamma, 2 - \gamma'; \xi, \eta), \\
& \xi^{1 - \gamma} \eta^{1 - \gamma'} F_4(\alpha + 2 - \gamma - \gamma', \beta + 2 - \gamma - \gamma'; 2 - \gamma, 2 - \gamma'; \xi, \eta).
\end{align}

The following reduction formulae can be found in \cite{Prudnikov} or \cite{Erdelyi}
\begin{align}
& F_4 \left( \alpha, \beta; \alpha, \beta; - \frac{x}{(1-x)(1-y)}, - \frac{y}{(1-x)(1-y)} \right) = \nn\\
& \qquad\qquad = \frac{(1-x)^\beta (1-y)^\alpha}{1-x y}, \label{e:redform1} \\
& F_4 \left( \alpha, \beta; \beta, \beta; - \frac{x}{(1-x)(1-y)}, - \frac{y}{(1-x)(1-y)} \right) = \nn\\
& \qquad\qquad = (1-x)^\alpha (1-y)^\alpha {}_2 F_1(\alpha, 1+\alpha-\beta; \beta; x y), \\
& F_4 \left( \alpha, \beta; 1+\alpha-\beta, \beta; - \frac{x}{(1-x)(1-y)}, - \frac{y}{(1-x)(1-y)} \right) = \nn\\
& \qquad\qquad = (1-y)^\alpha {}_2 F_1 \left(\alpha, \beta; 1+\alpha-\beta; - \frac{x(1-y)}{1-x} \right), \\
& {}_2 F_1 (2 \nu - 1, \nu; \nu; x) = (1 - x)^{1-2\nu}. \label{e:redform2}
\end{align}

\subsubsection{Integrals}

Here we present the list of integrals we use in the paper, which may be found in \cite{Prudnikov}.
\begin{enumerate}
\item[(i)] \begin{align}
& \int_0^\infty \D x \: x^{\alpha - 1} I_\lambda(a x) I_\mu(b x) K_\nu(c x) = \nn\\
& \qquad = \frac{2^{\alpha - 2} \Gamma \left( \frac{\alpha + \lambda + \mu - \nu}{2} \right) \Gamma \left( \frac{\alpha + \lambda + \mu + \nu}{2} \right)}{\Gamma(\lambda + 1) \Gamma(\mu + 1)} \cdot \frac{a^\lambda b^\mu}{c^{\alpha + \lambda + \mu}} \times \nn\\
& \qquad \qquad \times F_4 \left( \frac{\alpha + \lambda + \mu - \nu}{2}, \frac{\alpha + \lambda + \mu + \nu}{2}; \lambda + 1, \mu + 1; \frac{a^2}{c^2}, \frac{b^2}{c^2} \right),
\end{align}
valid for
\begin{equation}
\re (\alpha + \lambda + \mu) > | \re \nu |, \qquad |c| > |a| + |b|, \qquad \re c > |\re a| + |\re b|.
\end{equation}

\item[(ii)] \begin{align}
& \int_0^\infty \D x \: x^{\alpha - 1} K_\lambda(a x) K_\mu(b x) K_\nu(c x) = \nn\\
& \qquad = \frac{2^{\alpha - 4}}{c^\alpha} \left[ A(\lambda, \mu) + A(\lambda, -\mu) + A(-\lambda, \mu) + A(-\lambda, -\mu) \right], \label{e:KKK}
\end{align}
where
\begin{align}
A(\lambda, \mu) & = \left( \frac{a}{c} \right)^\lambda \left( \frac{b}{c} \right)^\mu \Gamma \left( \frac{\alpha + \lambda + \mu - \nu}{2} \right) \Gamma \left( \frac{\alpha + \lambda + \mu + \nu}{2} \right) \Gamma(-\lambda) \Gamma(-\mu) \times \nn\\
& \qquad \times F_4 \left( \frac{\alpha + \lambda + \mu - \nu}{2}, \frac{\alpha + \lambda + \mu + \nu}{2}; \lambda + 1, \mu + 1; \frac{a^2}{c^2}, \frac{b^2}{c^2} \right),
\end{align}
valid for
\begin{equation}
\re \alpha > | \re \lambda | + | \re \mu | + | \re \nu |, \qquad \re(a + b + c) > 0.
\end{equation}

\item[(iii)]
\begin{align} \label{e:I2K}
& \int_0^\infty \D x \: x^{\alpha - 1} K_\mu(c x) K_\nu(c x) = \nn\\
& \qquad = \frac{2^{\alpha - 3}}{\Gamma(\alpha) c^{\alpha}} \Gamma \left( \frac{\alpha + \mu + \nu}{2} \right) \Gamma \left( \frac{\alpha + \mu - \nu}{2} \right) \Gamma \left( \frac{\alpha - \mu + \nu}{2} \right) \Gamma \left( \frac{\alpha - \mu - \nu}{2} \right),
\end{align}
valid for
\begin{equation}
\re \alpha > | \re \mu | + | \re \nu |, \qquad \re c > 0.
\end{equation}

\item[(iv)] \begin{equation} \label{e:I1K}
\int_0^\infty \D x \: x^{\alpha - 1} K_\nu(c x) = \frac{2^{\alpha - 2}}{c^\alpha} \Gamma \left( \frac{\alpha + \nu}{2} \right) \Gamma \left( \frac{\alpha - \nu}{2} \right)
\end{equation}
valid for
\begin{equation}
\re \alpha > | \re \nu |, \qquad \re c > 0.
\end{equation}

\item[(v)]
\begin{align}
& \int_0^\infty \D x \: x^{\alpha - 1} \log x K_\nu(c x) = \frac{2^{\alpha - 3}}{c^\alpha} \Gamma \left( \frac{\alpha + \nu}{2} \right) \Gamma \left( \frac{\alpha - \nu}{2} \right) \times \nn\\
& \qquad \times \: \left[ \psi \left( \frac{\alpha + \nu}{2} \right)  + \psi \left( \frac{\alpha - \nu}{2} \right) - 2 \log \frac{c}{2} \right], \label{e:I1Klog}
\end{align}
valid for
\begin{equation}
\re \alpha > | \re \nu |, \qquad \re c > 0.
\end{equation}

\item[(vi)]
\begin{align}
& \int_0^\infty \D x \: x^{\alpha - 1} \log^2 x K_\nu(c x) = \frac{2^{\alpha - 4}}{c^\alpha} \Gamma \left( \frac{\alpha + \nu}{2} \right) \Gamma \left( \frac{\alpha - \nu}{2} \right) \times \nn\\
& \qquad \times \: \left[ \left( \psi \left( \frac{\alpha + \nu}{2} \right) + \psi \left( \frac{\alpha - \nu}{2} \right) \right) \cdot \left( \psi \left( \frac{\alpha + \nu}{2} \right) + \psi \left( \frac{\alpha - \nu}{2} \right) - 4 \log \frac{c}{2} \right) \right. \nn\\
& \qquad \qquad \qquad \left. + \: \psi' \left( \frac{\alpha + \nu}{2} \right) + \psi'\left( \frac{\alpha - \nu}{2} \right) + 4 \log^2 \frac{c}{2} \right], \label{e:I1Klog2}
\end{align}
valid for
\begin{equation}
\re \alpha > | \re \nu |, \qquad \re c > 0.
\end{equation}

\end{enumerate}

\subsection{\texorpdfstring{Triviality of $\< T^{\mu_1 \nu_1} J^{\mu_2} \mathcal{O} \>$}{Triviality of <TJO>}} \label{ch:vanish}

As our analysis shows, for any $d \geq 3$ and $\Delta_3$ satisfying unitarity bound the correlation functions $\lla T^{\mu_1 \nu_1} J^{\mu_2} \mathcal{O} \rra$ and $\lla T^{\mu_1 \nu_1} T^{\mu_2 \nu_2} J^{\mu_3} \rra$ are trivial, \textit{i.e.}, they are at most semi-local. The triviality of $\lla T^{\mu_1 \nu_1} T^{\mu_2 \nu_2} J^{\mu_3} \rra$ was proved in \cite{Costa:2011a} through a  position space analysis. Our results independently confirm this through calculations in momentum space. In this section we discuss the triviality of $\lla T^{\mu_1 \nu_1} J^{\mu_2} \mathcal{O} \rra$ as an example.

The tensor decomposition of the transverse-traceless part of the $\lla T^{\mu_1 \nu_1} J^{\mu_2} \mathcal{O} \rra$ correlator, the primary and secondary CWIs, and the transverse Ward identities are presented along with all 3-point functions in part II of the paper. Let us rewrite here the important data. The solution in terms of triple-$K$ integrals is
\begin{align}
A_1^{aI} & = \a_1^{aI} J_{3 \{000\}}, \\
A_2^{aI} & = 2 \a_1^{aI} J_{2 \{001\}} + \a_2^{aI} J_{1 \{000\}}.
\end{align}
The independent secondary CWIs are
\begin{align}
& \Lo_{2} A_1^{aI} + \Ro A_2^{aI} = 2 d \cdot \text{coefficient of } p_2^{\mu_1} p_3^{\mu_2} \text{ in } p_{1 \nu_1} \lla T^{\mu_1 \nu_1}(\bs{p}_1) J^{\mu_2 a}(\bs{p}_2) \mathcal{O}^I(\bs{p}_3) \rra, \\
& \Lo'_{1} A_1^{aI} + 2 \Ro' A_2^{aI} \nn\\
& \qquad = - 2 (d - 2) \cdot \text{coefficient of } p_2^{\mu_1} p_2^{\nu_1} \text{ in } p_{2 \mu_2} \lla T^{\mu_1 \nu_1}(\bs{p}_1) J^{\mu_2 a}(\bs{p}_2) \mathcal{O}^I(\bs{p}_3) \rra, \\
& \Lo_{2} A_2^{aI} = 4 d \cdot \text{coefficient of } \delta^{\mu_1 \mu_2} \text{ in } p_{1 \nu_1} \lla T^{\mu_1 \nu_1}(\bs{p}_1) J^{\mu_2 a}(\bs{p}_2) \mathcal{O}^I(\bs{p}_3) \rra
\end{align}
and the transverse Ward identities are
\begin{align}
p_1^{\nu_1} \lla T_{\mu_1 \nu_1}(\bs{p}_1) J^{\mu_2 a}(\bs{p}_2) \mathcal{O}^I(\bs{p}_3) \rra & = p_1^{\nu_1} \lla \frac{\delta T_{\mu_1 \nu_1}}{\delta A_{\mu_2}^a}(\bs{p}_1, \bs{p}_2) \mathcal{O}^I(\bs{p}_3) \rra, \label{e:tjoward1} \\
p_{2 \mu_2} \lla T_{\mu_1 \nu_1}(\bs{p}_1) J^{\mu_2 a}(\bs{p}_2) \mathcal{O}^I(\bs{p}_3) \rra & = 2 p_{2 \mu_2} \lla \frac{\delta J^{\mu_2 a}}{\delta g^{\mu_1 \nu_1}}(\bs{p}_2, \bs{p}_1) \mathcal{O}^I(\bs{p}_3) \rra. \label{e:tjoward2}
\end{align}
If $\beta_3 = \Delta_3 - \frac{d}{2} > 0$, the same reasoning as in section \ref{ch:tran_tjj} shows that the right-hand sides of (\ref{e:tjoward1}, \ref{e:tjoward2}) vanish in the zero-momentum limit. Then, by the results of section \ref{ch:allcases}, in the remaining cases the coefficient of $p_3^0$ in the series expansion of the right-hand sides of (\ref{e:tjoward1}, \ref{e:tjoward2}) is at most ultralocal. Assuming the limit $u = v$ can be taken, the secondary CWIs lead to the following equations
\begin{align}
0 & = \left( (d - \Delta_3)^2 - 4 \right) ( \Delta_3 + 2 v \epsilon ) \alpha_1^{aI} + (\Delta_3 - d - 2) \alpha_2^{aI}, \label{e:tjoseccwi1} \\
0 & = (d - \Delta_3 - 2)(2 + \Delta_3 + 2 v \epsilon) \alpha_1^{aI} - 2 \alpha_2^{aI}, \label{e:tjoseccwi2} \\
0 & = 2(d - 2 \Delta_3 - 2 v \epsilon) \alpha_1^{aI} - \alpha_2^{aI} = 0.
\end{align}
The only solution to these equations is trivial $\alpha_1^{aI} = \alpha_2^{aI} = 0$. The same analysis can be carried out in case when the limit $u = v$ cannot be taken.

Unlike in section \ref{ch:explicit}, there are no additional conditions following from the coefficients of $p_3^{2 \beta_3}$ or $p_3^{2 \beta_3} \log p_3$ in the series expansion in $p_3$ of the secondary CWIs \eqref{e:tjoseccwi1} - \eqref{e:tjoseccwi2}.  Recall that such additional constraints arise when the equations following from the coefficients of $p_3^{2 \beta_3}$ or $p_3^{2 \beta_3} \log p_3$ are more singular than the equations following from the zero-momentum limit. In our case, it turns out that all coefficients of $p_3^{2 \beta_3}$ or $p_3^{2 \beta_3} \log p_3$ on the left-hand sides of \eqref{e:tjoseccwi1} - \eqref{e:tjoseccwi2} can be written in terms of $l_{\frac{d}{2} + \epsilon \{ \frac{d}{2}, \frac{d}{2} - 1, - \Delta_3 + \frac{d}{2}\}}$, accounting for all possible singularities. One can check that $l_{\frac{d}{2} + \epsilon \{ \frac{d}{2}, \frac{d}{2} - 1, - \Delta_3 + \frac{d}{2}\}}$ cannot be more singular than $l_{\frac{d}{2} + \epsilon \{ \frac{d}{2}, \frac{d}{2} - 1, \Delta_3 - \frac{
d}{2}\}}$ assuming the unitarity bound $\Delta_3 \geq \frac{d}{2} - 1$.

\subsection{Identities with projectors} \label{ch:identities}

The projectors are defined as
\begin{align}
\pi^{\mu}_{\alpha} & = \delta^{\mu}_{\alpha} - \frac{p^{\mu} p_{\alpha}}{p^2}, \\
\Pi^{\mu \nu}_{\alpha \beta} & = \frac{1}{2} \left( \pi^{\mu}_{\alpha} \pi^{\nu}_{\beta} + \pi^{\mu}_{\beta} \pi^{\nu}_{\alpha} \right) - \frac{1}{d - 1} \pi^{\mu \nu}\pi_{\alpha \beta}, \\
\Pi^{\mu \nu \rho \sigma} & = \delta^{\rho \alpha} \delta^{\sigma \beta} \Pi^{\mu \nu}_{\alpha \beta},
\end{align}
One can find the following identities:
\begin{align}
p_\mu \pi^{\mu \nu} = p_\mu \Pi^{\mu \nu \rho \sigma} & = 0, \\
\delta_{\mu \nu} \pi^{\mu \nu} = \pi^\mu_\mu & = d - 1, \\
\Pi^{\mu \nu \rho}_{\ \ \ \ \rho} = \delta_{\rho \sigma} \Pi^{\mu \nu \rho \sigma} = \pi_{\rho \sigma} \Pi^{\mu \nu \rho \sigma} & = 0, \\
\Pi^{\mu \rho \nu}_\rho = \delta_{\rho \sigma} \Pi^{\mu \rho \nu \sigma} = \pi_{\rho \sigma} \Pi^{\mu \rho \nu \sigma} & = \frac{(d+1)(d-2)}{2(d-1)} \pi^{\mu \nu}, \\
\pi^\mu_\alpha \pi^\alpha_\nu & = \pi^\mu_\nu, \\
\Pi^{\mu \nu}_{\alpha \beta} \Pi^{\alpha \beta}_{\rho \sigma} & = \Pi^{\mu \nu}_{\rho \sigma}, \\
\Pi^{\mu \nu \rho}_{\ \ \ \ \alpha} \pi^{\alpha \sigma} & = \Pi^{\mu \nu \rho \sigma}, \\
\Pi^{\mu \nu}_{\alpha \beta} \Pi^{\alpha \rho \beta \sigma} & = \frac{d-3}{2(d-1)} \Pi^{\mu \nu \rho \sigma}.
\end{align}
Basic derivatives can be calculated directly.
Denoting
\begin{equation}
\partial_\mu = \frac{\partial}{\partial p^\mu}.
\end{equation}
we find
\begin{align}
\partial_\kappa \pi_{\mu \nu} & = - \frac{p_\mu}{p^2} \pi_{\nu \kappa} - \frac{p_\nu}{p^2} \pi_{\mu \kappa}, \\
\partial_\kappa \Pi_{\mu \nu \rho \sigma} & = - \frac{p_\mu}{p^2} \Pi_{\kappa \nu \rho \sigma} - \frac{p_\nu}{p^2} \Pi_{\mu \kappa \rho \sigma} - \frac{p_\rho}{p^2} \Pi_{\mu \nu \kappa \sigma} - \frac{p_\sigma}{p^2} \Pi_{\mu \nu \rho \kappa}, \\
\pi^\mu_\alpha \partial_\kappa \pi^\alpha_\nu & = - \frac{p_\nu}{p^2} \pi^\mu_\kappa, \\
\pi^{\mu \kappa} \partial_\alpha \pi^\alpha_\nu - \pi^{\mu \alpha} \partial_\alpha \pi^\kappa_\nu & = -(d-2) \frac{p_\nu}{p^2} \pi^{\mu \kappa} + \frac{p^\kappa}{p^2} \pi^\mu_\nu, \\
\Pi^{\mu \nu}_{\alpha \beta} \partial_\kappa \Pi^{\alpha \beta}_{\rho \sigma} & = - \frac{p_\rho}{p^2} \Pi^{\mu \nu}_{\kappa \sigma} - \frac{p_\sigma}{p^2} \Pi^{\mu \nu}_{\rho \kappa}, \\
\Pi^{\mu \nu}_{\kappa \beta} \partial_\alpha \Pi^{\alpha \beta}_{\rho \sigma} - \Pi^{\mu \nu \alpha}_{\ \ \ \ \beta} \partial_\alpha \Pi^{\kappa \beta}_{\rho \sigma} & = - \frac{1}{2} \frac{d - 1}{p^2} \left[ p_\rho \Pi^{\mu \nu}_{\kappa \sigma} + p_\sigma \Pi^{\mu \nu}_{\rho \kappa} \right] + \frac{p_\kappa}{p^2} \Pi^{\mu \nu}_{\rho \sigma}.
\end{align}
Analogous expressions with two derivatives are
\begin{align}
\pi^\mu_\alpha \partial^2 \pi^\alpha_\nu & = - \frac{2}{p^2} \pi^\mu_\nu, \\
p^\alpha \pi^\mu_\beta \partial_\alpha \partial_\kappa \pi^\beta_\nu & = \frac{p_\nu}{p^2} \pi^\mu_\kappa, \\
\Pi^{\mu \nu}_{\alpha \beta} \partial^2 \Pi^{\alpha \beta}_{\rho \sigma} & = - \frac{4}{p^2} \Pi^{\mu \nu}_{\rho \sigma}, \\
p^\gamma \Pi^{\mu \nu}_{\alpha \beta} \partial_\gamma \partial_\kappa \Pi^{\alpha \beta}_{\rho \sigma} & = \frac{p_\rho}{p^2} \Pi^{\mu \nu}_{\kappa \sigma} + \frac{p_\sigma}{p^2} \Pi^{\mu \nu}_{\rho \kappa}.
\end{align}

For the semi-local operators defined in \eqref{e:decompJ} and \eqref{e:decompT} we find
\begin{align}
\pi^\mu_\alpha \partial_\kappa j^{\alpha}_{\text{loc}} & = \frac{1}{p^2} \pi^\mu_\kappa r, \\
\pi^{\mu \kappa} \partial_\alpha j^\alpha_{\text{loc}} - \pi^{\mu \alpha} \partial_\alpha j^\kappa_{\text{loc}} & = \frac{d-3}{p^2} \pi^{\mu \kappa} r + \frac{1}{p^2} \pi^{\mu \kappa} p^\alpha \partial_\alpha r - \frac{p^\kappa}{p^2} \pi^{\mu \alpha} \partial_\alpha r, \\
\pi^\mu_\alpha \partial^2 j^\alpha_{\text{loc}} & = \frac{2}{p^2} \pi^{\mu \alpha} \partial_\alpha r, \\
p^\alpha \pi^\mu_\beta \partial_\alpha \partial_\kappa j^\beta_{\text{loc}} & = - \frac{2}{p^2} \pi^\mu_\kappa r + \frac{1}{p^2} \pi^\mu_\kappa p^\alpha \partial_\alpha r
\end{align}
and
\begin{align}
\Pi^{\mu \nu}_{\alpha \beta} \partial_\kappa t^{\alpha \beta}_{\text{loc}} & = \frac{2}{p^2} \Pi^{\mu \nu}_{\alpha \kappa} R^\alpha, \\
\Pi^{\mu \nu \kappa}_{\ \ \ \ \beta} \partial_\alpha t^{\alpha \beta}_{\text{loc}} - \Pi^{\mu \nu \alpha}_{\ \ \ \ \beta} \partial_\alpha t^{\kappa \beta}_{\text{loc}} & = \frac{d - 2}{p^2} \Pi^{\mu \nu \kappa}_{\ \ \ \ \alpha} R^\alpha + \frac{p^\beta}{p^2} \Pi^{\mu \nu \kappa}_{\ \ \ \ \alpha} \partial_\beta R^\alpha - \frac{p^\kappa}{p^2} \Pi^{\mu \nu \alpha}_{\ \ \ \ \beta} \partial_\alpha R^\beta, \\
\Pi^{\mu \nu}_{\alpha \beta} \partial^2 t^{\alpha \beta}_{\text{loc}} & = \frac{4}{p^2} \Pi^{\mu \nu \alpha}_{\ \ \ \ \beta} \partial_\alpha R^\beta, \\
p^\gamma \Pi^{\mu \nu}_{\alpha \beta} \partial_\gamma \partial_\kappa t^{\alpha \beta}_{\text{loc}} & = -\frac{4}{p^2} \Pi^{\mu \nu}_{\alpha \kappa} R^\alpha + \frac{2}{p^2} \Pi^{\mu \nu}_{\alpha \kappa} p^\beta \partial_\beta R^\alpha.
\end{align}

\bigskip

\section*{Acknowledgements}

We would like to thank Anatoly Dymarsky for discussions.
Research at the Perimeter Institute is supported by the Government of Canada
through Industry Canada and by the Province of Ontario through the Ministry of
Research \& Innovation. K.S.~acknowledges support from NWO via a VICI grant and from a grant of the John Templeton Foundation.  The opinions expressed in this publication are those of the authors and do not necessarily reflect the views of the John Templeton Foundation. A.B.~is supported through NWO via the VICI grant of K.S.


\newpage

\providecommand{\href}[2]{#2}\begingroup\raggedright\endgroup

\end{document}